\documentclass[aps,prb,amssymb,twocolumn,floatfix,showpacs]{revtex4-1}%
\usepackage{amsmath,amsfonts,amssymb}
\usepackage[english]{babel}
\usepackage{multirow}%

\usepackage{graphicx,color,fancyhdr,xspace}
\definecolor{darkgreen}{rgb}{0,0.5,0}
\definecolor{darkblue}{rgb}{0,0,0.2}
\definecolor{purple}{rgb}{0.35,0,0.35}
\definecolor{orange}{rgb}{1,0.5,0}

\RequirePackage[
   hyperindex,bookmarksnumbered,colorlinks, %%hypertex,pdfborder=0 0 0,
   plainpages,pdfstartview=FitH]{hyperref}
\hypersetup{linkcolor=blue,urlcolor=darkblue,citecolor=darkgreen}
\usepackage[pdfstartview=FitH]{hyperref} %% hypertex,

\usepackage{multirow}%
  \def\wrt{\emph{w.r.t.}\@\xspace}
  \def\eg{\emph{e.g.}\@\xspace}
  \def\ie{\emph{i.e.}\@\xspace}
  \def\cf{\emph{cf.}\@\xspace}

  \def\lhs{\emph{l.h.s.}\@\xspace}
  \def\rhs{\emph{r.h.s.}\@\xspace}

  \DeclareMathOperator*{\trace}{tr}

  \def\NS{\ensuremath{n_\mathrm{S}}\xspace}
  \def\TK{\ensuremath{T_\mathrm{K}}\xspace}
  \def\JH{\ensuremath{J_{H}}\xspace}

  \def\Ed{\ensuremath{\varepsilon_d}\xspace}
  \def\Ef{\ensuremath{\varepsilon_f}\xspace}
  \def\hc{\ensuremath{\mathrm{H.c.}}\xspace}
  \def\CS{\ensuremath{\mathrm{CS}}\xspace}

  \def\NK{\ensuremath{N_\mathrm{K}}\xspace}
  
  \def\EK{\ensuremath{E_\mathrm{K}}\xspace}
  \def\K{\ensuremath{\mathrm{K}}\xspace}

  \def\Etrunc{\ensuremath{E_\mathrm{trunc}}\xspace}

  \def\Atensor{$A$-tensor\xspace}
  \def\Atensors{$A$-tensors\xspace}

  \def\IREP{{IREP}\xspace}
  \def\IREPs{{IREPs}\xspace}

  \def\IROP{{IROP}\xspace}   %% irreducible operator set
  \def\IROPs{{IROPs}\xspace} %% irreducible operator sets

  \def\RLOs{{RLOs}\xspace}   %% raising and lowering operators
  \def\RLO{{RLO}\xspace}     %% raising or lowering operator

  \def\QSpace{\mbox{\textsf{QSpace}}\xspace}
  \def\QSpaces{\mbox{\textsf{QSpaces}}\xspace}

  \def\qMW{\ensuremath{q_\mathrm{MW}}\xspace}

  \def\myempty{myemptytoken}

  \newcommand{\QSP}[1]{\QSpace~(\ref{#1})\xspace}

  \newcommand{\EQ}[1]{Equation~(\ref{#1})\xspace}
  \newcommand{\Eq}[1]{Eq.~(\ref{#1})\xspace}
  \newcommand{\Eqs}[1]{Eqs.~(\ref{#1})\xspace}
  \newcommand{\Eqr}[2]{Eqs.~(\ref{#1}-\ref{#2})\xspace}
  \newcommand{\EQS}[2]{Eqs.~(\ref{#1}) and (\ref{#2})} % PRB style
  \newcommand{\Eqt}[1]{Eq.~\ref{#1}\xspace}

  \newcommand{\Fig}[1]{Fig.~\ref{#1}\xspace}
  \newcommand{\Figp}[2]{Fig.~\ref{#1}(#2)}
  
  \newcommand{\FIGP}[2]{Figure~\ref{#1}(#2)}

  \newcommand{\TBL}[1]{Table~\ref{#1}\xspace}
  \newcommand{\Tbl}[1]{Tbl.~\ref{#1}\xspace}
  \newcommand{\App}[1]{App.~\ref{#1}\xspace}
  \newcommand{\Sec}[1]{Sec.~\ref{#1}\xspace}
  \newcommand{\Secs}[1]{Secs.~\ref{#1}\xspace}

  \newcommand{\Secp}[1]{Sec.~\ref{#1}} % no \xspace here!
  \newcommand{\Eqp}[1]{Eq.~(\ref{#1})} % no \xspace here!
  \newcommand{\EQSp}[1]{Eqs.~(\ref{#1})} % no \xspace here!
  \newcommand{\FigP}[1]{Fig.~\ref{#1}} % no \xspace here!
  \newcommand{\TblP}[1]{Tbl.~\ref{#1}} % no \xspace here!
  \newcommand{\AppP}[1]{App.~\ref{#1}} % no \xspace here!

  \newcommand{\Pref}[1]{(\ref{#1})\xspace}

  \newcommand{\SU}[2][\myempty]{% KEEP THIS % (increase space, otherwise!)
     \ifx\myempty#1\ensuremath{\mathrm{SU}(#2)}\xspace
     \else\ensuremath{\mathrm{SU}(#2)_{\mathrm{#1}}}\xspace
     \fi
  }

  \def\Id{\ensuremath{\mathbf{1}}\xspace}

  \newcommand{\UA}[1]{% KEEP THIS % (increase space, otherwise!)
     \ifx\@empty#1\ensuremath{\mathrm{U}(1)}\xspace
     \else\ensuremath{\mathrm{U}(1)_{\mathrm{#1}}}\xspace
     \fi
  }

  \newcommand{\Sp}[1]{% KEEP THIS % (increase space, otherwise!)
     \ensuremath{\mathrm{Sp}(#1)}\xspace
  }

  \newcommand{\AS}{\ensuremath{%
     \UA{charge}\otimes\SU[spin]{2}}\xspace}

  \renewcommand{\SS}{\ensuremath{%
     \SU[spin]{2}\otimes\SU[charge]{2}}\xspace}
  \newcommand{\SC}{\ensuremath{%
     \SU[SC]{2}^{\otimes 2}}\xspace}

  \newcommand{\ASC}{\ensuremath{%
     \SU[spin]{2}\otimes\UA{charge}\otimes\SU[channel]{3}}\xspace}

  \newcommand{\SSP}{\ensuremath{%
     \SU[spin]{2}\otimes\Sp{6}}\xspace}

  \newcommand{\SSSS}{\ensuremath{%
     \SU[spin]{2} \otimes \SU[charge]{2} ^{\otimes3}}\xspace}

  \newcommand{\Sfour}{\ensuremath{\SU[SC]{2}^{\otimes4}}\xspace}

\begin{document}

\title{ Non-abelian symmetries in tensor networks: %%\\
a quantum symmetry space approach }

\author{A. Weichselbaum}
\affiliation{
  Physics Department, Arnold Sommerfeld Center for Theoretical Physics, and
  Center for NanoScience, Ludwig-Maximilians-Universit\"at, 80333 Munich,
  Germany
}

% 71.27.+a,   % Strongly correlated electron systems; heavy fermions (Electronic structure of bulk materials)
% 72.15.Qm,   % Scattering mechanisms and Kondo effect (Electronic transport in condensed matter)
% 73.21.La,   % Quantum dots (Electronic structure and electrical properties of ... low-dimensional structures)
% 75.40.Mg,   % Numerical simulation studies (Magnetic properties and materials)


\begin{abstract}
  A general framework for non-abelian symmetries is presented for
  matrix-product and tensor-network states in the presence of
  well-defined orthonormal local as well as effective basis sets.
  The two crucial ingredients, the Clebsch-Gordan algebra for
  multiplet spaces as well as the Wigner-Eckart theorem for
  operators, are accounted for in a natural, well-organized, and
  computationally straightforward way. The unifying
  tensor-representation for quantum symmetry spaces, dubbed
  \QSpace, is particularly suitable to deal with standard
  renormalization group algorithms such as the numerical
  renormalization group (NRG), the density matrix renormalization
  group (DMRG), or also more general tensor networks such as the
  multi-scale entanglement renormalization ansatz (MERA).
  In this paper, the focus is on the application of the
  non-abelian framework within the NRG. A detailed analysis is
  presented for a fully screened spin-3/2 three-channel Anderson
  impurity model in the presence of conservation of total spin,
  particle-hole symmetry, and \SU{3} channel symmetry. The same
  system is analyzed using several alternative symmetry scenarios.
  This includes the more traditional symmetry setting \SSSS, the
  larger symmetry \ASC, and their much larger enveloping
  symplectic symmetry \SSP. These are compared in detail,
  including their respective dramatic gain in numerical
  efficiency. In the appendix, finally, an extensive introduction
  to non-abelian symmetries is given for practical applications,
  together with simple self-contained numerical procedures to
  obtain Clebsch-Gordan coefficients and irreducible operators
  sets. The resulting \QSpace tensors can deal with any set of
  abelian symmetries together with arbitrary non-abelian
  symmetries with compact, \ie finite-dimensional, semi-simple Lie
  algebras.
\end{abstract}

\date{\today}
\maketitle

%%\onecolumngrid
%%\parskip 0.01in
%%\textwidth      4.5in %
  \def\tocname{Table of contents}
  %%\vspace{-0.3in}
  {\small{\tableofcontents}}
  %%\vspace{0.5in}
%%\textwidth      7in %
%%\par\twocolumngrid

%% \input paper.tex

\section{Introduction}

Numerical methods for strongly correlated quantum-many-body systems
are confronted with exponentially large Hilbert spaces. With a
limited number of exact analytical solutions at hand and with
perturbative treatments for low-energy or ground-state physics often
insufficient, a certain systematic treatment with respect the
Hilbert space is required. Besides quantum Monte Carlo approaches,
that explore quantum systems in a stochastic way  \cite{Foulkes01},
a systematic state space decimation is provided by renormalization
group (RG) techniques such as the density matrix renormalization
group (DMRG)\cite{White92} or the numerical renormalization group
(NRG)\cite{Wilson75}, both highly efficient for
quasi-one-dimensional systems, and since non-perturbative,
considered essentially exact.

Quantum-many-body Hilbert spaces are built from the direct product
of the state spaces of the participating individual particles. As
such particle statistics plays an essential role. While the focus of
this paper is on fermionic systems, generalizations to spin systems
are straightforward. The treatment of bosonic systems, on the other
hand, comes with the additional hurdle that even a single local
bosonic degree of freedom already has an infinite state space of its
own which must be truncated for numerical treatment. Nevertheless,
assuming that the bosonic state spaces can be properly categorized
in symmetry sectors, the complications deriving from their infinite
dimensionality are considered separate form the issues regarding the
description of pure symmetries of the Hamiltonian. In the case of
two-dimensional systems finally, more exotic types of particles
exist that are neither fermions nor bosons, but anyons. Much
attention has been paid to these recently within the framework of
tensor networks \cite{Bonderson07, Pfeifer10a, Pfeifer10,
Pfeifer11phd, Singh10, Singh10a}. While the treatment of particles
with non-abelian statistics is nicely complimentary to the work
presented here, this shall not be pursued any further in what
follows.

Methods such as the DMRG or the NRG then, are based on the same
algebraic structure of matrix product states (MPSs)
\cite{Rommer95,Wb09}. Initially introduced for one-dimensional
systems with MPS owing its name to this case, a wide range of
activity has emerged within recent years to generalize MPS to
tensor-networks for two- or higher-dimensional systems
\cite{Sandvik07,Murg07,Cirac09,Singh10}. While clearly
appealing from the point of view of area laws for
entanglement-entropy \cite{Bekenstein73,Wolf08,Eisert10}, tensor
network states (TNSs) often share the same disadvantage as linear
systems with periodic boundary conditions within the DMRG, namely
that state spaces become intrinsically non-orthogonal. Therefore
also the unique association of symmetry labels with each index in a
tensor is compromised. This, however, can be circumvented by
introducing an emerging extra-dimension, which is at the basis of
the recently developed multi-scale entanglement renormalization
ansatz (MERA) \cite{Vidal07_MERA,Vidal08}. Nevertheless, the
\emph{traditional} DMRG approach applied to 2D systems
\cite{Stoudenmire12} with open or cylindrical boundary conditions
yet with long-range interactions has continued to provide a highly
competitive, extremely well-controlled, even though numerically
expensive approach.

Within both, traditional DMRG as well as NRG, state spaces of entire
blocks are built iteratively by adding and merging one site at a
time. Clearly, the single index describing an effective basis for the
entire block or site can be chosen orthogonal. Moreover, the basis
states can be labeled in terms of the symmetries of the underlying
Hamiltonian. Operators written as matrix elements in this very same
basis therefore also share the same well-defined partitioning in
terms of symmetry sectors. By grouping symmetry state spaces
together, the Hamiltonian becomes block-diagonal, while general
operators usually obey well-defined selection rules between symmetry
sectors. Consequently, the sparsity of these operators due to
symmetry can be efficiently and exactly included in the numerical
description, such that usually only a few dense data blocks with
non-zero matrix elements remain, given the symmetry constraints.
While this well represents the advantage of implementing generic
abelian or point symmetries in a calculation, the presence of
non-abelian symmetries offers yet another strong simplification: many
of the non-zero matrix elements are actually \emph{not} independent
of each other, bearing in mind, for example, the Wigner-Eckart
theorem. Therefore going beyond abelian symmetries, non-abelian
symmetries allow to significantly \emph{compress} the non-zero blocks
in terms of multiplet spaces, \cite{McCulloch02,Toth08} while also
reducing their number. With the Clebsch-Gordan coefficient spaces
factorizing,\cite{Singh10} they can be split off systematically in
terms of a tensor-product and dealt with separately.

MPS is optimal for one-dimensional systems. When exploring systems
that are not strictly one dimensional but acquire \emph{width}, such
as ladders of several rungs in DMRG or multi-channel models in NRG,
the price to be paid for orthonormal state spaces is that one must
represent the system as a one-dimensional MPS nevertheless. This
introduces longer-range interactions to the mapped 1D system, with
the effect that the typically required dimensions of the state spaces
to be kept in a calculation, grow roughly exponentially with system
width. The number of symmetries then that (i) are available and (ii)
are also be exploited in practice, decides whether or not a
calculation is feasible. Abelian symmetries such as particle (charge
$Q$) or spin ($S_z$) conservation are usually implemented in DMRG
calculations. However, only very few groups have implemented
non-abelian symmetries, and these are also constrained to \SU{2}
symmetries only, \cite{McCulloch02} due to its complexity in the
actual implementation. General treatment of non-abelian symmetries
within the MERA, on the other hand, is currently under development.
\cite{Singh10,Singh11} NRG, in contrast, had been set up including
non-abelian \SU{2} spin symmetry from its very beginning,
\cite{Wilson75} dictated by limited numerical resources. So far,
however, only a very few isolated attempts including more complex
non-abelian settings exist within the NRG,
\cite{DeLeo05} %% Shimizu90,Allub95
while to our knowledge there exists no general realization yet of
arbitrary non-abelian symmetries in either method.

This paper focuses on the systematic description and implementation
of non-abelian symmetries of a given Hamiltonian within the
generalized MPS framework. This naturally also does include the
description of abelian symmetries where necessary, as they can be
trivially written in terms of Clebsch-Gordan coefficients. While the
focus within non-abelian symmetries belongs to \SU{N} and the
symplectic group \Sp{2n}, the generalization to other non-abelian
symmetries or also point groups is straightforward once their
particular Clebsch-Gordan coefficients are worked out. In contrast to
the well-known \SU{2} then, general non-abelian symmetries, such as
$\SU{N\ge 3}$, represent a significant increase in algorithmic
complexity, in that they can and routinely do exhibit inner and outer
multiplicity. The latter, for example, implies that in the
decomposition of the tensor-product of two irreducible
representations (\IREPs) into a direct sum of \IREPs, the same \IREP
may occur \emph{multiple} times. Nevertheless, this can be dealt with
properly on the algorithmic level, as will be shown in detail in this
paper.

While the presented non-abelian framework for general tensors is
straightforwardly applicable to traditional DMRG as well as NRG, the
paper focuses on the application within the NRG. Detailed results are
presented for a fully screened spin-$3/2$ Anderson impurity model
with \SU{3} channel-symmetry [\ie see Hamiltonian in
\Eq{eq:HKondoAM}{}]. This model has been suggested as the effective
microscopic Kondo model for iron impurities in gold or silver
\cite{Costi09}, historically the first system where Kondo physics was
observed experimentally. \cite{deHaas34,Kondo64} Being a true
three-channel system, this cannot be trivially rotated into a simpler
configuration of fewer relevant channels. The result is an extremely
challenging calculation within the NRG that requires non-abelian
symmetries for fully converged numerical results for reasonable
coarse-graining of the continuous bath. The non-abelian symmetries
present in the model considered are (i) particle-hole symmetry in
each of the three channels, $\SU[charge]{2}^{\otimes3}$, (ii) total
spin symmetry, \SU[spin]{2}, and (iii) channel symmetry,
\SU[channel]{3}. The non-abelian particle-hole \SU{2} symmetry,
however, does \emph{not} commute with the channel \SU{3} symmetry,
while the plain abelian charge \UA{} symmetry does commute. Overall,
this suggests a larger enveloping symmetry, which turns out to be the
symplectic symmetry \Sp{6} [for an introduction, see
\App{sec:Sp(2m)}{}]. With this, the following symmetry scenarios are
considered and compared in detail,
\begin{align*}
   & \SSSS \text{ ,} \\
   & \ASC  \text{ , and} \\
   & \SSP  \text{.}
\end{align*}
While the first setting represents a more traditional setup based on
multiple sets of plain \SU{2} symmetries only, the second setting
already includes the larger channel \SU{3} symmetry. Both of these
symmetries do not capture the full symmetry of the model, which
finally is achieved by using the enveloping \Sp{6} symmetry.

Due to the internal two-dimensional structure of the \SU{3} symmetry
based on the fact that \SU{3} has two commuting generators, \ie is of
rank 2, its multiplets have significantly larger internal dimension,
in practice, up to over a hundred. Therefore despite the reduction of
the particle-hole symmetry to a plain abelian symmetry, the second
setting with the \SU{3} channel symmetry allows to outperform the
more traditional setup based on \SU{2} symmetries only. Similarly,
with \Sp{6} a rank-3 symmetry, multiplets then easily reach
dimensions of several thousands there, which allows to reduce
multiplet spaces significantly further still. A detailed analysis of
this is provided in this paper, with a more general self-contained
introduction to non-abelian symmetries considered given in the
appendix [\cf \AppP{sec:SU123}].

From an NRG point of view,\cite{Wb07} a few essential steps are
required. These are (i) the evaluation of relevant operator matrix
elements required to construct the Hamiltonian, (ii) the generic
setup of an iteration, adding one site to the so-called Wilson chain,
and finally, for thermodynamical properties (iii) also the treatment
of the full thermal density matrix. \cite{Wb07} All of these steps
are simple in principle, yet come with the essential challenge to
have a flexible transparent framework for the treatment of
non-abelian symmetries in practice. In this paper, such a framework
is presented in terms of generalized contractions of tensors in the
presence of symmetry spaces, introduced as \QSpaces below.

The paper is thus organized as follows. Section II describes the MPS
implementation of non-abelian symmetries in terms of \QSpaces.
Section III describes the implications for calculating correlation
functions in the presence of irreducible operator sets. Section IV
gives a short review of the NRG together with specialties related to
non-abelian symmetries, such as calculating reduced density matrices.
This section also introduces the model Hamiltonian of a fully
symmetric 3-channel Anderson model. Section V then presents explicit
NRG results, followed by summary and outlook. Finally, also an
extended Appendix has been added to the paper. The latter is intended
to provide a more general pedagogical self-contained introduction to
non-abelian symmetries as they occur in fermionic lattice models,
together with their actual implementation in practice in terms of
\QSpaces.

\section{MPS implementation of non-abelian symmetries}

Consider some Hamiltonian $\hat{H}$ that is invariant under a set of
\NS symmetries,
\begin{align}
   \mathcal{S} \equiv
   \bigotimes_{\lambda=1}^{\NS} \mathcal{S}^\lambda
\text{,}\label{eq:allsymm}
\end{align}
that is, $[ \hat{H}, \hat{S}^\lambda_\alpha ]=0$, where $\alpha$
identifies the generator $\hat{S}^\lambda_\alpha$ for the simple
(non-abelian) symmetry $\mathcal{S}^\lambda$. To be specific, for
example, $\mathcal{S} = \SS \equiv \mathcal{S}^1 \otimes
\mathcal{S}^2$ with $\lambda \in \{1,2\}$ would stand for the
combination of spin and charge \SU{2} symmetry, respectively. The
tensor-product notation in \Eq{eq:allsymm} indicates that the
symmetries act independently of each other, that is $[
\hat{S}^\lambda_\alpha, \hat{S}^{\lambda'}_{\alpha'} ]=0$ for
$\lambda \neq \lambda'$.

Given the symmetries as in \Eq{eq:allsymm}, this allows to organize
the complete basis of eigenstates of $\hat{H}$ in terms of the
symmetry eigenbasis. Every state then belongs to a well-defined
irreducible multiplet $q^\lambda$ for each symmetry
$\mathcal{S}^\lambda$. The multiplet itself has an internal state
space structure that is described by the additional quantum labels
$q^\lambda_{z}$. For example, in the case of $\mathcal{S}^\lambda =
\SU{2}$, $q^\lambda$ ($q_z^\lambda$) corresponds to the spin
multiplet $S$ (the $S_z$ label), respectively.

Thus all states in a given vector space can be categorized using the
hierarchical label structure
\begin{equation}
   \vert qn; q_{z} \rangle \quad \text{(state-space label structure)}
\text{,}\label{eq:qnz-labels}
\end{equation}
where
\begin{enumerate}

\item[(i)] $q \equiv (q^{1},q^{2},\ldots,q^{\NS})$, to be
    referred to as \emph{q-labels} (quantum labels), references
    the irreducible representations (\IREPs) for each symmetry
    $\mathcal{S}^\lambda$, $\lambda=1,\ldots,\NS$. All states in
    given Hilbert space with the same q-labels are \emph{blocked}
    together, to be referred to as \emph{symmetry block} $q$.

\item[(ii)] Given a symmetry block $q$ then, the multiplet index
    $n_{(q)}$ identifies a specific multiplet within this space.
    It is therefore a plain index associated with given symmetry
    space $q$. Together with the q-labels, this forms the
    \emph{multiplet level} which is considered the topmost
    conceptual level. Using the composite notation $(qn)$ to
    identify an arbitrary multiplet, the subscript $q$ to the
    multiplet index $n_{(q)}$ is considered implicit and hence is
    dropped, for simplicity.

\item[(iii)] Finally, the set of labels $q_{z} \equiv (q^1_{z},
    q^2_{z}, \ldots, q^{\NS}_{z})$, to be referred to as
    \emph{z-labels}, resolves the internal structure of each
    multiplet in q. That is, for each \IREP $q^{\lambda}$,
    referring to the symmetry $\mathcal{S}^\lambda$ in $q$,
    $q^\lambda_z$ labels its internal \IREP space. As such, the
    z-labels are entirely defined by the symmetries considered.
    By construction, the eigenstates of the Hamiltonian $\hat{H}$
    are fully degenerate in the z-labels.

\end{enumerate}
Here the symmetry labels $q$ and $q_{z}$ describe the combined record
of labels derived from all symmetries considered. In practice, states
can mostly be treated on the higher multiplet level, while the lower
level in terms of the z-labels is split off and taken care of by
Clebsch-Gordan algebra and the coefficient spaces derived from it.

When non-abelian symmetries are broken, they are often reduced to
their abelian subalgebra. This can be easily implemented,
nevertheless, consistent with the presented framework. In particular,
in the abelian case, the non-abelian multiplet labels $q$ are absent,
while the abelian $q_z$ quantum numbers remain. Therefore the $q_z$
labels can be promoted to the status of q-labels, $q:=q_z$. As a
consequence, the concept of the actual $q_z$ labels becomes
irrelevant (therefore subsequently, the $q_z$ label space may simply
be set to zero, $q_z:=0$). The corresponding Clebsch Gordan
coefficients are all trivial scalars, \ie equal to 1. Yet these
``Clebsch Gordan coefficients for abelian symmetries'' do maintain an
important role, in that they take care of the proper addition rules
that come with abelian symmetries, resulting in $\langle q_1; q_2
\vert q \rangle = 1\cdot \delta_{q,q_1+q_2}$.

Given the MPS background of NRG or DMRG, states spaces are generated
iteratively, in terms of a product-space of a given effective state
space with a newly added local site. Operators, on the other hand,
are typically represented in local state spaces, and starting from
there, they can be written in terms of matrix elements in the
effective global state spaces. With this in mind, the implementation
of non-abelian symmetries within the MPS framework therefore is based
on the following two basic observations with respect to state space
and operator representations, respectively.
\begin{enumerate}

\item[(1)] State spaces: consider two distinct state spaces,
    $\vert Qn; Q_z \rangle$ and $\vert q l; q_z \rangle$ that,
    for example, represent a large effective state space and a
    small new local state space, respectively. Assuming that both
    state spaces all well-categorized in terms of \IREPs, then
    their tensor-product space can also be decomposed into a
    direct sum of new combined \IREPs $\vert \tilde{Q}\tilde{n};
    \tilde{Q}_z \rangle$ using Clebsch-Gordan coefficients
    (CGCs),
    \begin{eqnarray}
       \vert \tilde{Q}\tilde{n}; \tilde{Q}_z \rangle =
       \sum_{Qn; Q_z} \sum_{q l; q_z}
    && \bigl( A_{Q \tilde{Q}}^{[q]} \bigr) _{n\tilde{n}}^{[l]} \cdot
        C_{Q_z \tilde{Q}_z}^{[q_z]} \nonumber \\
    && \times \vert Qn; Q_z \rangle  \vert q l; q_z \rangle \text{.}
    \label{eq:CGstates}
    \end{eqnarray}
    Note that the Clebsch Gordan coefficients given by $C_{Q_z
    \tilde{Q}_z}^{[q_z]} \equiv \langle QQ_z; qq_z \vert
    \tilde{Q}\tilde{Q}_z \rangle$ (i) \textit{fully define} the
    internal multiplet space as specified by the Lie algebra, and
    (ii) determine the splitting, \ie which output multiplets
    $\tilde{Q}$ occur for given multiplets $Q$ and $q$. On the
    multiplet level, on the other hand, where $(A_{Q
    \tilde{Q}}^{[q]}) _{n\tilde{n}}^{[l]}$ combines the
    multiplets $Qn$ and $ql$ into the multiplet
    $\tilde{Q}\tilde{n}$ consistent with the splitting provided
    by the CGCs, the coefficients $(A_{Q \tilde{Q}}^{[q]})
    _{n\tilde{n}}^{[l]}$ may encode an arbitrary unitary
    transformation within the $\tilde{n}$ output space for each
    $\tilde{Q}$. The \rhs of \Eq{eq:CGstates} demonstrates, that
    the CGC spaces clearly factorize from the multiplet space
    $A_{Q \tilde{Q}}^{[q]}$ as a tensor product.

\item[(2)] Operators: the matrix elements of a specific
    irreducible operator set (\IROP) $\hat{F}^{q}$, \ie an \IROP
    that transforms according to multiplets $q$ for given
    symmetries [\cf App. \Eqp{eq:SymOp:irop}, or also
    \Secp{sec:irop+wet}] within some symmetry space $\vert
    Qn;Q_z\rangle$ can be written using the \emph{Wigner-Eckart
    theorem} as
    \begin{eqnarray}
       \langle Q'n';Q'_z \vert \hat{F}_{q_z}^{q} \vert Qn;Q_z\rangle
   &=& \bigl( F_{QQ'}^{[q]}\bigr)_{n n'}^{[1]} \cdot C_{Q_z Q'_z}^{[q_z]}
    \text{,}\label{eq:CGops} %% WET
    \end{eqnarray}
    with $C_{Q_z Q'_z}^{[q_z]}$ again the Clebsch-Gordan
    coefficients as in \Eq{eq:CGstates}. On the multiplet level,
    the \emph{reduced matrix elements} $(F_{QQ'}^{[q]})_{n
    n'}^{[1]} \equiv \langle Q'n' \Vert \hat{F}^{q} \Vert
    Qn\rangle$ refer to the \emph{single} irreducible operator
    set labeled by $q$, which is indicated by the superscript
    $[1]$. The Wigner-Eckart theorem thus allows to
    \emph{compactify} the operator matrix elements on the \lhs of
    \Eq{eq:CGops} as the tensor-product of reduced matrix
    elements and CGCs, as shown on the \rhs of \Eq{eq:CGops}.

\end{enumerate}
Therefore in both cases above, \emph{\ie in all tensor objects
relevant for a numerical calculation}, the CGC spaces factorize. This
allows to \emph{strongly} compress their size, and thus to
drastically improve on overall numerical performance. Moreover, note
that in both cases, \Eq{eq:CGstates} as well as \Eq{eq:CGops} the
underlying structure comprises tensors of rank-3 throughout. This
rank-3 structure holds for both, the reduced multiplet space as well
as the CGC spaces. Therefore, in either case, the final data
structure of either state space decomposition as well as reduced
operator sets is \emph{exactly} the same. It is implemented, in
practice, in terms of what will be referred to as \QSpace for general
tensors of arbitrary rank.

\subsection{General quantum space representation (\QSpaces) \label{sec:QSpacce}}

The generic representation, used in practice to describe all symmetry
related tensors $B$, is given by a listing of the following type,
\begin{equation}
   B\equiv\left\{\begin{tabular}[c]{c|c|c}%
      q-labels & reduced space $\left\Vert B\right\Vert $ & CGC
      spaces\\\hline\hline
      $\left\{  Q\right\}  _{1}$ & $B_{1}$ & $\{\mathrm{C}\}_{1}$\\
      $\left\{  Q\right\}  _{2}$ & $B_{2}$ & $\{\mathrm{C}\}_{2}$\\
      $\cdots$ & $\cdots$ & $\cdots$%
   \end{tabular}\right\}
\text{.}\label{eq:QSpace:Rep}%
\end{equation}
By notational convention, an actual operator $\hat{B}$ will be
written with a hat, while its representation in terms of matrix
elements in a specific basis will be written without the hat, hence
the corresponding \QSpace is referred to as \QSpace $B$.
Many explicit examples of \QSpaces are introduced and discussed in
detail in the appendix [\Sec{app:QSpaces}{}]. As an up-front
illustration, consider, for example, the general Hamiltonian of a
single spinful fermionic site in the presence of \SU{2} symmetry in
the spin (S) and charge sector (C), which can be written as the
\QSpace [see \Eq{QSpace:H0_cs}{}]
\begin{equation}
H \equiv \left\{
   \begin{tabular}[c]{cc|c|cc}
      $(S;C)$ & $(S^{\prime}; C^{\prime})$ & $\Vert H\Vert$ & \multicolumn{2}{|l}{CGC spaces}
   \\ \hline\hline
      $\tfrac{1}{2};0$ & $\tfrac{1}{2};0$ & $h_{\tfrac{1}{2},0}$ & $\Id^{(2)}$ & $1.$\\
      $0;\tfrac{1}{2}$ & $0;\tfrac{1}{2}$ & $h_{0,\tfrac{1}{2}}$ & $1.$ & $\Id^{(2)}$
   \end{tabular}
   \right\}
\text{.} \label{QSpace:H0:cs-ref}
\end{equation}
With every non-zero block listed as an individual row, one can see
that the only two reduced matrix elements $\Vert H\Vert$ free to
choose without compromising the $\SC \equiv \SS $ symmetry are the
parameters (numbers) $h_{1/2,0}$ and $h_{0,1/2}$. By definition, the
Hamiltonian is a scalar operator, therefore it is the only operator
within its \IROP, hence can be written as plain rank-2 \QSpace (the
third dimension for this \IROP would be a singleton dimension, hence
can be dropped). Being a scalar operator, the Hamiltonian is block
diagonal, which is reflected in equal symmetry sectors $(C;S)$ and
$(C';S')$ in each row for first and second dimension, respectively.
Moreover, in given case, the corresponding Clebsch-Gordan coefficient
(CGC) spaces also result in trivial identities, with $\Id^{(2)}$ the
two-dimensional identity. Note that the full set of CGC spaces in
each row needs to be interpreted as appearing in a tensor product
with the multiplet space, here the reduced matrix elements
$h_{1/2,0}$ or $h_{0,1/2}$ [\eg see \Eq{B-CG-tprod} below].

In general, the representation of a tensor $B$ of arbitrary rank-$r$
in the \QSpace in \Eq{eq:QSpace:Rep} [with \Eq{QSpace:H0:cs-ref} an
example for a rank-2 \QSpace{}], only lists the non-zero, \ie
relevant symmetry combinations. Having $r$ tensor dimensions, each of
its $r$ indices refers to its specific state space $\vert qn; q_{z}
\rangle _{i} \equiv \vert (q)_i n_i; (q_z)_i \rangle$ with
$i=1,\ldots,r$, and hence carries its own label structure as in
\Eq{eq:qnz-labels}. The q-labels $(q)_{i} \equiv \{q^{\lambda}\}_i$
already represent the \emph{combined} set of $\NS$ \IREP labels from
all symmetries $\mathcal{S}^\lambda$ for the state space at tensor
dimension $i$. In general, by convention, the internal order of the
q-labels $(q)_i$ \wrt $\lambda$ is fixed and follows the order of
symmetries used in \Eq{eq:allsymm}.

For a certain row $\nu$ of the \QSpace listing in \Eq{eq:QSpace:Rep}
then, the set of $r$ q-labels are grouped into
\begin{subequations}
\begin{equation}
   \{Q\}_{\nu} \equiv \{(q)_{1},\ldots,(q)_{r}\}_\nu
\text{.}\label{eq:QSpace:Q}
\end{equation}
The reduced matrix elements are stored in the dense rank-$r$ tensor
$B_{\nu}$ indexed by $n_{i}$ with $i=1,\ldots,r$. This is a plain
tensor, with the multiplet spaces possibly already rotated by
arbitrary unitary transformations and truncated. This is also
reflected in the fact that the indices $n_{i}$ are plain indices, \ie
carry no further internal structure.
Finally, for every one of the $\lambda=1\ldots \NS$ symmetries
included, the corresponding CGC space is stored in the sparse tensors
$\mathrm{C} _{\lambda,\nu}$, \emph{each of which is also of rank
$r$}. These CGC spaces are grouped into $\{\mathrm{C}\}_\nu$ in the
last column,
\begin{equation}
\{ \mathrm{C} \}_{\nu} \equiv
   \{ \mathrm{C}_{1;\nu},\ldots,\mathrm{C}_{\NS;\nu}\}
\text{.}\label{eq:QSpace:CGC}
\end{equation}
\end{subequations}%
As the q-labels $\{Q\}_{\nu}$ also define the z-labels, there is no
explicit need to store the z-labels $(q_z)_{i,\nu}$. The internal
running indices in $\mathrm{C}_{\lambda,\nu}$, however, are uniquely
associated with the z-labels. Note also the different index setting:
in contrast to \Eq{eq:QSpace:Q}, which contains a set of $r$
q-labels, \ie one for every dimension of the rank-$r$ tensor
$B$, \Eq{eq:QSpace:CGC} contains a set of \NS rank-$r$ CGC
spaces, \ie one for every symmetry.

In addition to the \QSpace listing in \Eq{eq:QSpace:Rep}, also the
type and order of symmetries considered is stored with a \QSpace, \cf
\Eq{eq:allsymm}, even though this is usually the same throughout an
entire calculation. Moreover, note that the row or record index $\nu$
in \Eq{eq:QSpace:Rep} is purely for convenience without any specific
meaning, as the order of records in a \QSpace can be chosen
arbitrarily. Nevertheless, it is required to refer to a specific
entry in a \QSpace.

For a given record $\nu$ in the \QSpace in \Eq{eq:QSpace:Rep} then,
the reduced space and the CGC spaces are to be interpreted as an
overall tensor-product,
\begin{equation}
  B_{\nu} \otimes \{ \mathrm{C} \}_{\nu} \equiv
  B_{\nu} \otimes \bigl(\bigotimes_{s=1}^{\NS}\mathrm{C}_{s;\nu}\bigr)
\text{,} \label{B-CG-tprod}
\end{equation}
while, of course, this is never explicitly done in practice. Yet
\Eq{B-CG-tprod} demonstrates the single most important motivation to
implement non-abelian symmetries in a numerical computation. By
splitting off the CGC spaces in terms of a tensor product, block
dimensions can be \emph{strongly} reduced for larger calculations
with several symmetries present. For the models analyzed in this
paper, for example, this was typically an average dimensional
reduction from plain abelian symmetries by a factor of $10$ up to
several hundreds. Considering that both NRG and DMRG scale like
$\mathcal{O}(D^{3})$ with $D$ the typical dimension of data blocks,
this is an enormous gain in efficiency. The factorized CGC spaces, on
the other hand, can be dealt with independently, as will be explained
in detail later. Assuming that usually the dimensions of the reduced
states spaces $B_\nu$ still exceed by far the typical dimensions
encountered for the CGC spaces, the latter bear little numerical
overhead. Only for larger-rank symmetries, such as the symmetry
\Sp{6} discussed later, multiplet dimensions can become large
themselves such that one needs to pay more attention to an efficient
treatment of their corresponding CGC spaces [see
\App{Sec:Sp6:applic}{}].

For \QSpaces where the CGC spaces in \Eq{eq:QSpace:Rep} exactly
correspond to the standard Clebsch-Gordan coefficients for each
symmetry, one may argue that actually similar to the z-labels, it is
not explicitly necessary to store the CGC spaces altogether, since
these are known. This is true, indeed, for these particular cases,
and CGC spaces may simply be referenced then. Nevertheless, the
explicit storage of the CGC spaces with a \QSpace as in
\Eq{eq:QSpace:Rep} has practical value. When combining \QSpaces
through contractions, \ie sum over shared indices, for example, quite
frequently \emph{intermediate} objects can arise that do have rank
different, in particular also larger than $3$ [\eg see the
intermediate objects indicated by the dashed boxes marked by $X$ in
\Fig{fig:matel}{}]. These then elude a description in terms of
standard rank-3 CGCs. In this case, the actual CGC spaces for
intermediate \QSpace are important, and even though they do not
necessarily resemble the interpretation of the original standard
rank-$3$ spaces of standard Clebsch-Gordan algebra anymore, these
spaces will be referred to as CGC spaces nevertheless, owing to their
origin.

Furthermore, for specific algorithms such as NRG and DMRG, on a
global level one typically deals with simple scalar operators such
as the Hamiltonian or a density matrix, apart from intermediate
steps where complex CGC structures can arise. Therefore the full
\emph{sequence} of contractions on the CGC level [\eg see
\Fig{fig:matel}{}] can be replaced by analytical expressions or sum
rules for Clebsch-Gordan coefficients.
In particular, in many situations the explicit knowledge of $3j$-
and $6j$-symbols, or more general $(3n)$-$j$ symbols, appears
sufficient \cite{McCulloch07, Koenig09, Fledderjohann11,
Pfeifer11phd, Singh11phd} with current applications in this
direction again mainly restricted at most to \SU{2}. If the
$(3n)$-$j$ symbols were known for arbitrary non-abelian symmetry,
the explicit storage of the CGC spaces with the \QSpaces would no
longer be required, indeed, and could be avoided altogether. Note,
however, that $(3n)$-$j$ symbols require \emph{specific}
contractions which must be implemented within the code dependent on
the context. While vast literature exists on $(3n)$-$j$ symbols,
this is limited to an overwhelming extent on the relatively simple
symmetry of \SU{2}, for which analytic expressions exist, indeed.
For arbitrary non-abelian symmetries, however, the $(3n)$-$j$
symbols may or may not be known \cite{Zodinmawia11}. For the \QSpace
as outlined in this paper, on the other hand, no special treatment
is required for specific contractions, and no explicit knowledge of
possibly symmetry dependent CGC sum rules is required. The \QSpace
approach solely relies on the correct construction of the standard
CGC spaces to start with, with the subsequent sums over CGC spaces
performed explicitly numerically and not analytically through
\emph{exactly} the same contraction as on the reduced multiplet
level, as discussed in more detail later.

Finally, the explicit inclusion of the CGC spaces allows to build in
strong consistency checks in the actual numerical implementation.
Imagine that the Hamiltonian is built by a sequence of complex
contractions. The Hamiltonian eventually must be a scalar operator,
\ie it is block diagonal in the symmetries and the CGC spaces reduce
to plain identities. This can simply be checked at the end of the
calculation, which thus provides a strong check of whether the
symmetries have been implemented correctly or not. At the stage of
intermediate contraction, however, the CGC spaces guarantee the
correct splitting and weight distribution between different emerging
symmetry sectors.

\subsection{\Atensors and Operators \label{Atensor:Operators}}

Consider the prototypical MPS scenario as in \Eq{eq:CGstates} that
takes some previously constructed state space $\vert i \rangle \equiv
\vert Qn;Q_z \rangle$ and adds a new local state space $\vert \sigma
\rangle \equiv \vert ql;q_z \rangle$, \eg a new physical site. The
state spaces are thus combined in a product-space described in terms
of the \IREPs $\vert j\rangle \equiv \vert \tilde{Q}\tilde{n};
\tilde{Q}_z \rangle$. Here the states $i$, $\sigma$, and $j$ are
introduced as notational shorthand for better readability. The
product space then is spanned by $\vert i\sigma \rangle \equiv \vert
\sigma \rangle \vert i \rangle $. The order of states in the latter
product emphasizes that state $\vert \sigma \rangle$ is typically
added \emph{after} and thus onto the existing state $\vert i
\rangle$, which is of particular importance for fermionic systems. In
general, the combined states \cite{Schollwoeck05}
\begin{equation}
   \vert j \rangle =\sum_{l\sigma} \vert i\sigma \rangle
   \underset{\equiv A_{ij}^{[\sigma]}}{\underbrace{\langle i\sigma|j\rangle }}
\text{,} \label{def:Atensor}%
\end{equation}
are described in terms of linear superpositions of the product space
$\vert i\sigma \rangle$ given by the coefficients
$A_{ij}^{[\sigma]}$, henceforth called \Atensor (rank-3) or
$A^{[\sigma]}$-matrices (rank-2). Without truncation,
$A_{ij}^{[\sigma]}$ denotes a full unitary matrix $U_{(i\sigma),j}$
where the round bracket indicates that the indices $i$ and $\sigma$
have been \emph{fused}, \ie combined into an effective single index.
The presence of symmetry and the proper categorization of state
spaces, however, imposes certain constraints on this unitary matrix,
as pointed out already with \Eq{eq:CGstates}. In particular, the
fully determined CGC spaces $C_{Q_z \tilde{Q}_z}^{[q_z]}$ factorize
from the \Atensor, allowing an arbitrary rotation in the reduced
multiplet space $A_{Q \tilde{Q}}^{[q]}$ only. For the specific case
then, that the reduced multiplet spaces are identical to partitions
of identity matrices with a clear one-to-one correspondence still of
input and output multiplets, the corresponding \Atensor will be
referred to as the \textit{identity \Atensor} [see \Fig{fig:omult}
later; for explicit examples, see App. \Eq{QSpace:A0_zs} or
\Eqp{QSpace:A0_cs}]. An identity \Atensor therefore represents the
full state space still without any state space truncation, and is
unique up to permutations in the combined output space. Its explicit
construction is a convenient starting point, in practice, when
merging new local state spaces with existing effective state spaces.

\begin{figure}[t!]
\begin{center}
\includegraphics[width=0.9\linewidth]{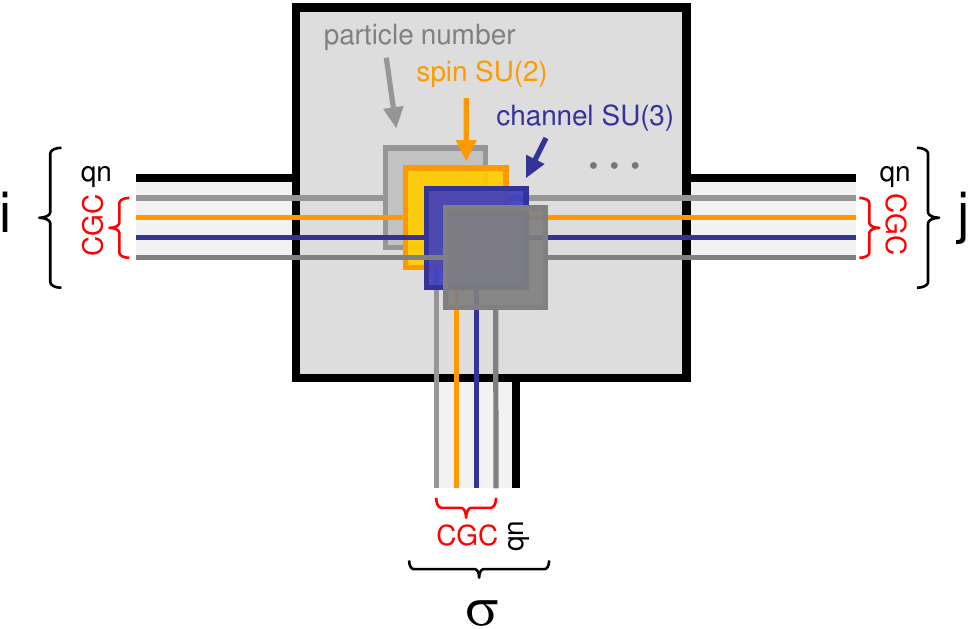}
\end{center}
\caption{
(Color online) Schematic depiction of a rank-3 \QSpace as an example
for a basic building block for an MPS or a tensor network, where
lines (boxes) represent indices (data spaces), respectively. Every
index is assumed to refer to a state space with similar physical background,
hence refers to the same global symmetries as in \Eq{eq:allsymm}, and has
the generic composite structure $\vert qn;q_z\rangle$ as in
\Eq{eq:qnz-labels}, where $q_z$ specifies the states within the CGC spaces.
The rank-3 \QSpace depicted can be interpreted in two entirely different
ways while sharing exactly the same underlying algebraic structure.
These are (i) the state space decomposition into \IREPs and (ii)
operator representation for a given \IROP in a given basis (see
text).
For the general interpretation of the \QSpace depicted, consider for
simplicity, a single row $\nu$ in \Eq{eq:QSpace:Rep}. The set
$\{{Q\}}_{\nu}$ defines the q-labels for all tensor dimensions (here
a total of three). With the q-labels fixed, the corresponding multiplet
index $n$ indexes the typically
large reduced multiplet space, indicated by the thick black
lines for each tensor dimension. The corresponding reduced rank-3
multiplet space $A_{\nu}$ is depicted by the large gray box in the
background. Moreover, with the q-labels fixed, this fixes the \IREPs
for every tensor dimension and every symmetry. The resulting sparse
CGC spaces are indicated by the small boxes around the center, with
one box for every symmetry, such as, for example, abelian particle
conservation, non-abelian spin \SU{2}, non-abelian channel \SU{3}, or
other. By construction, all CGC spaces share the same rank as the
underlying \QSpace. Therefore each CGC
space also has three lines attached, one for every tensor dimension.
In general, the CGC spaces refer to finite multiplet dimensions
for non-abelian symmetries, while for simpler symmetries, such
as abelian symmetries, the CGC spaces actually become trivial, \ie
scalars. These, nevertheless, are also interpreted as having the
same rank as the \QSpace using singleton dimensions throughout.
}\label{fig:qspace}%
\end{figure}

The entire construction of an \Atensor can be encoded compactly in
terms of a rank-3 \QSpace. Both coefficient spaces in
\Eq{eq:CGstates}, $C_{Q_z \tilde{Q}_z}^{[q_z]}$ as well as $A_{Q
\tilde{Q}}^{[q]}$, directly enter the \QSpace description in
\Eq{eq:QSpace:Rep}. A schematic pictorial representation of an
\Atensor is given in \Fig{fig:qspace}. There the states $i$ ($j$)
represent the open composite index to the left (right), respectively,
while $\sigma$ refers to the open composite index at the bottom.

As already argued with \Eq{eq:CGops}, an irreducible operator shares
exactly the same underlying CGC structure as an \Atensor. Thus also
its representation in terms of a \QSpace is \emph{completely
analogous}. Consider an \IROP set $\hat{F}^q \equiv
\{\hat{F}^q_{q_z}\}$, which transforms according to \IREP $q$. Here,
the composite index $\sigma \equiv (ql;q_z)$, for short, identifies
the specific operators in the \IROP set. As already indicated by the
superscript $[1]$ in \Eq{eq:CGops}, its associated multiplet index
$l$ has the trivial range $l=1$, since, by definition, the \IROP
represents a single \IREP on the operator level. With the states
$\vert i\rangle \equiv \vert Q'n';Q'_z \rangle$ and $\vert j\rangle
\equiv \vert Qn;Q_z\rangle$ now representing the same state space
within which the operator acts, with usually many multiplets and
different symmetries, the operator representation of the \IROP
$\hat{F}^{q}$ in the states $\langle i\vert$ and $\vert j \rangle$ is
evaluated using the Wigner-Eckart theorem in \Eq{eq:CGops}. Similar
to the \Atensor earlier, the resulting factorization of the CGC
spaces $C_{Q_z Q'_z}^{[q_z]}$ together with the remaining multiplet
space $F_{QQ'}^{[q]}$ of reduced matrix elements directly enter the
\QSpace description in \Eq{eq:QSpace:Rep}.

So even though an operator is usually considered a rank-2 object, the
fact that an \IROP consists of an operator \emph{set} indexed by
$\sigma$, adds a \emph{third} index to the \QSpace. In contrast to
the state interpretation of $\sigma$ for the \Atensor above, however,
here the ``index'' $\sigma$ has a different interpretation in that it
points to a specific operator in the \IROP set. By convention, the
operator index $\sigma$ will always be listed as third tensor
dimension in its \QSpace representation. Given the three-dimensional
representation of a general \IROP, therefore its entire construction
mimics the construction of an \Atensor in terms of a \QSpace. As a
consequence, \Fig{fig:qspace} exactly also resembles the \QSpace
structure of an \IROP. The states $i$ ($j$) used for the calculation
of the matrix element represent the open index to the left (right),
respectively, while the operator index $\sigma$ refers to the open
index at the bottom.

Scalar operators, finally, such as the Hamiltonian of the system or
density matrices, represent a special case, since there the \IROP set
contains just a single operator. Therefore the third index, \ie the
\emph{operator index}, becomes a singleton and hence can simply be
dropped [\eg see \Eq{QSpace:H0:cs-ref}{}]. Scalar operators therefore
are represented by rank-2 \QSpaces. They are block-diagonal in their
symmetries, and their CGC spaces are all equal to identity matrices,
with an example already given in \Eq{QSpace:H0:cs-ref}.

\subsection{Multiplicity \label{sec:multiplicty}}

For general non-abelian symmetries, frequently inner and outer
multiplicity occur. \cite{Elliott79,Alex11} Both are absent in
\SU{2}, yet do occur on a regular basis for \SU{N\ge3}. Inner
multiplicity describes the situation where for a given \IREP, several
states may share exactly the \emph{same} z-labels. Let $m^q_z$ denote
the number of times a specific z-label occurs within \IREP $q$. Then
the presence of inner multiplicity implies $m^q_z>1$ for at least one
z-label. Within such degenerate subspaces an arbitrary rotation is
allowed in principle. For global consistency, therefore the CGC
spaces must adopt a well-defined internal convention on how to deal
with inner multiplicity. This issue, however, is entirely contained
within the CGC algebra, which is explored in more detail in the App.
A [\eg see discussion following \Eqp{inner-mult}, and
\App{Sec:NumDeComp}{}]. On the level of a \QSpace, it is of no
further importance otherwise. Essentially, the only implication of
inner multiplicity is $q_z\to (q_z,\alpha_z)$ with $\alpha_z =
1,\ldots,m_z^{q}$ [\cf \Eq{inner-mult}{}], where $m_z^{q}$ depends on
the multiplet $q$. With this minor adjustment, it is assumed
throughout that the z-labels fully identify the internal multiplet
space. Note that, in practice, the extra label $\alpha_z$ is never
included explicitly. What is important, however, is a
\emph{consistent internal multiplet ordering} that respects
multiplicity [see \App{Sec:NumDeComp}{}].

Outer multiplicity, on the other hand, describes the situation where
in the state space decomposition of a product-space of two \IREPs,
$q_1$ and $q_2$, the \emph{same} output \IREP $q$ may appear multiple
times, the number of which is specified by $M_{q}^{[q_1,q_2]}$ [\cf
App. \Eqr{irrep-decomp}{def:clebsch} and discussion]. Therefore outer
multiplicity primarily also enters at the level of Clebsch-Gordan
coefficients, as it is based on pure symmetry considerations. In
contrast to inner multiplicity, however, outer multiplicity also
affects the reduced multiplet space, as will be elaborated upon in
what follows.

In the absence of outer multiplicity [\ie $M_{q}^{[q_1,q_2]} \le 1$
for all $q_1$, $q_2$, and $q$ of the symmetry, an example being
\SU{N\leq2}{}], all rows in the \QSpace in \Eq{eq:QSpace:Rep} must
have \emph{unique} $\{Q\}_{\nu}$. If this is not the case, then the
rows can be made unique by combining the rows with the same $\{Q\}$.
Assume, for example, $\{Q\}_{\nu}=\{Q\}_{\nu'}$ with $\nu\neq\nu'$:
clearly, the $\{Q\}$'s are already the same. Having the \textit{same}
symmetry labels, this refers to the same set of \IREPs, hence also
the CGC spaces of these records must be identical, up to a possible
global normalization factor which can be associated with the
multiplet space, instead. Furthermore, given
$\{Q\}_{\nu}=\{Q\}_{\nu'}$, the $A_{\nu}$ and $A_{\nu'}$ data blocks
do live in \emph{exactly} the same vector spaces for each individual
tensor dimension! Therefore $A_{\nu}$ and $A_{\nu'}$ can be simply
added up [here multiple contributions with the same $\{Q\}$ are
considered additive, consistent with general conventions regarding
sparse tensors; otherwise, say having given the same matrix element
twice with different values, would immediately lead to
contradictions].

\begin{figure}[t]
\begin{center}
\includegraphics[width=1\linewidth]{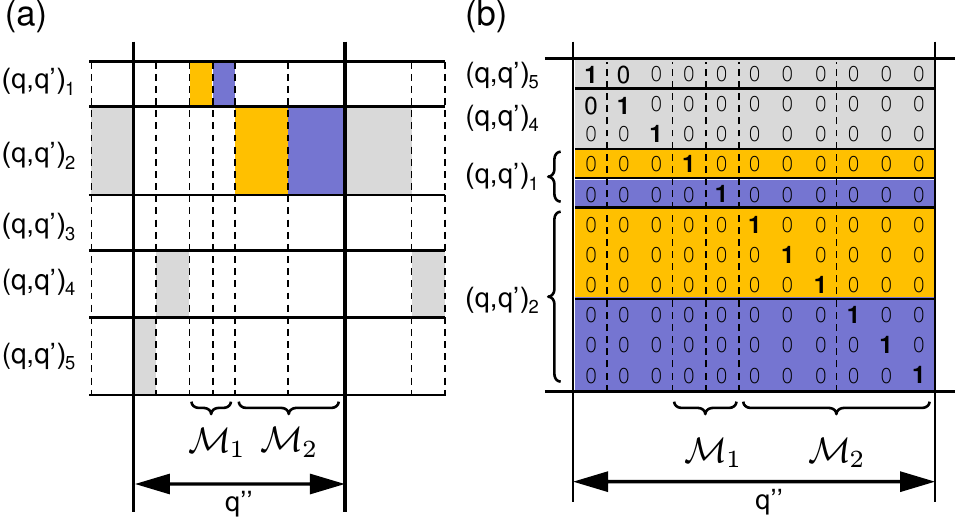}
\end{center}
\caption{
Effect of outer multiplicity on multiplet space ($A_\nu$) in terms of
an identity \Atensor -- Panel (a) Schematic depiction of the state
space decomposition of two input multiplet spaces with unique
symmetry combinations $(q,q')$ into combined multiplets $q''$ (rows
and columns, respectively). State spaces of the same symmetry are
grouped into blocks separated by solid lines (horizontally and
vertically). For simplicity, an identity \Atensor is depicted, for
which the individual sectors in $q''$ can be uniquely associated with
the $(q,q')$ they originate from. Hence each column, separated by
solid lines, has exactly one shaded block considered non-zero, with
all-zero blocks shown in white. Here vertical thin lines indicate
sub-blocks that originate from different $(q,q')$, yet are eventually
combined in the same block as they belong to the same symmetry $q''$
(separated by thick lines). Now, in the presence of outer
multiplicity a specific $(q,q')$ can contribute to the same $q''$
several times, as depicted schematically by the spaces
$\mathcal{M}_1$ and $\mathcal{M}_2$ for the rows $(q,q')_1$ and
$(q,q')_2$, respectively, both showing a multiplicity of $M_{q''}=2$.
Panel (b) depicts the \emph{enlarged} multiplet space for the output
multiplet $q''$ of panel (a) in order to accommodate the additional
multiplets arising from outer multiplicity. Being an identity
\Atensor, the entire block shown in panel (b) represents an identity
matrix (in contrast to an arbitrary \Atensor, which may have an
arbitrary unitary matrix in its place). The vertical lineup of
$(q,q')$ sectors is arbitrary, making the identity \Atensor unique up
to permutations. The identity matrix shown in the panel is sliced
into horizontal blocks as indicated, each of which is associated with
its own unique CGC space [not shown] as derived from the Lie algebra
of the symmetry under consideration. Each of these slices then
directly enters as a reduced multiplet space $A_\nu$ in a separate
row in the \QSpace as in \Eq{eq:QSpace:Rep}.
}\label{fig:omult}%
\end{figure}

In the presence of outer multiplicity, on the other hand, the
uniqueness of the q-labels $\{Q\}_{\nu}$ in the \QSpace in
\Eq{eq:QSpace:Rep} has to be relaxed. The reason for this is as
follows. Since outer multiplicity derives from the Clebsch-Gordan
algebra as in \Eq{def:clebsch}, the CGC spaces
\begin{align}
   C_{Q_z\tilde{Q}_{z}}^{[q_z]} \to
   C_{Q_z\tilde{Q}_{z},\alpha}^{[q_z]}
   \equiv \langle QQ_z; qq_z \vert \alpha\tilde{Q}, \tilde{Q}_z \rangle
\label{eq:CGCalpha}
\end{align}
acquire an additional label $\alpha=1,\ldots,M^{[Q,q]}_{\tilde{Q}}$
[different from the $\alpha_z$ used with inner multiplicity], where
$M^{[Q,q]}_{\tilde{Q}}$ indicates the outer multiplicity in
$\tilde{Q}$, given the product space of the \IREPs $Q$ and $q$. In
terms of a \QSpace object, one may therefore be tempted to enlarge
the CGC space from rank-3 to rank-4, with the dimension of the 4th
index being equal to $M^{[Q,q]}_{\tilde{Q}}$. This strategy alone,
however, does not capture the full picture since outer multiplicity
also \textit{enlarges} and thus effects the multiplet space $A_{\nu}$
of an \Atensor. By definition, outer multiplicity means that
\textit{different} multiplets with the same $q$ can emerge. The only
way they can be distinguished is through their Clebsch-Gordan
coefficients. Therefore rather than enlarging the CGC space in a
\QSpace, $M^{[Q,q]}_{\tilde{Q}}$ records with the \emph{same}
$\{Q\}_{\nu}$ are allowed, instead. These records have CGC spaces of
the same rank-3 dimensions, which, however, are clearly
distinguishable, as they are orthogonal to each other [\cf appendix
\Eqp{eq:clebsch:alpha:ortho}]. The $M^{[Q,q]}_{\tilde{Q}}$ sets of
Clebsch-Gordan coefficients arising from outer multiplicity are thus
spread over $M^{[Q,q]}_{\tilde{Q}}$ records within a \QSpace object.

The situation in the multiplet space for an identity \Atensor is
depicted schematically in \Fig{fig:omult}. In the absence of outer
multiplicity, each symmetry combination $(q,q')_{i}$ can only
contribute at most once to a given symmetry space $q''$ and gets its
space allocated, as depicted, for example, for $(q,q')_{4}$ in
\Figp{fig:omult}{a}, having only one non-zero block (shaded block)
within the $q''$ output multiplet. The symmetry combinations
$(q,q')_{1}$ and $(q,q')_{2}$, on the other hand, show outer
multiplicity, in that they result twice in the same multiplet $q''$,
\ie $M^{[q,q']_1}_{q''}=M^{[q,q']_2}_{q''}=2$.

For simplicity, in the absence of truncation and without any further
unitary rotation, the tensor-product on the multiplet level can be
represented as an identity \Atensor with a clear one-to-one
correspondence of input to output multiplets. This is depicted in
\Figp{fig:omult}{b} in terms of an identity matrix in the reduced
multiplet space. The identity matrix in panel (b) then is sliced
horizontally into blocks for each $(q,q')$ that contributes to $q''$.
In the presence of outer multiplicity, the state space for $q''$
needs to be \textit{enlarged} to accommodate the additional
multiplets. The slicing (horizontal solid lines) then also proceeds
for every output multiplet resulting from outer multiplicity, as
indicated in panel (b). As a result, $M_{q''}^{[q,q']}$ slices are
associated with exactly the same $Q\equiv \{q,q',q''\}$,
distinguishable only through their Clebsch-Gordan coefficients. These
slices directly enter as $A_\nu$ in separate rows in a \QSpace as in
\Eq{eq:QSpace:Rep}.

In summary, outer multiplicity requires an adaptation of the
multiplet space, which is naturally incorporated into a \QSpace by
allowing multiplet entries with the same $\{Q\}_{\nu}$ labels yet
with clearly distinguishable CGC spaces. That is, specific records
are also considered to refer to \emph{different} state spaces if
their CGC spaces are not exact copies (up to a global factor that can
be incorporated into the multiplet data) but rather orthogonal to
each other [see App. \Eq{eq:clebsch:alpha:ortho}{}]. In practice,
this is checked within a small numerical threshold ($\sim10^{-12}$)
accounting for numerical double precision noise. The great advantage
of this prescription is that then multiplicities fall completely in
line with the rest of the MPS algorithm without any specific further
treatment.

Finally, it is important to notice that the same concept of relaxing
the uniqueness of the $\{Q\}_{\nu}$ labels actually also can become
relevant for symmetries that \emph{do not} have intrinsic outer
multiplicity in its actual sense. Yet, in fact, through contractions
\emph{intermediate objects} can arise of rank larger than three [\eg
see the \QSpaces indicated by the dashed boxes marked by $X$ in
\Fig{fig:matel}{}], where records in a \QSpace with the same
$\{Q\}_{\nu}$ labels can also have \emph{incompatible} CGC spaces, in
the sense that they are \emph{not} the same up to overall factors. In
this case, also the uniqueness of the $\{Q\}_{\nu}$ must be relaxed
temporarily. For simplicity, this will also be referred to as outer
multiplicity.

\subsection{Contractions \label{sec:contractions}}

\begin{figure}[t]
\begin{center}
\includegraphics[width=0.6\linewidth]{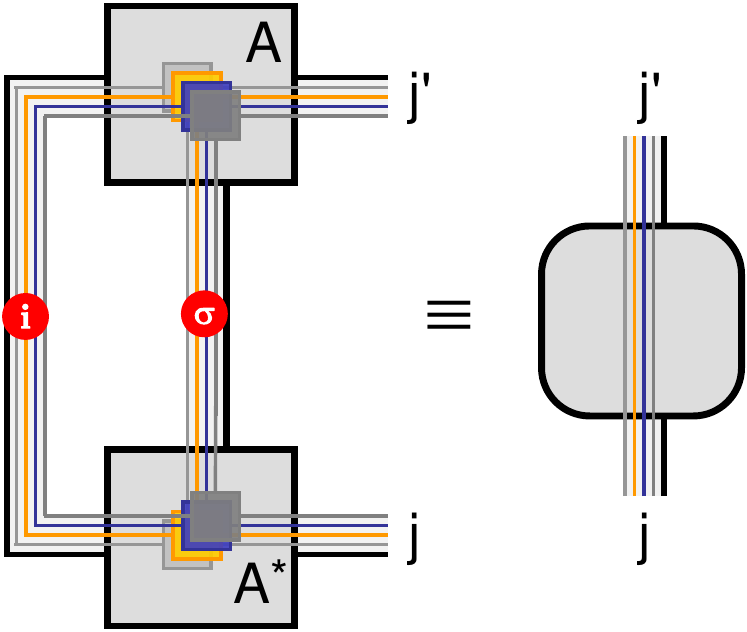}
\end{center}
\caption{
(Color online) Contraction of (i) an \Atensor or (ii) an irreducible
operator into a scalar. All indices specified are composite indices
of the type $\vert qn;q_z\rangle$. An \Atensor describes a
(truncated) basis transformation of the product-space of the new
local space $\vert \sigma\rangle$ with an effective previously
constructed basis $\vert i \rangle$, resulting in the combined state
space $\vert j' \rangle \equiv \sum_{i'\sigma'} A_{i'j'}^{[\sigma']}
\vert \sigma' \rangle \vert i'\rangle$, with the corresponding
bra-space $\langle j \vert \equiv \sum_{i\sigma}A_{ij}^{[\sigma]\ast}
\langle i \vert \langle \sigma\vert$ depicted in the lower part of
the figure. The result is the scalar identity operator, reflecting
the orthonormality condition \Eq{eq:Aortho}. An entirely different
interpretation of the same contraction pattern can be given when the
\Atensor is replaced by an \IROP $F^{\sigma}$. The contraction then
describes \Eq{eq:oprod:scalar2} and yields a scalar operator, with
its generic \QSpace representation schematically depicted to the
right.
} \label{fig:id-scalar}%
\end{figure}

The contraction of \QSpaces will be introduced in the following in
terms of a simple example, namely the orthonormalization condition on
the combined state space in a tensor-product space. Putting symmetry
labels aside for the sake of the argument, the \Atensor
$A_{ij}^{[\sigma]} \equiv \langle i\sigma|j \rangle $ in
\Eq{def:Atensor} combines the state spaces $\vert i\sigma \rangle $
into a combined (possibly truncated) orthonormal state space $\vert j
\rangle $. This directly leads to the standard orthogonality relation
for an \Atensor,
\begin{equation}
   \sum_{i\sigma} A_{ij}^{[\sigma]\ast} A_{ij'}^{[\sigma]}=\delta_{jj'}
\text{,}\label{eq:Aortho}%
\end{equation}
which is a simple example for the simultaneous contraction of two
tensors \wrt to two common indices, here $i$ and $\sigma$. By
construction, it is completely analogous in structure to the
orthogonality condition of CGCs as in App.
\Sec{eq:clebsch:alpha:ortho}.
Including symmetries, the contraction in \Eq{eq:Aortho} is depicted
in terms of \QSpaces in \Fig{fig:id-scalar}. Overall, indices are
represented by lines, and lines connecting two blocks such as the
indices $i$ and $\sigma$ are summed over, \ie contracted. In
practice, contraction of \QSpaces as defined in \Eq{eq:QSpace:Rep}
happens at several levels, since state indices are labeled by
composite indices that refer to a symmetry basis of the type $\vert
qn;q_{z} \rangle $. This implies for a contraction
$\sum_{i=i^{\prime}}$ of two \QSpace objects with respect to some
common state space $i$ and $i'$, that (i) the q-labels $q_{i}$ and
$q_{i'}$ of the \QSpaces as in \Eq{eq:QSpace:Rep} must be matched
for the indices $i$ and $i'$, respectively. For a given specific
match of rows $\nu$ and $\nu^{\prime}$ then, this is followed (ii)\
by the contraction of the corresponding reduced multiplet spaces,
and (iii) by exactly the same contraction of the CGC spaces, one for
each symmetry. This procedure derives from \Eq{B-CG-tprod}, since
the contraction of two tensors $B^{(1)}$ and $B^{(2)}$ for a given
match $\nu$ and $\nu'$, can be simply decomposed as the
\emph{sequential} contraction of its constituents, \ie the reduced
multiplet space and the corresponding CGC spaces,
\begin{eqnarray}
  &&\left[
      B^{(1)}_{\nu} \otimes \bigl(\bigotimes_{s=1}^{\NS}\mathrm{C}^{(1)}_{s;\nu}\bigr)
  \right] \cdot
  \left[
      B^{(2)}_{\nu'} \otimes \bigl(\bigotimes_{s=1}^{\NS}\mathrm{C}^{(2)}_{s;\nu'}\bigr)
  \right] \notag\\
&&=\bigl[ B^{(1)}_{\nu} \cdot B^{(2)}_{\nu'} \bigr] \otimes \bigl(\bigotimes_{s=1}^{\NS}
   \bigl[ \mathrm{C}^{(1)}_{s;\nu} \cdot \mathrm{C}^{(2)}_{s;\nu'} \bigr] \bigr)
\text{,} \label{contraction:oprod}
\end{eqnarray}
Here the multiplication ``$\,\cdot\,$'' is interpreted as
contraction \wrt to a certain subset of shared dimensions between
the tensors $B^{(1)}$ and $B^{(2)}$. Note that the rank of a \QSpace
and its index order are always shared by the multiplet space and CGC
spaces for consistency. Hence the overall contraction of the
\QSpaces is directly reflected in the elementary contraction of the
plain numerical tensors $A_{\nu}$ and $\{C\}_{\nu}$. That is, the
\textit{contraction pattern} depicted schematically in
\Fig{fig:id-scalar}, drawn in terms of boxes with connecting lines,
is exactly the same on all levels of the contraction.
By collecting the remaining \emph{non}-contracted q-labels, this
forms a new entry $\nu''$ in the resulting \QSpace, with the
(tensor) index order of the resulting tensor dimensions again being
the same for all $\{Q\}_{\nu''}$, $A_{\nu''}$, and $\{C\}_{\nu''}$
for consistency.

Finally, the resulting \QSpace is made unique in the $\{Q\}_{\nu''}$
labels as far as outer multiplicity permits. Records can only be
combined, \ie summed over, \textit{iff} the CGC spaces for given
records are all the same up to global factors which can be absorbed
into the multiplet data, instead (see \Sec{sec:multiplicty}). Outer
multiplicity plays no special role with contractions otherwise. Note
that independent of whether or not outer multiplicity is present,
when specifying a \textit{subset} of tensor dimensions within
$\{Q\}_{\nu}$ for contraction, the resulting \QSpace will, in
general, always have \emph{many} contributions to the same
$\{Q\}_{\nu''}$. For comparison, consider the completely analogous
case of regular square matrices of dimension $D>1$: a matrix element
$(M)_{ij}$ is uniquely identified in the overall index $(i,j)$, while
for example, the index $i$ is not sufficient as it refers to an
entire row of matrix elements. Moreover, when two matrices $M_{1}$
and $M_{2}$ are multiplied together,
\begin{equation}
   (B)_{jj'}=(M_{1} M_{2})_{jj'} =
   \sum_{i=i'} (M_{1})_{ji} (M_{2})_{i'j'}
\text{,}\label{eq:MMprod}
\end{equation}
the common index space (second index of $M_{1}$ and first index of
$M_{2}$) is summed over, \ie contracted. \textit{Every} match $i=i'$
results in a contribution. In particular, for some given $j$ and
$j'$, all $D$ matches $i=i'$ contribute and are summed up to the same
output space $(j,j')$. In the case of \QSpaces the situation is
\textit{exactly} analogous. All matches $i=i'$ in the q-labels
$q_{i}$ and $q_{i'}$ for the contracted index must be included. The
only real consequence of outer multiplicity is that in the resulting
\QSpace $B$ in \Eq{eq:MMprod} not necessarily all records with the
same $\{Q\}_{\nu}$ labels can be merged by adding them together. In
the specific case of the contraction in \Eq{eq:Aortho}, however, the
resulting \QSpace is simply the identity, and as such a scalar
operator with unique $\{Q\}_{\nu}$.

\subsection{Scalar operators \label{scalar-ops}}

Given the definition of an \Atensor in \Eq{def:Atensor}, the
contraction of the two \QSpaces $A$ and $A^{\ast}$ in
\Fig{fig:id-scalar} leads to the identity operator
$\mathbf{\hat{1}}^{(C)} \equiv \sum_j \vert j \rangle\langle j\vert$
in the possibly truncated combined space $C$ [\cf \Eq{eq:Aortho}{}].
Clearly, this also provides a strong check on the numerical
implementation of the symmetries. In particular, $\hat{\Id}^{(C)}$
represents a (trivial) example of a scalar operator, that can be
described as rank-2 \QSpace. The CGC spaces are all identity matrices
(up to overall factors that can be associated with the multiplet
space), and therefore the lines, that usually connect to the CGC
spaces within a \QSpace, can be directly connected through from $j$
to $j'$ on the \rhs of \Fig{fig:id-scalar}, with the CGC spaces
themselves no longer shown. In given case, due the orthonormality
condition in \Eq{eq:Aortho}, also the reduced multiplet space is
given by identity matrices. This actually also would allow to connect
through the thick black line on the \rhs of \Fig{fig:id-scalar}, and
thus also to skip the large remaining block on the \rhs for the
reduced multiplet space altogether.

\begin{figure}[tb]
\begin{center}
\includegraphics[width=0.7\linewidth]{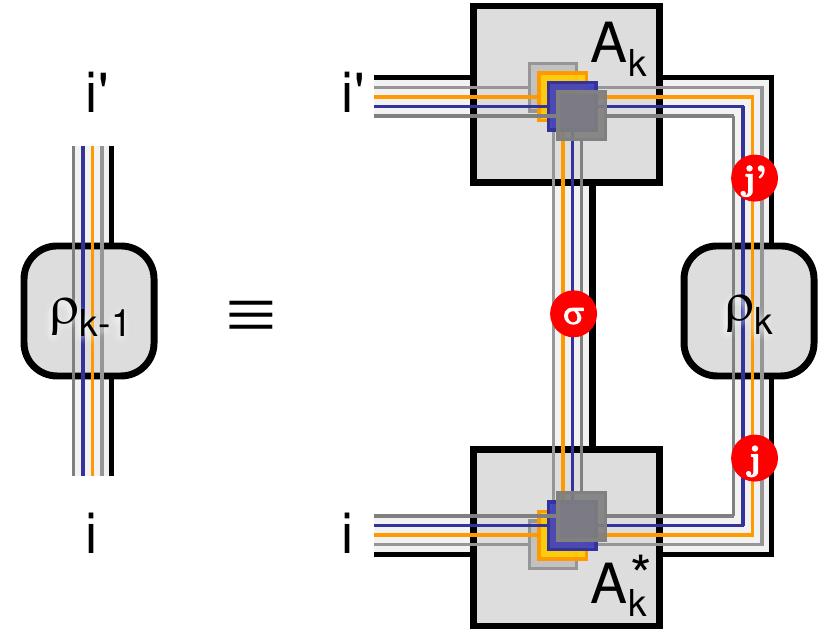}
\end{center}
\caption{ (Color online)
Backward update of density matrix $\rho_k$ given in the effective
basis $\vert j\rangle$ of a system up to and including site $k$
(right index) by tracing out the local state space $\vert
\sigma_k\rangle$ (middle index) given the basis transformation $A_k$ that
introduced site $k$. The result is the reduced density matrix
$\rho_{k-1}$ in the effective basis $\vert i\rangle$ of the system
up to and including site $k-1$.
}\label{fig:rhoupdate}%
\end{figure}

Figure \ref{fig:id-scalar}, however, allows yet an entirely different
interpretation. Remember that an irreducible operator set $\hat{F}^q$
has a completely analogous structure and interpretation in terms of
its internal CGC spaces when compared to an \Atensor (cf.
\Fig{fig:qspace}). Therefore it must hold that the
scalar-product-like contraction,
\begin{subequations}\label{eq:oprod:scalar}
\begin{equation}
   \hat{F}^{2} \equiv \hat{F}\cdot\hat{F}^{\dagger} \equiv
   \sum_{q_z} \hat{F}^q_{q_z}(\hat{F}^q_{q_z})^{\dagger}
%% \quad \text{(scalar operator)}
\label{eq:oprod:scalar1}%
\end{equation}
also results in a scalar operator (note that through the
Wigner-Eckart theorem, by convention, the state space associated with
the \emph{right} index of the operator $\hat{F}^q$ is combined with
the multiplet space $q$; \cf \App{sec:irop+wet}). With $\sigma\equiv
(q1;q_z)$ and the further sum through the operator (matrix)
multiplication, \Eq{eq:oprod:scalar1} shares exactly the same
contraction pattern as discussed in \Fig{fig:id-scalar} in the
context of the orthonormality of \Atensors earlier. Here the
resulting scalar operator, however, can have arbitrary positive
hermitian matrices in its multiplet space still, represented by the
large gray box on the \rhs of \Fig{fig:id-scalar}.
The reduction of \Eq{eq:oprod:scalar1} to a scalar operator is also
intuitively clear, given that the Hamiltonian itself is typically
constructed in terms of scalar operators of exactly this type [see,
for example, App. \Eq{H:tb-psiS} or \Eq{eq:Psi:CS4_scalar} given the
Hamiltonian in \Eq{H:tb-default}{}]. The notation in
\Eq{eq:oprod:scalar1} emphasizes that in the scalar product the
\textit{same} irreducible operator set $\hat{F}^q$ must be taken,
considering that the IROP $\hat{F}^q$ is \emph{different} from the
\IROP $(\hat{F}^\dagger)^q$. Nevertheless, since
$(\hat{F}^q_{q_z})^{\dagger} \sim (\hat{F}^{\dagger})^q_{-q_z}$, up
to possible signs originating from the definition of the CGC algebra
[\eg compare the \QSpaces in App. Tbls.~\ref{QSpace:psiS} and
\ref{QSpace:psiS-2} and accompanying discussion], these signs are
irrelevant in the scalar contraction. Hence it follows that also
\begin{equation}
   \tilde{F}^{2} \equiv \hat{F}^{\dagger}\cdot\hat{F}\equiv
   \sum_{q_z} (\hat{F}^q_{q_z})^{\dagger} \hat{F}^q_{q_z}
\label{eq:oprod:scalar2}%
\end{equation}
\end{subequations}%
is a scalar operator, yet different from \Eq{eq:oprod:scalar1}, as
indicated by the tilde on $\tilde{F}^{2}$. Similarly, note that if
the \Atensor had been contracted on the right instead of the left
index in \Fig{fig:id-scalar}, this also would have yielded a scalar
operator, namely a reduced density matrix up to normalization (\eg
\Fig{fig:rhoupdate} below using $\rho_k\equiv \Id$).

\begin{figure}[t]
\begin{center}
\begin{tabular}{l@{ }ll}
\raisebox{1.2in}{\large{(a)} } & & \includegraphics[width=0.6\linewidth]{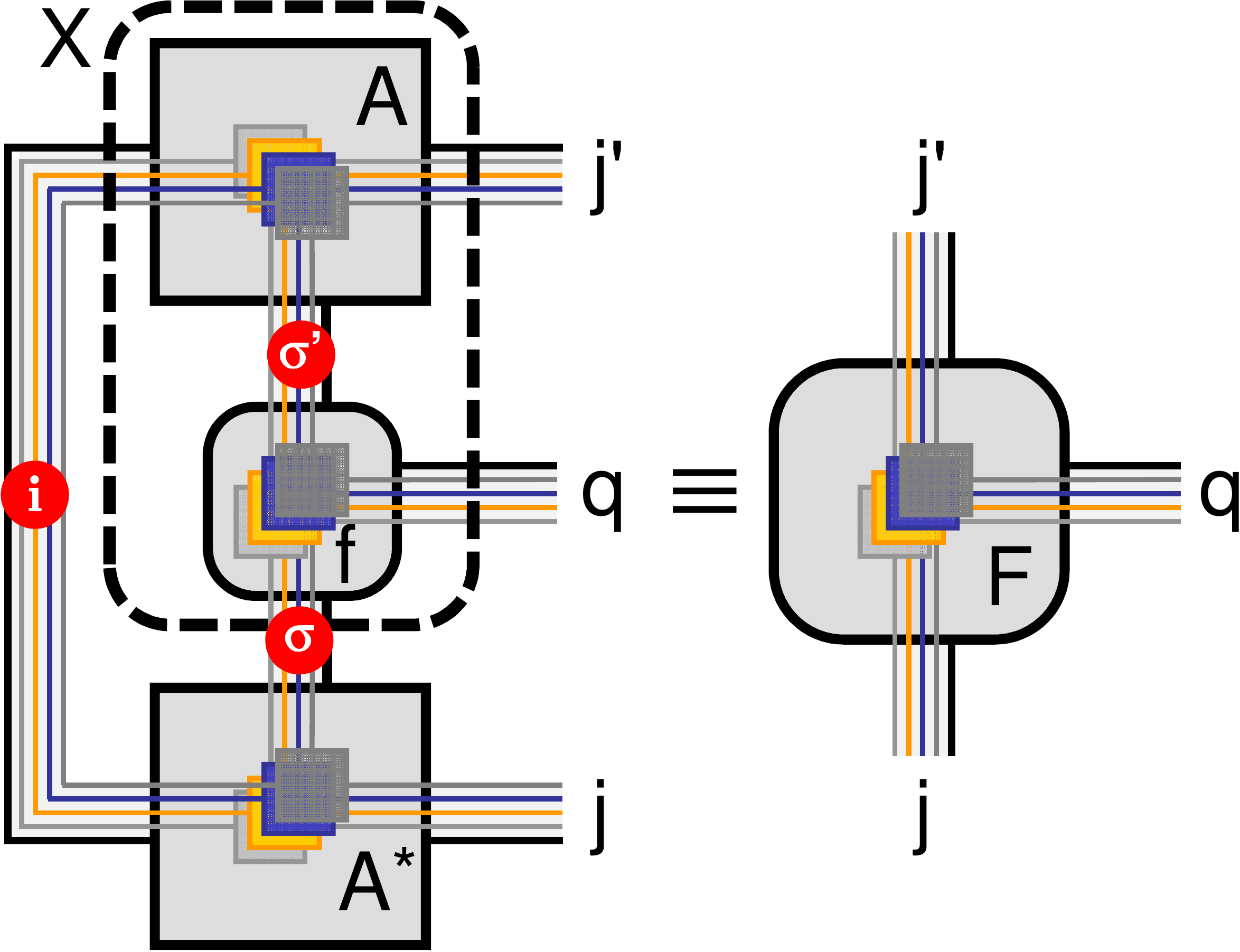} \\[5ex]
\raisebox{1.4in}{\large{(b)} } & & \includegraphics[width=0.65\linewidth]{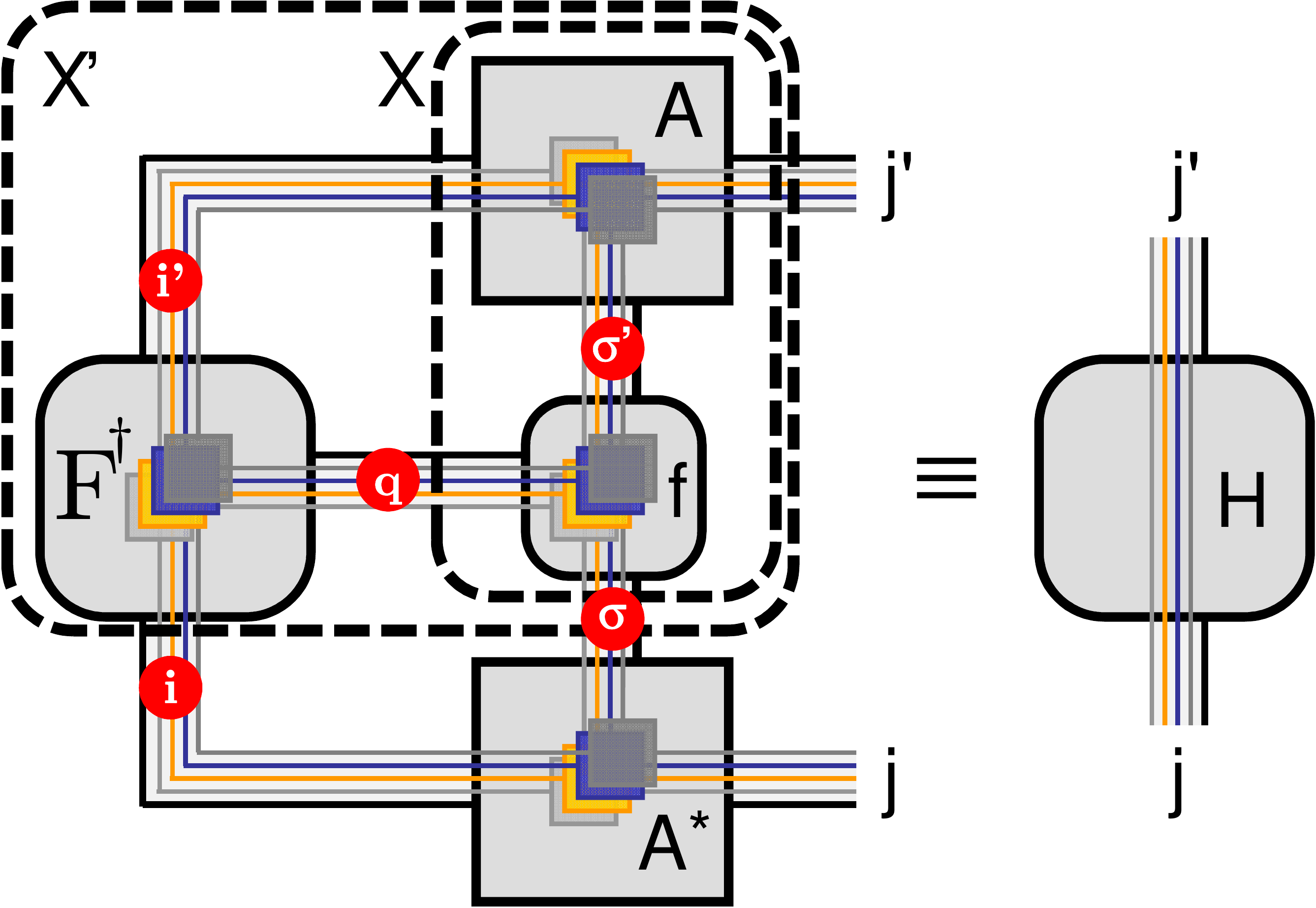}
\end{tabular}
\end{center}
\caption{ (Color online)
Typical evaluation of matrix elements given an \Atensor. The nested
dashed boxes $X^{(\prime)}$ indicate the sequential order of
contractions prior to the final contraction. In panel (a), the local
\IROP set $\hat{f}^q$ acts within the state space
$\vert\sigma\rangle$ of a given site. Its local matrix elements,
$\langle\sigma\vert \hat{f}^q \vert\sigma'\rangle$, are assumed to be
known and described in terms of the local rank-3 \QSpace $f$. The
local \IROP set is mapped into the larger effective space linked
through the \Atensor, $\vert j\rangle = \sum_{i\sigma}
A_{ij}^{[\sigma]} \vert\sigma\rangle \vert i\rangle$. The overall
result is the rank-3 \QSpace $F$ on the \rhs, \ie the desired matrix
elements $F_{jj'}^{[q]} \equiv \langle j\vert \hat{f}^q\vert
j'\rangle$. Panel (b) depicts a typical scalar nearest-neighbor
contribution to a Hamiltonian $\hat{H}\equiv
[\hat{f}^{q}]^\dagger_{k} \cdot [\hat{f}^{q}]_{k+1}$ of two
consecutive sites, say $k$ and $k+1$ using their respective
\Atensors. This contraction already uses an effective description of
the local operator $\hat{f}_k^\dagger$ at site $k$ in terms of the
\QSpace $F^\dagger$, obtained from the \Atensor $A_{(k)}$ at site $k$
as in panel (a) form the prior iteration. Using the \Atensor $A_{(k+1)}$
of site $k+1$, the overall contraction can be completed as indicated.
}\label{fig:matel}%
\end{figure}

\subsection{Operator matrix elements}

The typical calculation of matrix elements of operators for iterative
methods such as NRG or DMRG is depicted schematically in
\Fig{fig:matel}. While the complex many body states are generated
iteratively and described by \Atensors [\cf \Eq{def:Atensor}{}], an
elementary irreducible operator set $\hat{f}^{q}$, on the other hand,
usually operates locally within the state space $\vert \sigma
\rangle$ of a specific site. Therefore, the operator is described
initially in terms of the matrix elements $f_{\sigma\sigma'}^{[q]}
\equiv \langle \sigma\vert \hat{f}^q \vert \sigma'\rangle$. The link
to the many body states is given through the \Atensor that connects
given site to a generated effective state space $\vert i\rangle$,
$\vert j\rangle =\sum_{i\sigma} A_{ij}^{[\sigma]} \vert \sigma\rangle
\vert i\rangle$. The matrix elements of an \IROP in the combined
space $\vert j\rangle$ then become,
\begin{align}
   F_{jj'}^{[q]}
 & \equiv \langle j\vert \hat{f}^{q} \vert j'\rangle
 = \sum_{i\sigma,i'\sigma'}
   A_{ij}^{[\sigma]\ast} A_{i'j'}^{[\sigma']}
   \underset{\equiv \delta_{ii'} f_{\sigma\sigma'}^{[q]}}
   {\underbrace{\langle i\vert \langle \sigma\vert
      \hat{f}^q \vert \sigma'\rangle \vert i'\rangle
   }}\nonumber\\
&= \Bigl[\sum_{\sigma}
   A^{[\sigma]\dagger} \Bigl(\sum_{\sigma'} f_{\sigma\sigma'}^{[q]} A^{[\sigma']}
   \Bigr)\Bigr]_{jj'}
\text{,}\label{eq:Fmatel}%
\end{align}
It is exactly this procedure that is depicted in \Figp{fig:matel}{a}.
The matrix elements are calculated in a two-stage process. The sum in
the round brackets of \Eq{eq:Fmatel} (contraction of $\sigma'$) is
carried out first, leading to the temporary rank-4 tensor with open
indices $(i,j',\sigma,q)$ [box $X$ in \Figp{fig:matel}{a}]. This
rank-4 tensor then is contracted simultaneously in the indices $i$
and $\sigma$ with the $A^{\ast}$ tensor, providing the final result
shown on the \rhs of \Figp{fig:matel}{a}. Quite generally, for
contractions including several blocks as in \Fig{fig:matel}, these
are always done sequentially, adding one block at a time. This is
explicitly indicated in \Fig{fig:matel} by the (nested) dashed boxes,
with the final contraction connecting the remaining tensor to the
outer-most dashed box. Every individual contraction then follows the
multi-stage process over composite indices as described earlier in
\Sec{sec:contractions}.

The so obtained effective description $F_k^{[q]}$ of an operator
$\hat{f}^q$ acting on site $k$ using $A_k$ can be used then to
describe, for example, the typical scalar nearest-neighbor
contribution $[\hat{f}^{q}]^\dagger_{k} \cdot [\hat{f}^{q}]_{k+1}$ to
the Hamiltonian including site $k+1$. This operation is shown in
\Figp{fig:matel}{b}. In particular, one may use the identity \Atensor
$A_{k+1}^{\mathrm{Id}}$ for site $k+1$, such that the resulting
Hamiltonian is constructed in the full tensor-product space
$\vert\sigma\rangle_{k+1}\vert i\rangle_k$ of the system up to and
including site $k+1$. Here $\vert i\rangle_k$ describes the effective
space up to and including site $k$, whereas
$\vert\sigma\rangle_{k+1}$ describes the new local state space of
site $k+1$. This exactly corresponds the two-stage prescription used
within the NRG (and similarly also for the DMRG) to build the
Hamiltonian for the next iteration: (i) the tensor-product space
including the newly added site must be mapped into proper symmetry
spaces. This is taken care of by the construction of the identity
\Atensor $A_{k+1}^{\mathrm{Id}}$. (ii) The new Hamiltonian is built
using this identity \Atensor through contractions as shown in
\Figp{fig:matel}{b} [note that while the presence of outer
multiplicity in \QSpace $f$ is typically inherited by \QSpace $F$
through the basis transformation as in \Figp{fig:matel}{a}, the
internal contraction over the \IROP set index $q$ in
\Figp{fig:matel}{b} eventually leads to a scalar contribution to the
Hamiltonian, as discussed with \Eq{eq:oprod:scalar2}{}]. After
diagonalization and state space truncation in the combined state
space, the part of the resulting unitary matrix describing the kept
states can be contracted onto $A_{k+1}^{\mathrm{Id}}$, yielding the
actual final $A_{k+1}$.

\subsection{Density matrix and backward update \label{RHO:backward}}

Consider the density matrix $\hat{\rho}_k \equiv \sum_{jj'}
(\rho_k)_{jj'} \vert j\rangle \langle j'\vert$ given in the basis
$\vert j\rangle_{(k)}$, which is assumed to include all sites of a
system up to and including site $k$. With the local state space of
the last site $k$ described by $\vert\sigma_k\rangle$, tracing out
this last site from the density matrix $\rho_k$ corresponds to
contracting the $A_k$-tensor that connected site $k$ to the system's
MPS,
\begin{align}
    \hat{\rho}_{k-1} &= \trace_{\sigma_k} \bigl(\hat{\rho}_{k}\bigr)\nonumber\\
 &= \sum_{ij,i'j',\sigma}
    A_{ij}^{[\sigma_k]\ast} A_{i'j'}^{[\sigma_k]}\,(\rho_k)_{jj'}
    \vert i\rangle \langle i'\vert \nonumber\\
 &\equiv \sum_{ii'} \bigl(
      A^{[\sigma_k]\dagger} \rho_k A^{[\sigma_k]}
    \bigr)_{ii'} \vert i\rangle \langle i'\vert
\text{.}\label{def:backprop}%
\end{align}
\EQ{def:backprop} leads to a density matrix $\hat{\rho}_{k-1}$, which
now is written in the many-body basis $\vert i\rangle_{(k-1)}$ which
includes all sites up to and including site $k-1$. This
\textit{backward} update is a well-known operation within the NRG.
\cite{Hofstetter00,Wb07,Toth08,Wb11_rho} Its graphical depiction is
given in \Fig{fig:rhoupdate} [note that the sum over $i$ and $i'$ in
\Eq{def:backprop} connects to state spaces that are not yet
contracted; hence these correspond to open indices in
\Fig{fig:rhoupdate}{}].

The backward update of the density matrix in \Eq{def:backprop}
preserves its properties as a density matrix and as a scalar
operator. The former directly follows from the realization that the
orthonormality condition \Eq{eq:Aortho} with the \Atensor in the
last line of \Eq{def:backprop} is exactly equivalent to a complete
positive map.
Moreover, by tracing out part of a system such as a site that has
been added through a tensor product space and that itself can be
fully categorized using given symmetries, this procedure cannot break
symmetries by itself. This is to say, that the partial trace in
\Eq{def:backprop} preserves the property of a scalar operator.
However, the trace over CGC spaces adds important weight factors to
the reduced multiplet spaces, which are crucial, for example, to
preserve the overall trace of the density matrix during
back-propagation. While the contraction in \Fig{fig:rhoupdate} can be
easily performed, in practice, without the explicit knowledge of
these weights, their determination is straightforward and
instructive, nevertheless, as will be shown in the following.

The contraction in \Fig{fig:rhoupdate} clearly also holds for the CGC
spaces of every symmetry individually. Therefore it is sufficient to
focus on one specific symmetry. Let $i$ contain several multiplets
$q_i$, and consider, for simplicity, the special case where the local
state space $\sigma$ contains one specific multiplet $q_\sigma$ only.
In addition, also the reduced density matrix $\hat{\rho}_k$ is chosen
such that it only picks one very specific multiplet $q_j$. Focusing
on the Clebsch Gordan coefficients $C_{q_{i z} q_{j z}}^{[q_{\sigma
z}]} \equiv \langle q_i q_{i z}; q_\sigma q_{\sigma z} \vert q_j q_{j
z} \rangle$ for chosen symmetry then, which properly combine the
irreducible multiplets $q_{i}$ and $q_{\sigma}$ into the multiplet
$q_{j}$, the contraction in \Eq{def:backprop} with respect to the
fixed $q_{j}$ is given by
\begin{align}
& \sum_{q_{\sigma z^{{}}} q_{j z} q_{j'z}}
   \langle q_{i'} q_{i'z}; q_{\sigma} q_{\sigma z} \vert q_{j} q_{j'z} \rangle
   \langle q_{i } q_{i z}; q_{\sigma} q_{\sigma z} \vert q_{j} q_{j z} \rangle^{\ast}
   \!\cdot \delta_{q_{jz} q_{j'z}} \nonumber \\
&\quad = \sum_{q_{\sigma z}}
   \langle q_{i'} q_{i'z}; q_{\sigma} q_{\sigma z}\vert
   \Bigl(\sum_{q_{j z}} \vert q_{j} q_{j z} \rangle\langle q_{j} q_{j z} \vert\Bigr)
   \vert q_{i} q_{i z}; q_{\sigma} q_{\sigma z} \rangle
\nonumber\\
&\quad = f_{q_i q_{j}} \cdot \delta_{q_{i\!\phantom'} q_{i'}}
 \delta_{q_{i z\!\phantom'\!} q_{i'z}}
\text{,} \label{eq:rho-scalar}
\end{align}
where the $\delta_{q_{j z} q_{j'z}}$ in the first line comes from the
assumption that the initial $\hat{\rho}_k$ is a scalar. The last
identity follows from the fact that also $\hat{\rho}_{k-1}$ shall be
a scalar operator. Alternatively, the last equality can also be
understood as a general intrinsic completeness property of
Clebsch-Gordan coefficients. Either way, the remaining factor $f_{q_i
q_{j}}$ in the last line must be independent of the z-labels. The
factor $f_{q_i q_{j}}$ then, in a sense, reflects the weight of how
the \IREP $q_i$ together with the traced over \IREPs $q_\sigma$
contributes to the final total $q_j$. If, for example, for fixed
$q_i$ and the known set of $q_\sigma$ some final total $q_j$ cannot
be reached, then it holds $f_{q_i q_{j}}=0$ for this case.

From the scalar property of $\hat{\rho}_{k-1}$, \Eq{eq:rho-scalar}
can be further constrained to some specific $q_i=q_{i'}$. Also
summing over $q_{iz}=q_{i'z}$ then, the second line in
\Eq{eq:rho-scalar} becomes equal to $\trace \bigl(\sum_{q_{j z}}
\vert q_{j} q_{j z} \rangle\langle q_{j} q_{j z} \vert\bigr) =
d_{q_{j}}$, \ie the internal multiplet dimension of the \IREP
$q_{j}$. Together with the last line in \Eq{eq:rho-scalar}, it
follows,
\begin{equation}
   f_{q_i q_j} = \tfrac{d_{q_j}}{d_{q_i}}
\text{.}\label{rhoupd:fac}%
\end{equation}
as demonstrated, for example, for \SU{2} in [\onlinecite{Toth08}].
Note that \Eq{rhoupd:fac} holds in general for arbitrary symmetries,
and also in the presence of outer multiplicity. This follows by
recalling that one of the main assumptions that entered
\Eq{eq:rho-scalar} was to pick one specific multiplet $q_{j}$. This
single \IREP, however, may equally well also have been any of the
multiplets resulting from outer multiplicity, say multiplet $q_j\to
q_{j,\alpha}$, which nevertheless again leads to \Eq{rhoupd:fac}.

\begin{figure}[t!]
\begin{center}
\includegraphics[width=1\linewidth]{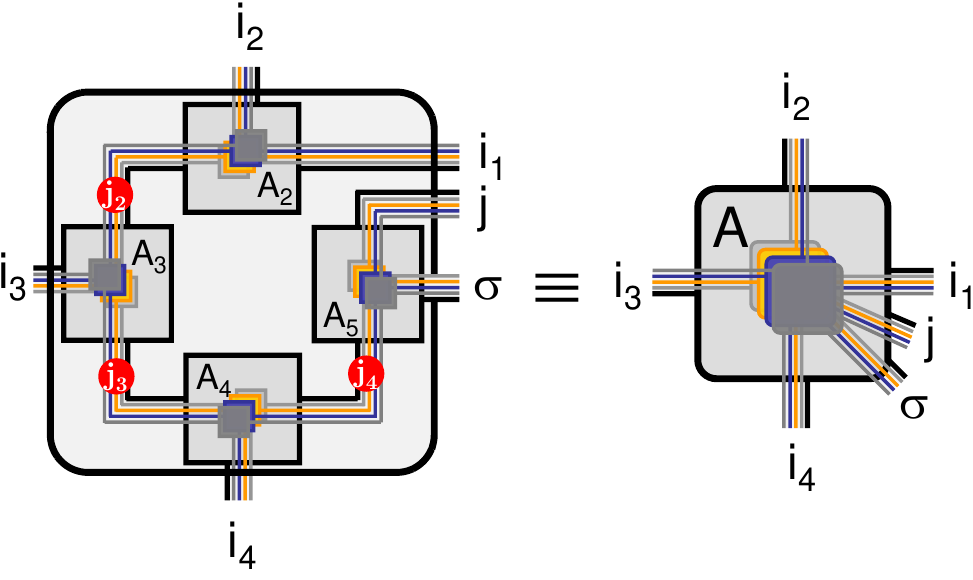}
\end{center}
\caption{(Color online)
  Generalized \Atensor that combines multiple state spaces, \ie four
  effective state spaces $\vert i_1\rangle, \ldots,\vert i_4 \rangle$
  together with one local degree of freedom $\vert \sigma\rangle$. Here
  it is assumed that all input state spaces describe proper orthonormal
  state spaces that act in different spaces, such that they can be
  combined into a simple product space. The index $j$, finally,
  represents the common global state space. In particular, it can be used
  to truncate the global Hilbert space to the state space of interest
  (within the DMRG, this may simply be the ground state, where the
  index $j$, being a singleton dimension, simply may be skipped then).
  While the general Clebsch-Gordan coefficients for the entire object
  may not be easily available (object to the right), the overall
  \Atensor can be built iteratively by adding one state space at a time
  (object to the left), starting, say, from $A_1$ which links the two
  state spaces $i_1$ and $i_2$ into the combined state space $j_2$ and
  hence allows to employ Clebsch-Gordan coefficients in the usual
  manner. The state space $j_2$ can then be combined with state spaces
  $i_3$, and so forth. Contraction of the intermediate indices
  $j_2,\ldots,j_4$, finally, leads to the generalized \Atensor to the
  right.
}\label{fig:Amult}%
\end{figure}

\section{ Implications for DMRG and beyond }

This section sketches strategies for using non-abelian symmetries in
the traditional DMRG \cite{White92,Stoudenmire12} with
generalizations to more general tensor networks. While the suggested
procedures eventually may be further optimized still, nevertheless,
they demonstrate the versatility of the presented \QSpace framework.
A particularly useful object in this context is the identity
\Atensor that was already introduced in \Sec{Atensor:Operators}.

\subsection{Generalized \Atensor for tensor networks}

The prototypical \Atensor as defined in \Eq{def:Atensor} combines
two physically distinct state spaces in terms of their
tensor-product space. One may be interested, however, in the case
where three or more state spaces need to be combined in the
description of a single combined state space, while nevertheless
also respecting symmetries. This situation, for example, occurs
regularly in the context of tree \cite{Shi06,Murg10} or tensor
network states
\cite{Vidal07_MERA,Sandvik07,Murg07,Cirac09,Singh10}. Let $m$
be the number of states spaces to be combined. Then this requires
the generalized Clebsch-Gordan coefficients $\langle q_{1} q_{1z};
\ldots; q_{m} q_{m,z} \vert q q_{z} \rangle$. Once known, in
principle they can be combined compactly into a generalized \Atensor
of rank $m+1$. The question is, how to obtain such a generalized
\Atensor in a simple manner, in practice.

For this, the \QSpace structure introduced in this paper proves very
useful. In particular, a generalized \Atensor can be obtained based
on the \emph{iterative pairwise addition} of individual state
spaces, which is a well-established procedure at every step. The
situation is depicted schematically in \Fig{fig:Amult}. To be
specific, \Fig{fig:Amult} considers four effective state spaces
$\vert i_\alpha \rangle \equiv \vert q_\alpha q_{\alpha,z}\rangle$
with $\alpha = 1,\ldots,4$, together with a local state space $\vert
\sigma\rangle \equiv \vert q_5 q_{5,z}\rangle$, thus having $m=5$.
This specific setting may correspond, for example, to the situation
in a tensor network state that describes a two-dimensional system
which, from the point of view of a specific site with state space
$\sigma$, has four effective states spaces to the top, bottom, left,
and right, respectively. Note, however, that here at least in
principle the state spaces $\vert i_\alpha\rangle$ with $\alpha =
1,\ldots,m$ are assumed to be \emph{physically different,
orthonormal} state spaces, such that their tensor-product space is a
well-defined meaningful Hilbert space. Starting with state spaces
$\vert i_1\rangle$ and $\vert i_2\rangle$ in \Fig{fig:Amult}, their
state space can be combined in terms of and identity \Atensor
$A_2^{\mathrm{Id}}$ in the usual fashion using standard
Clebsch-Gordon coefficients. The resulting state space $\vert
j_2\rangle$ then can be combined with state space $\vert i_3
\rangle$ using another identity \Atensor $A_3^{\mathrm{Id}}$, thus
obtaining $\vert j_3\rangle$. The procedure is repeated, for
example, until at the last step the local state space $\vert
\sigma\rangle$ is added, resulting in the full combined state space
$\vert j\rangle$, properly categorized in terms of symmetries. The
iteratively generated $m-1$ identity \Atensors $A_k^{\mathrm{Id}}$,
on the other hand, can be contracted into a single tensor of rank
$m+1$ by contracting the intermediate indices
$j_{2},\ldots,j_{m-1}$. This then results in the desired generalized
\Atensor, shown at the \rhs of \Fig{fig:Amult}.
Furthermore, in the context of DMRG or tensor network states, one is
typically interested in a single state, such as the ground state of
the system. In this case, the full combined state space $\vert
j\rangle$ is truncated to a single state. Thus the index $\vert
j\rangle$ becomes a singleton and as such can be dropped, for
simplicity. In general, by explicitly including the CGC spaces in
the \QSpace in \Eq{eq:QSpace:Rep}, generalized Clebsch-Gordan
coefficients can be easily obtained in terms of a generalized
\Atensor, which itself is constructed through a transparent
iterative procedure.

\subsection{Two-site treatment \label{sec:2site:treatment}}

\begin{figure}[t]
\begin{center}
\includegraphics[width=\linewidth]{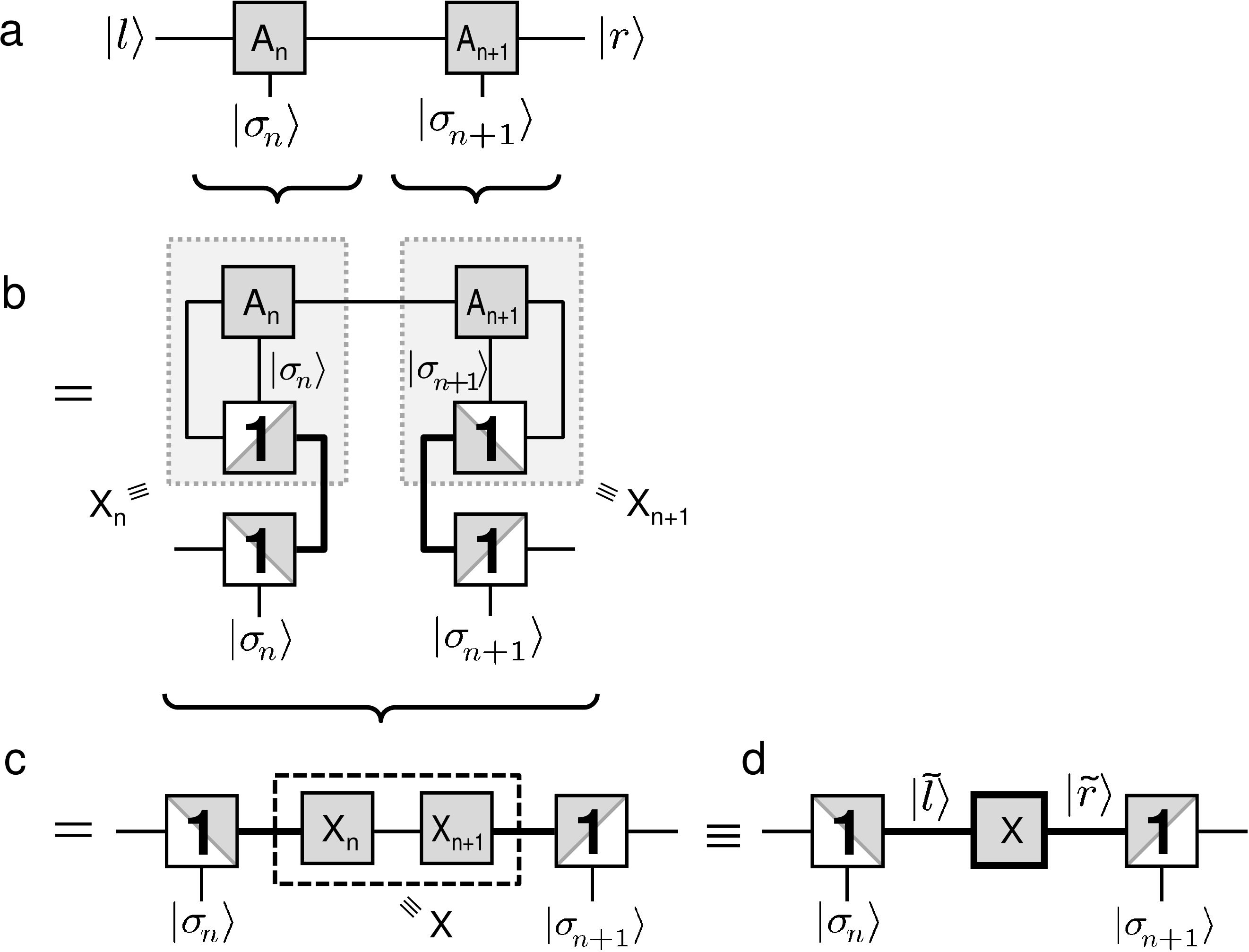}
\end{center}
\caption{
  DMRG treatment of a two-site setup using identity \Atensors. The
  internal CGC structure [\cf \Fig{fig:qspace}{}] is hidden in given
  case, for simplicity. Panel (a) Generic setup with orthonormalized
  state spaces for the left and right block of the system (open
  indices left and right), while explicitly considering the pair of
  intermediate sites $n$ and $n+1$. Panel (b) Insertion of an
  identity, \ie \emph{twice} the unitary identity \Atensor, for site $n$ and
  the left block (for site $n+1$ and the right block) allows to fuse
  the local state spaces with their respective environments.
  Contracting the \QSpaces $X_n$ and $X_{n+1}$ into $X \equiv X_n
  \cdot X_{n+1}$ (panel c), the setup in panel (d) is obtained.
  Overall, this allows to treat the more complex rank-4 two-site setup
  in panel (a) in terms of an intermediate rank-2 \QSpace $X$
  in panel (d) which is connected to two enlarged (fused) effective
  orthonormal state spaces $\tilde{l}$ and $\tilde{r}$
  (indicated by thick lines).
}\label{fig:A2toX}%
\end{figure}

A strategy for the treatment of a two-site setup common to the DMRG
is sketched in \Fig{fig:A2toX}. For this, consider the generic setup
of two adjacent sites $n$ and $n+1$ within an MPS setup with local
state spaces $\vert \sigma_n\rangle$ and $\vert\sigma_{n+1}\rangle$,
respectively (panel a). The state spaces $\vert l\rangle$ to the
left ($n'<n$) and $\vert r\rangle$ to the right ($n'>n+1$) are
assumed to be orthonormal and written in terms of proper multiplet
spaces. Using symmetries, this two-site configuration is considered
inefficient, however, since the local description of the Hamiltonian
fractures into many contributions. Therefore from a practical point
of view, it turns out advantageous even already on the level of
plain abelian symmetries, to transform the rank-4 two-site setup
[\cf \Figp{fig:A2toX}{b}] to an intermediate rank-2
bond-configuration [\cf \Figp{fig:A2toX}{d}]. Using identity
\Atensors, this can be done exactly even in the presence of complex
non-abelian symmetries.

In order to simplify the description of the two site setup, the left
state space and the local state space $\vert \sigma_n\rangle$ are
linked through an identity \Atensor into the combined
\emph{non-truncated} multiplet spaces $\vert s_n\rangle$. This
mapping which respects symmetries, corresponds to a unitary
transformation $U$. Therefore inserting $UU^\dagger = 1$, the
identity \Atensor needs to be inserted twice, as indicated to the
left of \Figp{fig:A2toX}{b}. Here the identity \Atensor is drawn
such that the two input spaces connect to the white triangle, while
the gray triangle solely links to the output space. This specific
depiction serves to emphasize the underlying \emph{unitary} mapping
in terms of CGCs from one basis to another. Nevertheless, for
simplicity, on the level of reduced matrix elements, \ie the
multiplet space, an identity matrix is maintained (\cf
\Sec{Atensor:Operators}{}). The original tensor $A_n$ can be
contracted now with the upper identity \Atensor in
\Figp{fig:A2toX}{b}, leading to the rank-2 \QSpace $X_n$. Exactly
the same treatment can be repeated for the right part of the system:
site $n+1$ is combined with the state space $\vert r\rangle$ for the
sites ($n'>n+1$) through their own identity \Atensor. The latter is
again inserted twice, and after contraction this leads to \QSpace
$X_{n+1}$. The two \QSpaces $X_n$ and $X_{n+1}$, finally, are
contracted into $X \equiv X_n \cdot X_{n+1}$ (panel c).

As seen in panel (d), the configuration resulting from this
transformation is such that sites $n$ and $n+1$ are now fully fused
without truncation through identity \Atensors with the left and
right part of the system, respectively. The original wave function,
on the other hand, is exactly encoded in the intermediate rank-2
\QSpace $X$. The enlarged tensor-product state-spaces [indicated by
thick lines in panels (b-d)] eventually connects to \QSpace $X$ in
panel (d). The wave-function encoded in \QSpace $X$ can be updated
then in the usual DMRG spirit, after rewriting all operators
relevant for the Hamiltonian within this local bond-configuration.
The resulting improved $\tilde{X}$ can be truncated then, followed
by an exact shift of the focus from sites $n$ and $n+1$ to the next
pair of sites, \eg $n+1$ and $n+2$. The last two steps are explained
in some more detail next.

\subsection{State space truncation}

\begin{figure}[t]
\begin{center}
\includegraphics[width=.65\linewidth]{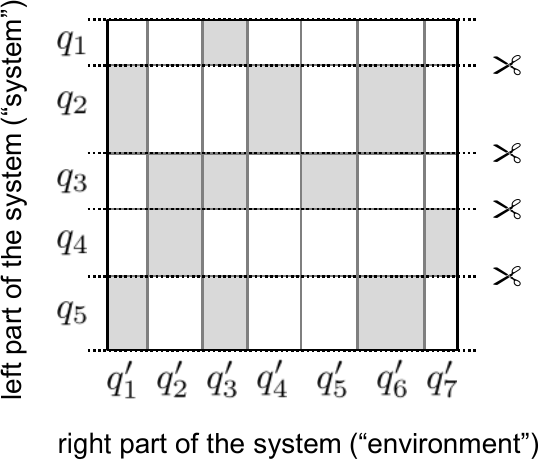}
\end{center}
\caption{
  State space truncation given the coefficient matrix $X$ in
  \Eq{eq:Psi:X} [\cf \Figp{fig:spectralSUN}{d}]. Here $X$ is
  schematically depicted in \emph{multiplet space} in terms of groups
  (blocks) of multiplets for both, the effective state space $\vert
  l\rangle$ for the left part (rows), as well as the effective state
  space $\vert r\rangle$ for the right part (columns) of the physical
  system analyzed. Blocks shaded in gray are considered non-zero,
  whereas white blocks are considered all-zero. For the purpose of
  truncation, it is sufficient to group the multiplets $q_i$ for the
  left part of the system, while arranging $(q'_i,q_\psi)$ as columns
  in multiplet space, and perform SVD for each block of rows for a
  specific $q_i$ in multiplet space only.
}\label{fig:DMRG:truncation}%
\end{figure}

Consider a wave function $\vert \psi \rangle$  written in the
effective local configuration of \Figp{fig:A2toX}{d},
\begin{equation}
   \vert \psi \rangle = \sum_{l,r} X_{lr} \vert l\rangle \vert r\rangle
\text{,}\label{eq:Psi:X}%
\end{equation}
having skipped the tildes, \ie $\vert \tilde{l}\rangle \to \vert
l\rangle$, and suppressing symmetry labels, for simplicity. Assume
this wave function $\vert \psi \rangle$ has a well defined global
symmetry described by the set of labels $q_\psi$. Now, both state
spaces, $\vert l\rangle$ as well as $\vert r\rangle$, represent
multiplet spaces that are grouped into blocks of states that belong
to the same symmetry multiplets. This is depicted schematically in
\Fig{fig:DMRG:truncation} for the matrix $X$ in multiplet space.
There white blocks are considered all-zero, while blocks shaded in
gray are considered non-zero. For a shaded block therefore, by
definition, its product space of the symmetries $q_i$ in $\vert
l\rangle$ (rows) and $q'_i$ in $\vert r\rangle$ (columns)
\emph{must} allow $q_\psi$ as a valid global multiplet.
Given the labeling in terms of multiplet labels $q_\psi$, moreover,
it is convenient to consider a \emph{full} single multiplet $\vert
\psi\rangle$, \emph{rather than picking a specific state} from the
internal space of multiplet $q_\psi$. Consequently, while the
coefficient space $X$ corresponds to a matrix, \ie a rank-2 object
in the multiplets $l$ and $r$ as depicted in
\Fig{fig:DMRG:truncation}, overall it is natural to consider the
\QSpace $X$ to have rank-3. For a single multiplet $q_\psi$, this
only affects the CGCs, while the multiplet space acquires a
singleton dimension.

One may trace out the right part of the system (together with the
third index regarding the CGC space of $\vert \psi \rangle$),
equivalent to calculating \mbox{$X\cdot X^\dagger$} in terms of
\QSpaces. Given a well defined global symmetry for $\vert \psi
\rangle$, the resulting reduced density matrix, is a scalar
operator, \ie block-diagonal in the multiplet spaces $q_i$,
corresponding to the blocks of rows in \Fig{fig:DMRG:truncation}.
From this, it follows that the eigenvalues and eigenvectors of the
reduced density matrix can be computed \emph{independently and thus
separately} for each block of rows to the same label $q_i$. When
calculating the reduced density matrix above, on the level of CGCs,
this corresponds to the situation already discussed in
\Sec{RHO:backward} in that the CGC spaces contract to identities.
Note, however, that the labels of the indices (state spaces) in
\Fig{fig:rhoupdate} acquire a somewhat altered interpretation here,
\ie $(i,\sigma)\to j$ becomes $(l,r)\to\psi$.
It follows from \Eq{rhoupd:fac} then, that the corresponding weight
factors for the reduced density matrix, that originate from the CGC
spaces, are \emph{completely independent} of the internal
dimensionality $d_{q'_i}$ of the multiplets $q'_i$ that represent
the multiplet space $\vert r\rangle$. Therefore for the purpose of
truncation, instead of explicitly calculating the reduced density
matrix, it is equally sufficient to use standard SVD decomposition
on the non-zero blocks in a row for a specific $q_i$ in
\Fig{fig:DMRG:truncation}. That is, truncation can be performed
fully on the level of multiplets only, temporarily putting aside the
CGC spaces. Consequently, SVD allows to rewrite the \QSpace
$X=U\cdot(SV^\dagger)$, where $U$ is a ``scalar operator''. That is,
all CGC spaces of $U$ are identities, and the newly generated
intermediate index inherits the symmetry labels of the multiplets
$l$, \ie is block-diagonal in $q_i$. In order to proceed to the next
DMRG iteration then, say sites $n+1$ and $n+2$, the resulting
\emph{truncated} \QSpace $\tilde{R}$ from $\tilde{X} =
U\cdot(\tilde{S}V^\dagger) \equiv U\tilde{R}$ can simply be
contracted onto $A_{n+1}$, whereas $U$ is contracted onto $A_n$.
Overall, this allows to truncate within properly orthonormalized
state spaces in the presence of arbitrary non-abelian symmetries,
which thus again reduces the dimension on the bond between sites $n$
and $n+1$.

Note furthermore, that the constraint to a single wave function with
well-defined multiplet label $q_\psi$ can be significantly relaxed.
It was already argued above, that it is convenient to consider the
\QSpace $X$ of rank-3, which thus keeps all states that constitute
the single multiplet $q_\psi$. However, this directly opens the door
towards the \emph{simultaneous} simulation of \emph{several}
multiplets $\{ \psi_k \}$ with possibly different multiplet spaces
$\{q_{\psi_k}\}$. Clearly, if each individual state $\psi_k$ belongs
to a well-defined overall symmetry multiplet, then the reduced
density matrix built from the scalar operator $\rho = \sum_k \vert
\psi_k\rangle\langle \psi_k\vert$ will still be block-diagonal in
the symmetry spaces. Therefore SVD can still be performed for every
individual block of rows $q_i$, while fusing the multiplet spaces
for $\psi$ with the multiplet spaces for the right part of the
system, \ie $q'_i \to (q'_i,q_{\psi})$.

\subsection{Wave function prediction}

Following the two-site update depicted in \Fig{fig:A2toX} above
together with subsequent truncation on the multiplet level, the
local description of the wave function $\psi$ can be carried over
exactly to the next bond: using the respective identity \Atensors
for sites $n+1$ towards the left and $n+2$ towards the right part of
the system, this then allows to switch to the bond-configuration
between sites $n+1$ and $n+2$, exactly as already discussed in
\Sec{sec:2site:treatment}.

\section{Correlations functions}

Correlations functions are usually calculated with respect to
operators whose transformation under given symmetries is known. This
is specifically so, as these operators often naturally derive from
the same fundamental building blocks, that also enter the
Hamiltonian. In the presence of non-abelian symmetries, a single
specific operator then that is not a scalar operator, is usually part
of a larger irreducible operator set. In practice, thus also its
correlation function is calculated \wrt the \emph{full} \IROP, for
simplicity, as will be explained in the following.

Consider, for example, the retarded Green's function
\begin{equation}
   G_{\sigma}(\omega) \equiv \langle d_{\sigma}^{\phantom\dagger}
   \Vert d_{\sigma}^{\dagger}\rangle_{\omega}
\label{G-specific}%
\end{equation}
that, in the time domain, creates a particle of preserved flavor
$\sigma$, and destroys it some time later. Clearly, a particle with
the same flavor must be destroyed later, otherwise the Green's
function is zero, \ie the Green's function is diagonal with respect to
symmetries. Now in the presence of symmetries, it must be possible to
write the operators $d_{\sigma}^{\dagger}$ as part of an irreducible
operator set, \eg some spinor (\IROP{}) $\hat{\psi}^q$ that
transforms according to \IREP $q$ with internal dimension $d_q$ (in
the case of plain abelian symmetries, it typically holds $d_q=1$, \ie
the operator $d_{\sigma}^{\dagger}$ is the only member of the
\IROP{}). Thus the calculation of the very specific correlation
function with respect to specific elements $d_{\sigma}$ and
$d_{\sigma} ^{\dagger}$ above can replaced by the Green's function
\begin{equation}
   G_{\psi^q}(\omega) \equiv
   \langle(\hat{\psi}^q)^{\dagger} \Vert \hat{\psi}^q\rangle_{\omega}
\text{.}\label{G-irrep}%
\end{equation}
To be clear, if $d_q>1$, this includes the scalar product of the
spinor components, and thus one is actually calculating the
\textit{same} Green's functions as in \Eq{G-specific} $d_q$ times,
\begin{align}
   G_{\psi^q}(\omega)
&= \sum_{q_z=1}^{d_q} \langle (\hat{\psi}_{q_z}^q)^{\dagger} \Vert
   \hat{\psi}_{q_z}^{q} \rangle_{\omega}
 = d_q \cdot \langle d_{(\sigma)}^{\ }\Vert d_{(\sigma)}^{\dagger
   }\rangle_{\omega} \nonumber \\
 & \Rightarrow G_{(\sigma)}(\omega) = \tfrac{1}{d_q} G_{\psi^q}(\omega)
\text{.}\label{eq:using-Gpsi}%
\end{align}
with $G_{(\sigma)}$ independent of $\sigma$ within its multiplet, as
implied by the round brackets. This apparent overhead, however, only
affects the CGC space, so this is negligible numerical overhead, yet
makes the calculation conceptually simple. Specifically, when
calculating matrix elements and their contribution to the Green's
function, eventually all indices can be fully contracted, so there is
no need for a special treatment of a specific z-label that represents
a peculiar $d_{\sigma}^{\dagger}$. Moreover, given the discussion of
scalar operators in \Sec{scalar-ops} earlier, one realizes that the
scalar product $\hat{\psi}^{\dagger} \cdot \hat{\psi}$ of the \IROP
$\hat{\psi}^q$ yields a scalar operator.

In the following, two explicit prototypical examples for correlation
functions are given that are used explicitly for the numerical
results presented in this paper. The first example is the spin-spin
correlation function or magnetic susceptibility $\chi_{d}\left(
\omega\right) $ defined at some site $d$. In the presence of spin
\SU{2} symmetry,
\begin{align}
   \chi_{d}(\omega)
&= \langle S_{x,d}\Vert S_{x,d}\rangle_\omega
 = \langle S_{y,d}\Vert S_{y,d}\rangle_\omega
 = \langle S_{z,d}\Vert S_{z,d}\rangle_\omega \nonumber \\
 & \equiv \tfrac{1}{3}
   \bigl\langle \mathbf{\hat{S}}_{d} \Vert\mathbf{\hat{S}}_{d}
   \bigr\rangle_\omega
\text{.}\label{eq:chi-SUN}%
\end{align}
Clearly, the local operator $\hat{S}_{d}^{2} \equiv
\mathbf{\hat{S}}_{d} \cdot\mathbf{\hat{S}}_{d}$ is a scalar operator,
with the corresponding spinor $\mathbf{\hat {S}}_{(d)}$ given by [\cf
App. \Eq{spin-irop}{}]
\[
  \mathbf{\hat{S}}\equiv%
  \begin{pmatrix}
     -\tfrac{1}{\sqrt{2}} \hat{S}_{+} \\
                          \hat{S}_{z} \\
     +\tfrac{1}{\sqrt{2}} \hat{S}_{-}
  \end{pmatrix}
\text{.}%
\]
The second example is the spectral function for a single spinful
channel in the presence of spin and particle-hole \SU{2} symmetry.
The spinor is given by [\cf App. \Eqp{eq:Psi:CS4}{}],
\[
   \hat{\psi}\equiv%
   \begin{pmatrix}
      s\hat{c}_{\uparrow}^{\dagger} \\
       \hat{c}_{\downarrow}\\
      s\hat{c}_{\downarrow}^{\dagger}\\
      -\hat{c}_{\uparrow}
   \end{pmatrix}
\text{.}%
\]
In the evaluation of the correlation function,
\begin{align*}
   \langle\hat{\psi}^{\dagger} \Vert \hat{\psi}\rangle_{\omega}
&= \sum_{q_z=1}^{d_q=4}
   \langle\hat{\psi}_{q_z}^{\dagger} \Vert \hat{\psi}_{q_z}^{\phantom\dagger}
   \rangle_{\omega} \\
&= \langle\hat{c}_{\uparrow}^{\phantom\dagger} \Vert \hat{c}_{\uparrow}^{\dagger}\rangle_{\omega}
 + \langle\hat{c}_{\downarrow}^{\dagger} \Vert \hat{c}_{\downarrow}\rangle_{\omega}
 + \langle\hat{c}_{\downarrow}^{\phantom\dagger} \Vert \hat{c}_{\downarrow}^{\dagger}\rangle_{\omega}
 + \langle\hat{c}_{\uparrow}^{\dagger} \Vert\hat{c}_{\uparrow}^{\phantom\dagger}\rangle_{\omega}
\end{align*}
the signs, including $s\equiv\pm1$, of the individual components are
irrelevant. Given the spin symmetry and the fact, that in the
presence of particle-hole symmetry spectral functions are symmetric
with respect to $\omega=0$, and in general $G_{B,B^{\dagger}}
(\omega) \equiv \langle B\Vert B^{\dagger} \rangle_{\omega} =
G_{B^{\dagger}, B}(-\omega)$, it follows that all four contributions
above describe exactly the same function, indeed, and therefore
\[
   G_{\sigma}(\omega) \equiv
   \langle\hat{c}_{\sigma}^{\phantom\dagger} \Vert \hat{c}_{\sigma}^{\dagger}\rangle_{\omega}
 = \tfrac{1}{4}\langle\hat{\psi}^{\dagger} \Vert \hat{\psi}\rangle_{\omega}
\text{.}%
\]

\section{The numerical renormalization group}

The non-abelian setup described above is straightforwardly applicable
to the NRG. \cite{Wilson75,Bulla08} Before doing so in detail, here a
brief reminder of the essentials of NRG is given, followed by the
introduction of the model Hamiltonian to be analyzed. By
construction, the NRG deals with so-called quantum impurity models --
an arbitrary small quantum system (the \emph{impurity}) that is in
contact with a macroscopic non-interacting typically fermionic bath.
Each part is simple to solve exactly on its own. The combination of
both, specifically in presence of interactions at the location of the
impurity, however, gives rise to strongly-correlated
quantum-many-body effects.

The systematic approach introduced by Wilson \cite{Wilson75} was a
logarithmic discretization in energy space of the continuum of the
bath (coarse graining), followed by an exact mapping onto a
semi-infinite so-called Wilson-chain, with the intact impurity space
coupled only to the very first site of this chain. Given the
half-bandwidth $W:=1$ of the bath, the discretization parameter
$\Lambda>1$, typically $\Lambda\gtrsim 1.7$, defines the logarithmic
discretization in terms of the intervals $\pm [\Lambda^{-m},
\Lambda^{-(m+1)}]$ with $m\ge0$ an integer, and energies taken
relative to the Fermi energy $\Ef \equiv 0$. Each of these intervals
is then described by a single effective fermionic state, with its
coupling and exact energy position chosen consistently \wrt the
hybridization of the original continuum model.
\cite{Yoshida90,Zitko09} The resulting discretized model is then
mapped onto the semi-infinite Wilson-chain (Lanczos
tridiagonalization).\cite{Demmel00} Hereby, the logarithmic
discretization of the non-interacting bath translates to an effective
tight-binding chain, with the hopping $t_{k}\sim\Lambda^{-k/2}$
between sites $k$ and $k+1$, decaying \textit{exponentially} in the
discretization parameter $\Lambda$. The latter then justifies the
essential renormalization group ansatz of the NRG in terms of
\textit{energy scale separation} -- large energies are considered
first, with approximate eigenstates at large energies
\textit{discarded} and considered unimportant for the description of
the still following lower energy scales. Thus each site of the Wilson
chain corresponds to an energy shell with a characteristic energy
scale $\omega_{k}\equiv \tfrac{a}{2}(\Lambda+1) \Lambda^{-k/2}$. Here
the constant $a$ of order 1. is chosen such, $t_{k-1}/\omega_k\to 1$ for
large $k$.\cite{Wb11_rho}

In practice, when considering the system up to site $k$, the
Hamiltonian of the rest of the system is ignored, equivalent to
assuming degeneracy in the state space of the remainder of the
system. With $\hat{H}_{k}$ the full Hamiltonian $\hat{H}$ including
the Wilson chain up to site $k$, its eigenstates $\vert
s\rangle_{k}$, $\hat{H}_{k} \vert s\rangle_{k}=E_{s}^{k} \vert
s\rangle_{k}$, and with $\vert e\rangle_{k}$ an arbitrary state of
the rest of the system following site $k$, then the essential spirit
of NRG after coarse graining of the bath can be condensed in the
following approximation,\cite{Anders05}
\begin{equation}
   H \vert se\rangle_{k} \simeq E_{s}^{k} \vert se\rangle_{k}
\text{,} \label{NRG-approx}%
\end{equation}
expressing energy scale separation, with $\vert se\rangle_{k} \equiv
\vert s\rangle_{k} \otimes \vert e\rangle_{k}$. The energies
$E_{s}^{k}$ are usually taken relative to the ground state energy
$E_{0}^{k}$ of iteration $k$, and rescaled by the energy scale
$\omega_k$. All of this will be referred to as rescaled energies, and
has the advantage that independent of the Wilson shell $k$, energies
are always of order $1$.

In this paper, the state space truncation at a given NRG iteration is
energy-based, \ie all states with $E_{s}^{k} \leq \EK$ are kept,
typically with $\EK \simeq 5 \ldots 7$ in rescaled energies. The
number of kept states \NK thus changes dynamically.
\cite{Bulla08,Zitko09,Wb11_rho}

\subsection{Full density matrix}

Within the NRG, \cite{Wilson75} a complete many-body basis set can be
formulated from the state space discarded at every iteration.
\cite{Anders05} Initially introduced for explicit time-dependence of
quantum quenches, they actually can also be used to improve on
existing calculations for thermodynamical quantities and expectation
values, \cite{Peters06} with a clean extension to arbitrary
temperatures using the full density matrix (FDM). \cite{Wb07} The
density matrix $\hat{\rho}\equiv e^{-\beta\hat {H}}/Z$ with
$\beta=1/k_{\mathrm{B}}T$, $k_{\mathrm{B}}$ the Boltzmann constant
and $T$ the temperature, obviously commutes with the Hamiltonian and
is a scalar operator in itself. Within the FDM-NRG approach,
\cite{Wb07} the density matrix
\begin{align}
   \hat{\rho} &=\tfrac{1}{Z} \sum_{k; qn;q_{z};e}e^{-\beta E_{qn}^k}
   \vert qn;q_{z};e \rangle_k{}\,_k\!\langle qn;q_{z};e \vert \text{,}%
\label{eq:rhoT}
\end{align}
can be constructed straightforwardly in terms of a \QSpace for every
Wilson shell $k$. Here $s\equiv (qn)$ stands for the multiplet label
for a given shell $k$. Note that the symmetry of the states $e$ is
irrelevant here, as this space is fully traced over. Given the usual
practice of NRG to rescale and shift energies at every iteration, all
of this, of course, must be undone before entering \Eq{eq:rhoT}
[given a general thermal density matrix, of course, \emph{all}
energies in \Eq{eq:rhoT} must be (i) at the same energy scale, \ie
\emph{non}-rescaled, and (ii) specified with respect to a
\emph{common} energy reference, \eg the overall ground state energy
of the Wilson chain]. \cite{Wb07}

By construction, all eigenenergies $E_{qn}^k$ are degenerate, \ie do
not depend on the z-labels. With the reduced density matrix being a
scalar operator, therefore the CGC spaces in the \QSpaces describing
\Eq{eq:rhoT} are all proportional to identity matrices, leading to
the overall normalization
\begin{equation}
  Z = \sum_{k;qn} d_{q} d_w^{N-k} e^{-\beta E_{qn}^k}
\text{,}\label{eq:Z:fdm}
\end{equation}
where $d_{q}$ is the internal dimension of multiplet $q$, and
$d_s^{N-k}$ reflects the degeneracy \wrt the rest of the Wilson chain
of final length $N$, with $d_w$ the state space dimension of a Wilson
site. \cite{Wb07}

\begin{figure}[t]
\begin{center}
\includegraphics[width=.9\linewidth]{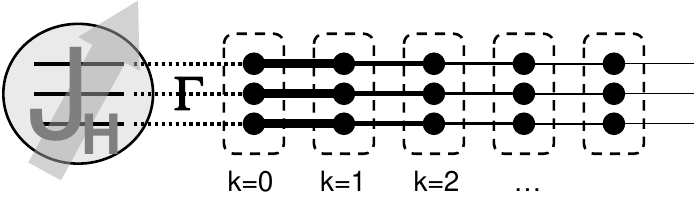}
\end{center}
\caption{
Schematic depiction of the fully screened Kondo-Anderson hybrid model
[\Eq{eq:HKondoAM} with $m=3$] in the NRG setup of a Wilson chain.
Three d-levels with onsite Hund's interaction of strength \JH couple
uniformly to their respective channel with hybridization $\Gamma$.
The semi-infinite Wilson chain for each channel represents a
tight-binding chain with exponentially decaying couplings, that
interacts with the other channels through the impurity only. For a
given NRG iteration, all terms in the Hamiltonian of the same energy
scale must be included simultaneously, leading to an extended Wilson
site [dashed boxes] of three spinfull fermionic levels with a state
space of $3^{4}=64$ states each.
}\label{fig:kondoAM}%
\end{figure}

\subsection{Model: symmetric three-channel system}

The historically first physical system where Kondo physics was
observed was that of \emph{Fe} impurities in \emph{Au}.
\cite{deHaas34,Kondo64} The effective microscopic model for this
material, however, is far from trivial. It was argued only very
recently in an extended study \cite{Costi09} that the physics of the
five d-orbitals of substitutional \emph{Fe} in \emph{Ag} or \emph{Au}
is dominated by 3-fold degenerate triplet space $t_{2g}$, with the
doublet space $e_{g}$ split-off by crystal fields and thus playing a
minor role. Together with the effective spin $3/2$ of the iron
impurity, this then results in an \SU{3} symmetric fully screened
3-channel Kondo model.

The actual model analyzed\cite{Costi09} is depicted schematically in
\Fig{fig:kondoAM}. It consists of $m=3$ spinful d-levels comprising
the impurity, that are interacting through the Hund's coupling of
strength \JH. Each of these impurity levels is coupled to its own
spinful bath channel with uniform hybridization $\Gamma$. This leads
to the Kondo-Anderson hybrid Hamiltonian,
\begin{subequations}\label{eq:HKondoAM}
\begin{equation}
   \hat{H} \equiv \hat{H}_{d} + \sum_{i=1}^{m=3} \sum_{p\sigma}
   \Bigl[ \sqrt{\tfrac{2\Gamma}{\pi}}
      \bigl(
         \hat{d}_{i\sigma}^{\dagger}\hat{c}_{ip\sigma}^{\phantom\dagger} + \hc
      \bigr)
    + \varepsilon_{p}
      \hat{c}_{ip\sigma}^{\dagger}\hat{c}_{ip\sigma}^{\phantom\dagger}
   \Bigr]
\text{,}\label{eq:HKondoAM-H}%
\end{equation}
where all energies will be given in context in units of the
half-bandwidth $W:=1$. The impurity is described by
\begin{equation}
   \hat{H}_{d} \equiv -\JH \mathbf{\hat{S}} \cdot \mathbf{\hat{S}}
%% + B\mathbf{\hat{S}}_{z}
\text{,}\label{eq:HKondoAM-d}%
\end{equation}
with the impurity spin
\begin{equation}
   \mathbf{\hat{S}}_{\alpha} \equiv
   \sum_{i}^{m}\mathbf{\hat{S}}_{i,\alpha} \equiv
   \sum_{i=1}^{m} \sum_{\sigma\sigma'}
   (\tfrac{1}{2}\tau_{\alpha})_{\sigma\sigma'}
   \hat{d}_{i\sigma}^{\dagger}
   \hat{d}_{i\sigma'}^{\phantom\dagger}
\label{eq:HKondoAM-S}%
\end{equation}
\end{subequations}
given in terms of the Pauli matrices $\tau_{\alpha}$ with
$\alpha\in\{x,y,z\}$. Here $\hat{d}_{i\sigma}^{\dagger}$
[$\hat{c}_{ip\sigma} ^{\dagger}$] creates a particle with spin
$\sigma \in\{\uparrow,\downarrow\}$ on d-level $i$ at energy $\Ed =
0$ [in bath channel $i$ at energy $\varepsilon_{p}$], respectively.
For $\JH \gtrsim \Gamma$, an effective spin-$3/2$ forms at the
impurity, leading to a symmetric fully-screened spin-$3/2$ system.
The resulting Kondo temperature \TK decays exponentially with
$\JH/\Gamma$, with \TK quickly becoming the smallest energy scale in
the system. In practice, choosing $\JH = 2\Gamma / (m+\tfrac{1}{2})$
leads to comparable Kondo temperatures $T_{K}$ for different $m$.
Compared to the standard Kondo Hamiltonian with $\mathbf{S} \cdot
\mathbf{s}$ coupling of the dot spin $\mathbf{S}$ with the lead spin
$\mathbf{s}$, the Hamiltonian in \Eq{eq:HKondoAM} in terms of
$\Gamma$ and \JH also allows for charge-fluctuations, while the
model maintains particle-hole symmetry.

In particular, the Anderson-like model in \Eq{eq:HKondoAM} has the
advantage that the impurity self-energy $\Sigma(\omega)$ can be
evaluated within the NRG in a simple fashion. From a more technical
point of view, this allows the straightforward calculation of an
improved spectral function from the self-energy. \cite{Bulla98} The
impurity Green's function [cf. \Eq{eq:using-Gpsi}{}]
\begin{align}
   G_{(i\sigma)}(\omega) &\equiv
   \langle \hat{d}_{(i\sigma)}^{\phantom\dagger} \Vert
   \hat{d}_{(i\sigma)}^{\dagger} \rangle_{\omega} \nonumber \\
   &\equiv G_{(i\sigma)}^{\prime}(\omega) -i\pi G_{(i\sigma)}^{\prime\prime}(\omega)
\text{,}
\end{align}
consisting of real and imaginary part, respectively, is constructed
within the NRG framework, as usual, from the spectral function
$A_{(i\sigma)}(\omega) \equiv -\tfrac{1}{\pi} \mathrm{Im}
G_{(i\sigma)}(\omega) \equiv G_{(i\sigma)}''(\omega)$. Subsequently,
the real part $G_{(i\sigma)}'(\omega)$ is obtained through
Kramers-Kronig transform of $A_{(i\sigma)}(\omega)$. \cite{Bulla08}
The calculation of the additional correlation function
$F_{(i\sigma)}(\omega)$ then,
\begin{align}
   F_{(i\sigma)}(\omega) \equiv %% -\tfrac{1}{\pi}\mathrm{Im}
   \bigl\langle[ \hat{d}_{(i\sigma)}, \hat{H}_{d}]\Vert\,\hat{d}_{(i\sigma)}^{\dagger}
   \bigr\rangle_{\omega} \text{,}
\label{eq:F4self}
\end{align}
obtained similarly from its spectral part $F_{(i\sigma)}''(\omega)
\equiv -\tfrac{1}{\pi} \mathrm{Im}F(\omega)$, allows to evaluate the
self-energy $\Sigma(\omega)$ at the impurity\cite{Bulla98}
\begin{align}
   \Sigma_{(i\sigma),\JH} \equiv \frac{F_{(i\sigma)}}{G_{(i\sigma)}}
\text{,}\label{eq:Sigma:def}
\end{align}
Note that, the commutator of the \IROP $\hat{d}_{(i\sigma)}$ with the
scalar Hamiltonian in \Eq{eq:F4self} again leads to an \IROP \wrt the
same \IREP $q$. Moreover, by symmetry, both $G_{(i\sigma)}$ and
$F_{(i\sigma)}$ are independent of $(i\sigma)$, as indicated by the
subscript bracket, and hence will be skipped altogether in the
following, for simplicity.

\subsubsection{Kondo limit from numerical perspective \label{sec:largeU}}

While the procedure to obtain the self-energy is straightforward for
an Anderson-like model, there is no simple way to do so for the plain
Kondo-like model with $\mathbf{S} \cdot \mathbf{s}$ interaction.
\cite{Bulla98} However, from the NRG point of view, the transition
from one to the other is straightforward. That is, knowing that the
Kondo temperature \TK decays exponentially with $\JH/\Gamma$, both,
\JH as well as $\Gamma$ can be taken \emph{much} larger than the
bandwidth $W:=1$ of the model, while keeping their ratio constant,
\begin{align}
   \JH,\ \Gamma\gg 1, \qquad \tfrac{\JH}{\Gamma}=\mathrm{const}
\text{.}\label{eq:large:JH}
\end{align}
This is a well-known procedure in the analytical Schrieffer-Wolff
transformation for the Anderson model into a Kondo model.
\cite{Schrieffer66} But, of course, exactly the same strategy can
also be pursued here within the NRG [see \Fig{fig:spectralSUN}
later]. For the local density of states at the impurity this leads to
a well-separated nearly discrete contribution to the spectral
function at $|\omega|\gg1$ far outside the bandwidth. For the
spectral range within the bandwidth, the actual spectral function for
the Kondo-model emerges. In particular, this procedure allows to
fully eliminate the free-orbital (FO) regime with strong
charge-fluctuations in the Anderson-like model right within the first
truncation step. From a numerical point of view, this is desirable as
the FO regime is typically the most expensive one. For example, for
the model discussed here using the symmetries below, using
energy-based truncation indicates that about a factor of $5\ldots 10$
more multiplets are required for the FO regime as compared to the
local moment (LM) or strong coupling (SC) regime at later NRG
iterations [\cf \Fig{fig:nrg_anal}{}]. Nevertheless, by maintaining
an Anderson-like description, the impurity self-energy remains easily
accessible numerically within the NRG, even though essentially the
correlation functions for the Kondo model are calculated.

\section{NRG results \label{sec:NRGresults}}

The model in \Eq{eq:HKondoAM} is a true three-channel system, in that
it is not possible to simply decouple a certain unitary superposition
of bath channels. Furthermore, within an NRG iteration, a site from
each channel must be included as they have the same coupling
strength, \ie energy scale, as schematically depicted in
\Fig{fig:kondoAM} [dashed boxes].

The non-abelian symmetries present in the system are,
\begin{itemize}
\item total spin symmetry: \SU[spin]{2},
\item particle-hole symmetry in each of the three channels:
    $\SU[charge]{2}^{\otimes3}$, and
\item channel symmetry: \SU[channel]{3}.
\end{itemize}
The latter symmetry \SU[channel]{3}, however, \emph{does not} commute
with particle-hole symmetry, while it does commute with the total
charge U(1)$_{\text{charge}}$, \ie the abelian subalgebra of
particle-hole symmetry [\cf App. \Eq{eq:CR:CT3}, and subsequent
discussion]. Having non-commutative symmetries, however, directly
suggests a larger enveloping symmetry, which in the present case is
the symplectic symmetry \Sp{6} [\ie \Sp{2m} with $m=3$, \cf
\App{sec:Sp(2m)}{}].

This allows to consider the following symmetry settings,
\begin{subequations}\label{eq:3level:syms}
\begin{eqnarray}
   && \SSSS\label{eq:SSSS-symm} \equiv \Sfour \text{,} \\
   && \ASC  \label{eq:ASC-symm} \text{, and } \\
   && \SSP  \label{eq:SSP-symm} \text{.}
\end{eqnarray}
\end{subequations}
All of these symmetry settings have been implemented, in practice,
and applied within the NRG framework, with results presented in the
following. The first setting in (\ref{eq:SSSS-symm}) represents a
more traditional NRG scenario based on a set of plain \SU{2}
symmetries. The second setting (\ref{eq:ASC-symm}) includes
\SU[channel]{3} together with the simple abelian symmetry
U(1)$_{\text{charge}}$ for total charge, while the last setting
(\ref{eq:SSP-symm}) represents the actual full symmetry of the model.

Even though the second setting in (\ref{eq:ASC-symm}) actually
includes an abelian component in terms of charge, it nevertheless
represents a stronger symmetry as compared to the first setting
(\ref{eq:SSSS-symm}). Since \SU[channel]{3} is a rank-2 symmetry with
two commuting z-operators, \ie generators of the Cartan subalgebra,
it possesses a \emph{two}-dimensional multiplet representation. This
results in much larger multiplets for setting (\ref{eq:ASC-symm}),
with the bare \SU{3} multiplet dimensions easily reaching up to 100
(\eg $d_q \le 125$ for the NRG run underlying \Fig{fig:SU3diags}, \cf
\AppP{sec:SU123}{}). As a consequence, this allows, on average,
smaller multiplet spaces and thus better numerical performance.

The first two symmetry settings in (\ref{eq:3level:syms}) emphasize
different symmetry aspects, yet allow to break certain symmetries
which, nevertheless, are present in the model Hamiltonian in
\Eq{eq:HKondoAM}. The first symmetry setup (\ref{eq:SSSS-symm})
strongly emphasizes particle-hole symmetry, while it does not use the
symmetric coupling of the levels to their respective channels. The
channel symmetry can thus be broken without reducing the symmetry
setting (\ref{eq:SSSS-symm}). The second symmetry setting
(\ref{eq:ASC-symm}), on the other hand, emphasizes the channel
symmetry (uniform $\Gamma$), while it allows to break the
particle-hole symmetry. Hence, in principle, a uniform level-shift
could be applied to the d-levels within this setting. Only the third
symmetry (\ref{eq:SSP-symm}) captures the full symmetry of the model,
as it combines channel symmetry with particle-hole symmetry into the
enveloping symmetry \Sp{6}. This is a rank-3 symmetry with multiplet
dimensions now easily reaching up to a several thousands (\eg see
\Tbl{NRG:Sp6:multiplets} for actual multiplets generated in a full
NRG run). A more detailed general discussion and comparison of all of
above symmetry setups in terms of their overall multiplet structure
and representation of a site with three spinful levels (\ie a Wilson
site) is given in \App{sec:SU123}.

\subsection{Spectral functions}

The spinor $\hat{\psi}^q$ to be used for fermionic hopping term as
well as for the calculation of spectral functions can be represented
for all symmetry settings by \IROPs with a well-defined multiplet
label $q$ [\cf \Eq{eq:using-Gpsi}{}]. For the first symmetry setting
in (\ref{eq:3level:syms}), $\Sfour$, the \IROP for the calculation of
spectral functions involves three 4-component spinors,
$\hat{\psi}^{[4]}_i$ for short, one for every channel $i=1,\ldots,
(m=3)$. The corresponding \IROP labels are
$q_{1}=(\tfrac{1}{2},\tfrac{1}{2},0,0)$,
$q_{2}=(\tfrac{1}{2},0,\tfrac{1}{2},0)$, and
$q_{3}=(\tfrac{1}{2},0,0,\tfrac{1}{2})$, respectively. The number of
components in the spinor derives from the two participating \SU{2}
multiplets $(S,C)=\tfrac{1}{2}$ for spin and one specific channel,
thus having $2\times 2=4$ operators in one specific \IROP
$\hat{\psi}^{[4]}_i$, indeed. With this, the hopping in the
Hamiltonian, for example, is given by $\hat{h}_{k,k+1} =
\sum_{i=1}^{m} \hat{\psi}_{k,i}^{[4]\dagger} \cdot
\hat{\psi}_{k+1,i}^{[4]}$. Note that this \emph{excludes} the
hermitian conjugate part, as this is already fully incorporated
through the particle-hole symmetry [see App.
\Eqp{eq:Fops3C:hopping:psi4}]. Furthermore, note that particle-hole
symmetry gives rise to intrinsic even-odd alternations for the
spinors along a chain [see \AppP{sec:particle-hole} for a detailed
discussion].

In contrast, the second symmetry setting in (\ref{eq:3level:syms}),
\ASC, leads to a single 6-component spinor, $\hat{\psi}^{[6]}$ for
short. Its \IROP multiplet label is given by
$q=(\tfrac{1}{2},-\tfrac{1}{2},0\ 1)$. This combines a 2-dimensional
\SU[spin]{2} multiplet $S=\tfrac{1}{2}$ and an abelian 1-dimensional
multiplet $C_z=-\tfrac{1}{2}$ with the 3-dimensional \SU[channel]{3}
multiplet $T=(0\ 1)$, resulting in the $2\times1 \times 3 =6$
operators in the multiplet. For comparison, here the hopping term in
the Hamiltonian in \Eq{eq:HKondoAM} is reduced to a total of two
terms, $\hat{h}_{k,k+1} = \bigl( \hat{\psi}_k^{[6]\dagger} \cdot
\hat{\psi}_{k+1}^{[6]} + \hc\bigr)$ [see App.
\Eqp{eq:Fops3C:hopping:psi6}].

Finally, for the third symmetry setting in (\ref{eq:3level:syms}),
\SSP, again a single spinor $\hat{\psi}^q$ is obtained, but now with
12 components, written as $\hat{\psi}^{[12]}$ for short. Its \IROP
label is given by $q=(\tfrac{1}{2},1\ 0\ 0)$, which combines the
2-dimensional \SU[spin]{2} multiplet $S=\tfrac{1}{2}$ with the
6-dimensional \Sp{6} multiplet $(1\ 0\ 0)$, \ie defining
representation of \Sp{6}. Overall, this again recovers the $2\times
6=12$ components of the spinor, indeed. For comparison, now the
hopping term in the Hamiltonian in \Eq{eq:HKondoAM} is reduced to the
\emph{single} term $\hat{h}_{k,k+1} = \bigl(
\hat{\psi}_k^{[12]\dagger} \cdot \hat{\psi}_{k+1}^{[12]} \bigr)$. The
scalar contraction of the spinor $\hat{\psi}^{[12]}$ with itself
recovers the original 12 terms in the fermionic hopping structure
between two sites in the Hamiltonian in \Eq{eq:HKondoAM}. Since
particle-hole symmetry is part of \Sp{6}, this again implies that (i)
the hermitian conjugate is already taken care of in the hopping term,
and (ii) that \Sp{6} again gives rise to the same intrinsic even-odd
alternations for the spinors along a chain, exactly analogous to what
has already been encountered for standard particle-hole symmetry.

\begin{figure}[t!]
\begin{center}
\includegraphics[width=\linewidth]{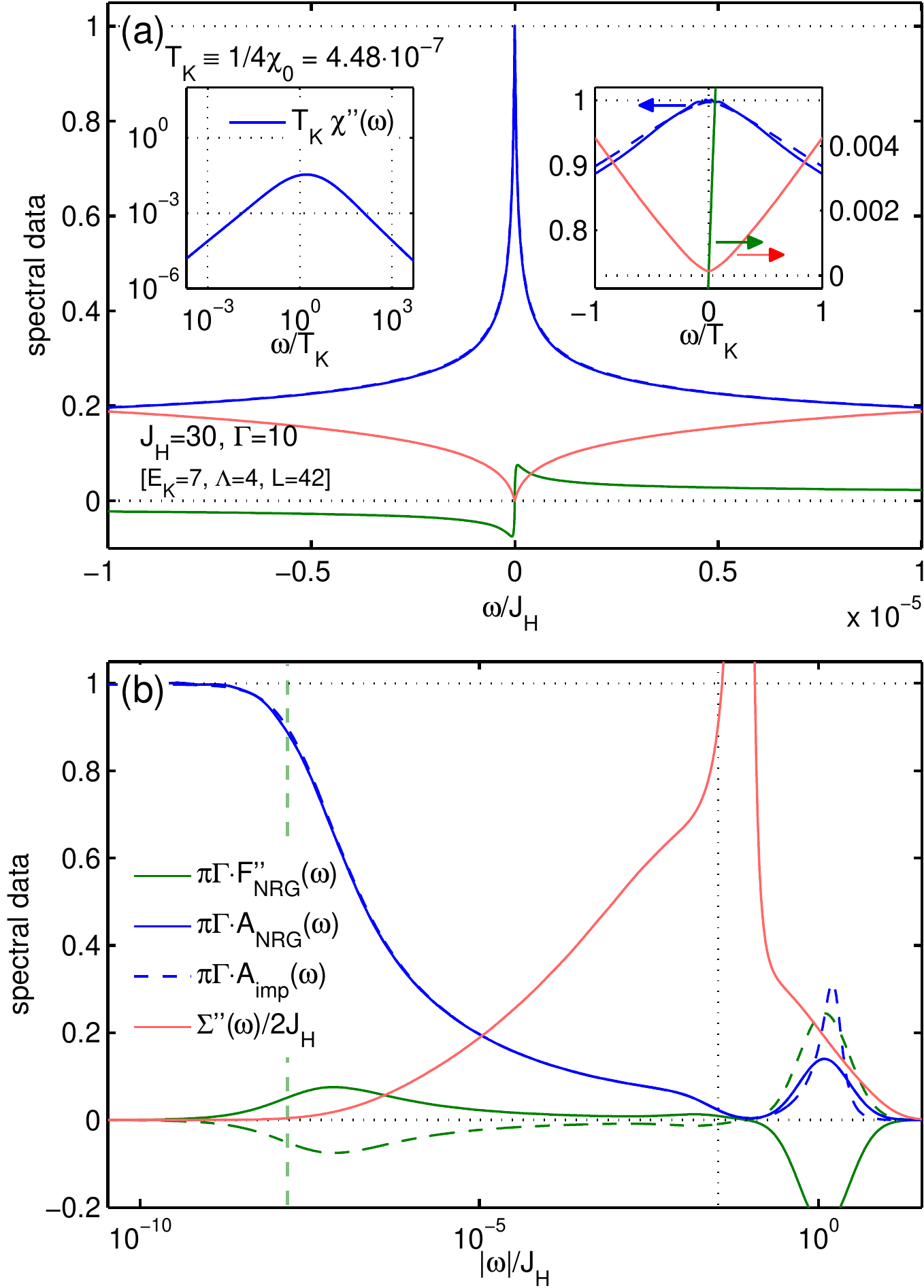}
\end{center}
\caption{ (Color online) \ASC analysis of the symmetric three-channel Anderson
model [\Eq{eq:HKondoAM} with model parameters specified in the lower
left of panel (a)]; the same data as in panel (a) is shown vs.
$\log(|\omega|)$ in panel (b) to zoom into the Kondo peak at small
frequencies with the legend for both panels shown in panel (b).
The spectral data $A_{\mathrm{NRG}}$ and the auxiliary
$F^{\prime\prime}_{\mathrm{NRG}}$ are shown together with the derived
self-energy $\Sigma''(\omega)$ and the improved spectral function
$A_{\mathrm{imp}}$ (see text). A zoom around $\omega=0$ is show in
the right inset of panel (a), with the left (right)\ axis belonging
to $A(\omega)$ [$\Sigma''(\omega)$ and $F''(\omega)$], respectively.
The spectral data for $A(\omega)$ and $\Sigma''(\omega)$ is symmetric
around $\omega=0$ and strictly positive, while $F''(\omega)$ is
antisymmetric. In panel (b) therefore the $\omega<0$ branch of
$F''(\omega)$ has been plotted in dashed lines, same color otherwise.
The left inset to panel (a) shows the spin-spin spectral data
$\chi''(\omega)$, with the resulting $\TK\equiv1/4\chi_0$
indicated in panel (b) and the left insets of both panels by the
vertical dashed line. }
\label{fig:spectralSUN}%
\end{figure}

The correlation functions calculated for the model in
\Eq{eq:HKondoAM} are presented in \Fig{fig:spectralSUN}, with the
model parameters indicated at the bottom left of panel (a). Panel (a)
shows the spectral data on a linear scale, while panel (b) shows the
same data vs. $\log(|\omega|)$ which therefore allows a logarithmic
zoom into the low energy regime. The legend shown with panel (b) also
applies to panel (a). The data in \Fig{fig:spectralSUN} was obtained
using the symmetry setting in (\ref{eq:ASC-symm}) including
\SU[channel]{3}. Note that having chosen an energy-based NRG
truncation with $\EK=7$, the spectral data for the other two symmetry
settings is identical, hence not shown. While the calculation is
somewhat more involved for the more traditional setup
(\ref{eq:SSSS-symm}), it becomes significantly more compact still
when finally including \Sp{6} as in (\ref{eq:SSP-symm}). Their
individual numerical efficiency will be discussed with
\Fig{fig:nrg_anal} below.

In \Fig{fig:spectralSUN}, the spectral function obtained from the NRG
is plotted as $A_{\mathrm{NRG}}(\omega) \equiv
G_{\mathrm{NRG}}''(\omega)$. The spectral data satisfies the Friedel
sum-rule to an excellent approximation, in that $\lim_{\omega\to 0}
\bigl(\pi\Gamma A_\mathrm{NRG}(\omega)\bigr) = 1$ [see right inset to
panel (a) for a zoom around $\omega=0$]. The self-energy
$\Sigma(\omega)$ was obtained by calculating the additional
correlation function $F_{\mathrm{NRG}}(\omega)$ [\Eq{eq:F4self}, to
be used in \Eq{eq:Sigma:def}{}]. The imaginary part $\Sigma''(\omega)
\equiv -\tfrac{1}{\pi}\mathrm{Im} \Sigma(\omega)$, plotted in
\Fig{fig:spectralSUN}, clearly approaches zero in a smooth parabolic
fashion at the Fermi energy, \ie $\lim_{\omega\to 0}
\Sigma(\omega)=\lim_{\omega\to 0}
\tfrac{d}{d\omega}\Sigma(\omega)=0$, as expected for a system who's
low-energy behavior corresponds to that of a Fermi liquid. This is
seen more clearly still in the zoom around $\omega=0$ in right inset
of \Figp{fig:spectralSUN}{a}, with the self-energy data associated
with the right axis. The self-energy $\Sigma(\omega)/\JH$ sharply
drops within $|\omega|\lesssim T_K$ from order 1. accurately down to
about $10^{-4}$ which is considered the NRG resolution limit.

The \emph{improved} spectral function $A_{\mathrm{imp}}(\omega)$
derived from the self-energy\cite{Bulla98} is also shown in
\Fig{fig:spectralSUN} [dashed red (black) line]. Within the Kondo
regime, the result closely follows the original
$A_{\mathrm{NRG}}(\omega)$, as demonstrated in the zoom in the right
inset of \Figp{fig:spectralSUN}{a} or also in panel (b). As expected
from the self-energy treatment, \cite{Bulla98} the improved spectral
function $A_{\mathrm{imp}}(\omega)$ allows clearly sharper resolution
for structures at finite frequencies, specifically so for larger
$\Lambda$. This can be observed, for example, for the hybridization
side peaks in \Figp{fig:spectralSUN}{b} at the energy of the Hund's
coupling \JH. Having chosen \JH much larger than the bandwidth [with
the bandwidth indicated by the vertical dotted line in panel (b)],
these hybridization side peaks essentially correspond to very narrow,
nearly discrete peaks that are much overbroadened through the
standard log-Gauss broadening of the NRG. \cite{Bulla08,Wb07} In
principle, these side peaks could be narrowed significantly further
by an adaptive broadening scheme. \cite{Freyn09} For the purposes of
this paper, however, this was irrelevant.

The dynamically generated exponentially small Kondo temperature \TK
for the system can be determined by taking the
full-with-at-half-maximum (FWHM) of the Kondo peak in the spectral
function. However, with NRG somewhat sensitive to broadening of the
underlying discrete data\cite{Bulla08} (see also supplementary
material in Ref.~\onlinecite{Wb07}), \TK is simply determined
therefore through the static magnetic susceptibility
$\chi_{0}=:1/4\TK$, \cite{Bulla98} where $\chi_{0}$ is obtained from
the impurity spin-spin correlation function $\chi(\omega) \equiv
\langle S_{(z),d} \Vert S_{(z),d}\rangle_\omega \equiv \chi'(\omega)
-i\pi\chi''(\omega)$ evaluated at $\omega=0$, with $S_{(z),d}$ the
total spin at the impurity [cf. \Eq{eq:chi-SUN}{}]. The resulting
spin-spin spectral function $\chi''(\omega)$ is shown in the left
inset to \Figp{fig:spectralSUN}{a}, together with the resulting
$\TK=4.4 \cdot 10^{-7}$ (in units of bandwidth). As expected,
$\chi''(\omega)$ shows a pronounced maximum around \TK. The value for
\TK is also indicated by the vertical dashed line in panel (b).

The NRG data presented in \Fig{fig:spectralSUN} clearly suggests
converged data, even without necessarily having to resort to
self-energy to get the low energy physics correct. \cite{Costi09} The
convergence is also supported by the analysis of the \emph{discarded
weight} \cite{Wb11_rho} which, inspired by DMRG, analyzes the decay
of the eigenspectrum of site-specific reduced density matrices built
from the ground state space a few iterations later. For given NRG
run, the discarded weight is estimated as $\varepsilon^{D}_{\chi=5\%}
= 3 \cdot10^{-11}$. This suggests good convergence, in agreement with
Ref. \onlinecite{Wb11_rho}. If, for example, an energy truncation of
$\EK=5,\ldots,6$ had been used, instead, NRG intrinsic parameter
dependent deviations of up to ten percents can still be seen \wrt to
the Friedel sum-rule or the agreement of $A_{\mathrm{NRG}}(\omega)$
with $A_{\mathrm{imp}}(\omega)$.

\begin{figure}[t!]
\begin{center}
\includegraphics[width=\linewidth]{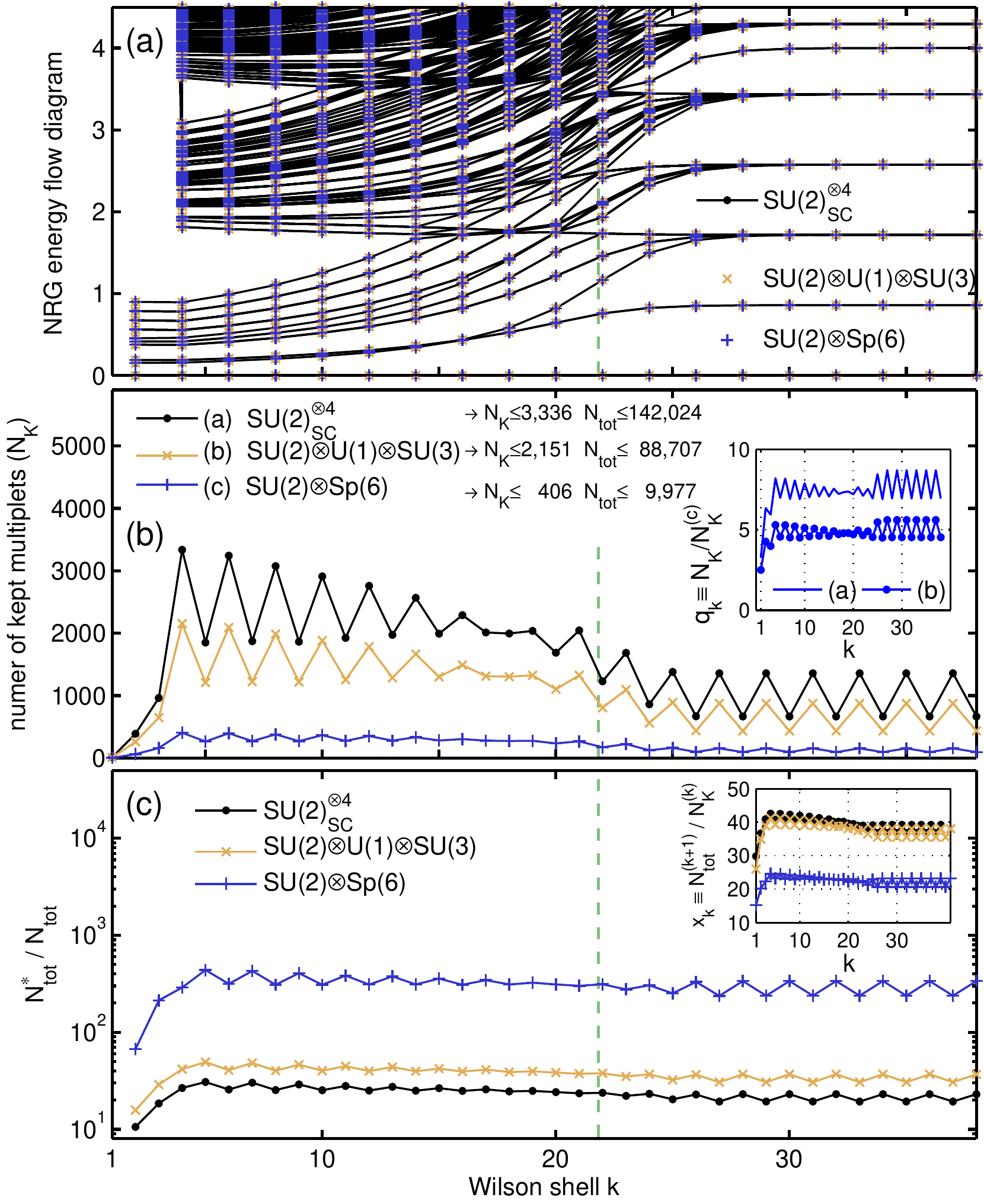}
\end{center}
\caption{ (Color online)
Comparison of the efficiency of the symmetry settings as outlined in
\Eq{eq:3level:syms} for the calculation of the spectral data in
\Fig{fig:spectralSUN} for the 3-channel model in \Eq{eq:HKondoAM}.
For a fair comparison, all calculations were performed using the same
energy-based truncation with $\EK=7$ for the same discretization
$\Lambda=4$ as in \Fig{fig:spectralSUN}. The vertical dashed lines
in all panels indicates the energy scale of \TK.
Panel (a) compares the energy flow diagrams resulting for even
iterations from the individual NRG runs, indicating perfect
consistency for all symmetry settings.
Panel (b) shows the number of kept multiplets for each
iteration. For each symmetry setting, at the top of the panel the
maximum dimension in the multiplet space over the entire NRG run is specified for
kept $(\NK)$ and total $(N_{\mathrm{tot}})$, \ie kept and discarded space,
respectively. The inset shows the ratio $q_k \equiv \NK/\NK^{(c)}$ of the
multiplets that needed to be
kept for the symmetry settings in \Eq{eq:SSSS-symm} and \Eq{eq:ASC-symm}
relative to the case when the full \Sp{6} is included [\Eq{eq:SSP-symm}{}].
Panel (c) shows the ratio $N_\mathrm{tot}^\ast/N_\mathrm{tot}$ of the actual
Hilbert-space dimension ($N_\mathrm{tot}^{\ast}$) at a given iteration,
which includes the internal multiplet dimensions, relative to the dimension
of the multiplet space ($N_\mathrm{tot}$). The inset shows the ratio
$x_k$ that describes the increase in the number of multiplets when adding
a new site prior to truncation.
}\label{fig:nrg_anal}%
\end{figure}

\subsection{Detailed comparison of symmetry settings}

An NRG specific technical comparison of the symmetry settings in
\Eq{eq:3level:syms} for the calculation in \Fig{fig:spectralSUN} is
presented in \Fig{fig:nrg_anal}. The underlying truncation had been
energy-based in all calculations ($\EK=7$), thus leading to a fair
comparison in terms of accuracy. With this, the physical properties,
and in particular the energy flow diagram \cite{Wilson75,Bulla98} in
\Figp{fig:nrg_anal}{a}, show perfect agreement using either symmetry
setting. Having sufficiently many states implies that for symmetries,
that are not explicitly and thus exactly included in the \QSpace
setup, their unintended breaking due to numerical double precision
noise does not play role.

\FIGP{fig:nrg_anal}{b} shows the number of kept multiplets for each
iteration $k$. Having chosen \JH and $\Gamma$ much larger than the
bandwidth [\cf \Figp{fig:spectralSUN}{a}], the free-orbital regime is
absent, with the transition from the local moment to the strong
coupling regime given by the energy scale of \TK [vertical dashed
line at $k\simeq 22$]. As expected from physical grounds, also the
local moment regime ($k<22$) requires a larger state space
(multiplet) dimension still for the same accuracy, \ie the same \EK,
as compared to the strong coupling regime $(k>22)$.

With the state space truncation based on the energy cutoff $\EK=7$,
the actual Hilbert state space dimension, \ie when including the
internal CGC space dimensions, is exactly the same for all symmetry
settings. In particular, the maximum total Hilbert state space
dimension per iteration that was diagonalized exactly for either
symmetry setting was $N_{\mathrm{tot}}^\ast \le 4,369,024$ or
$N_{\K}^\ast \le 68,266$ \wrt kept space only. These state spaces
could be strongly reduced to the effective and manageable multiplet
dimension as indicated at the top of \Figp{fig:nrg_anal}{b}, with
Wilson shell specific multiplet dimensions plotted in the panel.

\FIGP{fig:nrg_anal}{c} analyzes the actual reduction in multiplet
space due to presence of the CGS spaces in terms of the ratio of the
actual Hilbert space dimension $N_{\mathrm{tot}}^\ast$ relative to
the total multiplet dimension $N_{\mathrm{tot}}$ for each site along
the Wilson chain. Depending on the symmetry setting, on average, the
treatment of non-abelian symmetries allows to reduce the Hilbert
space dimension by at least a factor of 16, 20, or 300 for the
symmetries in (\ref{eq:3level:syms}), respectively. This demonstrates
an enormous numerical gain, considering that the numerical cost of
NRG roughly scales like $\mathcal{O}(N_\mathrm{tot}^3)$. Note that it
is exactly through the dimensional reduction to multiplet spaces,
that above NRG calculations had been feasible in practice, and this
within a few hours of runtime. In contrast, the plain abelian setting
simply would not have been able to deal with the underlying Hilbert
state space dimension using state of the art workstations [\cf App.
\TblP{tbl:3channel_Dtotal}].

Within the kept space, the multiplet dimension of the first two
settings in (\ref{eq:SSSS-symm}) and (\ref{eq:ASC-symm}) relative the
setting including the \Sp{6} are shown in the inset to
\Figp{fig:nrg_anal}{b} [(a) and (b), respectively]. This clearly
demonstrates a further reduction by a factor of about $5\ldots 8$
when including the full \Sp{6} symmetry. From the same inset, it is
also clear that the symmetry setting in \Eq{eq:ASC-symm} including
\SU[channel]{3} allows, on average, a 40\% further reduction of the
number of multiplets in the simulation as compared to the \Sfour
setting.

Furthermore, the inset to \Figp{fig:nrg_anal}{c} shows the ratio $x_k
\equiv N_\mathrm{tot}^{(k+1)} / \NK^{(k)}$ which indicates the
increase in total number of multiplets when adding a new site
\emph{prior} to truncation. While this factor shows a clear reduction
from the actual dimension of the local Hilbert space of a Wilson site
of $4^3=64$ states, the ratio $x_k$ is somewhat larger than what one
may naively expect, considering that, depending on the symmetry, a
Wilson site reduces to a total number of 4, 10 or 13 multiplets [see
App. \Tbl{64states:Sp6}, \Tbl{64states:SU123}, and
\TblP{64states:SU2x4}, respectively]. On the other hand, given
non-abelian symmetries, the combination of two multiplets typically
leads to clearly \emph{more} than just one overall multiplet. In this
sense, the major gain of using non-abelian symmetries is given by the
state space reduction demonstrated in \Figp{fig:nrg_anal}{b}. For the
first two symmetry settings in (\ref{eq:3level:syms}), the multiplet
space increase by adding a new site in terms of a product space
reduces the original abelian factor of 64 only modestly down to about
$38$. Only when using of the full \Sp{6}, this leads to a significant
further reduction of the ratio $x_k$ down to about 20, which thus
becomes nearly comparable in numerical cost to a two-channel
calculation with abelian symmetries, where a Wilson sites adds
$4^2=16$ states to the system.

The \SU{3} representations that are explicitly generated in the
calculation of \Fig{fig:spectralSUN} using \ASC are listed in
\Fig{fig:SU3diags} [\AppP{sec:SU123}]. The largest Clebsch-Gordan
space that is split off with respect to the \SU{3} sector only is the
$(4,4)$ representation with an internal multiplet dimension of 125.
In other words, by explicitly accounting for \SU{3} symmetries, in
the present case, a 125-fold degeneracy in the Hamiltonian had been
reduced to a \emph{single} multiplet, with the \SU{3} symmetry space
taken care of separately with minor computational overhead.
Nevertheless, the eigenstates in the \SU{3} setting still show
significant degeneracies. These can be entirely removed only by using
the full \Sp{6} symmetry, which allows to remove \emph{original
degeneracies in the Hamiltonian of several thousands}. Note that on
top of above symmetries, the spin \SU{2} multiplets present yet
another independent multiplet space that enters as a tensor product,
thus enlarging the overall symmetry space still further.

In terms of overall runtime on a state-of-the-art 8-core workstation,
this translated to about 6 hours of runtime for the \Sfour
symmetries, as compared to about 4.5 hours of runtime when including
\SU[channel]{3}. Using the full symmetry as in (\ref{eq:SSP-symm}),
on the other hand, took about $24\,\mathrm{hrs}$. While significantly
more efficient in terms of storage requirements [\cf
\Tbl{tbl:3channel_sparse}{}] thus facilitating calculations on
standard workstations, the huge CGC spaces in the last setting must
be dealt with carefully. As can be seen from \Figp{fig:nrg_anal}{b},
the total number of kept multiplets hardly reaches $400$, while the
\Sp{6} multiplets are fully comparable in terms of dimensionality,
with some multiplets even much larger internally than the actual
number of multiplets considered [\cf App.
\Tbl{NRG:Sp6:multiplets}{}]. While the sparse algebra had been
optimized by ourselves to also make use of the parallel shared memory
capacity [\cf \App{Sec:Sp6:applic}{}], in contrast, the full
multiplet spaces had access to the highly optimized shared BLAS
libraries. The latter benefitted the first two symmetry settings
(\ref{eq:SSSS-symm}) and (\ref{eq:ASC-symm}) in terms of overall
runtime. However, there is clearly room for further improvement in
dealing with the sparse algebra for larger rank symmetries as in
(\ref{eq:SSP-symm}).

\section{Summary and outlook}

A generic and transparent framework has been presented for the
implementation of non-abelian symmetries in tensor-networks in terms
of \QSpaces. For this, it was assumed that all participating state
spaces are strictly orthonormal and can be assigned proper
well-defined symmetry labels. Therefore the presented framework is
straightforwardly applicable to the traditional DMRG as well as to
the NRG. The latter was demonstrated in detail in this paper for an
\SU{3} symmetric 3-channel problem, which in the presence of
particle-hole symmetry can be further enlarged still to the
symplectic symmetry \Sp{6}. By reducing the actual state space to
the reduced multiplet space, while factorizing the Clebsch Gordan
coefficient space, this allows an efficient description of all
relevant tensors. While the explicit Clebsch Gordan algebra bears
little overhead for combinations of lower rank symmetries, the
average internal multiplet dimensions grow quickly with increasing
rank $r$ of a symmetry. In practice, one may roughly estimate that
the typical internal multiplet dimension grows like
$\mathcal{O}(10^r)$, for example, having $r=0,1,2,3$ for abelian,
\SU{2}, \SU{3}, and \Sp{6}, respectively. Starting with $r=3$, an
efficient sparse scheme on all CGC spaces becomes crucial. For
symmetries with rank larger than three, finally, it appears
desirable to develop general strategies and sum rules for the
contraction of extended complex networks of CGC spaces based on
$6n$-$j$ symbols.

A detailed self-contained general introduction to non-abelian
symmetries is given in \App{sec:basics101}, followed by many
explicit examples that arise in practice (\App{app:QSpaces}{}).
Several further highlights explained in detail in
\App{sec:num:implement} are: (i) a straightforward numerical recipe
for the general calculation of Clebsch Gordan coefficients based on
explicit product space decomposition in the presence of
multiplicity, (ii) a generic recipe for the determination of
irreducible operator sets, and last but not least, (iii) also a
general algorithm to get the framework for several symmetries
initialized from plain Fock space. The latter does not require any
initial detailed knowledge of specific symmetry labels other than
the general action of the underlying generators. These are known in
second-quantized form and thus also easily defined in Fock space.

While the work presented here is limited to situations where
effective state spaces are orthonormal and thus can be simply
categorized using well-defined symmetry labels, this eventually may
be relaxed to some extent. For example, orthonormal state spaces are
not straightforwardly applicable to two-dimensional systems with
two-dimensional tensor networks due to the presence of loops.
Nevertheless, the indices that connect tensors may be given a more
physical interpretation in terms of actual auxiliary physical state
spaces \cite{Banuls08,Saberi09}. This had been at the very basis of
the original AKLT construction \cite{AKLT88,Schollwoeck11} which
subsequently was generalized to two-dimensional (i)PEPS networks
\cite{Verstraete04_peps, Verstraete08}. With parity symmetry
\cite{Barthel09,Corboz10,Kraus10} and simple abelian symmetries such
as $Z_n$ \cite{Bauer11} successfully employed for two-dimensional
fermionic systems, iTEBD based algorithms \cite{Jordan08, Orus09}
may open the grounds to also use the non-abelian tensor framework as
described in this work in a widened context.

Finally, it is emphasized that also superoperators permit full
treatment in terms of symmetries using Clebsch-Gordan coefficient
spaces \cite{Mueller05,Buca12}. Note, for example, that the reduced
density matrices considered in this work are all block-diagonal \wrt
symmetries. As such they correspond to \emph{vectors} in
superoperator space with well-defined symmetry label, in that the
\emph{difference} of its quantum labels in regular operator space is
zero throughout. In general, this may also open the door to using
the \QSpace framework presented here to the simulation of
Liouvillian superoperators such as they occur, for example, in the
Lindblad equation for driven systems.

\section{Acknowledgements}

I want to thank Jan von Delft, Arne Alex, and Alan Huckleberry for
fruitful discussions on \SU{N} representation theory, and also G.
Zarand and I. Weymann for helpful discussions on the implementation
of non-abelian symmetries in context of conventional NRG
calculations. In addition, I want to thank the people above and also
Markus Hanl for their critical reading of the manuscript. This work
received support from the DFG (SFB-631, De-730/3-2, De-730/4-2,
SFB-TR12, WE\-4819/1-1).

\bibliographystyle{apsrev4-1}
\bibliography{D:/TEX/Lib/mybib}

\clearpage

\appendix
\addcontentsline{toc}{section}{Appendix}

\section{Non-abelian symmetries 101 \label{sec:basics101}}

The general more pedagogical introduction of non-abelian groups in
this appendix emerges from a practical numerical background of
treating quantum many-body phenomena. It does not claim to cover
non-abelian symmetries in every theoretical detail, yet requires
certain elementary concepts which will be briefly reviewed. The main
focus of this appendix then is on practical applications in quantum
lattice models. Specifically, this targets the numerical
renormalization group (NRG),\cite{Wilson75,Bulla08} density matrix
renormalization group (DMRG)\cite{White92,Schollwoeck05} or more
generally tensor networks,\cite{Singh10,Singh11} yet also exact
diagonalization, which itself may be formulated in a matrix product
state language. The appendix offers a general treatment of continuous
non-abelian symmetries, with modifications towards abelian,
point-groups, or discrete non-abelian symmetries straightforward.
Overall, this appendix should be self-contained, sufficient and
hopefully helpful to deal with general abelian and non-abelian
symmetries in numerical simulations.

The non-abelian symmetries of concern in this paper are continuous
symmetries. An element $\hat{G}$ of the corresponding \emph{Lie
group} $\hat{\mathcal{G}}$ can be parameterized by a set of $g$
continuous, independent, and real parameters $a_\sigma$,
\cite{Elliott79}
\begin{equation}
   \hat{G}(a_1,\ldots,a_g)
 = \exp\Bigl({i\sum\limits_{\sigma=1}^g a_\sigma \hat{S}_\sigma} \Bigr)
\text{,}\label{eq:gensymm}
\end{equation}
with $g$ the \emph{dimension} of the symmetry group. Infinitesimal
operations with $a_\sigma\ll1$ then define the set of $g$ generators
$\{\hat{S}_\sigma\}$, the number of which thus also reflects the
dimension of the group (note that the identity operator is a trivial
operation which therefore is never part of the set of generators).
For unitary symmetries, as considered throughout in this paper, the
generators in \Eq{eq:gensymm} are hermitian. Furthermore, when
dealing with exponentially large yet \emph{finite}-dimensional
quantum-many-body Hilbert spaces, the non-abelian symmetries also
must have finite-dimensional Lie algebras.

The commutator relations of the generators in \Eq{eq:gensymm},
\begin{equation}
   [\hat{S}_\sigma, \hat{S}_\mu]
 = \sum_\nu f_{\sigma\mu\nu}  \hat{S}_{\nu}
\text{,}\label{eq:Sop:CR}
\end{equation}
determine the tensor of the \emph{structure constants}
$f_{\sigma\mu\nu}$, which itself fully defines the underlying
\emph{Lie algebra}. The tensor $f_{\sigma\mu\nu}$ is antisymmetric in
that by construction $f_{\sigma\mu\nu} = -f_{\mu\sigma\nu}$, yet not
necessarily fully antisymmetric also \wrt to the last index $\nu$ [in
principle, it can be made fully antisymmetric using the
\emph{Cartan-Killing metric}, while distinguishes between co- and
contravariant indices in \Eq{eq:Sop:CR}; \cite{Pope06,Gilmore06} for
simplicity, however, this distinction is not made in this paper].
All generators are assumed to be connected to each other through
above commutator relations. That is, if a subgroup of generators
fully decouples in that it commutes with the rest of the generators,
then this subgroup forms a symmetry of its own. In this sense the
group of generators for a specific \emph{simple} symmetry is
irreducible.

A set of matrices $\{R_\sigma\}$, that obeys exactly the same
commutator relations as the generators (operators)
$\{\hat{S}_\sigma\}$ in \Eq{eq:Sop:CR}, allows a one-to-one
correspondence between the matrices $\{R_\sigma\}$ and the generators
of the symmetry. It is called a \emph{matrix representation} of the
Lie algebra. \label{text:Mrep} If the \emph{carrier space}, \ie the
vector space within which the matrix representation is defined, is
fully explored through repeated application of the individual
matrices of the representation, then this is called an
\emph{irreducible} matrix representation, to be denoted as
$\{I_\sigma\}$ henceforth. It is unique up to an overall similarity
transformation. Together with its carrier space it refers to an
irreducible representation (\IREP), specified by a unique label $q$.
If, on the other hand, part of the carrier space of a matrix
representation decouples, the representation is called
\emph{reducible}. This will be discussed in significantly more detail
later in the context of state space decomposition in \Secs{sec:irrep}
and \ref{sec:prodspaces}.

Consider an irreducible matrix representation $\{I^q_\sigma\}$ for
\IREP $q$ of dimension $d_q$. Its carrier space is spanned by the
\emph{multiplet} $\vert q \rangle \equiv \{\vert qq_z \rangle\}$,
where $q_z$ references the individual states within the multiplet $q$
(consider, for example, spin multiplets, where $\vert qq_z \rangle
\equiv \vert S,S_z\rangle$). The states $\vert qq_z \rangle$ forms an
irreducible space \wrt the action of the generators, in that for an
arbitrary symmetry operation $\hat{G}$ as in \Eq{eq:gensymm},
\begin{subequations}\label{eq:SymOps:gen}
\begin{equation}
   \hat{G} \vert qq_z \rangle
 = \sum_{q'_z} G_{q_z,q'_z}^{q}\vert qq'_z \rangle
\text{,}\label{eq:SymOp:irep}
\end{equation}
some linear superposition within the same multiplet space arises. The
coefficients $G_{q_z,q'_z}^{q}$ form a $d_q \times d_q$ dimensional
matrix, which represents the symmetry operation $\hat{G}$ within
multiplet $q$, and is given by $G^{q} \equiv \exp\left(i\sum_\sigma
a_\sigma I_\sigma^q\right)$ for some arbitrary but fixed values
$a_\sigma$.

Similar to the multiplet space $\vert qq_z \rangle$ of dimension
$d_q$, an \emph{irreducible operator} (\IROP) set $\hat{F}^q \equiv
\{\hat{F}^q_{q_z}\}$ can be defined in a completely analogous manner.
While it is not constrained to a specific carrier space, the \IROP
$\hat{F}^q$ consists of a set of $d_q$ \emph{operators} that are
associated with multiplet $q$. As such, it can be written as a vector
of operators, \ie a generalized spinor. For a given symmetry
operation $\hat{G}$ then, the \IROP transforms analogously to
\Eq{eq:SymOp:irep}, which for an operator implies
\begin{equation}
   \hat{G} \hat{F}^q_{q_z} \hat{G}^{-1}
 = \sum_{q'_z} G_{q_z,q'_z}^{q} \hat{F}^q_{q'_z}
\text{.}\label{eq:SymOp:irop}
\end{equation}
\end{subequations}
On the level of infinitesimal operations, $|a_\sigma|\ll 1$, in
contrast to the plain action of generators on a ket-state as in
\Eq{eq:SymOp:irep}, \Eq{eq:SymOp:irop} shows that the transformation
of an \IROP directly translates to \emph{commutator relations} [\lhs
of \Eq{eq:SymOp:irop}{}] with the generators of the symmetry,
instead.

The practical relevance of above general statements will be discussed
in much detail in what follows, together with many examples relevant
in actual numerical calculations.

\subsection{Simple example: rotational symmetry}

A simple and well-known example of a non-abelian symmetry is the
rotational group SO(3) in real space in three dimensions. An arbitrary
rotation can be written as $G = e^{i S}$ with $S$ an arbitrary
hermitian, yet fully complex three-dimensional \emph{matrix} (hence
no hats). The latter is required for $G$ to be real. Consequently,
this leaves three real parameters $(a_x,a_y,a_z)$, with $S=a_x S_x +
a_y S_y + a_z S_z$.
The generators \cite{Elliott79}
\begin{equation}
   S_x = \begin{pmatrix} 0 & 0 & 0 \\ 0 & 0 &-i \\  0 & i & 0  \end{pmatrix},
   S_y = \begin{pmatrix} 0 & 0 & i \\ 0 & 0 & 0 \\ -i & 0 & 0  \end{pmatrix},
   S_z = \begin{pmatrix} 0 &-i & 0 \\ i & 0 & 0 \\  0 & 0 & 0  \end{pmatrix}
\label{eq:rep-O3}
\end{equation}
represent infinitesimal rotations around the $x$, $y$, and $z$-axis,
respectively. The SO(3) symmetry therefore has dimension $g=3$, and
its Lie algebra is defined by,
\begin{equation}
   [\hat{S}_\sigma, \hat{S}_\mu]
 = i\sum_\nu \varepsilon_{\sigma\mu\nu} \hat{S}_\nu
\text{,}\label{eq:CR:O3}
\end{equation}
with $\sigma,\mu,\nu \in \{x,y,z\}$ and $\varepsilon_{\sigma\mu\nu}$
the Levi-Civita tensor, having switched to general operator notation
[operator $\hat{S}$ (with hat) rather matrix $S$]. Being generators
of SO(3), the matrix representation in \Eq{eq:rep-O3} already
represents a 3-dimensional \IREP. As it is the simplest non-trivial
\IREP for SO(3), it is also called its \emph{defining representation}.
By combining state spaces that share this symmetry then, many other
\IREPs can be generated, including, for example, the (trivial) scalar
representation of dimension 1.

With respect to continuous functions $f(x,y,z)$ in three-dimensional
space, the generators of infinitesimal rotations are given by the
differential operator described by the angular momentum operator
$\hat{\mathbf{L}} = \hat{\mathbf{r}} \times \hat{\mathbf{p}}$ with
$\hat{\mathbf{p}} \sim \mathbf{\nabla}_{\mathbf{r}}$. By
construction, its three components $\hat{L}_{i}$ also obey exactly
the same Lie algebra as the generators in \Eq{eq:CR:O3}. The same
also holds for the spin algebra \SU{2} in complex space, which
describes the symmetry for spinful particles such as electrons if
rotational spin symmetry is not broken, \ie in the absence of an
external magnetic field. Hence the rotational group SO(3) is
isomorphic to the spin \SU{2}. In contrast to SO(3), however, the
defining representation of \SU{2} is \emph{two}-dimensional [\cf
\Eq{eq:pauli-def}{}], and hence also allows half-integer spin
multiplets, which are entirely absent in SO(3). Having essentially
twice as many multiplets in \SU{2} as compared to SO(3), \SU{2} is
thus called a double cover or 2:1 cover of SO(3).

\subsection{\SU{2} spin algebra}

In this paper, the setup and notation for non-abelian symmetries is
generalized from \SU{2}. Therefore the symmetry \SU{2} will be
recapitulated in some more detail, introducing the semantics used for
the general treatment of non-abelian symmetries. In this sense, the
semantics used in this paper is somewhat more inclined towards the
physics background, rather than strictly adhering to the mathematical
language of Lie algebras. The latter, nevertheless, will be indicated
in context.

Similar to the SO(3) symmetry, an arbitrary unitary transformation in
two-dimensional complex space is given by $G=e^{i S}$ with $S$ an
arbitrary two-dimensional hermitian matrix. This again has three
independent real parameters $(a_x,a_y,a_z)$, such that $S=a_x S_x +
a_y S_y + a_z S_z$. Here $S_{\sigma}=\tfrac{1}{2} \tau_{\sigma}$,
with $\sigma\in\{x,y,z\}$, is given by the standard Pauli spin
matrices $\tau_{\sigma}$,
\begin{equation}
   \tau_{x} = \begin{pmatrix}  0 & 1  \\ 1 &  0 \end{pmatrix},\
   \tau_{y} = \begin{pmatrix}  0 & -i \\ i &  0 \end{pmatrix},\
   \tau_{z} = \begin{pmatrix}  1 & 0  \\ 0 & -1 \end{pmatrix}
\text{,}\label{eq:pauli-def}
\end{equation}
For \SU{2}, this represents the smallest non-trivial matrix
representation, therefore this also becomes its defining
representation. The commutator relations of the matrices
$\tau_{\sigma}$ are \emph{exactly} the same as for SO(3) in
\Eq{eq:CR:O3}, since \SU{2} also refers to the same rotational
symmetry. Therefore, the generators for \SU{2} will again also be
denoted by the operators $\{\hat{S}_\sigma\}$ with $\sigma \in
\{x,y,z\}$ in what follows.

For a general irreducible representations of \SU{2}, \eg a spin
multiplet, the usual choice of basis is such that the z-component of
the spin operator, $\hat{S}_{z}$, becomes diagonal in its matrix
representation $S_{z}$, while the other two operators $\hat{S}_{x}$
and $\hat{S}_{y}$ remain non-diagonal (due to their non-commuting
properties, only one spin component can be fully diagonalized, given
the freedom of a similarity transformation for the whole
representation). Using the notation $\vert qq_{z}\rangle \equiv \vert
S,S_z\rangle$ for general spin multiplets, the multiplet label $q$
(q-label) then can take the values $q=0, \tfrac{1}{2}, 1,
\tfrac{3}{2}, 2, \ldots$ with the internal multiplet label (z-label)
spanning the $2q+1$ values $q_{z} \in \{ -q,-q+1,\ldots,+q \}$. The
raising and lowering operators (\RLOs) are defined as
\cite{Sakurai94}
\begin{align}
   \hat{S}_{\pm}  &  \equiv \hat{S}_{x} \pm i\hat{S}_{y}
\text{,}
\label{eq:SU2-PM}
\end{align}
such that $\hat{S}_{-} \equiv (\hat{S}_{+})^{\dagger}$, with the
commutator relations
\begin{subequations}\label{spin:CR}
\begin{align}
   &[ \hat{S}_{z}, \hat{S}_{\pm} ]  = \pm\hat{S}_{\pm} \label{spin-cr:zpm} \\
   &[ \hat{S}_{+},\hat{S}_{-} ]  \equiv [ \hat{S}_{+}, \hat{S}_{+}^{\dagger} ]
   =2\hat{S}_{z}
\text{.}\label{spin-cr:pm}%
\end{align}
\end{subequations}
For spin multiplets $\vert qq_{z} \rangle$ then, it holds
\cite{Sakurai94}
\begin{align}
   &\hat{S}_{z} \vert qq_{z} \rangle =q_{z} \vert qq_{z}\rangle \nonumber\\
   &\hat{S}_{\pm } \vert qq_{z} \rangle =
   \sqrt{q(q+1) -q_{z}(q_{z}\pm1)}\ \vert q,q_{z}\pm1 \rangle
\text{.}\label{eq:Spm:state}
\end{align}

While the operator set $\{\hat{S}_{x},\hat{S}_{y},\hat{S}_{z}\}$
generates the \SU{2} symmetry group, this set itself \emph{does not}
yet represent an irreducible operator (\IROP), in that it does not
yet transform according to a specific symmetry multiplet. For this, a
specific linear superposition of the original operators as in
\Eq{spin-cr:pm} is required. In particular, the transformation of an
\IROP set under given symmetry is completely analogous to the
transformation of the symmetry eigenstates in \Eq{eq:Spm:state}. As
indicated with \Eq{eq:SymOp:irop}, the major difference is that the
action of a generator $\hat{S}_\sigma$ applied onto a state is simply
replaced by the \emph{commutator} of the generator with an operator.
For example, for an \IROP $\hat{F}^q$ given in terms of the set of
operators $\{\hat{F}^q_{q_ z}\}$ which transform like the (state)
multiplet $q$, it follows for consistency with \Eq{eq:Spm:state},
\begin{subequations}\label{eq:CR-SpinOps}
\begin{align}
  &[ \hat{S}_{z}, \hat{F}^q_{q_z} ] = q_z \hat{F}^q_{q_z} \label{spin-cr:zF} \\
  &[ \hat{S}_{\pm}, \hat{F}^q_{q_z} ] = \sqrt{q(q+1)-q_z(q_z\pm1)}\cdot\hat{F}^q_{q_z\pm1}
\label{spin-cr:pmF}
\end{align}
\end{subequations}
This allows, for example, to complete the operator $\hat{S}_{z}$ into
an \textit{irreducible} spin operator set as follows. Clearly, $[
\hat{S}_{z}, \hat{S}_{z} ] = 0\cdot \hat{S}_{z}$, which implies that
the operator $\hat{S}_{z}$ has z-label $q_z=0$, \ie $\hat{S}^q_{0}
\equiv \hat{S}_{z}$ with $q$ still unknown. Applying the \RLOs yields
the operators corresponding to $q_z=\pm1$,
\begin{align*}
   \underset{ =[\hat{S}_{\pm}, \hat{S}_{z} ] = \mp\hat{S}_{\pm} }
   {\underbrace{[\hat{S}_{\pm}, \hat{S}^q_{0} ]}}
   \!\! = \sqrt{q(q+1)+0}\cdot\hat{S}^q_{\pm1}
\text{.}
\end{align*}
With the further application of \RLOs
yielding zero, \ie $[ \hat{S}_{+}, \hat{S}_{+} ] = [ \hat{S}_{-},
\hat{S}_{-} ] = 0$, the operator space is thus exhausted. The
\emph{irreducible} spin operator set therefore has three members $q_z
\in \{-1,0,+1\}$, and thus transforms like a spin multiplet
$q=\max(q_z)=1$,
\begin{equation}
   \hat{S}^{1} \equiv \{  \hat{S}^1_{q_{z}} \} \equiv
   \left(\begin{array}{l}
      \hat{S}_{1,+1} \\ \hat{S}_{1,0}\\ \hat{S}_{1,-1}%
   \end{array}\right)
 = \begin{pmatrix}
     -\tfrac{1}{\sqrt{2}}\hat{S}_{+} \\
      \hat{S}_{z}\\
     +\tfrac{1}{\sqrt{2}}\hat{S}_{-}
   \end{pmatrix}
\text{.} \label{spin-irop}%
\end{equation}
Note that the signs and prefactors are crucial for consistency with
the Wigner-Eckart theorem later.

In above derivation, the z-operator in \Eq{spin-cr:zF} allowed to
directly determine the z-label $q_z$. The \RLOs in \Eq{spin-cr:pmF},
on the other hand, served to explore the multiplet space, in that
they generated the remaining operators $\hat{F}^q_{q'_z}$ with proper
well-defined prefactors. In given case of spin \SU{2}, these factors
are known [\cf \rhs of \Eq{spin-cr:pmF}{}]. In situations, where they
may not be known right away, they can nevertheless be determined in a
straightforward manner. For simplicity, in the absence of inner
multiplicity for given multiplet, for canonical raising or lowering
operator $S_\pm$ (see \Sec{sec:Spm}) the combined application of
$\hat{S}_\pm^{\phantom\dagger}$ followed by $\hat{S}_\pm^\dagger$
onto an operator of given multiplet $q$ results in the same operator,
\ie
\begin{eqnarray}
   [\hat{S}_\pm^\dagger,
   [\hat{S}_\pm^{\phantom\dagger}, \hat{F}^q_{q_z}]]
 = a_\pm^2 \hat{F}^q_{q_z} \text{,}\nonumber
\end{eqnarray}
from which the prefactor $a_\pm^2$ can be easily determined. The
analogous situation for a state space multiplet $\vert qq_z \rangle$
is $\hat{S}_\pm^\dagger \hat{S}_\pm^{\phantom\dagger} \vert
qq_z\rangle = a_\pm^2 \vert qq_z\rangle$, with $a_\pm^2 \geq0$ since
$\hat{S}_\pm^\dagger \hat{S}_\pm^{\phantom\dagger}$ is a positive
operator; in case of spin \SU{2}, this exactly reflects the prefactor
on the \rhs of \Eq{spin-cr:pmF}, \ie $a_\pm^2=q(q+1)-q_z(q_z\pm1) \ge
0$. Therefore if the application of $\hat{S}_\pm$ results in a new
operator component in the multiplet, \ie $a_\pm^2>0$, then this
operator is exactly given by
\begin{eqnarray}
   \hat{F}^q_{q'_z}
 = \tfrac{1}{\sqrt{a_\pm^2}} [\hat{S}_\pm^{\phantom\dagger}, \hat{F}^q_{q_z}]
\text{.}\label{eq:Fqqz:norm}
\end{eqnarray}
This already contains the correct normalization and sign, with the
the latter strictly determined by the outcome of the commutator. The
z-label $q_z$ can be derived directly from the structure constants of
the underlying Lie algebra, \ie \Eq{spin-cr:zF}.
For a more general discussion on \IROPs and their general
decomposition also in the presence of inner multiplicity for the
\IROP multiplet $q$, see \Sec{sec:irop+wet}.

\subsection{Generators and symmetry labels}

Symmetries $\mathcal{S}$ within a quantum mechanical framework are
described by a set of generators $\hat{S}_{\sigma}$ that leave the
Hamiltonian $\hat{H}$ of the system invariant. Therefore it must hold
for all generators of the symmetries considered that
\begin{equation}
   [ \hat{S}_{\sigma}, \hat{H} ] = 0
\text{.}\label{eq:SHcommute}
\end{equation}
Thus by definition, the Hamiltonian is a scalar operator.
The generators of independent symmetries $\mathcal{S}$ and
$\mathcal{S}'$ commute trivially, by definition, as they operate in
independent symmetry sectors. Therefore, for simplicity, a single
specific non-abelian symmetry $\mathcal{S}$ is considered in the
following, also referred to as \emph{simple} non-abelian symmetry, a
prototypical example being \SU{N}.

Therefore let $\mathcal{S}$ be a simple non-abelian symmetry. By
construction then, its set of generators $\{\hat{S}_{\sigma}\}$ is
fully connected via the structure constants in \Eq{eq:Sop:CR}, \ie is
irreducible but not necessarily an \IROP yet [\eg see previous
discussion for \SU{2}]. With the symmetry reflected in the unitary
transformation $\hat{G}=e^{i\varepsilon\hat{S}_{\sigma}}$ with
hermitian $\hat{S}_{\sigma}$ [\cf \Eq{eq:gensymm}{}], it follows that
for infinitesimal $\varepsilon\ll 1$, the invariance of the
Hamiltonian under this unitary transformation, \ie
$\hat{U}\hat{H}\hat{U}^{\dagger}=\hat{H}$, is trivially equivalent to
\Eq{eq:SHcommute}.

In order to ensure maximally independent generators, all operators in
$\{\hat{S}_{\sigma}\}$ can be taken orthogonal with respect to each
other and specifically also with respect to the identity matrix
(which is always excluded from the set of generators $\{S_\sigma\}$).
This requires a scalar or inner product for matrices, which is
provided by
\begin{align}
   \langle A, B \rangle &\equiv \trace\bigl(A^\dagger B\bigr)
\text{,}\label{eq:fro:scalar}
\end{align}
together with the resulting Frobenius norm $\Vert A \Vert^2 = \langle
A, A \rangle = \trace\bigl(A^\dagger A\bigr)$. For the generators of
the symmetry, thus one requires
\begin{subequations}\label{conv:Sortho}
\begin{align}
  & \trace \bigl( S_{\sigma}^{\dagger} S_{\sigma'}^{\phantom\dagger}\bigr)
  = a_{\sigma^{{}}} \delta_{\sigma\sigma'} \label{conv:Sortho:S}\\
  & \trace\bigl( S_{\sigma} \bigr)
  = \trace\bigl( \Id^{(\dagger)} S_{\sigma} \bigr) = 0
\text{,}\label{conv:Straceless}%
\end{align}
\end{subequations}
The generators in \Eq{conv:Straceless} are understood as
finite-dimensional matrix representations of the operators
$\hat{S}_{\sigma}$ in some specific carrier space, here the defining
representation. Moreover, the orthogonality \wrt to the identity in
the last equation implies that all generators $S_{\sigma}$ are
traceless.
Note that if $\mathbf{1}$ had been amongst the generators,
it would form a subgroup of its own, and hence can be split
off as a $\mathrm{U}(1)$ factor. This is exactly what
distinguishes, for example, the unitary group $\mathrm{U}(N)$
from the special unitary group $\SU{N}$.

\subsubsection{Z-operators (Cartan subalgebra)}

For a given simple non-abelian symmetries, it is always possible to
identify a maximal set of mutually commuting hermitian generators
which form the so-called \emph{Cartan subalgebra} of the symmetry's
Lie algebra. These can be fully diagonalized simultaneously (together
with the Hamiltonian), and hence can be considered diagonal. They
shall be referred to as the \emph{z-operators} [as they generalize
the concept of the operator $S_z$ for \SU{2}{}],
\begin{equation}
   [ \hat{S}_{z^{{}}}, \hat{S}_{z'} ] = 0
\qquad \text{(z-operators).}\label{eq:CR:Szz}%
\end{equation}
For a given Hamiltonian $\hat{H}$ then, this implies that every
eigenstate $\hat{H} \vert n\rangle =E_{n}\vert n\rangle $, in
addition, can also be labeled with its respective set of symmetry
labels $\vert n\rangle \to \vert qn; q_z\rangle$, leading to
\begin{equation}
  \hat{H} \vert qn; q_z\rangle =E_{qn}\vert qn; q_z\rangle
\text{.}\label{eq:Hsym:eig}%
\end{equation}
Here $q$ identifies the multiplet, \ie a set of states $q_{z}$ that
are connected in an irreducible manner through all of the generators
of the symmetry. While the index $n$ originally identified all states
in given Hilbert space, it is now sufficient that it labels the
multiplet within the space of multiplets that share the same $q$. The
composite index $(qn)$ then is referred to as \emph{multiplet index}.
Similarly, also the eigenenergies $E_{qn}$ in \Eq{eq:Hsym:eig}
acquire symmetry labels. These, however, are independent of $q_{z}$
since, by construction, the states within a symmetry multiplet are
degenerate in energy. More generally, with $q_{z}$ entirely
determined by symmetry for a given multiplet $q$, they can easily be
generated and thus omitted where convenient.

Given a specific multiplet $qn$, the labels $q_{z}$ are equal to the
eigenvalues of the z-operators,
\begin{equation}
   \hat{S}_{z} \vert qn;q_{z} \rangle = q_{z} \vert qn;q_{z} \rangle
\qquad\text{(z-labels),}\label{Zop:diag}%
\end{equation}
which will be referred to as \emph{z-labels}. If more than one
z-operator is associated with given symmetry $\mathcal{S}$, say a
total of $r$ z-operators, where $r$ thus defines to the \emph{rank}
of the symmetry, then the z-label structure associated with a
multiplet also consist of a collective set of $r$ z-labels (note that
$r$ needs to be differentiated here from the rank $r$ of a tensor or
\QSpace as used in the main text). For example, the symmetry group
\SU{N} has rank $r=N-1$. Therefore the rank of \SU{2} is 1, \eg a
single label $q$ suffices to identify a state within an \SU{2} spin
multiplet. \SU{3}, on the other hand, already acquires a
two-dimensional label structure for $q_z$, and thus also for $q$.

Note that the z-labels in \Eq{Zop:diag} for the states of a specific
multiplet $q$ may not necessarily be unique, in that the
\textit{same} $q_{z}$ may occur multiple times. Let $m_{z}$ describe
how often a specific z-label occurs within given multiplet $q$. Then
the presence of $m_{z}>1$ for at least one z-label is called
\textit{inner multiplicity}. It is then necessary to introduce an
extra label $\alpha$ that uniquely identifies the state within this
degeneracy,
\begin{align}
    \vert qn; q_z\rangle &\rightarrow \vert qn; q_z \alpha_z \rangle
    \text{,} \quad \text{(inner multiplicity)}
\label{inner-mult}
\end{align}
with $\alpha_z \in \{1,\ldots,m_z\}$. While inner multiplicity is
absent for \SU{N\leq2}, it occurs on a regular basis for \SU{N\geq3}.
The situation for \emph{outer} multiplicity is analogous (see
\Secp{sec:irrep}).

The label for the entire multiplet $q$ (to be referred to
collectively as \emph{q-labels}) is in principle arbitrary, yet must
be unique to identify the multiplet. Since for a continuous symmetry
infinitely many \IREPs exist, it is natural that the q-labels inherit
the $r$-dimensional label structure of the z-labels. In particular,
it is possible to construct a set of $r$ \emph{scalar} operators,
called Casimir operators, %% (see \Secp{sec:SUN:casimir}),
that define a unique set of $r$ constants for each multiplet. In
practice, however, the q-labels are derived from $q \equiv
\max\{q_z\}$, \ie by the z-labels corresponding to the \emph{maximum
weight state} (see \Secp{sec:maxweight}) which in principle can be
related to the constants derived from the Casimir operators.
\cite{Elliott79}

\subsubsection{Raising and lowering operators (roots) \label{sec:Spm}}

While for an arbitrary unitary element $\hat{G}$ of the symmetry
\emph{hermitian} $\{\hat{S}_{\sigma}\}$ are required, on the level of
generators, in principle, arbitrary linearly-independent linear
superpositions within the space of generators $\hat{S}_{\sigma}$ can
be taken. Using such a reorganized set of generators, instead, this
still preserves \Eq{eq:SHcommute}, yet alters the structure constants
$f_{\sigma\mu\nu}$ for given symmetry $\mathcal{S}$. This freedom is
used in the following to define canonical raising and lowering
operators, which are non-hermitian, in general.

Consider the action of a generator $\hat{S}_{\sigma}$ onto a symmetry
eigenstate $\vert qn;q_{z} \rangle $. The z-operators are special, in
that they are diagonal and hence return the same state, yet weighted
by the eigenvalue $q_{z}$. The remaining generators, however, are
non-diagonal, hence change the state and thus explore the multiplet
space. In general, these generators can be reorganized such that all
of them represent proper raising or lowering operators (\RLOs), with
the canonical commutator relations,
\begin{equation}
   [ \hat{S}_{z}, \hat{S}_{\sigma} ]
   = f_{z\sigma\sigma}\hat{S}_{\sigma} \equiv f_{z\sigma} \hat{S}_{\sigma}
\text{,} \label{eq:CR:Sz-pm}
\end{equation}
with no summation over $\sigma$. The action of these canonical \RLOs
in z-label space, in the literature also referred to as \emph{root
space}, then defines the canonical form. By definition, the canonical
\RLOs $\{\hat{S}_{\pm}\}$ of a specific Lie algebra are expected to
have the property that their application onto a symmetry eigenstate
in the multiplet with well-defined z-labels will generate another
eigenstate of the z-operators, yet with raised or lowered, \ie
\emph{well-defined different} z-labels. This is exactly what is
expressed through the commutator relations in \Eq{eq:CR:Sz-pm}. In
particular, the structure constants take the simple form, where a
non-zero contribution can only arise if the last two indices in
$f_{z\sigma\sigma'}$ are identical, hence the shortcut notation
$f_{z\sigma}$ in the last term in \Eq{eq:CR:Sz-pm}. By construction,
$f_{z\sigma}$ is fully antisymmetric. Note that \Eq{eq:CR:Sz-pm} also
can be interpreted as an eigenvalue equation for the generators of
the group. Since the z-operators $\hat{S}_z$ are symmetric, the
resulting eigenvalue problem is always well-defined with real
eigenvalues $f_{z\sigma}$.

As a specific example, \Eq{eq:CR:Sz-pm} was already encountered for
\SU{2} in \Eq{spin-cr:zpm}. Here it states more generally that the
commutator of an arbitrary generator $\hat{S}_{\sigma}$ with a
z-operator yields the very same operator $\hat{S}_{\sigma}$ up to the
scalar structure factor $f_{z\sigma}$. This factor can be zero,
\eg when $\hat{S}_{\sigma}$ refers to another z-operator as in
\Eq{eq:CR:Szz}, therefore $f_{zz'}=0$. For every z-operator,
however, there must exist \emph{at least} one \RLO $\hat{S}_{\sigma}$
with $f_{z\sigma} \neq 0$, since otherwise the group of
generators would be reducible.

With \Eq{eq:CR:Sz-pm}, the application of a generator
$\hat{S}_{\sigma}$ onto a symmetry eigenstate $\vert qn;q_{z}
\rangle$ yields
\begin{align}
    \hat{S}_{z} \cdot \hat{S}_{\sigma} \vert qn;q_{z} \rangle
 &= \underset{
        \overset{({\tiny \ref{eq:CR:Sz-pm}})}{=}
        f_{z\sigma}\hat{S}_{\sigma}
    }{\underbrace{[\hat{S}_{z},\hat{S}_{\sigma}]}}
    \!\!\vert qn;q_{z} \rangle
  + \hat{S}_{\sigma} \cdot
    \underset{=q_{z} \vert qn;q_{z} \rangle}{\underbrace{
    \hat{S}_{z} \vert qn;q_{z} \rangle }} \nonumber\\
 &= (q_{z}+f_{z\sigma}) \hat{S}_{\sigma} \vert qn;q_{z} \rangle
\text{.}\label{op-trafo-0}%
\end{align}
If $\hat{S}_{\sigma}$ is an \RLO with $f_{z\sigma} \neq 0$, the state
$\hat{S}_{\sigma} \vert qn;q_{z} \rangle$ is again a symmetry
eigenstate, yet with a uniform shift in the z-labels,
\begin{equation}
    q_{z} \rightarrow q_{z'} \equiv q_{z}+f_{z\sigma }
\text{.}\label{eq:Sop:zshift}
\end{equation}
Therefore the action of an \RLO $\hat{S}_{\sigma}$ in root space is
generic, \ie independent of the specific multiplet $q$ or the state
$q_z$ under consideration. Nevertheless, the \RLO may annihilate the
state, \ie $\hat{S}_{\sigma} \vert qn;q_{z} \rangle = 0$, which is
essential to obtain a finite-dimensional multiplet space.
Furthermore, \Eq{eq:Sop:zshift} allows to pair up raising and
lowering operators. That is, if $\hat{S}_{\sigma}$ is a raising
operator, then with
\begin{equation}
   [ \hat{S}_{z}, (\hat{S}_{\sigma})^\dagger ] =
 - [ \hat{S}_{z}, \hat{S}_{\sigma} ]^\dagger =
 - f_{z\sigma} (\hat{S}_{\sigma})^\dagger
\text{,}\label{eq:SpSz:rel}
\end{equation}
the operator $(\hat{S}_{\sigma})^\dagger$ changes the z-labels
exactly in the opposite direction as $\hat{S}_{\sigma}$ in
\Eq{eq:Sop:zshift}. In this sense, $(\hat{S}_{\sigma})^\dagger \equiv
\hat{S}_{-\sigma}$ represents the corresponding lowering operator.
The actual definition of what is a raising or lowering operator is
not entirely unique, as it depends on the specific underlying sorting
scheme of the z-labels adopted in root space. This does not matter,
however, as long as the sorting is done consistently throughout.
\cite{Pope06,Gilmore06}

In the presence of inner multiplicity a few complications arise. Most
importantly, an \RLO usually will generate a superposition in the
$m_{z'}$-fold degenerate state space in the resulting $q_{z'}$,
\begin{equation}
   \hat{S}_{\sigma} \vert qn; q_{z} \alpha_z\rangle
 = \sum_{\alpha_{z'}=1}^{m_{z'}}
   s_{q_{z} \alpha_z; q_{z'} \alpha_{z'}}^{[q \sigma ]}
   \vert qn;q_{z'}  \alpha_{z'}\rangle
\label{stateop:rise-lower}%
\end{equation}
with some coefficients $s_{q_{z} \alpha_z; q_{z'} \alpha_{z'}}^{[q
\sigma ]}$. As a consequence, the application of a raising operator
$\hat{S}_{\sigma}$ followed by its complimentary lowering operator
$\hat{S}_{\sigma}^{\dagger}$ onto a symmetry eigenstate,
\begin{equation}
   \hat{S}_{\sigma}^\dagger \hat{S}_{\sigma}^{\phantom{\dagger}} \vert qn;q_{z} \alpha_z\rangle
 = \sum_{\alpha_{z'}=1}^{m_{z}}
   s_{q_{z}; \alpha_z\alpha_{z'}}^{[q\sigma]}
   \vert qn;q_{z} \alpha_{z'}\rangle
\label{eq:SpSm-state}
\end{equation}
with some other coefficients $s_{q_{z}; \alpha_z
\alpha_{z'}}^{[q\sigma]}$, \emph{does} return to the same symmetry
labels $q_z$, yet \emph{not necessarily to the same state}. If the
resulting state in \Eq{eq:SpSm-state} does not replicate the initial
state $\vert qn;q_{z} \alpha_z \rangle$ up to an overall factor, then
this allows to explore the other states in the degenerate subspace at
$q_z$. This is relevant for the decomposition of state spaces, where
the resulting state as in \Eq{eq:SpSm-state} needs to be
orthonormalized in a consistent fashion with respect to the already
explored states of the multiplet including the state $\vert qn;q_{z}
\alpha_z \rangle$ (see \Secp{Sec:NumDeComp}, for more detail on the
numerical implementation).

While all z-operators are required, \eg for the definition of the
z-labels, it is usually not required to explicitly construct all of
the \RLOs, as some of these operators can be generated through a
product of a smaller set of \RLOs. As will be seen below in the case
of \SU{N} or \Sp{2n}, the number of actually required \RLOs can
always be reduced to the rank of the symmetry, \ie the number of
z-operators. This minimal set of \RLOs will be referred to as
\emph{simple} \RLOs, consistent with their general notation in the
literature as \emph{simple roots} of the symmetry. In a sense, these
simple \RLOs are the ones that induce the smallest shifts in the
z-labels. \cite{Pope06,Gilmore06} Again, their definition is not
entirely unique, depending on conventions such as normalization of
generators or what specific sorting scheme is applied to the
z-labels. The simple \RLOs still fully generate and connect the state
spaces of any \IREP. The underlying intuitive reason is that an
$r$-dimensional z-label structure only requires $r$ linearly
independent vectors to explore its space (for a rigorous proof, see
for example Refs.~\onlinecite{Pope06,Gilmore06}). Therefore given $r$
z-operators $\{ Z_{1},\ldots,Z_{r} \} $, it is sufficient to choose a
specific subset of $r$ raising operators $\{ S_{1+},\ldots,S_{r+}\}
$, with the corresponding lowering operators $S_{i-}\equiv
(S_{i+})^\dagger$. This reduction to simple \RLOs is very useful in
practice, yet does not restrict the non-abelian treatment in any way.

\subsubsection{Maximum-weight state \label{sec:maxweight}}

Consider some multiplet $q$ of internal dimension $d_q$ for a given
non-abelian symmetry group $\mathcal{S}$ of rank $r$. Then each of
the $d_{q}$ states carries a set of $r$ z-labels. When depicted
graphically as points in $r$-dimensional space, this is called the
\emph{weight diagram} for the multiplet [for \SU{3}, for example, a
collection of weight diagrams generated in an actual NRG run is shown
in \FigP{fig:SU3diags}{}]. Since the z-operators are traceless, the
values of the z-labels are naturally centered around the origin, \ie
$q_{z}=0$. Inner multiplicity, if present, decreases as a function of
distance $\vert q_{z}\vert $ to the origin, such that the outermost
points in a weight diagram always refer to unique states without any
remaining multiplicity. By choosing a \emph{lexicographic} ordering
in the $r$ z-labels, \cite{Elliott79} the \textit{maximum weight}
(MW) is defined by
\begin{equation}
    \qMW \equiv \max \bigl\{ q_z \bigr\}
\text{.}\label{MWlabel}
\end{equation}
The state with $q_z=\qMW$ is called the maximum-weight state. This
state is guaranteed to be unique to the multiplet for non-abelian
symmetries, \cite{Elliott79,Pope06,Gilmore06} hence can be used as
label for the entire multiplet, \ie $q=\qMW$. While the state space
$|q_z|=\max(|q_z|)$ will not be unique, in general, since it refers
to several states at the circumference of the weight diagram, $\max
\{q_z\} $ does provide a unique set of z-labels. This underlines the
importance of lexicographic ordering.

As an example, consider the well-known spin \SU{2}. The states within
the multiplet $q$ are labeled by $\vert qq_{z} \rangle $ where
$q_{z}=-q,\ldots,+q$ identifies each state within the multiplet. This
results in a one-dimensional weight diagram, with the multiplet
itself labeled by the maximum weight states, $q=\max(q_z)$, indeed.

Clearly, the q-labels for a multiplet themselves are also not
entirely unique and hence depend on convention. In particular, if the
rank of a symmetry group $\mathcal{S}$ is $r>1$, the order of the
z-operators themselves is a priori arbitrary. Hence there is a
certain freedom in the order of the z-labels, which in return affects
the definition of the maximum weight state. Given a certain order in
the z-operators then, the lexicographic sorting of sets of z-labels
is typically done in \emph{reverse} order, \ie starting with the
\emph{last} of the $r$ label for a given $q_z$. Moreover, having
identified \qMW, this still leaves the freedom to use a linearly
independent transform of \qMW as label for the entire multiplet for
consistency with literature. For example, for \SU{3} [\Sp{6}{}] this
is discussed with \Eq{SU3-labels} [\Eq{eq:MW:Sp6}{}], respectively.

\subsection{Example \SU{N}}

\subsubsection{Defining representation}

The symmetry \SU{N} is defined as the unitary symmetry of an
$N$-dimensional space. The defining representation, \ie the \IREP
with smallest non-trivial dimension, is therefore given by $N\times
N$ dimensional matrices. Since according to \Eq{conv:Straceless} all
generators are traceless, only $N-1$ diagonal z-operators exist, the
diagonals of which form an $N$-dimensional orthogonal vector space
that is also orthogonal to the diagonal of the identity matrix. The
raising (lowering) operators are chosen as $N\times N$ matrices with a
single entry of $1.$ anywhere in the upper (lower) triangular space,
respectively, away from the diagonal.
For this, let
\begin{subequations}\label{eq:DREP:basis}
\begin{equation}
   \vert e_{i} \rangle \equiv (0,\ldots,0,1_{(i)} ,0,\ldots,0)^T
\text{,}\label{eq:def:ei}
\end{equation}
with $i \in \{ 1,\ldots,N \} $
be the $N$-dimensional cartesian column basis vectors, and
\begin{equation}
   E_{ij} \equiv \vert e_{i} \rangle \langle e_{j} \vert
\text{,}\label{eq:def:Eij}
\end{equation}
\end{subequations}
the matrices of the related operator basis, which also contain just a
single entry of $1$ in their $N\times N$ dimensional matrix space,
\ie $(E_{ij})_{i'j'} = \delta_{ii'} \delta_{jj'}$. Then the
generators can be written as follows,
\begin{subequations}\label{eq:SUN-gen}
\begin{align}
   S_{i\neq j}^{\SU{N}}  &= E_{ij}
   =\left\{\begin{array}[c]{l}%
      \text{raising operator for }i<j\\
      \text{lowering operator for }i>j
      \end{array}
   \right. \label{eq:SUN-gen-pm} \\
   S_{_{z,k<N}}^{\SU{N}}  &= \bigl( \sum_{i=1}^{k} E_{ii} \bigr) -k E_{k+1,k+1}
\text{.}\label{eq:SUN-gen-z}%
\end{align}
\end{subequations}
These matrices are orthogonal as in \Eq{conv:Sortho}, while the
(arbitrary) normalization was chosen such that, for convenience, all
entries are integers. The choice of generators for \SU{N} in
\Eq{eq:SUN-gen} guarantees canonical \RLOs, and thus simplifies the
group's commutator relations \wrt z-operators exactly the way as
indicated in \Eq{eq:CR:Sz-pm}. This can be easily seen by observing
that for a diagonal operator of the type $(\hat{Z})_{ij} = z_{i}
\delta_{ij}$, the matrix elements of the commutator with an arbitrary
operator $(\hat{S})_{ij}=s_{ij}$ is given by
\[
   [ \hat{Z}, \hat{S} ]_{ij} = s_{ij}(z_{i}-z_{j})
\text{,}
\]
that is, existing non-zero matrix elements in $\hat{S}$ are weighted
by differences in diagonal elements of $\hat{Z}$, while there cannot
arise any new matrix elements unequal zero in $[\hat{Z},\hat{S}]$ as
compared to $\hat{S}$. Clearly, if $\hat{S}_\pm$ only has a single
non-zero entry as for the operators in \Eq{eq:SUN-gen-pm}, it follows
$[\hat{Z}, \hat{S}_\pm] = \mathrm{const} \cdot \hat{S}_\pm$, in
agreement with \Eq{eq:CR:Sz-pm}.

From \Eq{eq:SUN-gen-pm} above, a total of $\tfrac{1}{2}N(N-1)$
different raising operators arise. However, not all of these are
required to fully explore the multiplet space. Consider, for example,
the subset of $r=N-1$ raising operators
\begin{equation}
    \{S_{+}^{\SU{N}}\}_r \equiv \{\hat{S}_{12}^{\SU{N}}\!\!,\
    \hat{S}_{23}^{\SU{N}}\!\!, \ldots, \hat{S}_{N-1,N}^{\SU{N}} \}
\text{,}\label{eq:SUN:Sp}
\end{equation}
which thus matches the rank $r$ of the symmetry group \SU{N} and thus
also the number of z-operators. From repeated application of these
operators, it is easily seen that the remaining raising operators not
contained in \Eq{eq:SUN:Sp} can be generated. For example,
$\hat{S}_{13}^{\SU{N}}$ is generated by $\hat{S}_{12}^{\SU{N}} \cdot
\hat{S}_{23}^{\SU{N}}$. Therefore, above minimal set of $r$ raising
operators with their hermitian conjugate set of lowering operators is
sufficient, indeed, to explore all multiplet spaces.

\subsubsection{The symmetry \SU{3}}

\begin{figure}[tb]
\begin{center}
\includegraphics[width=1\linewidth]{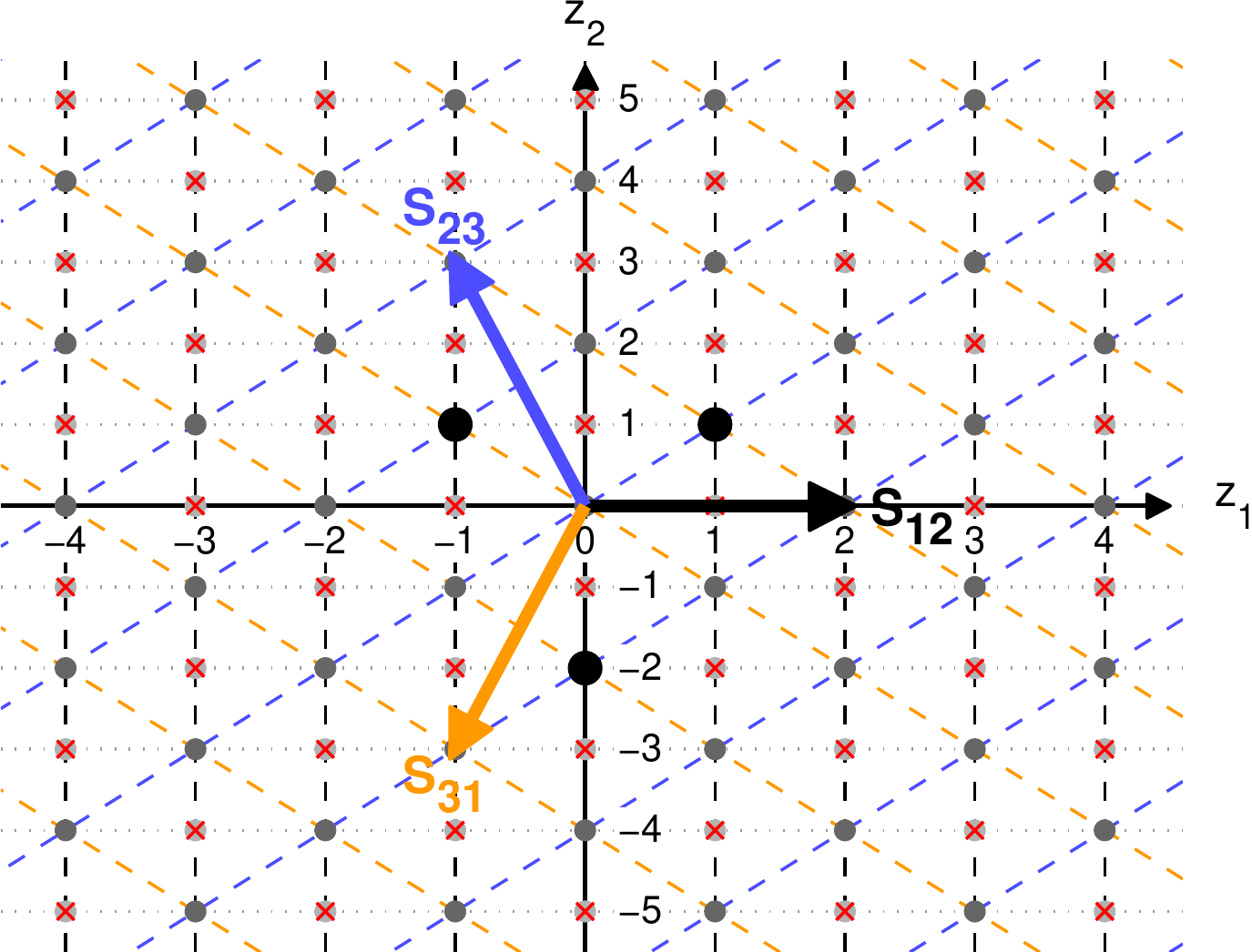}
\end{center}
\caption{(Color online)
Root space $(z_1,z_2)$ for \SU{3}. The three large black dots
depict the weight diagram of the three-dimensional defining
representation of \SU{3}. Large arrows indicate the shifts in
z-labels due to the action of the raising operators $S_{12}$,
$S_{23}$, and the lowering operator $S_{31}$, while dashed lines
close to orthogonal to these arrows indicate lines of constant
$S_{12}$, $S_{23}$, and $S_{31}$ (color match with corresponding
arrows). Dark symbols indicate accessible z-labels, while light
crossed-out symbols are not accessible within \SU{3} (see text). }
\label{fig_SU3}%
\end{figure}

The defining representation for \SU{3} is chosen as in
\Eq{eq:SUN-gen}, with the z-operators given by,
\begin{equation}
   Z_1 \equiv \left(\begin{array}{rrr}
       1 & \phantom{-}0 & 0 \\ 0 & -1 & 0 \\ 0 & 0 &  \phantom{-}0  \end{array}\right),\
   Z_2 \equiv \left(\begin{array}{rrr}
       1 & \phantom{-}0 & 0 \\ 0 &  1 & 0 \\ 0 & 0 & -2 \end{array}\right)
\text{.}\label{eq:SU3:zops}
\end{equation}
Their diagonals can be collected as rows into a matrix,
\begin{equation}
   z=\left(\begin{array}{rrr}
      1  & -1 &  0 \\
      1  &  1 & -2
   \end{array}\right)
\text{,}\label{eq:SU3:zmat}
\end{equation}
the columns of which give the z-labels $(z_1,z_2)$ for the three
states in the defining representation (see large black dots in
\Fig{fig_SU3}{}). This represents the weight diagram for the defining
representation.

The canonical commutator relations as in \Eq{eq:CR:Sz-pm} yield the
structure constants $(f)_{z,\sigma} \equiv f_{z\sigma}$ for
$z\in\{1,2\}$ and $\sigma\in\{12, 23, 31\}$,
\begin{equation}
    f = \left(\begin{array}{rrr}
      2  & -1 & -1  \\
      0  &  3 & -3
   \end{array}\right)
\text{.}\label{eq:SU3:roots}
\end{equation}
The columns in \Eq{eq:SU3:roots} thus define the roots, \ie the shift
in z-labels when applying either $S_{12}$, $S_{23}$, or $S_{31}$,
respectively. These vectors (roots) are depicted in \Fig{fig_SU3} by
large thick arrows. Clearly, the three points in the weight diagram
of the defining representation can be connected by these roots,
equivalent to (repeatedly) applying raising or lowering operators.

With the convention, that z-labels are lexicographically sorted
starting with the last z-label, \ie sorting \wrt $z_2$ first and then
$z_1$, the three states in the defining representation are already
properly sorted from largest to smallest [left to right in
\Eq{eq:SU3:zops}{}]. Furthermore, \Eq{eq:SU3:roots} shows that
$S_{12}$ and $S_{23}$ correspond to positive roots, since
$(2,0)>(0,0)$ and also $(-1,3)>(0,0)$. As their application makes
z-labels \emph{larger}, they represent raising operators, indeed,
while $S_{31}$ is a lowering operator, all in agreement with
\Eq{eq:SUN-gen-pm}. The third raising operator thus would be $S_{13}$
with root $(1,3)$ which, however, is not a simple root and hence can
be dropped.

Finally, \SU{3} still contains well-known \SU{2} subalgebras. That
is, for example, by using $S_{12}$ as a raising operator for the
$(x,y)$ subspace together with its corresponding z-operator
$[S_{12},S_{12}^\dagger] =: 2S_z^{(12)} = Z_1$ while keeping the $y$
component abelian, this shows that every \emph{line of points} in the
$(z_1,z_2)$ plane in \Fig{fig_SU3} parallel to $S_{12}$ must
correspond to a proper \SU{2} multiplet. The same also holds for the
two remaining permutations of $(x,y,z)$ using $S_{23}$ or $S_{31}$
for the \SU{2} subspace, instead. These \SU{2} subalgebras clearly
obey the standard commutator relations for \SU{2}.

\subsubsection{Symmetry labels for \SU{3}
\label{sec:SU3:symlabels}}

The q-labels for a given \IREP within \SU{3} are derived from its
maximum-weight labels $\qMW \equiv \max\{(z_{1},z_{2})\}$. With the
z-labels additive through tensor products (see latter in the
appendix), the z-labels of \emph{arbitrary} multiplets must be
integer multiples of the z-labels of the defining representation.
This immediately excludes the z-labels (points) in \Fig{fig_SU3} that
are crossed out. In particular, with the columns of \Eq{eq:SU3:zmat}
being linearly dependent, one may therefore use the columns of
\begin{equation}
   \tilde{z} = \left(\begin{array}{rr}
      1  &  0 \\
      1  &  2
   \end{array}\right)
\text{,}\label{eq:SU3:ztilde}
\end{equation}
as basis for the maximum weight labels, for consistency with
literature. \cite{Elliott79,LiE92} Given $\qMW$, the actual label of
the multiplet then is determined by
\begin{equation}%
   q \equiv (q_{1},q_{2}) \equiv \tilde{z}^{-1} \cdot \qMW
 = \left(\begin{array}{rr}
      1  &  0 \\
     -\tfrac{1}{2}  &  \tfrac{1}{2}
   \end{array}\right) \qMW
\text{.} \label{SU3-labels}%
\end{equation}
This prescription makes the q-labels independent of the specific
normalization conventions chosen for the z-operators. Furthermore,
with $\tilde{z} = z \cdot (1,0,-1)^T$ and the vectors in the columns
in \Eq{eq:SU3:ztilde} being positive by the adopted sorting scheme,
this guarantees plain positive integers for the multiplet labels $q$.
These labels also lie dense, in the sense that any $(q_1,q_2)$ with
$q_i\ge0$ results in a valid multiplet.

The defining representation with $\qMW=(1,1)$ has the q-label
$(1,0)$. Its weight diagram together with many further examples for
multiplets, as generated, in practice, from an actual NRG run using
\SU{3}, are presented in \FigP{fig:SU3diags}. Note, however, that
weight diagrams are mainly a matter of presentation of multiplets,
while in practice a listing of z-labels suffices to describe the
multiplet space.

\subsection{Decomposition into irreducible representations\label{sec:irrep}}

The generators of a specific symmetry group $\mathcal{S}$ represent
an irreducible finite set of operators $\{\hat{S}_\sigma\}$, assumed
to act in the full Hilbert space of a given physical system. Within
(small) subspaces of the system, finite dimensional matrix
representations can be constructed that obey exactly the same
commutator relations as the generators in terms of their structure
constants in \Eq{eq:Sop:CR}. As such, a given matrix representation
$\{R_{\sigma}\}$ inherits all the properties of the generators. In
particular, the matrix representation has the same number of
operators as $\{\hat{S}_\sigma\}$ with a one-to-one correspondence in
the symmetry label $\sigma$. Therefore the z-operators as well as the
\RLOs share exactly the same interpretation within the
$D$-dimensional carrier space of $\{R_{\sigma}\}$.

Consider some arbitrary matrix representation $\{R_{\sigma}\}$ that
may have emerged, for example, from a tensor product space. As it
operates in a $D$-dimensional carrier space, all of its matrices
share the same dimension $D\times D$. Assume a well-defined symmetry
eigenstate within this space is available, to be called \textit{seed
state}, with a typically easy example being a maximum weight state.
Then repeated application of \RLOs from the set $\{R_{\sigma}\}$
generates a (sub)space which eventually describes a full symmetry
multiplet, \ie an \IREP. By construction this subspace already
diagonalizes the z-operators. Thus the z-labels are known, which also
provides the q-labels for the multiplet, \eg by simply taking the
maximum weight labels, $q_1=\qMW$.

If this multiplet $q_1$ with $d_{q_1}$ symmetry eigenstates spans the
entire $D$-dimensional carrier space, then the matrix representation
$\{R_{\sigma }\}$ is already irreducible. If only a subspace of the
$D$-dimensional carrier space was generated, \ie $d_{q_1}<D$, the
matrix representation $\{R_{\sigma }\}$ is reducible. Multiplet $q_1$
then defines a fully separated space, given the symmetry operations
in $\{R_{\sigma}\}$. Combining the orthonormal state space of
multiplet $q_1$ as columns into a matrix $V_1$, the matrix
representation $\{R_{\sigma}\}$ can be cast into the space of
multiplet $q_1$, $R_{\sigma} \to I^{[q_1]} \equiv V_1^\dagger
R_{\sigma} V_1$, which thus constructs the irreducible matrix
representation $I^{[q_1]}$ for \IREP $q_1$.

In case the $D$-dimensional vector space is not exhausted yet, above
procedure can be repeated with another seed state within the
remainder of the vector space, generating further irreducible
multiplets $q_2,\,q_3,\ldots$, until the $D$-dimensional vector space
is fully exhausted. By combining the state spaces of the multiplets
thus generated, the resulting unitary matrix $V\equiv
[V_1,V_2,\ldots]$ allows to block-decompose the original matrix
representation $\{R_{\sigma }\}$ in terms of its irreducible
representations,
\begin{equation}
   V^\dagger R_{\sigma}V = \bigoplus_{q} M_{q} I_{\sigma}^{[q]}
\text{.} \label{irrep-decomp}
\end{equation}
where $q$ runs through all \IREPs $I^{[q]}$. Note that a given \IREP
may be generated multiple times in the decomposition, which is
indicated by the \textit{outer multiplicity} $M_q \in
[0,1,2,\ldots]$. The presence of outer multiplicity therefore refers
to the situation that $M_{q}>1$ for at least one $q$ in the
decomposition. In this case, also inner multiplicity may occur, which
refers to non-uniqueness of z-labels within an irreducible multiplet
[\cf \Eq{inner-mult}{}], both of which are specifically relevant, for
example, for \SU{N>2} or \Sp{2n>2}.

As seen from above construction, the matrix representation $I^{[q]}$
of \IREP $q$ is tightly connected to the symmetry multiplet $q$. In
general, $I^{[q]}$ is unique only up to a global similarity
transformation, as this does not affect commutator relations. By
using its related multiplet state space, however, this space (i) can
be chosen such that it diagonalizes all z-operators, and (ii) can put
into a well-defined order as provided, for example, by the
lexicographic ordering in the z-labels used to define the maximum
weight state. Based on this basis, the matrix representation
$I^{[q]}$ can be determined uniquely. This procedure on obtaining
unique irreducible matrix representations will be adopted throughout.

The decomposition in \Eq{irrep-decomp}, finally, can be done fully
numerically along the same lines as already sketched above.
Particular attention, however, must be paid to issues related to
inner and outer multiplicity for overall consistency. This will be
discussed in more detail in \Sec{Sec:NumDeComp}.

\subsection{Tensor product spaces\label{sec:prodspaces}}

Consider two irreducible matrix representations $I^{[q_1]}$ and
$I^{[q_2]}$ of some non-abelian symmetry group $\mathcal{S}$, with
their matrix elements written in the basis of the symmetry
eigenstates $\vert q_1 q_{1z} \rangle$ and $\vert q_2 q_{2z} \rangle$
of the two \IREPs $q_1$ and $q_2$, respectively. The two multiplets
are assumed to live in different spaces, so they can be joined
through a tensor product, \ie $\vert q_1 q_{1z} \rangle \vert q_2
q_{2z} \rangle \equiv \vert q_1 q_{1z}; q_2 q_{2z} \rangle$. Then the
generators of the symmetry in the combined space are defined in an
\textit{additive} fashion, which derives from the origin of the
generators in infinitesimal symmetry operations, \cf \Eq{eq:gensymm},
\begin{equation}
    R_{\sigma}^{\mathrm{tot}} \equiv
    I_{\sigma}^{[q_1]} \otimes \Id^{[q_2]}
  + \Id^{[q_1]} \otimes I_{\sigma}^{[q_2]}
\text{.}\label{eq:add-sym-op}%
\end{equation}
Note that the additivity of the symmetry generators \textit{directly}
also implies the additivity of z-labels for non-abelian symmetries in
general. And even if the non-abelian part of the \SU{N} symmetry is
broken, \eg reduced to an abelian symmetry with quantum labels
$q_{z}$, these are, of course, still additive.

By construction, the tensor product representation $\{R_{\sigma}
^{\mathrm{tot}}\}$ in \Eq{eq:add-sym-op} is also a representation of
the symmetry, as it obeys the same commutator relations within the
combined system as the \IREPs $I_{\sigma}^{[q_i]}$ within their
individual space,
\begin{align}
   [ R_{\sigma^{{}}}^{\mathrm{tot}}, R_{\sigma'}^{\mathrm{tot}} ]
&= \underset{ =\sum\limits_{\mu} f_{\sigma\sigma'\mu} R_{\mu}^{[q_1]}
   }{\underbrace{ [ R_{\sigma^{{}}}^{[q_1]}, R_{\sigma'}^{[q_1]} ] }}
   \otimes\Id^{[q_2]}
 + \Id^{[q_1]} \otimes
   \underset{ =\sum\limits_{\mu} f_{\sigma\sigma'\mu} R_{\mu}^{[q_2]}
   }{\underbrace{ [R_{\sigma^{{}}}^{[q_2]}, R_{\sigma'}^{[q_2]}] }} \nonumber \\
&= \sum_{\mu} f_{\sigma\sigma'\mu} R_{\mu}^{\mathrm{tot}}
\text{.}\label{tprod-rep:cr}%
\end{align}
The product representation $\{R_{\sigma}^{\mathrm{tot}}\}$, however,
is typically reducible. The resulting decomposition into \IREPs is
done exactly the same way as in \Eq{irrep-decomp}.

The unitary transformation that rotates the product space $\vert
q_{1} q_{1z};q_{2}q_{2z} \rangle $ into its combined symmetry
multiplets $\vert q q_{z}; (\alpha)\rangle$ is given by the
Clebsch-Gordan coefficients (CGC),
\begin{equation}
   \vert q q_{z}; (\alpha) \rangle
 = \sum_{q_{1z},q_{2z}} \vert q_{1}q_{1z};q_{2}q_{2z} \rangle
   \underset{
     \equiv C_{q_{1z} q_{z}; (\alpha)}^{[q_{2z}]}
   }{\underbrace{
     \langle q_{1}q_{1z};q_{2} q_{2z}| q q_{z}; (\alpha) \rangle
   }}
\text{,} \label{def:clebsch}%
\end{equation}
with the shorthand notation $C_{q_{1z} q_{z}; (\alpha)} ^{[q_{2z}]}$
for CGCs, consistent with the MPS tensors in the main body of the
paper [\cf \Eqp{eq:CGCalpha}]. Note that the CGCs implicitly also
carry the multiplet labels $q_1$, $q_2$, and $q$.
The index $\alpha$ has been added to account for possible
outer-multiplicity [\cf \Eqt{irrep-decomp}{}], in that for input
multiplets $q_{1}$ and $q_{2}$ the same output multiplet $q$ can
appear $M^{[q_1,q_2]}_{q}$ times, therefore $\alpha=1, \ldots,
M^{[q_1,q_2]}_{q}$ for a given $q$. If outer multiplicity is absent,
the index $\alpha$ can be omitted, hence the round brackets around
$\alpha$ in \Eq{def:clebsch} or \Eq{eq:clebsch:alpha:ortho}.

As outer multiplicity also refers to \emph{different} multiplets and
hence state spaces, the Clebsch-Gordan coefficients, reflecting a
unitary transformation, obey the general orthogonality condition,
\begin{equation}
   \sum_{q_{1z} q_{2z}}
   C_{q_{1z}q_{z};(\alpha)}^{[q_{2z}]}
   C_{q_{1z}q'_{z};(\alpha')}^{[q_{2z}]}
 = \delta_{q_{z}, q'_{z}} \cdot \bigl(\delta_{\alpha,\alpha'}\bigr)
\text{.}\label{eq:clebsch:alpha:ortho}%
\end{equation}
This holds within the same output multiplet $q$, whereas the overlap
between different output multiplets is strictly zero. While outer
multiplicity is intrinsically connected to the underlying symmetry
and hence to CGCs, in addition, this also affects the output
multiplet space which must accommodate the additional multiplets [\eg
see \QSpace discussion in the main text; note also that
\Eq{eq:clebsch:alpha:ortho} is completely analogous in structure to
the orthogonality relation of \Atensors as in \Eq{eq:Aortho}{}].

\subsection{Irreducible operator sets and Wigner-Eckart theorem
\label{sec:irop+wet}}

Consider a set of generators $\{\hat{S}_{\sigma}\}$ of some symmetry
group $\mathcal{S}$ that a Hamiltonian $\hat{H}$ commutes with. Then
all energy eigenstates of the Hamiltonian can be categorized with
well-defined quantum-labels, as indicated in \Eq{eq:Hsym:eig}. In
order to maintain an effective book keeping of the quantum labels
when calculating matrix elements of operators, it must be possible to
similarly categorize the operators themselves. Typically, the
operators of interest are closely related to the Hamiltonian, \ie
consist of operators that also appear in the Hamiltonian or are
composites thereof, such as creation, annihilation, occupation, spin
operators, etc. Since the Hamiltonian can be properly constructed
within the given symmetry, so can be its constituents.

An \emph{irreducible operator set} is constructed in a completely
analogous fashion as an irreducible state space, with an explicit
example already derived for the spin operator in \Eq{spin-irop} using
\Eq{eq:CR-SpinOps}. Consider the generic setup of a set of generators
$\{\hat{S}_{\sigma}\}$ including \RLOs. Then
irreducible state multiplets can be generated through iterative
application of these operators,
\begin{equation}
   \hat{S}_{\sigma} \vert qq_{z} \rangle
 = s_{q_{z}q_{z}^{\prime}}^{[q \sigma]} \vert qq_{z}^{\prime} \rangle
\text{,}\label{eq:irrep-state}%
\end{equation}
as in \Eq{stateop:rise-lower}, while ignoring inner multiplicity for
the sake of the argument and having dropped the energy multiplet
index $n$ for simplicity. Given an operator $\hat{F}$, on the other
hand, its transformation according to a symmetry is fully reflected
in its \emph{commutator relations} with the generators of the
symmetry. This is easily motivated through infinitesimal symmetry
operations as in \Eq{eq:SymOp:irop}. The commutator relations, on the
other hand, also emerge naturally when analyzing the effect of a
generator of the symmetry acting onto a symmetry state $\vert qq_{z}
\rangle$ that already also has the operator $\hat{F}$ applied to it,
\begin{equation}
   \hat{S}_{\sigma} \cdot \hat{F} \vert qq_{z}^{\prime}\rangle
 = [ \hat{S}_{\sigma} , \hat{F} ] \vert qq_{z}^{\prime}\rangle
 + \hat{F} \cdot \hat{S}_{\sigma} \vert qq_{z}^{\prime}\rangle
\text{.}\label{eq:op-trafo}
\end{equation}
The second term on \rhs clearly describes the symmetry properties of
the state $\vert qq_{z} \rangle$, while the first term yields the
transformation properties of the operator $\hat{F}$ which are
independent of the carrier space. This is similar to what has already
been seen in \Eq{op-trafo-0} for the combined action of two
generators.

Now \emph{iff} an operator set $\hat{F}^q \equiv \{ \hat{F}^q_{q_{z}}
\}$ transforms exactly the same way as the state space of \IREP $q$
in \Eq{eq:irrep-state}, that is
\begin{equation}
   [ \hat{S}_{\sigma}, \hat{F}^q_{q_{z}} ]
 = s_{q_{z}q_{z}^{\prime}}^{[q\sigma]} \hat{F}^q_{q_{z}^{\prime}}
\text{,}\label{def:irrep-op}
\end{equation}
then the operator set $\hat{F}^q$ is called an \textit{irreducible
operator} (\IROP) set that transforms like the multiplet $q$. It
carries the symmetry labels $(q,q_{z})$ the same way as an
irreducible state multiplet does.

\subsubsection{\IROP decomposition \label{sec:irop-decomp}}

In the case that a specific member of the \IROP set is already known,
then \Eq{def:irrep-op} allows to generate the full \IROP set exactly
the same way as a state multiplet can be generated. Again, the
maximum weight label determined the multiplet $q$ that the \IROP
represents. This was exactly the procedure adopted, for example, to
obtain the spin operator in \Eq{spin-irop}. Furthermore,
\Eq{spin-irop} also serves as a simple demonstration that the space
of generators itself clearly also can be cast into a single \IROP.
The corresponding multiplet is called the \emph{regular
representation} then.

More generally, it is instructive to realize that irreducible
operator sets (\IROPs) and symmetry multiplets (\IREPs) can be
treated on a nearly equal footing. In particular, the notion of
proper orthonormalization of state spaces can be directly applied
also to \IROP sets, up to a global normalization factor. This is
motivated by the observation, that given a scalar multiplet $\vert
0\rangle$ for which $\vert F^q_{q'_z}\rangle \equiv F^q_{q'_z} \vert
0\rangle \neq 0$, \ie does not vanish, then $\vert F^q_{q'_z}\rangle$
represents the multiplet \emph{vector space} for \IREP $q$. With
proper overall normalization of the \IROP $F^q$, it follows
\[
  \delta_{q_z,q'_z} = \langle F^q_{q_z} \vert F^q_{q'_z}\rangle
  = \langle 0\vert F^{q\dagger}_{q_z} F^q_{q'_z}\vert 0 \rangle
\text{.}
\]
The last equation also holds, if the scalar multiplet $\vert 0
\rangle$ is replaced by an arbitrary other symmetry eigenstate $\vert
q q_z\rangle$. For matrix representations of \IROPs and operators
more generally, this motivates the scalar or inner product for two
matrices as in \Eq{eq:fro:scalar}. Thus equipped with scalar product
and norm for matrices, an \IROP decomposition can be done
\emph{exactly } the same way as the multiplet decompositions for
symmetry multiplets starting from a specific symmetry eigenstate
(\IROP component). This is important, in particular, in the presence
of inner multiplicity in the multiplet of an \IROP for consistency
with the Wigner-Eckart theorem.

\subsubsection{Wigner Eckart theorem}

It follows from \EQS{eq:op-trafo}{def:irrep-op}, that the states
resulting from the \IROP $\{ \hat{F}^{q_{1}}_{q_{1z}} \}$ applied to
a multiplet $\vert q_{2} q_{2z} \rangle $,
\begin{align*}
  & \hat{S}_{\sigma} \cdot \hat{F}^{q_{1}}_{q_{1z}}  \vert q_{2}q_{2z} \rangle \\
  &=\underset{
      =s_{q_{1z}^{{}} q_{1z}^{\prime}}^{[q_{1}\sigma]}
       \hat{F}^{q_{1}}_{q_{1z}^{\prime}}
    }{\underbrace{ [ \hat{S}_{\sigma}, \hat{F}^{q_{1}}_{q_{1z}}  ]
    }}
    \vert q_{2}q_{2z} \rangle + \hat{F}^{q_{1}}_{q_{1z}} \cdot
    \underset{
       =s_{q_{2z}^{{}}q_{2z}^{\prime}}^{[q_{2}\sigma]} \vert q_{2}q_{2z}^{\prime} \rangle
    }{\underbrace{\hat{S}_{\sigma}\left\vert q_{2}q_{2z}\right\rangle }}
\text{,}%
\end{align*}
transforms exactly the same way under given symmetry as a tensor
product of two state multiplets,
\begin{align*}
   & \hat{S}_{\sigma} \cdot \vert qq_{z} \rangle_{1} \vert qq_{z}\rangle_{2} \\
   &=\underset{
      =s_{q_{1z}^{{}}q_{1z}^{\prime}}^{[q_{1}\sigma]} \vert qq_{z}^{\prime}\rangle _{1}
   }{\underbrace{
      \hat{S}_{1\sigma} \vert qq_{z} \rangle _{1}
   }} \otimes \vert qq_{z} \rangle_{2}
   + \vert qq_{z} \rangle _{1} \otimes
   \underset{
     =s_{q_{2z}^{{}} q_{2z}^{\prime}}^{[q_{2}\sigma]} \vert qq_{z}^{\prime}\rangle_{2}
   }{\underbrace{\hat{S}_{2\sigma} \vert qq_{z}\rangle_{2}}}
\text{,}%
\end{align*}
using \EQS{eq:add-sym-op}{eq:irrep-state}. Therefore the action of an
\IROP $\hat{F}^{q_{1}}$ onto the state space of an \IREP $q_2$ shares
exactly the same algebraic structure in terms of symmetries like the
product space of the two multiplets $q_1$ and $q_2$.

This motivates the Wigner-Eckart theorem. With the definition of the
Clebsch-Gordan coefficients in \Eq{def:clebsch}, it is thus clear
that up to scalar factors depending on the normalization of the
operator set, the \emph{same} CGCs also apply for the state space
decomposition arising out of $\hat{F}^{q_{1}}_{q_{1z}} \vert
q_{2}q_{2z} \rangle$. In particular, it follows for the matrix
elements of the operator \wrt a given state space,
\begin{align}
   \langle q q_{z};(\alpha) \vert \hat{F}^{q_{1}}_{q_{1z}} \vert q_{2}q_{2z}\rangle &
   \equiv \langle q q_{z};(\alpha) \vert \cdot
   \bigl( \hat{F}^{q_{1}}_{q_{1z}}  \vert q_{2}q_{2z}\rangle \bigr) \nonumber\\
&= \langle q;(\alpha)\Vert \hat{F}^{q_{1}} \Vert q_{2}\rangle \cdot
   C_{q_{1z} q_{z}; (\alpha)}^{[q_{2z}]}
\text{,}\label{WET:motiv}%
\end{align}
where, again, $\alpha$ accounts for possible outer multiplicity.
$\langle q;(\alpha)\Vert\hat{A}_{q_{1}}\Vert q_{2}\rangle$ is called
the \textit{reduced matrix element}. It is entirely independent of
the z-labels, \ie the internal structure of the \IREPs$q_{1}$,
$q_{2}$, and $q$.

The first line in \Eq{WET:motiv} specifies the adopted convention for
matrix elements given the Wigner-Eckart theorem: the operator is
acting to the \emph{right} ket-state, the symmetry labels of which
are combined. The resulting object is contracted with the bra-states.
This is important for consistency, since the \IROP $\hat{F^q}$ is
subtly different from the \IROP $(\hat {F}^{\dagger})^q$. Therefore
one \emph{must} be careful with expressing a matrix element through
$\langle q q_{z} \vert \hat{F} \vert q_{2}q_{2z}\rangle = \langle
q_{2}q_{2z} \vert \hat{F}^{\dagger} \vert q q_{z}\rangle^{\ast}$.
Even though usually $(\hat{F^q_{q_z}})^\dagger \sim
(\hat{F}^\dagger)^q_{-q_z}$, further signs may bee needed to ensure
for consistency within the Clebsch-Gordan coefficients [\eg see
discussion around \Eqt{eq:IROP-psiS2} later].

\subsection{Several independent symmetries\label{sec:multsym}}

A physical system often exhibits several symmetries. Each of the
$\lambda = 1,\ldots,\NS$ symmetries $\mathcal{S}^{\lambda}$ is
completely described by its own set of generators $\{
\hat{S}_{\sigma}^{\lambda} \}$. As these symmetries act independently
of each other, this implies that their generators must commute,
\begin{equation}
   [ \hat{S}_{\sigma^{{}}}^{\lambda}, \hat{S}_{\sigma^{\prime}}^{\lambda^{\prime}} ]
   =0 \qquad \text{for } \lambda\neq \lambda^{\prime}
\text{.}\label{eq:multSymm:CR}
\end{equation}
This allows to assign independent quantum labels $(q^{\lambda}
q_{z}^{\lambda})$ with respect to each individual symmetry [\cf
discussion following \Eqp{eq:allsymm} in the main paper]. On the
multiplet level, the symmetries are given compactly by the combined
\emph{q-labels}, $q \equiv (q^{1}, q^{2}, \ldots, q^{\NS})$, while
similarly their \emph{z-labels} are given by $q_{z} \equiv
(q_{z}^{1},q_{z} ^{2}, \ldots, q_{z}^{\NS})$. Here the elementary
multiplet labels $q^{\lambda}$ and $q_{z}^{\lambda}$ can already
consist of a set of labels themselves, the number of which is
determined by the rank $r$ of the respective symmetry
$\mathcal{S}^{\lambda}$ [\cf \Eq{Zop:diag}{}].

When a non-abelian symmetry is broken, it it is reduced to simpler
subalgebras. In particular, it may be reduced to its abelian core of
z-operators (Cartan subalgebra). For example, consider the rotational
spin \SU{2} symmetry. This symmetry can be broken by applying a
magnetic field. The system still maintains, however, a continuous
rotational symmetry around the axis of the magnetic field, leaving
the $q_{z}$ symmetry intact, while the multiplet label $q$ becomes
irrelevant. Similarly, if particle-hole symmetry (see later) is
broken, only the abelian quantum number of total charge [\ie the
z-label] remains.

Abelian symmetries therefore fit seamlessly into the general
non-abelian framework outlined in this paper. With the multiplet
label $q$ irrelevant, the $q_z$ are promoted to the status of a
q-label, instead, with no z-labels remaining [with all multiplets
being one-dimensional, the z-labels are no longer required, \ie can
be set to zero, for simplicity]. This then allows to write the
abelian symmetry in terms of trivial scalar Clebsch-Gordan
coefficients. The latter, nevertheless, are important as they account
for the proper addition rules \wrt the abelian z-labels,
\begin{equation}
   \langle q_{1(z)}q_{2(z)} \vert q_{(z)}\rangle
 = \delta_{q_{(z)},q_{1(z)}+q_{2(z)}}
\text{.}\label{clebsch:abelian}%
\end{equation}

\subsection{Symmetries in physical systems}

In the following, several examples of symmetries in simple physical
systems will be given, with the associated spinors and irreducible
operator sets explained in detail. In particular, this concerns
fermionic systems with spin or particle-hole symmetry.

For the model Hamiltonians in strongly correlated electron systems,
correlation through interaction plays an important role, while the
terms describing interaction typically preserve certain underlying
global symmetries. Since the arguments of demonstrating symmetries of
a specific Hamiltonian, however, are rather similar, in general, it
suffices to consider a simple non-interacting Hamiltonian. Simple
issues related to interactions are discussed with \Eq{U:ph-symmetric}
below.

For simplicity, therefore much of the following discussion will be
exemplified in terms of the Hamiltonian of a plain spinful fermionic
tight-binding chain,
\begin{equation}
   \hat{H} = \sum_{k} t_k \
   \underset{\equiv \hat{h}_{k,k+1} }{\underbrace{
   \sum_{\sigma}\bigl(
      \hat{c}_{k\sigma}^{\dagger} \hat{c}_{k+1,\sigma}^{\phantom\dagger} + \hc
   \bigr) }}
\text{,}\label{H:tb-default}
\end{equation}
where $\hat{c}_{k\sigma} ^{\dagger}$ creates a particle at site $k$
with spin $\sigma \in \{\uparrow, \downarrow\}$. The Hamiltonian in
\Eq{H:tb-default} has spin-independent hopping amplitudes $t_k$,
hence possesses spin-\SU{2} symmetry, \SU[spin]{2} in short.
Furthermore, it is particle-hole symmetric, implying particle-hole
\SU{2} symmetry, also called charge-\SU{2} symmetry, or
\SU[charge]{2} in short.

\subsubsection{\SU{2} spin symmetry}

Using the two-dimensional spinor
\begin{equation}
   \hat{\psi}_{S,k} \equiv
   \begin{pmatrix} \hat{c}_{k\uparrow} \\ \hat{c}_{k\downarrow} \end{pmatrix}
\label{spinor2S}
\end{equation}
for each site $k$, the Hamiltonian in \Eq{H:tb-default} can be
rewritten as
\begin{equation}
   \hat{H} = \sum_{k} t_k \bigl(
   \hat{\psi}_{S,k}^{\dagger} \cdot \hat{\psi}_{S,k+1}^{\phantom\dagger}
   + \hc \bigr)
\text{,}\label{H:tb-psiS}
\end{equation}
where the sum over $\sigma$ was incorporated in the scalar product of
the vector of operators in $\hat{\psi}_{S,k}$. Clearly, the
two-dimensional scalar product is invariant under an arbitrary
unitary two-dimensional transformation $U$, \ie $\psi_{k}^{\dagger}
\psi_{k+1}^{{}} = (U\psi_{k})^{\dagger} (U\psi_{k+1})$, thus
exhibiting spin-\SU{2} symmetry. The spinor in \Eq{spinor2S} is
defined in a site specific manner. When concentrating on a single
site, therefore the index $k$ can be dropped for convenience.

The generators of spin-\SU{2} symmetry are constructed in terms of
the two-dimensional defining representation of $\{S_\sigma\} \equiv
\{S_{+}, S_{z}, S_{-} \}$ [\cf \Eq{eq:SU2-PM}{}]. These can be
written as operators (distinguished by the hat) through second
quantization in the full Hilbert space,
\[
   \hat{S}_{\sigma} =
   \hat{\psi}_{S}^{\dagger}S_{\sigma} \hat{\psi}_{S}^{\phantom\dagger}
\text{,}
\]
which up to prefactors leads to the spin \IROP $\hat{S}^{1} \equiv \{
-\tfrac{1}{\sqrt{2}} \hat{S}_{+}; \hat{S}_{z}; +\tfrac{1}{\sqrt{2}}
\hat{S}_{-}\}$, already derived in \Eq{spin-irop}. The raising
operator, for example, is given by
\begin{equation*}
    \hat{S}_{+} = \hat{\psi}_{S}^{\dagger}
    \left( \begin{array}[c]{cc}
       0 & 1 \\
       0 & 0
    \end{array} \right)
    \hat{\psi}_{S}
  = \hat{c}_{\uparrow}^{\dagger} \hat{c}_{\downarrow}^{\phantom\dagger}
\text{,} %%\label{op:S+}
\end{equation*}
which flips a down-spin to an up-spin for given site. Similarly, the
z-operator is given by
\begin{align*}
    \hat{S}_{z}
 &= \hat{\psi}^{\dagger} (\tfrac{1}{2} \tau_{z}) \hat{\psi}
  = \tfrac{1}{2}(\hat{c}_{\uparrow}^{\dagger} \hat{c}_{\uparrow}^{\phantom\dagger}
     - \hat{c}_{\downarrow}^{\dagger} \hat{c}_{\downarrow}^{\phantom\dagger})
  \equiv
    \tfrac{1}{2} ( \hat{n}_{\uparrow}-\hat{n}_{\downarrow} )
\text{.} %%\label{op:Sz}%
\end{align*}

Furthermore, $
    [ \hat{S}_{+}, \hat{c}_{\downarrow}^{\dagger} ]
  = [ \hat{c}_{\uparrow}^{\dagger} \hat{c}_{\downarrow}^{\phantom\dagger},
      \hat{c}_{\downarrow}^{\dagger} ]
  = \hat{c}_{\uparrow}^{\dagger}
$ shows that the spinor $\hat{\psi}_{S}^{\dagger}$ already represents
an \IROP for the $q=\tfrac{1}{2}$ multiplet of \SU[spin]{2},
\begin{subequations}\label{eq:IROP-psiS2}
\begin{equation}
     (\hat{\psi}^\dagger_S)^{[1/2]} =
     \begin{pmatrix}
        \hat{c}^\dagger_{\uparrow} \\
        \hat{c}^\dagger_{\downarrow}
     \end{pmatrix}
\text{.}\label{eq:IROP-psiS2-1}
\end{equation}
This is already properly sorted \wrt z-labels, in that the second
component correspond to the lower $q_z=-\tfrac{1}{2}$ element of the
multiplet, since $[ \hat{S}_{z}, \hat{c}_{\downarrow}^{\dagger} ] =
(-\tfrac{1}{2}) \cdot \hat{c}_{\downarrow}^{\dagger}$.

In contrast, the \IROP for the spinor $\hat{\psi}_{S}$, \ie
\emph{without} the dagger, is similar, yet has subtle differences.
In particular, with %
$ [ \hat{S}_{+}, \hat{c}_{\uparrow}^{\phantom\dagger} ] =
  [ \hat{c}_{\uparrow}^{\dagger} c_{\downarrow}^{\phantom\dagger},
    \hat{c}_{\uparrow}^{\phantom\dagger} ] =-\hat{c}_{\downarrow}^{\phantom\dagger}
$, the role of spin within the multiplet is reversed, \ie $q_z \to
-q_z$, while also an additional sign is acquired,
\begin{equation}
     (\hat{\psi}_S)^{[1/2]}=
     \begin{pmatrix}
       -\hat{c}_{\downarrow} \\
        \hat{c}_{\uparrow}
     \end{pmatrix}
\text{.}\label{eq:IROP-psiS2-2}
\end{equation}
\end{subequations}
This extra sign is important in context of the Wigner-Eckart theorem
in \Eq{WET:motiv}, where the particular order of first applying, \ie
combining an operator with the ket-state is directly related to the
order in the Clebsch-Gordan coefficients. This is convention, of
course, but consistency is crux.

In terms of the proper \IROPs in \Eqs{eq:IROP-psiS2}, finally, the
Hamiltonian in \Eq{H:tb-default} can be written in either \IROP
while, however, one must not mix them,
\begin{subequations}\label{H:tb:IROP}
\begin{eqnarray}
   \hat{H}
&=& \sum_{k} t_k \Bigl( \bigl[
   (\hat{\psi}_S)^{[1/2]} \bigr]^\dagger \cdot
   (\hat{\psi}_S)^{[1/2]}
   + \hc \Bigr) \qquad\label{H:tb:IROP-1} \\
&=& \sum_{k} t_k \Bigl(
   (\hat{\psi}^\dagger_S)^{[1/2]} \cdot \bigl[
   (\hat{\psi}^\dagger_S)^{[1/2]} \bigr]^\dagger
   + \hc \Bigr)
\text{.}\label{H:tb:IROP-2}
\end{eqnarray}
\end{subequations}
The second line is essentially the same as the spinor expression in
\Eq{H:tb-psiS}, yet with the difference, that here the underlying
\IROP structure has been pointed out explicitly.

\subsubsection{\SU{2} particle-hole symmetry
for spinful system \label{sec:particle-hole}}

The particle-hole symmetry \SU[charge]{2} of the Hamiltonian in
\Eq{H:tb-default} can be made apparent in a similar way as for the
spin symmetry above. Consider the spinor in the charge sector,
\[
   \hat{\psi}_{C,k\sigma} \equiv
   \begin{pmatrix}
      \hat{c}_{k\sigma} \\
      s_{k}^{\phantom\dagger} \hat{c}_{k,-\sigma}^{\dagger}
   \end{pmatrix}
\]
with alternating phases $s_{k}=(-1)^{k}$ along the chain in
\Eq{H:tb-default}. Again, the Hamiltonian can be written as sum over
\emph{scalar products} in the spinors,
\begin{align*}
   \sum_{\sigma}\hat{\psi}_{C,k\sigma}^{\dagger} \cdot
   \hat{\psi}_{C,k+1,\sigma}^{\phantom\dagger}
&= \sum_{\sigma}
   \bigl(
     \hat{c}_{k\sigma}^{\dagger} \hat{c}_{k+1,\sigma}^{\phantom\dagger}
   - \hat{c}_{k,-\sigma}^{{}} \hat{c}_{k+1,-\sigma}^{\dagger} \bigr) \\
&= \hat{h}_{k,k+1}
\text{,}
\end{align*}
suggesting another underlying \SU{2} symmetry. Note that the
alternating sign $s_k$ is crucial to recover the correct hopping
structure in \Eq{H:tb-default}. Given the spinor in the charge
sector, the raising operator becomes
\[
   \hat{\psi}_{C,k\sigma}^{\dagger}
    \left( \begin{array}[c]{cc}
      0 & 1\\
      0 & 0
   \end{array}  \right)
   \hat{\psi}_{C,k\sigma}
 = s_{k} \hat{c}_{k\sigma}^{\dagger} \hat{c}_{k,-\sigma}^{\dagger}
\]
which, up to a sign, is the \emph{same} for both spins. It is
therefore sufficient in the charge sector to consider a spinor for
one specific $\sigma$ in $\hat{\psi}_{C,k\sigma}$ only. Therefore,
again concentrating on a single site and hence dropping the site
index $k$, now with fixed $\sigma=\uparrow$, the spinor in the charge
sector is given by,
\begin{equation}
   \hat{\psi}_{C} \equiv
   \begin{pmatrix}
      \hat{c}_{\uparrow}\\
      s \hat{c}_{\downarrow}^{\dagger}%
   \end{pmatrix}
\text{.}\label{spinor2C}%
\end{equation}
The associated raising operator becomes
\begin{equation}
   \hat{C}_{+}=s\hat{c}_{\uparrow}^{\dagger} \hat{c}_{\downarrow}^{\dagger}
\text{,}\label{op:C+}%
\end{equation}
which now creates a pair of particles with opposite spin, while the
z-operator is
\begin{align}
   \hat{C}_{z}  &= \hat{\psi}_{C}^{\dagger} (\tfrac{1}{2}\tau_{z})\hat{\psi}_{C}^{\phantom\dagger}
 = \tfrac{1}{2} (
      \hat{c}_{\uparrow}^{\dagger} \hat{c}_{\uparrow}^{\phantom\dagger} -
      \hat{c}_{\downarrow}^{\phantom\dagger} \hat{c}_{\downarrow}^{\dagger}
   )\nonumber\\
   &  \equiv\tfrac{1}{2} (\hat{n}_{\uparrow}+\hat{n}_{\downarrow}-1)
\text{.}\label{op:Cz}%
\end{align}
With $\hat{n} \equiv \hat{n}_{\uparrow}+\hat{n}_{\downarrow}$, the
z-operator $\hat{C}_{z}$ counts the total charge on given fermionic
site relative to half-filling. With
\begin{subequations}\label{eq:Cp:CR}
\begin{eqnarray}
   [ \hat{C}_{+}, \hat{c}_{\uparrow} ]
 &=& [ s\hat{c}_{\uparrow}^{\dagger} \hat{c}_{\downarrow}^{\dagger} , \hat{c}_{\uparrow} ]
 =  -s\hat{c}_{\downarrow}^{\dagger}
\label{eq:Cp:CR-1} \\
 {[} \hat{C}_{+}, \hat{c}_{\downarrow} ]
 &=& [ s\hat{c}_{\uparrow}^{\dagger} \hat{c}_{\downarrow}^{\dagger} , \hat{c}_{\downarrow} ]
 =   s\hat{c}_{\uparrow}^{\dagger}
\text{,}\label{eq:Cp:CR-2}
\end{eqnarray}
\end{subequations}%
this allows to construct the $q=\tfrac{1}{2}$ \IROPs for
\SU[charge]{2},
\begin{subequations}\label{eq:IROP-psiC2}
\begin{eqnarray}
     (\hat{\psi}_C)^{[1/2]} =
     \begin{pmatrix}
       s\hat{c}_{\downarrow}^{\dagger}\\
       -\hat{c}_{\uparrow}
     \end{pmatrix} \label{eq:IROP-psiC2-1}
\\
     (\hat{\psi}^\dagger_C)^{[1/2]}=
     \begin{pmatrix}
       s\hat{c}_{\uparrow}^{\dagger} \\
        \hat{c}_{\downarrow}
     \end{pmatrix}
\text{,}\label{eq:IROP-psiC2-2}
\end{eqnarray}
\end{subequations}
again associating the lower component with the $q_z=-\tfrac{1}{2}$
element of the $q=(1/2)$ multiplet [\cf \Eqs{eq:IROP-psiS2}{}]. An
irrelevant overall minus sign has been applied to the spinor in
\Eq{eq:IROP-psiC2-1} for later convenience. With this, the hopping
term in the Hamiltonian in \Eq{H:tb-default} can be rewritten in
terms of the scalar products
\begin{eqnarray}
   \hat{h}_{k,k+1} &=& \bigl[
     (\hat{\psi}_{Ck})^{[1/2]}\bigr]^\dagger \cdot
     (\hat{\psi}_{C,k+1})^{[1/2]} \nonumber\\
 &+& \bigl[
     (\hat{\psi}^\dagger_{Ck})^{[1/2]}]^\dagger \cdot
     (\hat{\psi}^\dagger_{C,k+1})^{[1/2]}
\end{eqnarray}
The spinors in the charge sector do mix spin components, which
essentially also requires full spin symmetry [see later discussion of
symplectic group $\Sp{2m}$ in \Sec{sec:Sp(2m)}{}]. More importantly,
the construction of the \SU[charge]{2} symmetry allows it to fully
commute with the spin-\SU{2} symmetry introduced earlier,
\begin{align}
  & [ \hat{S}_{z}, \hat{C}_{z}]
     =\tfrac{1}{4} [ \hat{n}_{\uparrow}-\hat{n}_{\downarrow}, \hat{n}-1 ]  = 0
\nonumber\\
  & [ \hat{S}_{z}, \hat{C}_{+}]
    = \tfrac{s}{2}
       \underset{ =\hat{c}_{\uparrow}^{\dagger} \hat{c}_{\downarrow}^{\dagger} }
       {\underbrace{
       [ \hat{c}_{\uparrow}^{\dagger} \hat{c}_{\uparrow}^{\phantom\dagger},
         \hat{c}_{\uparrow}^{\dagger} \hat{c}_{\downarrow}^{\dagger}]
       }}
  - \tfrac{s}{2}
    \underset{=\hat{c}_{\uparrow}^{\dagger}\hat{c}_{\downarrow}^{\dagger}}
    {\underbrace{
    [ \hat{c}_{\downarrow}^{\dagger}\hat{c}_{\downarrow}^{\phantom\dagger},
      \hat{c}_{\uparrow}^{\dagger}\hat{c}_{\downarrow}^{\dagger} ]
    }} = 0
\nonumber\\
  & [ \hat{S}_{+}, \hat{C}_{z}]
    = \tfrac{1}{2}
      \underset{=-\hat{c}_{\uparrow}^{\dagger} \hat{c}_{\downarrow}^{\phantom\dagger}}
      {\underbrace{
      [ \hat{c}_{\uparrow}^{\dagger} \hat{c}_{\downarrow}^{\phantom\dagger},
        \hat{c}_{\uparrow}^{\dagger} \hat{c}_{\uparrow}^{\phantom\dagger} ]
      }}
  + \tfrac{1}{2}
    \underset{=\hat{c}_{\uparrow}^{\dagger} \hat{c}_{\downarrow}^{\phantom\dagger}}
    {\underbrace{
    [ \hat{c}_{\uparrow}^{\dagger} \hat{c}_{\downarrow}^{\phantom\dagger},
      \hat{c}_{\downarrow}^{\dagger} \hat{c}_{\downarrow}^{\phantom\dagger}]
    }}=0
\nonumber\\
  & [\hat{S}_{+},\hat{C}_{+}]
    = s [ \hat{c}_{\uparrow}^{\dagger} \hat{c}_{\downarrow}^{\phantom\dagger},
          \hat{c}_{\uparrow}^{\dagger} \hat{c}_{\downarrow}^{\dagger}
    ] = sc_{\uparrow}^{\dagger}\hat{c}_{\uparrow}^{\dagger}=0
\text{.}\label{eq:CS-commute}
\end{align}
That is, the two symmetries act completely independent of each other
and thus can coexist simultaneously, written as the overall symmetry
$\SU[spin]{2} \otimes \SU[charge]{2}$.

If interactions are present in the system, such as local Coulomb
interaction $U\hat{n}_{\uparrow}\hat{n}_{\downarrow}$, then the
particle-hole symmetric regime requires a specific altered onsite
energy relative to the chemical potential. With $\hat
{n}_{\sigma}^{2} = \hat{n}_{\sigma}$, and $\hat{n} \equiv
\hat{n}_{\uparrow} + \hat{n}_{\downarrow}$, it follows that
$\hat{n}_{\uparrow}\hat {n}_{\downarrow} = \tfrac{1}{2}
(\hat{n}-1)^{2} + \tfrac{1}{2} (\hat{n}-1)$, and therefore
\begin{equation}
   \varepsilon_{d} \hat{n} + U \hat{n}_{\uparrow}\hat{n}_{\downarrow}
 = (\varepsilon_{d} + \tfrac{U}{2})
   \underset{=\hat{C}_{z}}{\underbrace{(\hat{n}-1) }}
   + \tfrac{U}{2} \bigl( \hat{n}-1 \bigr)^2
   + \mathrm{const}
\text{.}\label{U:ph-symmetric}%
\end{equation}
The first term on the \rhs is proportional to the $\hat{C}_{z}$
operator, which thus acts like a magnetic field for \SU[spin]{2}.
Therefore for full particle-hole symmetry to hold, this term must be
zero, which requires $\varepsilon _{d}=-\tfrac{U}{2}$. In particular,
in the absence of interaction, this implies $\varepsilon_{d}=0$. The
actual interaction term, \ie the second term on the \rhs in
\Eq{U:ph-symmetric}, also resembles $\hat{C}_{z}$. Yet it is
quadratic, and for this it also holds, $\hat{C}_z^2 = \hat{C}_x^2 =
\hat{C}_y^2$. Therefore, this term can actually be written as
$\hat{C}^2$ which itself, like spin $\hat{S}^2$ for \SU[spin]{2},
represents the Casimir operator for \SU{2}, and thus is compatible
with $\SU[spin]{2}\otimes\SU[charge]{2}$.

The actual \IROP for particle creation and annihilation given $\SC
\equiv \SU[spin]{2} \otimes \SU[charge]{2}$ symmetry can be generated
using above symmetry operations. This generates a four-dimensional
spinor. As it turns out, the resulting \IROP is the combination of
the two \IROPs generated in the spin symmetric case in
\Eqs{eq:IROP-psiS2} as well as in the particle-hole symmetric case in
\Eqs{eq:IROP-psiC2},
\begin{equation}
   \hat{\psi}_{\CS}^{\bigl[\tfrac{1}{2},\tfrac{1}{2}\bigr]} \equiv
   \begin{pmatrix}
      s \hat{c}_{\uparrow}^{\dagger} \\
        \hat{c}_{\downarrow} \\
      s \hat{c}_{\downarrow}^{\dagger} \\
      - \hat{c}_{\uparrow}
   \end{pmatrix}
\text{.} \label{eq:Psi:CS4}
\end{equation}
The multiplet labels $\bigl[\tfrac{1}{2},\tfrac{1}{2}\bigr]$ will be
derived with \Eq{eq:Psi:CS4:zlabels} below. The signs for the
individual components in above \IROP have been properly adjusted,
considering that the raising operator in the charge sector itself,
\Eq{op:C+}, carries the alternating sign $s_{(k)}=(-1)^k$. For
example, commuting $\hat{C}_{+}$ onto the fourth component, yields
the third component of the spinor $\hat{\psi}_{\CS}$ [\cf
\Eq{eq:Cp:CR-1}{}], while commuting $\hat{S}_{+}$ onto the third
component yields the first component, and so on. Again, keeping track
of the alternating sign $s_{k}=(-1)^{k}$ is crucial to recover the
hopping structure in \Eq{H:tb-default},
\begin{align}
   & \bigl(\hat{\psi}_{\CS,k}\bigr)^{\dagger} \cdot
     \hat{\psi}_{\CS,k+1}^{\phantom\dagger} \nonumber\\
   &=- \underset{=-\hat{c}_{k+1,\uparrow}^{\dagger}\hat{c}_{k\uparrow}^{\phantom\dagger}}
      {\underbrace{\hat{c}_{k\uparrow}^{\phantom\dagger} \hat{c}_{k+1,\uparrow}^{\dagger}}}
     + \hat{c}_{k\downarrow}^{\dagger}\hat{c}_{k+1,\downarrow}^{\phantom\dagger}
     - \underset{=-\hat{c}_{k+1,\downarrow}^{\dagger} \hat{c}_{k\downarrow}^{\phantom\dagger}}
      {\underbrace{\hat{c}_{k\downarrow}^{\phantom\dagger} \hat{c}_{k+1,\downarrow}^{\dagger}}}
     + \hat{c}_{k\uparrow}^{\dagger} \hat{c}_{k+1,\uparrow}^{\phantom\dagger}
      \nonumber\\
   & = \hat{h}_{k,k+1}
\text{,}\label{eq:Psi:CS4_scalar}%
\end{align}
The full tight-binding Hamiltonian simply becomes $\hat{H} = \sum_{k}
t_k (\hat{\psi}_{\CS,k})^{\dagger} \hat{\psi}_{\CS,k+1}
^{\phantom\dagger}$ where the hermitian conjugate term has been
incorporated already in the spinor structure. This also reflects the
irrelevance of taking the hermitian conjugate version of the \IROP in
\Eq{eq:Psi:CS4} as this results in essentially the same object after
properly reordering of its components and taking care of signs.

With \Eq{eq:Psi:CS4_scalar} being a scalar product in a
four-dimensional spinor space, one may be tempted to think that a
plain tight binding chain actually has a non-abelian symmetry with a
defining representation of dimension 4. This cannot be the symmetry
\SU{4}, however, since \SU{4} has rank-3 and thus requires
\emph{three} commuting abelian z-operators. The symmetries discussed
here, however, only have \emph{two} abelian z-operators, namely total
spin and total charge. The symmetry that appears compatible with this
scenario, at second glance, is the symplectic symmetry \Sp{4} [see
\Sec{sec:Sp(2m)} below]. Nevertheless, even the latter can be
excluded, since raising and lowering operators are severely
constrained by the fact that the creation and annihilation operators
appear in \emph{pairs} for the \emph{same} fermionic particle in the
\IROP of \Eq{eq:Psi:CS4}. Consequently, quadratic operators of the
type $(\hat{c}_{\sigma})^{\dagger} \hat{c}_{\sigma}^{\dagger} =
(\hat{c}_{\sigma}^{\dagger})^\dagger \hat{c}_{\sigma} = 0$ are
immediately excluded. With this, the symmetry of the spinor in
\Eq{eq:Psi:CS4} has to remain the product of two symmetries, \ie $\SC
\equiv \SU[spin]{2}\otimes\SU[charge]{2}$.

Having determined the \IROP $\hat{\psi}_{\CS}$ in \Eq{eq:Psi:CS4} by
repeated application of \RLOs $\hat{S}_{\pm}$ and $\hat{C}_{\pm}$,
the z-labels for each of the four components, on the other hand, can
be determined through the z-operators $\hat{C}_{z} \equiv
\tfrac{1}{2} \bigl( \hat{c}_{\uparrow}^{\dagger} \hat{c}_{\uparrow}
^{\phantom\dagger} - \hat{c}_{\downarrow}^{\phantom\dagger}
\hat{c}_{\downarrow}^{\dagger} \bigr)$ and $\hat{S}_{z} \equiv
\tfrac{1}{2} \bigl( \hat{c}_{\uparrow}^{\dagger}
\hat{c}_{\uparrow}^{\phantom\dagger} - \hat{c}_{\downarrow}^{\dagger}
\hat{c}_{\downarrow}^{\phantom\dagger} \bigr)$, resulting in the
z-labels $q_z \equiv (C_{z},S_{z})$, respectively. The results are
summarized in the following table.
\begin{equation}
  \begin{tabular}[c]{lc}
  {[} z-operator, \IROP component ] & $(C_{z},S_{z})$ \\\hline\\[-2ex]
  $\left. \begin{array}[c]{c}
     {[} \hat{C}_{z}, s \hat{c}_{\uparrow}^{\dagger} ] = +\tfrac{1}{2}(sc_{\uparrow}^{\dagger}) \\
     {[} \hat{S}_{z}, s \hat{c}_{\uparrow}^{\dagger} ] = +\tfrac{1}{2}(s \hat{c}_{\uparrow}^{\dagger})
  \end{array}
  \hspace{0.48in}\right\}$ & $(+\tfrac{1}{2},+\tfrac{1}{2})$
\\[3ex]
  $\left. \begin{array}[c]{c}
     {[} \hat{C}_{z}, \phantom{s}\hat{c}_{\downarrow} ] =
      - \tfrac{1}{2}(\phantom{s}\hat{c}_{\downarrow}) \\
     {[} \hat{S}_{z}, \phantom{s}\hat{c}_{\downarrow} ] =
      +\tfrac{1}{2}(\phantom{s}\hat{c}_{\downarrow}^{\phantom\dagger})
  \end{array}
  \hspace{0.48in}\right\}$ & $(-\tfrac{1}{2},+\tfrac{1}{2})$
\\[3ex]
  $\left. \begin{array}[c]{c}
     {[} \hat{C}_{z}, s \hat{c}_{\downarrow}^{\dagger} ] =
         +\tfrac{1}{2}(s \hat{c}_{\downarrow}^{\dagger}) \\
     {[} \hat{S}_{z}, s \hat{c}_{\downarrow}^{\dagger} ]
       = -\tfrac{1}{2}(sc_{\downarrow}^{\dagger})
  \end{array}
  \hspace{0.48in}\right\}$ & $(+\tfrac{1}{2},-\tfrac{1}{2})$
\\[3ex]
  $\left. \begin{array}[c]{c}
     {[} \hat{C}_{z}, -\hat{c}_{\uparrow} ] = -\tfrac{1}{2}(-\hat{c}_{\uparrow}) \\
     {[} \hat{S}_{z}, -\hat{c}_{\uparrow} ] = -\tfrac{1}{2}(-\hat{c}_{\uparrow})
  \end{array}
  \hspace{0.40in}\right\}$ & $(-\tfrac{1}{2},-\tfrac{1}{2})$
\\[3ex]\hline\\[-2ex]
  \end{tabular}
  \label{eq:Psi:CS4:zlabels}%
\end{equation}
These z-labels demonstrate that both the charge and spin multiplet
contained in $\hat{\psi}_{\CS}$ corresponds to a $q=\tfrac{1}{2}$
multiplet. The maximum weight state has the z-labels $(\tfrac{1}{2},
\tfrac{1}{2})$, which thus labels the spinor, as was already
indicated in \Eq{eq:Psi:CS4}.

Similarly, the local state space of a fermionic site must be
organized consistent with the $\SU[spin]{2} \otimes \SU[charge]{2}$
symmetry above. The local state space consists of the empty state
$\left\vert 0\right\rangle$, the singly occupied states $\left\vert
\uparrow\right\rangle$ and $\left\vert \downarrow\right\rangle$, and
the doubly occupied state $s \left\vert \uparrow\downarrow
\right\rangle$. Note that the sign in the last state is crucial, as
it is generated by the raising operator $\hat{C}_{+}$ acting on the
empty state $\left\vert 0 \right\rangle$. In summary,
\begin{equation}
   \begin{tabular}[c]{l|rr|rr}
   state & $C$ & $C_{z}$ & $S$ & $S_{z}$ \\ \hline
   $\left\vert 0\right\rangle$ & $\tfrac{1}{2}$ & $-\tfrac{1}{2}$ & $0$ & $0$ \\
   $\left\vert \uparrow\right\rangle \equiv \hat{c}_\uparrow^{\dagger}\vert 0\rangle$
      & $0$ & $0$ & $\tfrac{1}{2}$ & $+\tfrac{1}{2}$ \\
   $\left\vert \downarrow\right\rangle \equiv \hat{c}_\downarrow^{\dagger}\vert 0\rangle$
      & $0$ & $0$ & $\tfrac{1}{2}$ & $-\tfrac{1}{2}$ \\
   $s \left\vert \uparrow\downarrow \right\rangle \equiv
    s \hat{c}_{\uparrow}^{\dagger} \hat{c}_{\downarrow}^{\dagger}\vert 0\rangle$
      & $\tfrac{1}{2}$ & $+\tfrac{1}{2}$ & $0$ & $0$.\\[0.5ex]\hline
   \end{tabular}
   \ \label{fstates:SC4}%
\end{equation}
Therefore the local four-dimensional state space of fermionic site is
spanned by the two multiplets, $q\equiv(C,S) \in \{(0, \tfrac{1}{2}),
(\tfrac{1}{2}, 0) \}$.
If particle-hole symmetry is broken yet particle number still
preserved, then $2C_{z}$ from the middle column describes the total
number of particles relative to half-filling [\cf \Eq{op:Cz}{}].

With the ordering convention of state labels being $\vert
C,C_{z};S,S_{z} \rangle$ and $\sigma \in \{ \uparrow,\downarrow \}
\equiv \{ +1,-1 \} $, the non-zero matrix elements of the 4-component
spinor in \Eq{eq:Psi:CS4} can be calculated. For example,
\begin{align*}
   +s\sigma
 &= \langle -\sigma \vert \hat{c}_{\sigma} \cdot s \vert \uparrow\downarrow \rangle
  = s\langle \uparrow\downarrow \vert s \cdot s\hat{c}_{\sigma}^{\dagger} \cdot
    \vert -\sigma\rangle^{(\ast)}
\\
 & \equiv s \langle
    \tfrac{1}{2}, \tfrac{+1}{2};0,0 \vert
    \hat{\psi}_{(\tfrac{1}{2}\tfrac{+1}{2};\tfrac{1}{2}\tfrac{\sigma}{2})} \vert
    0,0;\tfrac{1}{2},\tfrac{-\sigma}{2} \rangle
\\
 &= s\underset{=1\text{ (charge)}}{\underbrace{
     \langle\tfrac{1}{2},\tfrac{+1}{2}|\tfrac{1}{2},\tfrac{+1}{2};0,0\rangle
   }}
   \underset{=\tfrac{+\sigma}{\sqrt{2}}\text{ (spin)}}{\underbrace{
     \langle0,0|\tfrac{1}{2},\tfrac{\sigma}{2};\tfrac{1}{2},\tfrac{-\sigma}{2}\rangle
   }}
   \langle\tfrac{1}{2},0 \Vert\psi\Vert 0,\tfrac{1}{2}\rangle
\\
 & \Rightarrow \langle\tfrac{1}{2},0 \Vert \psi \Vert 0,\tfrac{1}{2}\rangle
 = \sqrt{2}
\text{.}%
\end{align*}
The order inversion of the matrix element in the first line was used
since the spinor $\hat{\psi}_{\CS}$ in \Eq{eq:Psi:CS4} contains the
creation operator $s\hat{c}_{\sigma}^{\dagger}$ and not its hermitian
conjugate. The overall complex conjugation $\langle \cdot \rangle
^{(\ast)}$, however, is irrelevant since all matrix elements are
real, hence the notation of putting the asterisk in brackets.

The second non-zero reduced matrix element can be calculated in a
similar fashion,
\begin{align*}
   1  &= \langle 0 \vert \hat{c}_{\sigma} \vert \sigma\rangle
       = s \langle \sigma \vert \cdot s\hat{c}_{\sigma}^{\dagger} \cdot
         \vert 0\rangle ^{(\ast)} \\
      &\equiv
         s \langle 0,0;\tfrac{1}{2},\tfrac{\sigma}{2} \vert
         \hat{\psi}_{(\tfrac{1}{2}\tfrac{+1}{2};\tfrac{1}{2}\tfrac{\sigma}{2})} \vert
         \tfrac{1}{2},\tfrac{-1}{2};0,0 \rangle \\
      &= s \underset{=+\tfrac{1}{\sqrt{2}}\text{ (charge)}}{\underbrace{
         \langle0,0|\tfrac{1}{2},\tfrac{+1}{2};\tfrac{1}{2},\tfrac{-1}{2}\rangle}}
         \underset{=1\text{ (spin)}}{\underbrace{
           \langle\tfrac{1}{2},\tfrac{\sigma}{2}|\tfrac{1}{2},\tfrac{\sigma}{2};0,0\rangle
         }}
         \langle 0,\tfrac{1}{2} \Vert\psi\Vert \tfrac{1}{2},0\rangle \\
      &  \Rightarrow\langle0,\tfrac{1}{2} \Vert \psi\Vert \tfrac{1}{2},0\rangle
      =  s \sqrt{2}%
\end{align*}
Overall, this leads to the reduced matrix elements in the charge-spin
sectors $(C,S)\in \{ (0,\tfrac{1}{2}), (\tfrac{1}{2},0) \}$
\begin{equation}
    \psi_{\CS}^{[1/2,1/2]}
  = \left(\begin{array}[c]{cc}
       0 & s\sqrt{2}\\
       \sqrt{2} & 0
    \end{array}\right)
\text{.}\label{eq:Psi:CS}%
\end{equation}
Note that although the spinor in \Eq{eq:Psi:CS4} has four components,
\ie is of rank-3, on the reduced multiplet level in \Eq{eq:Psi:CS}
the spinor becomes a two-dimensional object as expected from an
\IROP. The further internal structure is entirely taken care of by
rank-3 Clebsch Gordan coefficients, which have been omitted in
\Eq{eq:Psi:CS} [for a full description of $\psi_{\CS}^{[1/2,1/2]}$
including Clebsch-Gordan coefficients in terms of a \QSpace see
\TblP{QSpace:F3_cs}].

The operator in \Eq{eq:Psi:CS} is non-hermitian. In the context of
two-site hopping, however, this nevertheless leads to a hermitian
term in the Hamiltonian, as required. Indicating the local symmetry
eigenspace for site $k$ by $\left\vert \sigma\right\rangle _{k}$, the
matrix elements of the hopping term in the tensor-product basis
$\left\vert \sigma_{k+1} \right\rangle \left\vert
\sigma_{k}\right\rangle $ (in this order, assuming site $k+1$ is
added after site $k$) are given by
\begin{equation}
   \langle \sigma_{k} \vert \langle \sigma_{k+1} \vert
   \hat{\psi}_{k}^{\dagger} \hat{\psi}_{k+1}^{\phantom\dagger}
   \vert \sigma_{k+1}^{\prime}\rangle \vert \sigma_{k}^{\prime} \rangle
 = \psi_{k}^{\dagger}\otimes[z\psi]_{k+1}
\text{,} \label{psi_zpsi}%
\end{equation}
where the $\psi$'s to the right without the hat denote the matrix
elements in the local $\vert \sigma\rangle$ basis. Note, that
fermionic signs apply, when $\hat{\psi}_{k}^{\dagger}$ is moved, for
example, to the left of $\langle \sigma_{k+1}\vert$, such that the
tensor-product on the \rhs of \Eq{psi_zpsi} contains $[z\psi]_{k+1}$
rather than $\psi_{k+1}$, where $\hat{z}_{k} \equiv (-1)^{\hat
{n}_{k}}$ is diagonal in $\vert \sigma_{k}\rangle $ and adds signs
corresponding to the number of particles in $\vert \sigma_{k} \rangle
$. Note that with the particle number being related to $\hat{C}_{z} =
\tfrac{1}{2} (\hat{n}-1)$, the operator $\hat{z}$ is well-defined in
terms of the symmetry labels. It is a scalar operator, since
$(-1)^{\hat{n}-1} = (-1)^{(\hat{n}-1)^2}=(-1)^{4\hat{C}_z^2}$, hence
does not alter the Clebsch-Gordan content of the operator
$\hat{\psi}$ but rather acts on the multiplet level only. For the
hopping $\hat{\psi}_k^\dagger \hat{\psi}_{k+1}^{\phantom\dagger}$
between two nearest-neighbor sites, \Eq{eq:Psi:CS} finally leads to
\begin{equation*}
   H_{k,k+1}=\begin{pmatrix}
      0 & \sqrt{2}\\
      s_{k}\sqrt{2} & 0
   \end{pmatrix} \otimes
   \underset{=\begin{pmatrix}
      0 & -s_{k+1}\sqrt{2}\\
      \sqrt{2} & 0
   \end{pmatrix}}{\underbrace{
   \left(\begin{array}{rr}
     -1 & 0\\
      0 & 1
   \end{array}\right)
   \begin{pmatrix}
      0 & s_{k+1}\sqrt{2}\\
      \sqrt{2} & 0
   \end{pmatrix}}}
\text{,}
\end{equation*}
written as a plain tensor product on the level of the multiplet
spaces of two fermionic sites. For the sake of the argument, the
product space here is not yet described in terms of proper combined
symmetry multiplets of sites $k$ and $k+1$.

With $s_{k}=(-1)^{k}$, $H_{k,k+1}$ in the last equation clearly
yields a hermitian object for all iterations. For example, for even
$k$, the hopping term is given by
\begin{equation*}
   H_{k,k+1}^{[k\text{ even}]} =
   \begin{pmatrix}
      0 & \sqrt{2}\\
      \sqrt{2} & 0
   \end{pmatrix} \otimes
   \begin{pmatrix}
      0 & \sqrt{2}\\
      \sqrt{2} & 0
   \end{pmatrix}
\end{equation*}
similar in structure to a hermitian object of the type
$\tau_{x}\otimes \tau_{x}$ in terms of Pauli matrices, while for odd
$k$,
\begin{equation*}
   H_{k,k+1}^{[k\text{ odd}]} =
   \begin{pmatrix}
      0 & \sqrt{2}\\
      -\sqrt{2} & 0
   \end{pmatrix} \otimes
   \begin{pmatrix}
      0 & -\sqrt{2}\\
      \sqrt{2} & 0
   \end{pmatrix}
\end{equation*}
similar in structure to the hermitian $(i\tau_{y}) \otimes (
i\tau_{y}) =-\tau_{y} \otimes \tau_{y}$. Hence for every even (odd)
iteration, one has a $\tau_{x} \otimes \tau_{x}$ ($\tau_{y} \otimes
\tau_{y}$) structure, respectively, a prescription that is periodic
with every pair of iterations. This intrinsic even-odd behavior is
not specifically surprising, considering, for example, that two
particles are needed to return to the same charge quantum numbers
related to particle hole symmetry.

In summary, using \Eq{eq:Psi:CS4_scalar}, the hopping in the
Hamiltonian is given by
\begin{equation}
   \hat{H} = \sum_{k} t_{k}
   \hat{\psi}_{k}^{\dagger} \hat{\psi}_{k+1}^{\phantom\dagger}
\text{.}\label{eq:Htb:CS}
\end{equation}
In a sense, the net effect of incorporating spin \SU{2} was to
eliminate the spin index, while incorporation of particle-hole \SU{2}
eliminates the hermitian conjugate term in the Hamiltonian. Together
they reduce the four terms initially required for a single hopping in
\Eq{H:tb-default} to the single scalar term $\hat{\psi}_{k}^{\dagger}
\hat{\psi}_{k+1}^{\phantom\dagger}$.

\subsubsection{Particle-hole \SU{2} symmetry
for several channels}

The alternating sign in the raising operator
$\hat{C}_{k,+}=s\hat{c}_{\uparrow }^{\dagger}
\hat{c}_{\downarrow}^{\dagger}$ in \Eq{op:C+} defines the doubly
occupied states as $\vert \tfrac{1}{2};\tfrac{+1}{2}\rangle =
\hat{C}_{k,+} \vert \tfrac{1}{2};\tfrac{-1}{2} \rangle = s
\hat{c}_{\uparrow}^{\dagger}c_{\downarrow}^{\dagger} \vert 0\rangle$;
for even sites, $s=+1$, therefore $\vert \tfrac{1}{2}; \tfrac{+1}{2}
\rangle = \hat{c}_{\uparrow}^{\dagger} \hat{c}_{\downarrow}^{\dagger}
\vert 0 \rangle$. For odd sites, on the other hand, $\vert
\tfrac{1}{2}; \tfrac{+1}{2} \rangle = - \hat{c}_{\uparrow}^{\dagger}
\hat{c}_{\downarrow}^{\dagger} \vert 0 \rangle =
\hat{c}_{\downarrow}^{\dagger} \hat{c}_{\uparrow }^{\dagger} \vert 0
\rangle$. In practice, for consistency, usually a certain
well-defined fermionic order is adopted. Above raising operator
$\hat{C}_{k,+}$ thus suggests that it may be useful to \emph{reverse
the fermionic order of every other site} for the local state space
included there.

Fully reversing the fermionic order of a given site $k$ with several
fermionic channels $i=1,\ldots,m$ implies for the \emph{matrix
elements} for particle creation or annihilation operators,
\begin{align*}
   c_{k,i\sigma}  & \rightarrow
   \tilde{c}_{k,i\sigma} \equiv z_k c_{k,i\sigma}
\\
   c_{k,i\sigma}^{\dagger}  & \rightarrow
   \tilde{c}_{k,i\sigma}^{\dagger} \equiv
   -z_k c_{k,i\sigma}^{\dagger}
\text{.}
\end{align*}
This transformation is equivalent to a unitary transformation local
to site $k$. Similar to \Eq{psi_zpsi}, $\hat{z}_{k} \equiv
(-1)^{\hat{n}_{k}}$ with $\hat{n}_{k} \equiv \sum_{i\sigma}
\hat{n}_{k,i\sigma}$ and $\hat{n}_{k,i\sigma} \equiv
\hat{c}_{k,i\sigma}^\dagger \hat{c}_{k,i\sigma}^{\phantom\dagger}$
again takes care of fermionic signs for the full multi-level site
$k$. Being a scalar operator, $\hat{z}_{k}$ is independent of the
fermionic order.

Now consider the effect of flipping the fermionic order for the odd
sites in the tight-binding chain that carry the sign $s=-1$, assuming
particle-hole symmetry in every channel. For a specific channel, this
(i) takes away the sign in the raising operator $\hat {C}_{k,+}$, and
(ii) implies, for example, for the 4-component spinor in
\Eq{eq:Psi:CS4} for a single channel,
\begin{equation}
   \hat{\psi}_{\CS,k\text{ odd}} \equiv
   \begin{pmatrix}
      -\hat{c}_{\uparrow}^{\dagger}\\
       \hat{c}_{\downarrow}\\
      -\hat{c}_{\downarrow}^{\dagger}\\
      -\hat{c}_{\uparrow}
   \end{pmatrix} \rightarrow
   \begin{pmatrix}
      +\hat{z}\hat{c}_{\uparrow}^{\dagger}\\
       \hat{z}\hat{c}_{\downarrow}\\
      +\hat{z}\hat{c}_{\downarrow}^{\dagger}\\
      -\hat{z}\hat{c}_{\uparrow}
   \end{pmatrix}
   \equiv \hat{z}_k \hat{\psi}_{\CS,k\text{ even}}
\text{,}\label{eq:psi4:EO}
\end{equation}
having intermittently dropped the index $k$ for readability.
Therefore, up to the local operator $\hat{z}_{k}$ which assigns
fermionic signs to the full Hilbert space of a local site, the matrix
elements of the spinor for the odd sites are \emph{exactly} the same
as the matrix elements of the spinor for even sites. Therefore with
$\hat{\psi}$ taken as the spinor for even sites in the chain, the
required spinor for odd sites becomes $\hat{z}\hat{\psi}$. Together
with the additional fermionic signs in the nearest-neighbor hopping
term as already encountered in \Eq{psi_zpsi}, the hopping structure
$\hat{h}_{k,k+1}$ of the tight-binding Hamiltonian in \Eq{eq:Htb:CS}
becomes,
\begin{align*}
   \psi_{k}^{\dagger} \otimes [z \cdot (z\psi)]_{k+1}
&= \psi_{k}^{\dagger} \otimes \psi_{k+1}^{\phantom\dagger} \text{ \ for }k\text{ even}
\\
   (z\psi)_{k}^{\dagger} \otimes [ z \cdot \psi]_{k+1}
&= (z\psi)_{k}^{\dagger} \otimes (z\psi)_{k+1} \text{ for }k\text{ odd.}
\end{align*}
This result generalizes to any number of channels with particle-hole
symmetry. As such it much simplifies the structure and thus the
treatment of the two different kinds of spinors for even and odd
sites, respectively, that had been required initially.

\subsubsection{Symmetric three-channel system
\label{sec:3channel}}

Consider the generalization of the spinful one-channel setup in
\Eq{H:tb-default} to a spinful three-channel system,
\begin{equation}
  \hat{H} = \sum_k t_k \cdot
  \underset{ \equiv \hat{h}_{k,k+1} }{\underbrace{
     \sum_{i=1}^{m=3} \sum_{\sigma}
     (\hat{c}_{k,i\sigma}^{\dagger} \hat{c}_{k+1,i\sigma}^{\phantom\dagger} + \hc)
  }}
\text{,}\label{eq:H3C}%
\end{equation}
where $\hat{c}_{k,i\sigma}^{\dagger}$ creates a particle at site $k$
in channel $i\ $with spin $\sigma$. This model is relevant for the
system analyzed in the main body of the paper where the specific
number of three channels, for example, originates from the underlying
orbital band structure in terms of a partially filled $d$-shell. The
Hamiltonian in \Eq{eq:H3C} can also be complemented with interaction
terms that are compatible with the symmetries discussed in the
following. This can include onsite interaction $U$ at half-filling
[\cf \Eq{U:ph-symmetric}{}], or uniform local Hund's coupling \JH
[\eg see \Eqp{eq:HKondoAM-d}{}]. Here, however, the focus of the
discussion is on symmetries, for which the Hamiltonian in \Eq{eq:H3C}
suffices.

The Hamiltonian in \Eq{eq:H3C} possesses \SU{2} spin symmetry, \SU{2}
particle-hole symmetry in each channel, and also \SU{3} channel
symmetry, while not all of these symmetries necessarily are
independent of, \ie commute with each other. All of these symmetries
can be defined within the Hilbert space of a local site, hence again
focusing the discussion on a single site $k$ in the following, while
dropping the site index $k$, for simplicity. For each of the three
channels, the associated spinful fermionic level is represented by
the four states as in \Eq{fstates:SC4}, leading to a total of
$4^{3}=64$ state for a given site.

The total spin-\SU{2} symmetry of a site is described by the
generators
\begin{align}
   \begin{array}[c]{l}
   \hat{S}_{+} =\sum\limits_{i}\hat{S}_{i+}
 = \sum\limits_{i}
   \hat{c}_{i\uparrow}^{\dagger} \hat{c}_{i\downarrow}^{\phantom\dagger}
\\
   \hat{S}_{z} =\sum\limits_{i}\hat{S}_{iz}
 = \tfrac{1}{2} \sum\limits_{i} (\hat{n}_{i\uparrow}-\hat{n}_{i\downarrow})
   \text{,}
\end{array}
\label{eq:3C:Sops}
\end{align}
where $\hat{S}_{i\sigma}$ represents the spin operators for the
fermionic level $i$, with $\hat{S}_{z} = \tfrac{1}{2} [ \hat{S}_{+},
\hat{S}_{+}^{\dagger} ]$, as expected for \SU{2}.

The particle-hole symmetry exists for every channel $i$, and is
described by the \SU{2} symmetry,
\begin{subequations}
\begin{equation}
   \begin{array}[c]{l}
   \hat{C}_{i+}=s\hat{c}_{i\uparrow}^{\dagger}\hat{c}_{i\downarrow}^{\dagger} \\
   \hat{C}_{iz}=\tfrac{1}{2} [\hat{C}_{i+},\hat{C}_{i+}^{\dagger}]
 = \tfrac{1}{2} \bigl(\hat{n}_{i}-1 \bigr)
 \text{,}
\end{array}\label{eq:3C:Cops}
\end{equation}
which includes the same sign-factor $s_{k}=(-1)^{k}$ as in \Eq{op:C+}
to correctly represent the hopping structure in the Hamiltonian
\Eq{eq:H3C}. The total charge relative to half-filling is given by
(up to a factor of $2$)
\begin{equation}
   \hat{C}_{z}\equiv\sum_{i} \hat{C}_{iz}
   \text{.}\label{eq:3C:Cz}
\end{equation}
\end{subequations}
Finally, the channel symmetry is given by the minimal set of two
raising operators $\{\hat{T}_{+},\hat{U}_{+}\} \equiv \{
\hat{S}_{12}, \hat{S}_{23}\}$ together with the z-operator
$\{\hat{T}_{z},\hat{Y}\} \equiv \{ \hat{Z}_{1}, \hat{Z}_{2} \}$ as
introduced through \Eqs{eq:SUN-gen} in \Eq{eq:SU3:zops},
\begin{equation}
  \begin{array}[c]{ll}
    \hat{T}_{+} = \sum\limits_{\sigma}\hat{c}_{1\sigma}^{\dagger}
       \hat{c}_{2\sigma}^{\phantom\dagger} \text{,}
  & \hat{T}_{z}= \sum\limits_{\sigma}
   \bigl(\hat{n}_{1\sigma} - \hat{n}_{2\sigma} \bigr) \text{,}
  \\
    \hat{U}_{+}=\sum\limits_{\sigma}\hat{c}_{2\sigma}^{\dagger}
       \hat{c}_{3\sigma}^{\phantom\dagger} \text{,}
  & \hat{Y}= \sum\limits_{\sigma}
    \bigl( \hat{n}_{1\sigma} + \hat{n}_{2\sigma}-2\hat{n}_{3\sigma} \bigr)
    \text{.}
  \end{array}
\label{sop3C:TUY}%
\end{equation}
Here the notation for the generators of \SU{3} has been changed to
another notation frequently also found in literature, so these
generators can be better distinguished from the generators for spin
and particle-hole symmetry. In particular, the operators
$\hat{T}_{+}$ and $\hat{T}_{z}$ generate an \SU{2} subalgebra, that
is linked to the full \SU{3} symmetry through the generators
$\hat{U}_{+}$ and $\hat{Y}$. The normalization of the z-operators,
however, has been chosen consistent with \Eqs{eq:SUN-gen}, such that
plain integer matrix elements arise.

The spin symmetry clearly commutes with the particle-hole symmetry in
each channel, which follows from the previous one-channel discussion
in \Eqs{eq:CS-commute}. Therefore it remains to analyze the
compatibility of the \SU{3} channel symmetry. All z-operators clearly
commute. For the \SU{3} raising operators, it follows with respect to
the spin symmetry,
\begin{align}
    \bigl[  \hat{T}_{+},\hat{S}_{+}\bigr]
 &= \sum\limits_{\sigma,i}
    \bigl[ c_{1\sigma}^{\dagger} \hat{c}_{2\sigma}^{\phantom\dagger},
    \hat{c}_{i\uparrow}^{\dagger} \hat{c}_{i\downarrow}^{\phantom\dagger}\bigr]
\nonumber\\
 &= \sum\limits_{\sigma} (\delta_{\sigma\uparrow}-\delta_{\sigma \downarrow}) \cdot
    \hat{c}_{1\uparrow}^{\dagger} \hat{c}_{2\downarrow}^{\phantom\dagger} = 0
\label{eq:CR:ST3}%
\end{align}
with a similar expression for $\hat{U}_{+}$ instead of $\hat{T}_{+}$
with a shift in the channel indices. Note that in order for the \rhs
to vanish, the sum over the spin $\sigma$ is essential which shows
the importance of the summation over $\sigma$ in \Eqs{sop3C:TUY}. As
a consequence, the \SU{3} channel symmetry in \Eq{sop3C:TUY} commutes
with the \SU{2} spin symmetry, indeed.

The compatibility of the \SU{3} channel symmetry with the \SU{2}
particle-hole symmetry, however, cannot be established, since
\begin{align}
    \bigl[ \hat{T}_{+}, \hat{C}_{i+} \bigr]
 &= \sum_{\sigma} \bigl[
      c_{1\sigma}^{\dagger} \hat{c}_{2\sigma}^{\phantom\dagger},
      s\hat{c}_{i\uparrow}^{\dagger} \hat{c}_{i\downarrow}^{\dagger}
    \bigr]
 \nonumber\\
 &= s\delta_{i2} (
      \hat{c}_{1\uparrow}^{\dagger} \hat{c}_{2\downarrow}^{\dagger}
     -\hat{c}_{1\downarrow}^{\dagger} \hat{c}_{2\uparrow}^{\dagger}
    ) \neq 0
\label{eq:CR:CT3}%
\end{align}
cannot be made to vanish for all channels $i$ at the same time.
Therefore the non-abelian channel and particle-hole symmetries cannot
coexist independently of each other. Nevertheless, the generators of
each individual symmetry do commute with the Hamiltonian, \emph{which
thus suggests a larger symmetry}, with \Eq{eq:CR:CT3} already
indicating one of the additional generators. As it turns out, this
symmetry is \Sp{2m} with $m$ the number of channels. This symmetry
will be introduced and discussed in the next section.

By reducing the non-abelian particle-hole symmetry to its abelian
conservation of total charge, however, this abelian symmetry does
commute with the \SU{3} channel symmetry,
\begin{align*}
    \sum_{i} \bigl[ \hat{T}_{+}, \hat{C}_{iz} \bigr]
 &= \sum_{\sigma,i\sigma^{\prime}} \tfrac{1}{2} \bigl[
    c_{1\sigma}^{\dagger}\hat{c}_{2\sigma}^{\phantom\dagger},
    c_{i\sigma^{\prime}}^{\dagger}
    \hat{c}_{i\sigma^{\prime}}^{\phantom\dagger} \bigr]
 \\
 &= \sum_{i} \tfrac{1}{2} (\delta_{i2}-\delta_{i1}) \sum_{\sigma}
    \hat{c}_{1\sigma}^{\dagger} \hat{c}_{2\sigma}^{\phantom\dagger}
  = 0
\text{.}%
\end{align*}
In order to get a commuting abelian charge symmetry, the z-operators
for the channel-specific particle-hole symmetry must be summed over
all channels $i$. With all commuting symmetries combined, this leads
to the overall symmetry \ASC, consisting of the \SU{2} total spin
symmetry in \Eq{eq:3C:Sops}, the abelian total charge of the system
in \Eq{eq:3C:Cz}, and the channel \SU{3} symmetry in \Eq{sop3C:TUY}.

A more conventional symmetry setup can be obtained by giving up the
channel \SU{3} symmetry. Bearing in mind that the channel-specific
\SU{2} particle-hole symmetries commute with total spin, this also
allows the symmetry setup \SSSS.

The symmetry combinations above can be motivated also by a simple
counting argument with respect to conserved abelian quantum numbers.
Note that the preserved abelian quantum numbers in the Hamiltonian
\Eq{eq:H3C} are the particle number in each of the three channels
together with the total spin $S_z$. This results in a total of four
z-operators, and thus four z-labels. Now, by including non-abelian
flavors, the number of z-operators clearly cannot increase, but will
remain the same. Total spin has one z-operator, the channel \SU{3}
symmetry has two z-operators, and the channel-specific particle-hole
symmetries have three z-operators, which combined results in
$1+2+3=6$ z-operators. This set of z-operators therefore \emph{cannot
be independent} of each other, as already seen in the earlier
discussion. Yet, in fact, both of the alternative symmetry setups
above do have a total of four z-operators. For \ASC, these are
$1+1+2$ from spin, charge, and channel symmetry, respectively, while
for \SSSS these are $1+3$ from spin and each channel.

For the symmetry setting \ASC then, the hopping term in the Hamiltonian in
\Eq{eq:H3C} is described by a 6-component \IROP
$\hat{\psi}_k^{[6]}$ (annihilation operators for spin-up and spin-down
combined), that can be obtained, for example, numerically
as described in \Secp{sec:fock2sym}. This leads to
\begin{subequations}\label{eq:Fops3C:hopping}
\begin{equation}
  \hat{h}_{k,k+1} =
  \hat{\psi}_n^{[6]\dagger}\cdot \hat{\psi}_k^{[6]} +\hc
\label{eq:Fops3C:hopping:psi6} %%\text{.}
\end{equation}
In contrast, for the second symmetry setting \SSSS, the \IROPs
$\hat{\psi}_{k,i}^{[4]}$ required for the hopping term are already
exactly the 4-component spinors in \Eq{eq:Psi:CS4}, \ie one for each
individual channel, $i=1,\ldots,3$. The hopping in the Hamiltonian is
thus described by
\begin{equation}
\hat{h}_{k,k+1} =
  \sum_{i=1}^{m=3} \hat{\psi}_{k,i}^{[4]\dagger} \hat{\psi}_{k+1,i}^{[4]}
\text{,}\label{eq:Fops3C:hopping:psi4}
\end{equation}
\end{subequations}%
\ie \emph{without} the hermitian conjugate part as this is already
included through particle-hole symmetry. Furthermore, note that
particle-hole symmetry also acquires even-odd alternations for the
spinors along a chain [see \AppP{sec:particle-hole}].

\subsection{The symplectic group \Sp{2m}
\label{sec:Sp(2m)}}

All Hamiltonians considered in this paper are time-independent, hence
obey time-reversal symmetry. Time-reversal symmetry then is described
by an anti-unitary operator
$\hat{T}=\hat{\Sigma}_y\hat{K}$,\cite{Sakurai94} that includes a
standard unitary operation $\hat{\Sigma}_y$ together with the
operator $\hat{K}$, which stands for complex conjugation [the
notation of $\hat{\Sigma}_y$ has been chosen for latter convenience;
see \Eq{def:Sp2m:Yop} below]. The time-reversal operator obeys
$\hat{T}^2=\pm 1$, where for spin-half particles, such as electrons
as considered throughout in this paper, it holds $\hat{T}^2=-1$. The
latter is important for the symmetry \Sp{2m}, since it implies that
the unitary $\hat{\Sigma}_y$ must be antisymmetric. This follows
simply by looking at the matrix elements of the time-reversal
operator for arbitrary states $\vert a\rangle$ and $\vert b\rangle$
in some real basis $i$,
\begin{align*}
   \langle a \vert \hat{T}b \rangle  = \sum_{i,j} a_i^\ast
   (\Sigma_y)_{ij} b_j^\ast
\text{,}
\end{align*}
yet it also holds,
\begin{align*}
   \langle a \vert \hat{T}b \rangle  =
   \langle \underset{ -1 }{\underbrace{ \hat{T}^2 }} b \vert \hat{T}a \rangle  =
   -\sum_{i,j} b_j^\ast (\Sigma_y)_{ji} a_i^\ast
\text{.}
\end{align*}
As this applies for arbitrary states $\vert a\rangle$ and $\vert
b\rangle$, this shows that, given $\hat{T}^2=-1$,
the unitary $\hat{\Sigma}_y$ must be antisymmetric, indeed.

Since a time-independent Hamiltonian obviously commutes with the
time-reversal operator, it follows that all eigenstates of the
Hamiltonian can also be written as eigenstates of the the
time-reversal operator $\hat{T}$. As a consequence, all unitary
symmetry operations $\hat{G} = \exp(i\sum_\sigma a_\sigma
\hat{S}_\sigma)$ can be \emph{constrained} to unitaries which also
leave the time-reversal operator invariant. That is,
\begin{align}
 & \hat{T} \overset{!}{=} \hat{G} \hat{T} \hat{G}^{-1}
 = \hat{G} \hat{\Sigma}_y
   \underset{ =\hat{G}^{T}\hat{K} }{\underbrace{ \hat{K} \hat{G}^{\dagger} }}
   \nonumber \\
 & \Rightarrow\quad \hat{\Sigma}_y = \hat{G} \hat{\Sigma}_y \hat{G}^{T}
\text{.}\label{eq:T:time-reversal}
\end{align}
For the generators $\hat{S}_\sigma$ of a symmetry group this implies
(\eg by expansion of the exponential in $\hat{G}$ to first order in
$a_\sigma$), that
\begin{align}
   \hat{S}_\sigma\hat{\Sigma}_y + \hat{\Sigma}_y\hat{S}^{T}_\sigma = 0
\text{,}\label{eq:T:S-constraint}
\end{align}
This exactly corresponds to the definition of the Lie algebra
\Sp{2m}. Having a unitary, \ie non-singular, yet also antisymmetric
$\hat{\Sigma}_y$, this requires a global Hilbert space of \emph{even}
dimension $N$, since
$\mathrm{det}(\Sigma_y)=\mathrm{det}(\Sigma_y^T)=(-1)^N
\mathrm{det}(\Sigma_y) \neq 0$. While this argument holds on the
entire Hilbert space, for a specific symmetry subspace (carrier
space) of an irreducible representation of \Sp{2m} this is not
necessarily the case. Specifically, there are \IREPs with \emph{odd}
dimensions, a simple example being the scalar representation with
dimension $1$. Within such an irreducible representation, a
non-singular antisymmetric $\Sigma_y$ does not exist. This is not a
problem, however, since the existence of $\Sigma_y$ is required only
globally, and also in the \emph{defining} representation, which thus
has to be of even dimension.

Consider such a matrix representation of \Sp{2m} of even dimension,
which allows to explicitly construct the non-singular antisymmetric
$\Sigma_{y}$. In this case, an arbitrary matrix $S_{\left(
\sigma\right) }$ within the space of the generators of the symmetry
can be written as a tensor-product with a two-dimensional space,
which itself can be expanded in terms of the Pauli matrices
$\tau_{\sigma}$ [\cf \Eq{eq:pauli-def}{}],
\begin{equation}
   S_{(\sigma)} \equiv
   \sum_{x=0}^{3} \tau_{x} \otimes
   \mathrm{S}_{x}^{(\sigma)}
\text{,}\label{eq:Sp2m:Srep0}
\end{equation}
where $x\in\{0,1,2,3\}\equiv\{0,x,y,z\}$ and $\tau_{0}\equiv1^{\left(
2\right) }$ the two-dimensional identity matrix. Here the same letter
$S$ is used left and right in \Eq{eq:Sp2m:Srep0}, as their
interpretation is related. Nevertheless, they refer to different
objects. So in order to distinguish them, the generators on the \lhs
are written with Greek-letter subscripts ($\sigma$), while their
decomposition $\mathrm{S}_{x}^{\left( \sigma\right) }$ is denoted in
roman font with roman or numeric subscripts. Moreover, for
readability, the index $\sigma$ referring to a specific generator
will be skipped in the following where not explicitly required (hence
the $\sigma$ has been put in brackets).

Now, with representations of a symmetry unique up to similarity
transformation, one is free to choose the form of the matrix representation of
the operator $\hat{\Sigma}_{y}$ in Eq.~(\ref{eq:T:S-constraint}). In the
two-dimensional (block) space described by the Pauli matrices then,
$\Sigma_{y}$ is chosen as follows, \cite{Pope06,Gilmore06}%
\begin{align}
   \Sigma_{y} &=
   \tau_{y}\otimes 1^{(m)}
    \equiv
   \begin{pmatrix}
      0^{(m)}
   & -i 1^{(m)} \\
      i 1^{(m)}
   & 0^{(m)}
   \end{pmatrix}
\text{,}\label{def:Sp2m:Yop}%
\end{align}
where the last term explicitly denotes the tensor block-decomposition
of $m\times m$ matrices, with $0^{(m)}$ [$1^{(m)}$] an $m\times m$
dimensional zero [identity] matrix, respectively. This $\Sigma_{y}$
fulfills the minimal requirement that it is (i) unitary and (ii)\
antisymmetric. Using the $\Sigma_{y}$ in \Eq{def:Sp2m:Yop} in the
defining equation for \Sp{2m}, \Eq{eq:T:S-constraint}, and the fact
that the generators $S_{\left( \sigma\right) }$ in \Eq{eq:Sp2m:Srep0}
shall refer to hermitian operators to start with, this implies for
the decomposition $\mathrm{S}_{x}$, that $\mathrm{S}_{0}\equiv
i\mathrm{A}\ $is a purely imaginary and antisymmetric matrix, while
the remaining $S_{x}$ for $x=(1,2,3)$ must be real symmetric
matrices. In summary, this allows to rewrite the matrix
block-decomposition in \Eq{eq:Sp2m:Srep0} in the form,
\cite{Pope06,Gilmore06}
\begin{equation}
   S=\begin{pmatrix}
      i\mathrm{A}+\mathrm{S}_{3} & \mathrm{S}_{1}-i\mathrm{S}_{2}\\
      \mathrm{S}_{1}+i\mathrm{S}_{2} & i\mathrm{A}-\mathrm{S}_{3}%
   \end{pmatrix}
   \equiv
   \begin{pmatrix}
      \mathrm{C} & \mathrm{D}^{\dagger}\\
      \mathrm{D} & -\mathrm{C}^{T}
   \end{pmatrix}
\text{,}\label{eq:Sp2m:Srep}%
\end{equation}
where $\mathrm{C}\equiv i\mathrm{A}+\mathrm{S}_{3}$ ($\mathrm{D}%
\equiv\mathrm{S}_{1}+i\mathrm{S}_{2}$) is an arbitrary hermitian
(symmetric) $m\times m$ matrix, respectively. The resulting number of
free parameters is $m^{2}+m$ for the matrix $\mathrm{D}$ (where the
$+m$ comes from the fact that the diagonal can be fully complex), and
$m^{2}$ for the hermitian matrix $\mathrm{C}$. The total number of
free parameters of the $(N\equiv2m)$-dimensional matrices therefore
is,
\begin{equation}
   g=m\left(  2m+1\right)  \equiv\tfrac{N}{2}\left(  N+1\right)
\text{.}\label{eq:Sp2m:dim}%
\end{equation}
In case of the defining representation, by construction, this also
corresponds to the dimension of the symmetry group \Sp{2m}. For
comparison, for example, the orthogonal group $O(N)$ has dimension
$\tfrac{N}{2}\left( N-1\right) $.

Setting the off-diagonal block-matrix $\mathrm{D}$ in
\Eq{eq:Sp2m:Srep} to zero, and using arbitrary hermitian \textit{yet
also traceless} matrices $\mathrm{C}$, this directly demonstrates
that $SU(m)$ is contained as a subalgebra within \Sp{2m}. This
subalgebra $SU(m)$ has rank $m-1$, \ie has $m-1$ z-operators. Now,
the full \Sp{2m} symmetry also includes the \textit{tracefull}
hermitian matrix $\mathrm{C}$. This introduces the remaining $m$-th
z-operator, $Z_{m}=\tau_{z}\otimes1^{(m)}$. With a total of $m$
z-operators, \Sp{2m} therefore has rank $m$, with the z-operators
given by
\begin{subequations}\label{eq:Sp2m:zops}
\begin{equation}
   Z_{k} \equiv
   \tau_{z}\otimes\mathrm{Z}_{k}^{(m)}
\text{,}\label{eq:Sp2m:Zk}
\end{equation}
where
\begin{equation}
  \mathrm{Z}_{k}^{(m)} = \left\{
     \begin{array}[c]{ll}
       (\mathrm{Z}_{k}^{(m)})^{\SU{m}} & k=1,\ldots,m-1\\
       1^{(m)} & k=m
     \end{array}
  \right.
\text{,}\label{eq:Sp2m:Zm}
\end{equation}
\end{subequations}%
with $\bigl(\mathrm{Z}_{k}^{(m)}\bigr)^{\SU{m}}$ the standard
$m\times m$ dimensional z-operators for \SU{m}. By construction, all
of these z-operators can be considered diagonal, as they form a
mutually commuting set of matrices.

Leaving the space of strictly hermitian generators, the canonical
\RLOs from the \SU{m} subalgebra are given by
\begin{equation}
   S_{ij} \equiv
   \begin{pmatrix}
      \mathrm{S}_{ij} & 0\\
      0 & -\mathrm{S}_{ij} ^{T}%
   \end{pmatrix}\text{,} \qquad (i\neq j)
\label{eq:Sp2m:SUm}%
\end{equation}
with $\mathrm{S}_{ij} \equiv E_{ij}$ given by the non-symmetric
matrices in \Eq{eq:def:Eij}. This encodes both, raising and lowering
operators, depending on $i<j$ or $i>j$, respectively.
Having $(m^2-1)+1=m^2$ generators from the \SU{m} subalgebra together
with $Z_{m}$, the remaining $m\left( m+1\right) $ operators are split
equally into complimentary raising and lowering operators. The
corresponding canonical \RLOs can be chosen as follows,
\cite{Pope06,Gilmore06}
\begin{align}
   \tilde{S}_{ij}^{\pm} &\equiv
   \tfrac{1}{2}\left(\tau_{x}\pm i\tau_{y}\right)
   \otimes \mathrm{\tilde{S}}_{ij}
\text{,} \qquad (\text{all }i,j)
\label{eq:Sp2m:Sij}
\end{align}
with the \emph{symmetric} matrices $\mathrm{\tilde{S}}_{ij} \equiv
\tfrac{1}{2} (E_{ij}+E_{ji})$. %% keep diagonal in S{i=j} equal to 1!
Here the tilde serves to differentiate the \RLOs from the \SU{m}
subalgebra in \Eq{eq:Sp2m:SUm}. Having symmetric
$\mathrm{\tilde{S}}_{ij}$, \ie $\mathrm{\tilde{S}}_{ij} =
\mathrm{\tilde{S}}_{ji}$, this describes a total of
$\tfrac{1}{2}m\left( m+1\right) $ raising operators. Complemented by
$\tfrac{1}{2}m\left( m+1\right) $ lowering operators, indeed, this
completes the group of generators for the Lie algebra \Sp{2m}.

Using the canonical representation for \SU{m} together with above
extension to \Sp{2m}, this provides the canonical representation for
\Sp{2m} as in \Eq{eq:CR:Sz-pm}. For example, with
\begin{equation}
   (\vec{z}_i)_k \equiv z_{k,i}^{(m)} \equiv (\mathrm{Z}_{k}^{(m)})_{ii}
\label{eq:zik:vecs}
\end{equation}
referring to the $i$-th diagonal matrix element of the diagonal
matrices $\mathrm{Z}_{k}^{(m)}$, it follows
\[
   \mathrm{Z}_{k}^{(m)}\mathrm{\tilde{S}}_{ij} =
   \mathrm{\tilde{S}}_{ij}\mathrm{Z}_{k}^{(m)} =
   \underset{\equiv (\vec{z}_i+\vec{z}_j)_k }{\underbrace{
   \bigl( z_{k,i}^{(m)} + z_{k,j}^{(m)} \bigr) }
   }\cdot\mathrm{\tilde{S}}_{ij}
\text{,}
\]
and thus
\begin{subequations}\label{eq:Sp2m:[Z,S]}
\begin{align}
   \bigl[  Z_{k},\tilde{S}_{ij}^{\pm}\bigr]  & =
   \bigl[
      \tau_{z}\otimes\mathrm{Z}_{k}^{(m)},
      \tfrac{1}{2} (\tau_{x}\pm i\tau_{y}) \otimes\mathrm{\tilde{S}}_{ij}
   \bigr] \nonumber \\
   & = \underset{=\pm (\tau_{x}\pm i\tau_{y})}{\underbrace{
      \tfrac{1}{2} [ \tau_{z}, \tau_{x}\pm i\tau_{y} ] }}
   \otimes\bigl(  (\vec{z}_i+\vec{z}_j)_k \mathrm{\tilde{S}}_{ij}\bigr)
   \nonumber \\
   & =\pm (\vec{z}_i+\vec{z}_j)_k \cdot \tilde{S}_{ij}^{\pm}
\text{,}\label{eq:Sp2m:[Z,S2]}
\end{align}
(no summation over $i$ or $j$). Similarly, for the \RLOs
$\mathrm{S}_{ij}$ from the \SU{m} subalgebra, with
\[
   \bigl[  \mathrm{Z}_{k}^{(m)}, \mathrm{S}_{ij} \bigr] =
   (\vec{z}_i - \vec{z}_j)_k  \cdot \mathrm{S}_{ij}
\]
it follows,
\begin{align}
   \bigl[  Z_{k}, S_{ij} \bigr] & =
   \Bigl[  \tau_{z} \otimes \mathrm{Z}_{k}^{(m)},
      \begin{pmatrix}
         \mathrm{S}_{ij} & 0\\
         0 & - \mathrm{S}_{ij} ^{T}
      \end{pmatrix}
   \Bigr]  \nonumber\\
&= \begin{pmatrix}
      \bigl[  \mathrm{Z}_{k}^{(m)},\mathrm{S}_{ij} \bigr]  & 0 \\
      0 & +\bigl[  \mathrm{Z}_{k}^{(m)}, \mathrm{S}_{ij} ^{T}\bigr]
   \end{pmatrix} \nonumber\\
   & =(\vec{z}_i - \vec{z}_j)_k \cdot S_{ij}
\text{,}\label{eq:Sp2m:[Z,S1]}
\end{align}
\end{subequations}%
since $\left[ \mathrm{Z},\mathrm{S}^{T}\right] =-\left[ \mathrm{Z},
\mathrm{S}\right] ^{T}$. This confirms that the z-operators together
with the raising and lowering operators are in the expected canonical
form, indeed.

\subsubsection{Internal multiplet ordering}

The block-decomposition of \Eq{eq:Sp2m:Srep0} is not yet ordered \wrt
to the \RLOs, \ie the z-labels [here, by definition, it is assumed
that a raising (lowering) operator leads to a larger (smaller)
z-label in root space which directly links to the underlying sorting
implemented in root space].
The starting point, however, is correct: (i) The ($D=2m$) dimensional
first state $\vert e_1 \rangle $ [\cf \Eq{eq:def:ei}{}] does
represent the maximum weight state, indeed, and (ii) by applying the
$m-1$ lowering operators from the \SU{m} subalgebra, this iteratively
demotes the MW-state through the states $\vert e_2 \rangle$, \ldots,
$\vert e_m \rangle$. So far the state order is correct.

However, the next lower state is obtained by the $m$-th lowering
operator, \ie the one that links to the full \Sp{2m} symmetry. This
will generate the state $\vert e_D \rangle$, which thus is not in
order. Through another sequence of lowering operators from the \SU{m}
subalgebra, finally this proceeds through the states $\vert e_{D-1}
\rangle $, \ldots, $\vert e_{D-m+1} \rangle $ with additional
alternating signs. The full sequence of normalized states thus
obtained starting from the MW-state, can be collected as columns into
a unitary matrix $U$,
\begin{subequations}\label{eq:Sp2m:Usort}
\begin{equation}
   U\equiv
   \begin{pmatrix}
      1^{(m)} & 0 \\
      0 & \Sigma^{(m)}
   \end{pmatrix}
\text{,}\label{eq:Sp2m:resort:U}
\end{equation}
with the $m\times m$ dimensional matrix $\Sigma^{(m)}$
\begin{equation}
   \Sigma^{(m)}\equiv
   \begin{pmatrix}
      \cdot & \cdot & \cdot & \text{\reflectbox{$\ddots$}}\\
      \cdot & \cdot & +1 & \cdot\\
      \cdot & -1 & \cdot & \cdot\\
      +1 & \cdot & \cdot & \cdot
   \end{pmatrix}
\text{,}\label{eq:Sp2m:Sigma(m)}
\end{equation}
\end{subequations}%
to be distinguished from $\Sigma_y$ in \Eq{def:Sp2m:Yop} associated
with time-reversal symmetry. The unitary $U$ in \Eq{eq:Sp2m:resort:U}
maps the basis into the correct order \wrt to sorted z-labels, as is
assumed throughout this paper. Therefore this basis convention will
be used henceforth, which requires $U$ to be applied to all
generators.

The transformation of an arbitrary symmetry operation $S$ in
\Eq{eq:Sp2m:Srep} then leads to $ S \rightarrow U^{\dagger}SU$, that
is
\begin{equation}
   \begin{pmatrix}
      \mathrm{C} & \mathrm{D}^{\dagger}\\
      \mathrm{D} & -\mathrm{C}^{T}
   \end{pmatrix}
\rightarrow
   \begin{pmatrix}
      C & \left(  \Sigma^{T}D\right)  ^{\dagger}\\
      \Sigma^{T}D & -C^{t}
   \end{pmatrix}
\text{.} \label{eq:Sp2n:gen}%
\end{equation}
In $\Sigma^{T}D$, $\Sigma^{T}$ flips the order of the rows in $D$
with alternating signs, starting with $+1$ on the new first row. The
transformation $C^{t} \equiv \Sigma^{T}C^{T}\Sigma$ in the lower
right block, finally, corresponds to inversion of $C$ \wrt its center
with alternating checker-board like minus signs applied, starting
with plus signs along the regular matrix diagonal. With $C$
hermitian, when taken real, $C^{t}$ is equivalent to transposition
\wrt the \emph{minor} diagonal, \cite{Gilmore06} thus indicated by
superscript lowercase $t$ [this is in contrast to the standard
transposition $(\cdot)^T$ around the regular diagonal].

All generators inherited from the \SU{m} subalgebra thus become
\begin{equation}
   S \rightarrow
   \begin{pmatrix}
      S_{i} & 0\\
      0 & -S_{i}^{t}%
   \end{pmatrix}
\text{.}\label{eq:Sp2n:SUm}
\end{equation}
In particular, all z-operators have the diagonal in the lower-right
diagonal flipped to reverse order. The \emph{simple} \RLOs from the
\SU{m} subalgebra now have two strictly positive entries $+1$ at the
first upper subdiagonal at symmetric positions \wrt the \emph{center}
of the matrix. The remaining simple raising operator completing the
\Sp{2m} algebra (see below) is given by the matrix
$\tilde{\mathrm{S}}_{mm} = E_{mm} \to E_{mm}\Sigma = +E_{m1}$ in the
upper right block, thus naturally completing the set of simple
raising operators of the type
\begin{equation}
   S^{+}_{(\alpha=1)}=
   \begin{pmatrix}
      0 & 1 & \ \cdot\  & \ \cdot\  & \ \cdot\ & \ \cdot\ \\
      \cdot & 0 & 0 & \cdot & \cdot & \cdot\\
      \cdot & \cdot & 0 & \ddots & \cdot & \cdot\\
      \cdot & \cdot & \cdot & \ddots & 0 & \cdot\\
      \cdot & \cdot & \cdot & \cdot & 0 & 1\\
      \ \cdot\  & \ \cdot\  & \cdot & \cdot & \cdot & 0
   \end{pmatrix}
\text{,}\label{eq:Sp2n:SP}
\end{equation}
with $\alpha=1,\ldots,m$ indicating the position of the entries of
$1$ moving towards the center of the first upper off-diagonal.

\subsubsection{Multiplet labels for \Sp{2m} for $m=3$}

With the \RLOs defined to have at most two matrix elements exactly
equal to $1$, the canonical commutator relations in
\Eqs{eq:Sp2m:[Z,S]} directly depict the diagonal elements of the
z-matrices. As already indicated in \Eq{eq:zik:vecs}, these diagonals
can be combined as \emph{rows} into an $r\times D$ matrix $z_{k,i}$,
to be referred to as $z$-matrix, with $r=m$ being the rank of the
symmetry and $D=2m$ the dimension of the defining matrix
representation. The vectors $\vec{z}_i$ in \Eq{eq:zik:vecs} thus
refer to the \emph{columns} in the $z$-matrix, and therefore directly
reflect the $q_z$-labels, \ie the root space.

For \Sp{6}, this $3\times 6$ dimensional $z$-matrix reads
\begin{equation}
   z=\underset{ \equiv \tilde{z} }{\underbrace{
   \left(\begin{array}{rrr}
      1  & -1 &  0 \\
      1  &  1 & -2 \\
      1  &  1 &  1
   \end{array}\right| }}
   \left.\begin{array}{rrr}
       0 &  1 & -1 \\
       2 & -1 & -1 \\
      -1 & -1 & -1
   \end{array}\right)
\text{.}\label{eq:Sp6:zmat}
\end{equation}
By construction, all matrix elements are integers, for simplicity.
The z-labels of the defining representation are \emph{directly}
specified by the columns $\vec{z}_i$ of the $z$-matrix. Moreover,
since the z-labels are additive for tensor-product spaces, this
implies that the z-labels for \emph{arbitrary} \IREPs also contain
integers only.

Consequently, the root space is fully spanned by simple linear
integer combinations of the vectors $\vec{z}_i$. Furthermore, also
the action of the \RLOs themselves can be expressed as simple shifts
in root space [\cf \Eqs{eq:Sp2m:[Z,S]}{}]. While in the defining
representation, the z-labels in the carrier space are clearly unique,
they are not linearly independent. In particular, it is sufficient to
focus the discussion on the linearly independent subset of the
vectors $\vec{z}_i$ in terms of the leading $3\times 3$ block
$\tilde{z}$ of the $z$-matrix in \Eq{eq:Sp6:zmat}.

In terms of the three column vectors $\vec{z}_i$ in $\tilde{z}$, the
simple roots are given (i) by the simple roots of \SU{m}, which (ii)
is complemented by one further root involving $\vec{z}_3$,
\begin{equation}
   \begin{array}{lllrrrrll}
      \vec{\alpha}_{1}  &= \vec{z}_{1}-\vec{z}_{2} &= ( &2, &0,&\phantom{-}0 &)^T &
      \hat{=}& S_{12} \\
      \vec{\alpha}_{2}  &= \vec{z}_{2}-\vec{z}_{3} &= (&-1,&3,&0 &)^T &
      \hat{=}& S_{23} \\
      \vec{\alpha}_{3}  &=2\vec{z}_{3}             &= (&0,&-4,&2 &)^T &
      \hat{=}& \tilde{S}_{33}^+
   \end{array}
\text{,}\label{eq:Sp2m:roots}
\end{equation}
where the correspondence with the raising operators indicated in the
last column follows from \Eq{eq:Sp2m:[Z,S]}.
Having $\vec{\alpha}_{i}\cdot\vec{\alpha}_{j}\le 0$ for $i\neq j$
together with taking smallest integer combinations derived from the
action of \RLOs in \Eq{eq:Sp2m:[Z,S]}, this suggests \emph{simple}
roots. \cite{Pope06,Gilmore06}

Similar to \SU{N}, the convention on the sorting of the z-labels is
chosen lexicographic, yet as always, starting from the last z-label.
In this sense, the vectors $\vec{\alpha}_{i}$ in \Eq{eq:Sp2m:roots}
are greater than $(0,0,0)^T$, hence positive. The corresponding
operators thus \emph{increase} the z-labels, \ie correspond to
raising operators, indeed. Moreover, having reduced the symmetry to
its simple roots, equivalently, this also defines the set of
\emph{simple} \RLOs that are sufficient to fully explore multiplet
spaces.
Note that above convention on the sorting of the z-labels is already
also consistent with the state order in the defining representation
in \Eq{eq:Sp6:zmat}: the z-labels strictly decrease, starting from
the MW-state (the very left column) all the way to the last state
represented by the very right column.

In principle, the z-labels of the MW-state already could be used as
labels for the entire multiplet. However, using the vectors
$\hat{z}_i$ as (non-)orthogonal basis that spans the root space, also
$\vec{q} \equiv \tilde{z}^{-1} \max\left\{ \vec{z}\right\}$ could be
used as multiplet label, instead. The latter has the advantage that
it guarantees that the multiplet labels are strictly positive
integers or zero. For consistency with literature, however, the
multiplet labels for \Sp{2m} are still modified somewhat further, and
thus finally derived from the MW-state as follows,
\begin{subequations}\label{eq:Sp2m:qlabel}
\begin{equation}
   \vec{q} \equiv \underset{ \equiv Q }{\underbrace{ M\tilde{z}^{-1} }}
   \cdot\max\left\{ \vec{z}\right\}
\label{eq:MW:Sp6}
\end{equation}
where the matrix $M$,
\begin{eqnarray}
   M \equiv
   \left(\begin{array}{rrr}
      1 & -1 & \phantom{-}0\\
      0 & 1 & -1\\
      0 & 0 & 1
   \end{array}\right)
\text{,}\label{eq:Sp2m:qlabel:M}
\end{eqnarray}
has been added as a further minor modification for consistency with
standard literature \cite{LiE92} which further ensures that the
multiplet labels lie dense, \ie with $q=(q_1,q_2,q_3)$ any $q_i\ge0$
will result in a valid multiplet. Overall,
\begin{equation}
   Q \equiv \left(\begin{array}{rrr}
      1 & 0 & \phantom{-}0\\
      -\tfrac{1}{2} & \tfrac{1}{2} & 0\\
      0 & -\tfrac{1}{3} & \tfrac{1}{3}
   \end{array}\right)
\text{.}\label{eq:Sp2m:qlabel:Q}
\end{equation}
\end{subequations}
For example, when applied from the right to the $z$-matrix in
\Eq{eq:Sp6:zmat}, all resulting matrix elements (z-labels) are either
$\pm1$ or 0. In particular, the MW-state of the defining
representation of \Sp{2\cdot3} has the q-labels $(1,0,0)$.

\subsubsection{Construction of \Sp{2m} for $m$-channel setup}

Given the three-channel setup in the previous section with $m=3$, the
resulting defining representation for \Sp{2m} is
($2m=6$)-dimensional. As seen from the earlier introduction of this
model in \Sec{sec:3channel}, this contains an \SU{3} subalgebra,
together with a third z-operator, namely total particle conservation.
This subalgebra of a total of $9$ generators can now be completed by
$6$ raising operators together with their hermitian conjugates, \ie
their corresponding lowering operators. This leads to a total of $21$
generators, consistent with the dimension of the group \Sp{2\cdot3}.

Using a sorted z-label space, this requires that the unitary $U$ in
\Eq{eq:Sp2m:resort:U} is applied to all generators of the defining
representation, as well as to the initial spinor $\hat{\psi}^{[2m]}
\equiv ( \hat{c}_{1\uparrow}, \ldots, \hat{c}_{m\uparrow},
\hat{c}_{1\downarrow}^{\dagger}, \ldots,
\hat{c}_{m\downarrow}^{\dagger} )^T$ derived from \Eq{eq:Sp2m:Srep0}.
In case of $m=3$, the properly sorted $6$-dimensional spinor (\IROP)
spinor becomes,
\begin{equation}
   \hat{\psi}^{[6]}_{(\uparrow)}\equiv%
   \begin{pmatrix}
   \hat{c}_{1\uparrow}\\
   \vdots\\
   \hat{c}_{m\uparrow}\\
   +\hat{c}_{m\downarrow}^{\dagger}\\
   -\hat{c}_{m-1,\downarrow}^{\dagger}\\
   \vdots\\
   (-1)^{m-1}\hat{c}_{1\downarrow}^{\dagger}%
   \end{pmatrix}
\label{eq:Sp2m:psi}
\end{equation}
This naturally generalizes particle-hole symmetry in the presence of
channel symmetry. The symmetry preserving hopping term in
\Eq{eq:H3C}, for example, can now be written as scalar contraction
$\hat{h}_{k,k+1} = \sum_\sigma \bigl( \hat{\psi}^{[6]}_{k\sigma}\bigr
)^\dagger \cdot \hat{\psi}^{[6]}_{k+1,\sigma}$. Note that if, in
addition, also \SU[spin]{2} is present, this would further double the
dimension of the \IROP in \Eq{eq:Sp2m:psi} to a set of 12 operators,
such that the hopping term in \Eq{eq:H3C} can be written as
\emph{single} scalar contraction $\hat{h}_{k,k+1} =
\bigl(\hat{\psi}^{[12]}_k\bigr)^\dagger \cdot
\hat{\psi}^{[12]}_{k+1}$.

All generators are given in second quantization by the quadratic form
$\hat {S}_{\sigma}\equiv\hat{\psi}^{\dagger}S_{\sigma}\hat{\psi}$,
with $S_{\sigma}$ a $2m$-dimensional generator from the defining
representation. Specifically, the remaining $\tfrac{1}{2} m(m+1)$
raising operators for the $m$-channel setup in \Eq{eq:Sp2m:Sij} that
complete \Sp{2m} are given by
\begin{align}
   \tilde{S}_{ij}^{+}=\tfrac{1}{2}\left(  \tau_{x} + i\tau_{y}\right)
   \otimes \bigl(\mathrm{\tilde{S}}_{ij} \Sigma \bigr)\equiv
   \begin{pmatrix}
      0 & \mathrm{\tilde{S}}_{ij} \Sigma \\
      0 & 0
   \end{pmatrix}
\text{,}\label{eq:Sp2m:SopsX}
\end{align}
which leads to
\begin{equation}
   \hat{S}_{ij}^{+} \equiv
   \hat{\psi}^{\dagger}\tilde{S}_{ij}^{+}\hat{\psi} = \tfrac{1}{2}\bigl(
   \hat{c}_{i\uparrow}^{\dagger}\hat{c}_{j\downarrow}^{\dagger} +
   \hat{c}_{j\uparrow}^{\dagger}\hat{c}_{i\downarrow}^{\dagger}\bigr)
\text{.}\quad(\text{all }i,j)
\label{eq:mchannel:Spij}
\end{equation}
This generates a pair of particles, the nature of which originates
from the underlying general particle-hole symmetry. With $\bigl\{
\hat{\psi}_{i},\hat{\psi}_{j}^{\dagger}\bigr\} =\delta_{ij}$ for
$\nu=1,\ldots,2m$, and therefore
\begin{eqnarray}
   \bigl[  \hat{S}_{\sigma},\hat{S}_{\sigma^{\prime}}\bigr] &\equiv&
   \bigl[
      \hat{\psi}_{i}^{\dagger}\left(  S_{\sigma}\right)  _{ij}\hat{\psi}_{j},
      \hat{\psi}_{i'}^{\dagger}\left(  S_{\sigma'}\right)_{i'j'}\hat{\psi}_{j'}
   \bigr] \nonumber \\ &=&
   \hat{\psi}^{\dagger}
   \left[  S_{\sigma},S_{\sigma^{\prime}}\right]  \hat{\psi}
\text{,}\label{eq:CR:mat2Sops}
\end{eqnarray}
the commutator relations within the matrix representations of the
defining representation earlier directly carry over to the quadratic
second-quantized operators as in \Eq{eq:mchannel:Spij}.

\section{Numerical implementation \label{sec:num:implement}}

Tensor-product spaces are an essential ingredient to numerical
renormalization group techniques such as NRG or DMRG. State spaces
are enlarged iteratively by adding a small local state space at a
time, \ie a physical site with a few degrees of freedom. With respect
to the description of strongly-correlated entangled quantum many-body
states, this leads to a description which is well-known as matrix
product states (MPS). Both, the existing state space (iteratively
constructed itself) as well as the newly added state-space, have
finite dimension and well-defined symmetry labels. New
representations can therefore only emerge through the tensor product
of the two spaces. In particular, all iteratively constructed quantum
many body states strictly derive from the \IREPs of the elementary
sites. With operators usually acting locally, these are also
expressed in the symmetries of the local basis. Furthermore, the
local state space of a site is usually small. For example, a
fermionic site has the four states described in \Eq{fstates:SC4}.
Therefore the \IREPs present within the local state space are usually
just the smallest non-trivial \IREPs, often just the defining
representation itself. For identical sites, the local symmetry space
can be setup once and for all at the beginning of the calculation.

Having identified and labeled all symmetries on the local site level,
this sets the stage for generic iterative algorithms such as NRG or
DMRG. The remainder is a large exercise on tensor-product spaces. By
construction, the iteratively combined spaces are finite, yet as they
grow rapidly, they are eventually truncated on the multiplet level
while leaving the symmetry content of the individual multiplets, \ie
the CGC spaces, fully intact.

\subsection{Tensor product decomposition of symmetry spaces
\label{Sec:NumDeComp}}

The decomposition of the tensor-product space of two \IREPs into
irreducible multiplets has already been discussed more generally in
\Secs{sec:irrep} and \ref{sec:prodspaces}. In the actual numerical
implementation, however, in particular the presence of inner and out
multiplicity must be taken care of meticulously for overall
consistency. This will be discussed in the following.

Similar to \Sec{sec:prodspaces}, consider a specific arbitrary
non-abelian symmetry group $\mathcal{S}$ whose Clebsch-Gordan
coefficients may not necessarily be easily accessible analytically
for arbitrary multiplets. Assume two of its \IREPs, $q_1$ and $q_2$,
with dimensions $d_{q_1}$ and $d_{q_2}$, respectively, are known
together with their irreducible representations of the generators
$I_{\sigma}^{[q_1]}$ and $I_{\sigma}^{[q_2]}$, specifically the
z-operators (Cartan subalgebra) and the simple \RLOs (simple roots).
In practice, these representations either refer to small \IREPs such
as the defining representation, or have been generated through prior
iterative calculations. As in \Eq{eq:add-sym-op}, consider their
tensor-product,
\begin{equation}
    R_{\sigma}^{\mathrm{tot}} \equiv
    I_{\sigma}^{[q_1]} \otimes \Id^{[q_2]}
  + \Id^{[q_1]} \otimes I_{\sigma}^{[q_2]}
\text{,} \label{eq:add-sym-op2}%
\end{equation}
resulting in matrices of dimension $D = d_{q_1} d_{q_2}$. Clearly the
commutator relations are preserved, and the z-labels are
\textit{additive} under this operation [\cf \Secp{sec:prodspaces}].

In order to determine the decomposition into \IREPs, a tempting route
may be through the construction of the group's Casimir operators in
the combined state space and their simultaneous diagonalization
together with the z-operators. However, in the presence of outer or
inner multiplicity, subspaces exist that are fully degenerate in
Casimir operators as well as in the z-operators. In this case, for
overall consistency a \emph{unique deterministic} algorithm must be
constructed that (i) separates multiplets in the presence of outer
multiplicity, and (ii) fixes a choice of basis for degenerate spaces
within a multiplet in the presence of inner multiplicity. Moreover,
the explicit construction of the Casimir operators bears some efforts
of its own. In practice, therefore a more straightforward approach
has been adopted, instead, as will be explained in the following.

The main hurdles in the decomposition of the tensor-product in
\Eq{eq:add-sym-op2} into \IREPs is the possible occurrence of outer
and inner multiplicity. The strategy employed here to deal with this
situation is based on the uniqueness and accessibility of the MW
(maximum-weight) states as introduced in \Sec{sec:maxweight}. For
this, throughout the procedure below, the \emph{same} lexicographic
sorting scheme of the z-labels, used to obtain the MW-state in
\Eq{MWlabel}, is employed to order all states within an \IREP. The
sorting is descending, such that the MW-state appears \textit{first}
within a multiplet.

\begin{figure}[tb]
\begin{center}
\includegraphics[width=0.8\linewidth]{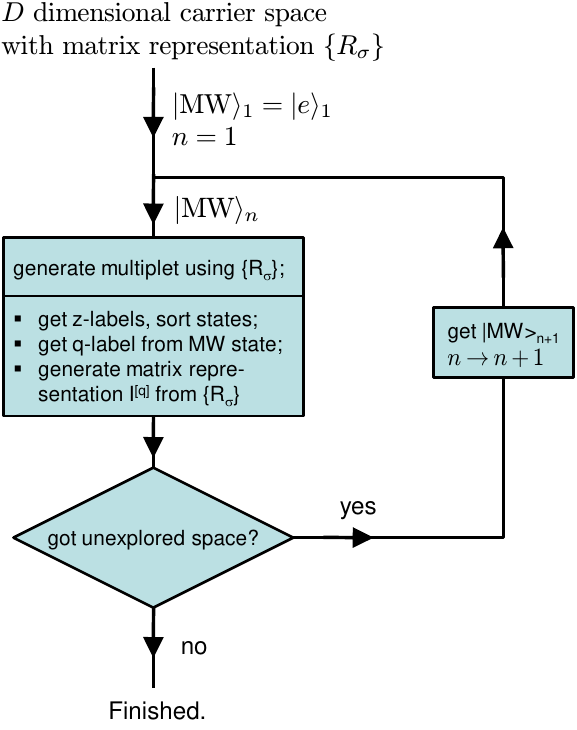}
\end{center}
\caption{ Schematic procedure of state space decomposition
of given $D$ dimensional vector space with known set of generators
$\{R_{\sigma}^{(\mathrm{tot})}\}$.
}\label{fig:statedecomp}%
\end{figure}

Since the z-labels are additive, it also follows for a tensor product
of two such representations that the first state automatically also
represents a MW-state of some multiplet,
\begin{equation}
    \vert\mathrm{MW}\rangle_{1} \equiv \vert e_{1} \rangle
\label{eq:decomp:MW1}
\text{,}%
\end{equation}
where the vectors $\vert e_{k} \rangle$ [\cf \Eq{eq:def:ei}{}] form
the cartesian basis for the $D$-dimensional space of the
representation $R_{\sigma}^{\mathrm{tot}}$ in \Eq{eq:add-sym-op2}.
Given that the MW-state of a representation is guaranteed to be
unique, \cite{Elliott79,Pope06,Gilmore06} the state $\vert
\mathrm{MW} \rangle_{1}$ is already a proper symmetry eigenstate, \ie
an eigenstate of all z-operators. This was always double checked, in
practice, as a safety measure.
The further procedure then is schematically depicted in the work flow
diagram in \Fig{fig:statedecomp}: starting with $\vert
\mathrm{MW}\rangle_{n=1}$,
\begin{enumerate}

\item the symmetry eigenstate $\vert \mathrm{MW}\rangle _{n}$ is
    used as the \textit{seed state} to sequence its complete
    \IREP (the \textit{current} multiplet). This is done by
    repeatedly applying an \emph{arbitrary but fixed order of
    simple lowering operators only} to the current set of vectors
    in the multiplet. Therefore starting with the MW-state $\vert
    \mathrm{MW}\rangle_{n}$ and adding the newly acquired
    symmetry states one at a time, this introduces a well-defined
    state order, independent of whether their z-labels are
    degenerate or not. In the presence of inner multiplicity, it
    is important to notice, however, that it is not guaranteed
    that a newly acquired state is automatically orthogonal to
    the already existing states within the current multiplet.
    Therefore, a newly acquired state, if it represents a new
    vector space component, must be orthonormalized with respect
    to the existing states. This is repeated, until the current
    multiplet space is exhausted.

\item The states in the multiplet thus generated, by
    construction, already have well-defined z-labels (this again
    was double-checked, in practice); the states are sorted with
    respect to these labels in descending lexicographic order
    while keeping subspaces that are degenerate in the z-labels
    in their \emph{original order} in order to remain
    deterministic. Within this order, the first state defines the
    label for the generated multiplet, \ie $q=\qMW$. In addition,
    the matrix representation in \Eq{eq:add-sym-op2}, when cast
    into the current \IREP space results in the newly generated
    irreducible matrix representation $\hat{I}_{\sigma}^{[q]}$.

\item If the $D$-dimensional vector space is not fully exhausted
    yet, a new seed state is determined by finding the
    \textit{smallest} $k$ for which $\vert e_{k}\rangle$ exhibits
    a new vector component \wrt the symmetry states already
    collected. Having started with $k=1$ above, it follows $k>1$.
    After proper orthonormalization with respect to the
    previously explored space, this state becomes the next seed
    state. If it already does represent a MW-state, which is
    typically the case in that it is destroyed by all raising
    operators, then $\vert \mathrm{MW}\rangle_{n+1}$ has been
    found. Otherwise repeatedly apply simple raising operators on
    the current seed state until the unique new maximum weight
    state $\vert \mathrm{MW}\rangle_{n+1}$ is reached. Continue
    with (1), setting $n\rightarrow n+1$.

\item If on the other hand, the $D$-dimensional vector space in
    \Eq{eq:add-sym-op2} is already fully exhausted, the
    decomposition of the tensor-product space into $n$
    irreducible representations is completed, and the procedure
    terminates.

\end{enumerate}

Note that no explicit reference to z-labels has been made, except for
step (2). That step, however, is actually not required right away for
the decomposition, with its results only relevant for subsequent
calculations. By construction, therefore this procedure is
deterministic and does not dependent on dealing with degeneracies in
the z-labels or inner and outer multiplicities. The MW-states are
accessible by keeping \IREPs sorted in their z-labels throughout.
They represent the entry point in sequencing its \IREP, which
\textit{guarantees} that inner and outer multiplicities are dealt
with in a consistent fashion. Finally, note that the choice of the
seed states $\vert \mathrm{MW}\rangle_{n}$, \ie starting with $+\vert
e\rangle_k$, also provides the sign convention.

The resulting unitary transformation into the irreducible symmetry
subspaces directly determines (i) the Clebsch Gordan coefficients,
and (ii) the matrix-representations of the newly generated \IREPs.
With only a few Clebsch-Gordan coefficients usually unequal zero and
of order $1$, small numbers below a numerical noise threshold for
double-precision ($10^{-12}$) are neglected, \ie set to zero.
Moreover, a non-zero Clebsch-Gordan coefficient can typically be
expressed as a rational number, or the square root of a rational
number, an efficient approximation of which can be found through
continued fraction techniques. Therefore if an excellent fractional
approximation was found within the same accuracy of $10^{-12}$, this
rational approximation also was used, instead.

\subsection{State space initialization and operator compactification
\label{sec:fock2sym}}

In the presence of several symmetries, a given state space is
represented by a certain set of multiplet combinations. For a single
fermionic site in the presence of spin-symmetry and particle-hole
symmetry, this still can be easily characterized by hand [\cf
\Eq{fstates:SC4}{}]. The situation, however, can quickly become more
involved. For example, for a spinful three-channel calculation with
\SU{3} channel symmetry as in \Eqp{eq:HKondoAM-H}, a site is
represented by $4^{3}=64$ states (4 fermionic states for each of the
3 channels). If for example, particle number, spin symmetry and
channel symmetry is preserved, then this system exhibits \ASC
symmetry, as discussed in \Sec{sec:3channel}. Given these symmetries,
the 64-dimensional Hilbert space of a site cannot be decomposed into
a tensor product of convenient smaller units with already
well-defined \ASC symmetry labels themselves. For the channel
symmetry it is essential, of course, that all three channels are
present, while it is also essential for the spin symmetry that both
spin species are present. Therefore in the example above, the
64-dimensional space of site already appears as the smallest building
block. It can be reduced to a set of irreducible multiplet
combinations, of course, but explicit determination can quickly
become tedious if done by hand, while the problem can be tackled
completely generally and straightforwardly on a numerical level.

In order to get started numerically, a simple and natural starting
point is the Fock space representation. While this usually does not
represent the symmetry eigenbasis, of course, nevertheless all
generators of the symmetries present, in particular its raising,
lowering, and z-operators, are known in second-quantized form and can
be equally constructed in Fock space.

The z-operators typically have a simple form. In particular, for
3-channel setup mentioned above, the z-operators are already all
diagonal in the Fock space, \cf \Eqs{eq:3C:Sops}, (\ref{eq:3C:Cz}),
or (\ref{sop3C:TUY}). This thus already provides the z-labels. Next,
note that the order of the states \wrt to their z-labels is important
for consistency with the Clebsch Gordan coefficients later, which
suggests using the same lexicographic order as for the determination
of the MW-states. Sorting the states in this order and applying the
same prescription for state space decomposition as explained in
\Sec{Sec:NumDeComp}, this suffices to fully identify all symmetry
multiplets within the given $D$-dimensional Hilbert space.

\subsubsection{Compactifying operators using Wigner-Eckart theorem}

Irreducible operator sets can be equally constructed starting from
the Fock space representation of a \emph{seed operator} that is part
of some irreducible operator set. This seed operator is typically
known, yet can be completed to an \IROP set, by using the \RLOs in
Fock space representation and numerically evaluating the commutators
in \Eq{def:irrep-op} (see also subsequent discussion in
\Sec{sec:irop-decomp}{}). Using the same unitary transformation that
brings the Fock space into the correct symmetry eigenbasis as
described above, the \IROP set is rotated into the space of symmetry
eigenstates. With this, however, this \IROP set is still represented
in the \emph{fully expanded} multiplet space, \ie this space still
references both multiplet labels and their corresponding z-labels on
the same flat level. However, through the Wigner-Eckart theorem,
\Eq{WET:motiv},
\begin{equation}
    \langle q q_{z} \vert \hat{F}^{q_{1}}_{q_{1z}} \vert q_{2}q_{2z}\rangle
  = \langle q \Vert \hat{F}^{q_{1}} \Vert q_{2}\rangle \cdot
    C_{q_{1z} q_{z}(\sigma)}^{[q_{2z}]}
\text{,}\label{eq:op-compact}
\end{equation}
many of the matrix elements can be related to each other through
Clebsch Gordan coefficients. The \IROP set can therefore be
\textit{compactified} as a tensor-product of reduced matrix-elements
$\langle q\Vert \hat{A}_{q_{1}} \Vert q_{2}\rangle$ in the multiplet
space times the CGC space $C_{q_{1z} q_{z}(\sigma)} ^{[q_{2z}]}$.

The CGC spaces are known from a separate numerical calculation, \eg
they can be generated by several iterations of tensor-product
decompositions starting from the defining representation. Therefore,
the final compactification in \Eq{eq:op-compact} of the fully
expanded matrix elements of the \IROP also serves as a major
consistency check. The first non-zero matrix-element $\langle q q_{z}
| \hat{F}^{q_{1}} _{q_{1z}} | q_{2}q_{2z}\rangle$ for the already
known multiplet spaces $(q;q_{1},q_{2})$ can be used to determine the
reduced matrix element $\langle q\Vert \hat{F}^{q_{1}} \Vert
q_{2}\rangle$, with its corresponding Clebsch Gordan coefficient
known. This, however, immediately predicts the existence of a set of
other non-zero matrix elements within the same multiplet spaces
$(q;q_{1},q_{2})$. These matrix elements must exist and agree within
numerical noise. The matched matrix elements are marked and
considered taken care of. If the same value of a matrix element
occurs several times within the multiplets $(q;q_{1},q_{2})$ for the
same z-labels, the first one that matches is taken. This check is
thus not entirely unique, but nevertheless a strong one, and
sufficient to obtain the space of reduced matrix elements. Finally,
having identified all non-zero matrix elements, the multiplet matrix
element space $\langle q\Vert \hat{F}^{q_{1}} \Vert q_{2}\rangle$ are
stored together with their referenced CGC space in terms of a \QSpace
as discussed in the main text.

\section{Example \QSpaces \label{app:QSpaces}}

\QSpaces represent an efficient numerical description of tensors of
arbitrary rank in the presence of arbitrary quantum symmetries [\cf
\Eqp{eq:QSpace:Rep}]. This includes both abelian and non-abelian
symmetries, with the extension to further symmetries such as point
symmetries being straightforward. The \QSpaces are decomposed into a
set of reduced multiplet spaces together with their respective CGC
(Clebsch Gordan coefficient) spaces. In the following several
elementary examples of \QSpaces are given as they appeared in
practice. Elementary \QSpaces typically have rank-2 (such as scalar
operators with identity CGC spaces) or rank-3 (\IREPs and \IROPs with
reference to standard rank-3 CGC spaces), while combinations of these
through subsequent algebraic operations can easily result in
higher-rank intermediate objects.

The notation regarding the elementary data arrays will be as follows.
Plain matrices of dimension $m\times n$ will be written as $a=\left[
a_{11},\ldots ,a_{1n};\ \ldots;\ a_{m1},\ldots,a_{mn}\right] $, \ie
$m$ rows of equal length $n$ separated by semicolons. The commas
within a row are considered optional. In order to deal with $m\times
n\times k$ dimensional rank-3 objects, the notation $\left\{
a_{1},a_{2},\ldots,a_{k}\right\} $ is used, which shall indicate that
the matrices $a_{1},\ldots,a_{k}$, all of the same dimension $m\times
n$, are concatenated along the third dimension. Trailing singleton
dimensions will be considered implicit if required, \eg a scalar such
as $1.$ can stand for an arbitrary rank-r object in that a number
also represents a $1\times 1\times\ldots \times1$ object. Identity
matrices of dimension $n$ will be denoted by $\Id^{(n)}$.

\subsection{Fermionic site with \AS symmetry}

Consider the state space of a single fermionic site with the four
states: empty $\vert 0\rangle $, singly occupied
$\vert\uparrow\rangle$ and $\vert\downarrow\rangle$, and double
occupied $\vert\uparrow\downarrow \rangle$. The symmetries considered
are particle conservation \UA{charge}, and full spin symmetry
\SU[spin]{2}. The z-operators are $\hat{C}_{z} \equiv \tfrac{1}{2} (
\hat{n}_{\uparrow} + \hat{n}_{\downarrow} - 1)$ and $\hat{S}_{z}
\equiv \tfrac{1}{2} ( \hat{n}_{\uparrow} - \hat{n}_{\downarrow})$,
with the corresponding quantum labels $C_{z}$ for charge and $S$ for
total spin. For consistency with later, here the charge is treated as
the reduction of the non-abelian particle-hole symmetry to its
abelian part, which also reduces the set of symmetry operations to
the z-operator $\hat{C}_{z}$ only [hence the factor $\tfrac{1}{2}$].
Consequently, the z-label of the underlying non-abelian symmetry is
promoted into a q-label, while the CGC space becomes trivial ($1.$)
with internal \emph{multiplet} dimension of $1$. In order to stress
the difference between the original z-label which can become
negative, and the \SU{2} q-labels of multiplets which are positive,
by definition, the q-labels are therefore written as ($+C_z,S$),
emphasizing the origin of the q-label $C_z$ being derived from a
z-operator.

\subsubsection{Symmetry space and operators of one site}

The states $\left\vert 0\right\rangle $, $\left\vert \uparrow
\right\rangle$, $\left\vert \downarrow \right\rangle $, and
$\left\vert \uparrow\downarrow \right\rangle$ already represent the
correct symmetry eigenstates [\cf \Eq{fstates:SC4}{}],
\begin{equation}
  \begin{tabular}[c]{ll|c}
    \multicolumn{2}{l|}{multiplet space} & dimension \\
    \multicolumn{2}{l|}{$\vert C_{z};S\rangle $} & $d_{C_{z}}\times d_{S}=d_{\mathrm{tot}}$
  \\[0.5ex]\hline
    $\left\vert -\tfrac{1}{2};0 \right\rangle $ & $\equiv \vert 0\rangle$ & $1\times1=1$\\
    $\left\vert +\tfrac{1}{2};0 \right\rangle $ &
       $\equiv \left\vert \uparrow\downarrow \right\rangle$ & $1\times1=1$\\
    $\left\vert \ \,\ 0; \tfrac{1}{2} \right\rangle$ &
       $\equiv \{ \left\vert\uparrow\right\rangle, \left\vert\downarrow\right\rangle \}$ &
    $1\times2=2$
  \end{tabular}\label{site:ZSbasis}%
\end{equation}
The matrix elements of a generic Hamiltonian in this basis can be
written as \QSpace [see definition in \Eqp{eq:QSpace:Rep}],
\begin{equation}
   H \equiv \left\{
   \begin{tabular}[c]{cc|c|cc}
      $\left(  C_{z};S\right)  $ & $\left(  C_{z}^{\prime};S^{\prime}\right)  $ &
      $\Vert H\Vert$ & \multicolumn{2}{|l}{CGC spaces} \\ \hline\hline
      $-\tfrac{1}{2};0$ & $-\tfrac{1}{2};0$ & $h_{-\tfrac{1}{2},0}$ & $1.$ & $1.$\\
      $+\tfrac{1}{2};0$ & $+\tfrac{1}{2};0$ & $h_{+\tfrac{1}{2},0}$ & $1.$ & $1.$\\
      $0;\tfrac{1}{2}$ & $0;\tfrac{1}{2}$ & $h_{0,\tfrac{1}{2}}$ & $1.$ &
      $\Id^{\left(  2\right)  }$%
   \end{tabular}
\quad\right\}\label{QSpace:H0_ez}%
\end{equation}
The Hamiltonian is a scalar operator, hence its rank as an \IROP can
be reduced from three to two, as it is the only operator in its
irreducible set. Consequently, all CGC spaces reduce to the identity,
as reflected in the last two columns of the \QSpace
\Pref{QSpace:H0_ez}. Each of the remaining two indices explicitly
refers to symmetry states, hence the \QSpace requires the two sets of
q-labels $q\equiv\left( C_{z};S\right)$ and $q^{\prime}\equiv\left(
C_{z}^{\prime}; S^{\prime}\right)$ referring to the first (second)
index shown in the first (second)\ column, respectively. With the
Hamiltonian preserving the symmetries, it must be block-diagonal, \ie
$q=q^{\prime}$ for all records in \Pref{QSpace:H0_ez}. Both of the
symmetry spaces $\left( \pm \tfrac{1}{2};0\right) $ have a single
state only, therefore the corresponding entries in the multiplet
space $h_{\pm 1/2,0}$ are $1\times1$ dimensional blocks, \ie numbers.
The last symmetry multiplet $\left( 0;\tfrac{1} {2}\right) $ has two
states owing to the \SU{2} symmetry [see \Pref{site:ZSbasis}{}]. By
means of the Wigner Eckart theorem, the space of reduced matrix
elements, $h_{0,1/2}$, is therefore again a number while the CGC
space becomes a 2-dimensional identity matrix. Therefore the most
general representation of a scalar operator for a single fermionic
level in the presence of \AS symmetry is given by the three numbers
$\{h_{-1/2,0},h_{+1/2,0},h_{0,1/2}\}$ in the multiplet space. The
remaining matrix elements are constrained due to symmetry.

\begin{table*}[ptb] %% ptb
\caption{ Example \QSpaces in the presence of \AS symmetry for a
single fermionic site ($\hat{\psi}_{S}^{\dagger}$ and
$\hat{\psi}_{S}$), and for the combination of the state space of two
sites (\Atensor{}). Having the two symmetries of abelian \UA{charge}
and non-abelian \SU[spin]{2}, the respective CGC spaces C (trivial)
and S appear in the right columns. The record index $\nu$ in the
first column, as well as the explicit specification of the dimensions
of the reduced multiplet space and the combined CGC spaces are just
added for better clarity. For comparison, \Tbl{QSpaces:CS} shows how
the \QSpaces \Pref{QSpace:psiS} and \Pref{QSpace:A0_zs} are modified
for the case that the abelian charge symmetry also becomes a
non-abelian \SU[charge]{2} particle-hole symmetry.}

\begin{subequations}
\begin{align}
   \psi_{S}^{\dagger} \equiv \left\{
   \begin{tabular}[c]{c|ccl|cc|ccc}
      record & & & & \multicolumn{2}{|l|}{reduced matrix elements} & \multicolumn{2}{|l}{CGC spaces} & \\
      index $\nu$ & $(C_{z};S)$ & $(C_{z}^{\prime};S^{\prime})$ &
      $(C_{z}'';S'')$ &
      $\Vert\psi_{S}^{\dagger}\Vert$, & dimension & C & S & dimension
   \\ \hline\hline
      1. & $0;\tfrac{1}{2}$ & $\tfrac{-1}{2};0$ & $\tfrac{+1}{2};\tfrac{1}{2}$ &
      $1.$ & $1\times 1\times 1$ & $1.$, & $\{  [1;0],[0;1]\}$ &
      $2\times 1\times 2$
   \\[1ex]
      2. & $\tfrac{+1}{2};0$ & $0;\tfrac{1}{2}$ & $\tfrac{+1}{2};\tfrac{1}{2}$ &
      $\sqrt{2}$ & $1\times 1\times 1$  & $1.$, & $\tfrac{1}{\sqrt{2}}\{[0\ 1], [-1\ 0]\}$ &
      $1\times 2\times 2$
   \end{tabular}
   \right\}\label{QSpace:psiS}%
\end{align}

\begin{align}
   \psi_{S} \equiv \left\{
   \begin{tabular}[c]{c|ccl|cc|llc}
      record & & & & \multicolumn{2}{|l|}{reduced matrix elements} & \multicolumn{2}{|l}{CGC spaces} & \\
      index $\nu$ & $(C_{z}; S)$ & $(C_{z}^{\prime};S^{\prime})$ &
      $( C_{z}^{\prime\prime};S^{\prime\prime} )$ &
      $\Vert\psi_{S}\Vert$, & dimension & C & S & dimension
   \\ \hline\hline
      1. & $0;\tfrac{1}{2}$ & $\tfrac{+1}{2};0$ & $\tfrac{-1}{2};\tfrac{1}{2}$ &
      $1.$ & $1\times 1\times 1$ & $1.$, & $\{ [1;0],[0;1]\}$  & $2\times 1\times 2$
   \\[1ex]
      2. & $\tfrac{-1}{2};0$ & $0;\tfrac{1}{2}$ & $\tfrac{-1}{2};\tfrac{1}{2}$ &
      $-\sqrt{2}$ & $1\times 1\times 1$  & $1.$, &
      $\tfrac{1}{\sqrt{2}}\{[0\ 1],[-1\ 0]\}$ & $1\times 2\times 2$
   \end{tabular}
   \right\}\label{QSpace:psiS-2}%
\end{align}
\end{subequations}

\begin{quote}
\QSpaces $\hat{\psi}_{S}^{\dagger} = \{ \hat{c}_{\uparrow}^{\dagger},
\hat{c}_{\downarrow}^{\dagger} \}$ and $\hat{\psi}_{S} = \{
-\hat{c}_{\downarrow }; \hat{c}_{\uparrow} \}$ representing \IROPs
for a single spinful fermionic level [\cf \EQSp{eq:IROP-psiS2}]. Note
that the \IROP $\hat{\psi}_{S}^{\dagger}$ is interpreted differently
compared to the \IROP $(\hat{\psi}_{S})^\dagger$, hence
$\hat{\psi}_{S}^{\dagger} \neq (\hat{\psi}_{S})^\dagger$ [\eg note
the sign in multiplet space $\Vert \hat{\psi}_{S} \Vert$ in the
second record of $\hat{\psi}_{S}$ or the reverted signs in the
q-labels for $(C_{z}'';S'')$ associated with the \IROP in the third
column; see text].
\end{quote}%

\begin{align}
   A \equiv \left\{ \begin{tabular}[c]{c|ccc|ll|ccc}
   record & site 1 & site 2 & combined & multiplet space &  &
   \multicolumn{2}{|l}{CGC Spaces} & \\
   index $\nu$ & $(C_{z};S)$ & $(C_{z}^{\prime};S^{\prime})$ &
   $(C_{z}^{\prime\prime}; S^{\prime\prime})$ & $\Vert A\Vert$ & dimension & C & S & dimension
\\ \hline\hline
   1. & $\tfrac{-1}{2};0$ & $\tfrac{-1}{2};0$ & $-1;0$ & $1.$ & $1 \times1
   \times1$ & $1.$ & $1.$ & \multicolumn{1}{l}{$1 \times1 \times1$}%
\\[0.5ex]\hline
   2. & $\tfrac{-1}{2};0$ & $0;\tfrac{1}{2}$ & $\tfrac{-1}{2};\tfrac{1}{2}$ &
   $\{ [1],[0] \}  $ & $1 \times1 \times2$ & $1.$ &
   $\{ [1\ 0],[0\ 1] \}  $ & \multicolumn{1}{l}{$1 \times2 \times2$} \\
   3. & $0;\tfrac{1}{2}$ & $\tfrac{-1}{2};0$ & $\tfrac{-1}{2};\tfrac{1}{2}$ &
   $\{ [0],[1] \}  $ & $1 \times1 \times2$ & $1.$ &
   $\{ [1;0],[0;1] \}  $ & \multicolumn{1}{l}{$2 \times1 \times2$}
\\[1ex]\hline
   4. & $\tfrac{-1}{2};0$ & $\tfrac{1}{2};0$ & $0;0$ & $\{ [1],[0],[0] \}$
   & $1 \times1 \times3$ & $1.$ & $1.$ & \multicolumn{1}{l}{$1 \times1 \times1$} \\
   5. & $0;\tfrac{1}{2}$ & $0;\tfrac{1}{2}$ & $0;0$ & $\{ [0],[1],[0] \}$
   & $1 \times1 \times3$ & $1.$ &
   $[  0\ \tfrac{-1}{\sqrt{2}};\ \tfrac{1}{\sqrt{2}}\ 0 ]$ &
   \multicolumn{1}{l}{$2\times2 \times1$} \\
   6. & $\tfrac{1}{2};0$ & $\tfrac{-1}{2};0$ & $0;0$ & $\{ [0],[0],[1] \}$ &
   $1 \times1 \times3$ & $1.$ & $1.$ & \multicolumn{1}{l}{$1 \times1 \times1$}
\\[1ex] \hline
   7. & $0;\tfrac{1}{2}$ & $0;\tfrac{1}{2}$ & $0;1$ & $1.$ & $1 \times1 \times1$
   & $1.$ &$
   \begin{array}[c]{c}\bigl\{
     {[} 1\ 0;0\ 0 ], \\
     {[} 0\ \tfrac{1}{\sqrt{2}};\ \tfrac{1}{\sqrt{2}}\ 0 ], \\
     {[} 0\ 0;0\ 1 ]
   \bigr\}\end{array}
   $ & \multicolumn{1}{l}{$2 \times2 \times3$}
\\[5ex]\hline
   8. & $0;\tfrac{1}{2}$ & $\tfrac{1}{2};0$ & $\tfrac{1}{2};\tfrac{1}{2}$ &
   $\{ [1],[0] \}$ & $1 \times1 \times2$ & $1.$ & $\{ [1;0],[0;1] \}$ &
   \multicolumn{1}{l}{$2 \times1 \times2$} \\
   9. & $\tfrac{1}{2};0$ & $0;\tfrac{1}{2}$ & $\tfrac{1}{2};\tfrac{1}{2}$ &
   $\{ [0],[1] \}$ & $1 \times1 \times2$ & $1.$ &
   $\{ [1\ 0],[0\ 1] \}  $ & \multicolumn{1}{l}{$1 \times2 \times2$}
\\[1ex]\hline
   10. & $\tfrac{1}{2};0$ & $\tfrac{1}{2};0$ & $+1;0$ & $1.$ & $1 \times1 \times1$ &
   $1.$ & $1.$ & \multicolumn{1}{l}{$1 \times1 \times1$}%
   \end{tabular}
   \right\}\label{QSpace:A0_zs}%
\end{align}

\begin{quote}
\QSpace of identity \Atensor combining two fermionic sites. Site 1
with symmetries $(C_{z}; S)$ and site 2 with symmetries
$(C_{z}^{\prime}; S^{\prime})$ are combined into the global symmetry
$(C_{z}^{\prime\prime} ;S^{\prime\prime})$.
The specific order of the records is irrelevant and hence arbitrary.
Here, the records have been sorted with respect to the combined
quantum labels $q^{\prime\prime }\equiv (C_{z}'';S'')$, where groups
with the same $q^{\prime\prime}$ are indicated by horizontal lines
for clarity. The dimensions in the last (third) index are therefore
the same within a group that shares the same $(C_{z}'';S'')$.
\end{quote}
\label{QSpaces:AS}
\end{table*} %% table*

As an example for a non-scalar \IROP, consider the spinor of particle
creation operators $\hat{\psi}_{S}^{\dagger} = \{
\hat{c}_{\uparrow}^{\dagger}, \hat{c}_{\downarrow}^{\dagger} \}$ that
encodes \SU{2} spin symmetry [\cf \Eq{eq:IROP-psiS2}], with its
\QSpace representation shown in \Pref{QSpace:psiS}. The z-labels of
the \IROP set $\hat{\psi }_{S}^{\dagger}$ are determined through the
z-operators $\hat{C}_{z}$ and $\hat{S}_{z}$ acting on the components
of $\hat{\psi}_{S}^{\dagger}$,
\begin{align*}
   {[}\hat{C}_{z},\hat{c}_{\sigma}^{\dagger} ]  &= \tfrac{+1}{2} \cdot
   \hat{c}_{\sigma}^{\dagger}\\
   {[}\hat{S}_{z},\hat{c}_{\sigma}^{\dagger}]  &= \tfrac{\sigma}{2} \cdot
   \hat{c}_{\sigma}^{\dagger}
\text{,}%
\end{align*}
with $\sigma\equiv\left\{ \uparrow,\downarrow\right\} \equiv\left\{
+1,-1\right\} $. The \IROP$\hat{\psi}_{S}^{\dagger}$ is therefore
identified with the multiplet $q^{\prime\prime}\equiv\left(
C_{z}^{\prime\prime };S^{\prime\prime}\right) =\left(
\tfrac{+1}{2};\tfrac{1}{2}\right)$, as indicated in the third column
of \Pref{QSpace:psiS}. The \QSpace representation of $\hat{\psi}
_{S}^{\dagger}$ derives from the matrix-elements
\[
   \psi_{S}^{\dagger} \to \langle C_{z}S
   | \cdot \Bigl( (\hat{\psi}_{S}^{\dagger}) ^{\left( \tfrac{+1}{2};\tfrac{1}{2}\right)}
   | C_{z}^{\prime}S^{\prime}\rangle \Bigr)
\]
using the Wigner Eckart theorem as in \Eq{WET:motiv}.

The operator index in the \QSP{QSpace:psiS} is listed third, by
convention. The two non-zero matrix elements of each
$\hat{c}_{\sigma}^{\dagger}$ within the four-dimensional space of
single fermionic site implies a total of four non-zero matrix
elements in $\psi_{S}^{\dagger}$, all having norm $1$, with one
matrix-element being negative. These matrix elements can be directly
identified in \QSP{QSpace:psiS}. Since the reduced matrix elements
$\Vert\psi_{S}^{\dagger}\Vert$ and the CGC spaces are to be
interpreted as tensor product, the $\sqrt{2}$ factors in the last
line cancel. With $\hat{\psi}_{S}^{\dagger}$ representing
non-hermitian operators, the first column $q\equiv\left(
C_{z};S\right) $ is in general different form the second column
$q^{\prime}\equiv\left( C_{z}^{\prime};S^{\prime}\right) $. Moreover,
since $\hat{\psi}_{S}^{\dagger}$ creates one particle, the first
column, for example, cannot contain the empty state
$(\tfrac{-1}{2};0)$, while the second column cannot contain the
double occupied state $(\tfrac{+1}{2};0)$.

In contrast, the \QSpace representation of the \IROP $\psi_{S}$, \ie
without the dagger, is shown in \Pref{QSpace:psiS-2}. Note that for
$\hat{\psi}_{S} \equiv \{ -\hat{c}_{\downarrow}; \hat{c}_{\uparrow})$
to be an irreducible operator as compared to
$\hat{\psi}_{S}^{\dagger} = \{ \hat{c}_{\uparrow}^{\dagger},
\hat{c}_{\downarrow}^{\dagger} \}$, the reverse order in spin and the
minus sign in the first component is essential [see discussion along
with \EQSp{eq:IROP-psiS2}]. In terms of the \QSP{QSpace:psiS-2}, this
leads to the extra minus signs in the \emph{multiplet space} of the
second row. Moreover, the z-labels of the operator $\hat{\psi}_{S}$
itself flipped sign \wrt $\hat{\psi}_{S}^\dagger$ as expected as it
removes a particle rather than adding one [see the multiplet labels
$q'' \equiv (C_{z}''; S'')$ in the third column of
\Pref{QSpace:psiS-2}{}]. This is to emphasize that the application of
the Wigner Eckart theorem must be performed consistently, \ie
switching sides in the application of an operator as in $\langle
C_{z}S|\cdot(\psi^{\dagger}|C_{z}^{\prime}S^{\prime}\rangle
)=(\psi|C_{z}S\rangle)^{\dagger}\cdot|C_{z}^{\prime}S^{\prime}\rangle$
must be dealt with carefully.

\subsubsection{Identity A-tensor for two fermionic sites}

Consider the combination of two fermionic sites. Alluding to
\FigP{fig:qspace}, let site 1 (2) be described by $\left\vert
i\right\rangle $ ($\left\vert \sigma\right\rangle $), respectively,
both representing a 4-dimensional state space $\left\{ \left\vert
0\right\rangle ,\left\vert \uparrow\right\rangle, \left\vert
\downarrow \right\rangle ,\left\vert \uparrow\downarrow\right\rangle
\right\}$ of their own. The decomposition of the combined space in
terms of the overall symmetry \AS is fully described by the rank-3
\QSP{QSpace:A0_zs}.

\begin{table*}[ptb]
\caption{ Example \QSpaces in the presence of $\SC \equiv \SS$
symmetry for a single fermionic site ($\hat{\psi}_{CS}$), and for the
combination of the state space of two sites (\Atensor{}). The CGC
spaces for \SU[charge]{2} and \SU[spin]{2} are indicated by the C and
S, respectively. The record index $\nu$ as well as the explicit
specification of the dimensions are just added for clarity. For
comparison, \Tbl{QSpaces:AS} shows the same \QSpaces for the case
where the particle-hole symmetry is reduced to abelian charge
conservation. }
\begin{align}
   \psi_{\CS} \equiv \left\{
   \begin{tabular}[c]{c|ccc|cc|ccc}
   record & & & & \multicolumn{2}{|c|}{red. matrix elements} & CGC spaces & & combined \\
   index $\nu$ & $\left( C;S\right)  $ & $\left(  C^{\prime};S^{\prime}\right)$ &
   $\left( C_{z}^{\prime\prime};S^{\prime\prime}\right)$ &
   $\Vert\hat{\psi}_{\CS}\Vert$ & dimension & C & S & dimension
\\ \hline\hline & & & & & & & & \\[-2ex]
   1. & $0;\tfrac{1}{2}$ & $\tfrac{1}{2};0$ & $\tfrac{1}{2};\tfrac{1}{2}$ &
   $s\sqrt{2}$ & $1\times1\times1$ & \multicolumn{1}{|c}{$
   \begin{array}[c]{c}
      \tfrac{1}{\sqrt{2}} \bigl\{ [ 0\ 1 ]; [-1\ 0 ]  \bigr\} \\
   %% (1\times2\times2)
   \end{array}
   $} & \multicolumn{1}{c}{$
   \begin{array}[c]{c}
      \bigl\{ [ 1;0 ]; [ 0;1 ] \bigr\} \\
   %% (2\times1\times2)
   \end{array}
   $} & $2\times2\times4$
\\[2ex] %%\hline
   2. & $\tfrac{1}{2};0$ & $0;\tfrac{1}{2}$ & $\tfrac{1}{2};\tfrac{1}{2}$ &
   $\sqrt{2}$ & $1\times1\times1$ & \multicolumn{1}{|c}{$
   \begin{array}[c]{c}
      \bigl\{ [ 1;0 ]; [0;1 ] \bigr\} \\
   %% (2\times1\times2)
   \end{array}
   $} & \multicolumn{1}{c}{$
   \begin{array}[c]{c}
      \tfrac{1}{\sqrt{2}} \bigl\{ [0\ 1 ]; [-1\ 0 ] \bigr\} \\
   %% (1\times2\times2)
   \end{array}
   $} & $2\times2\times4$
%%\\[1ex]\hline
   \end{tabular}
   \right\}
\label{QSpace:F3_cs}%
\end{align}
\quote{\QSpace of spinor $\hat{\psi}_{\CS}$ defined in
\Eq{eq:Psi:CS4} with the reduced matrix elements for given symmetries
already calculated in \Eq{eq:Psi:CS}. The operator index within the
\IREP is listed third, as usual. The alternating sign $s$ required
with particle-hole symmetry appears with the reduced matrix elements
in the first record only [\cf \Eq{eq:Psi:CS}; note that the same sign
$s$ is also picked up by the double occupied state, \cf
\Eq{fstates:SC4} or \Pref{QS:CZ}{}].
}%

\begin{align}
   A \equiv \left\{
   \begin{tabular}[c]{c|ccc|ll|ccc}
   record & site 1 & site 2 & combined & \multicolumn{2}{|l}{multiplet space} &
   \multicolumn{2}{|l}{CGC spaces} & combined \\
   index $\nu$ & $(C;S)$ & $(C^{\prime};S^{\prime})$ &
   $(C^{\prime\prime}; S^{\prime\prime})$ & $\Vert A \Vert$ &
   dimension & C & S & dimension
\\ \hline\hline & & & & & & & & \\[-2ex]
   1. & $0;\tfrac{1}{2}$ & $0;\tfrac{1}{2}$ & $0;0$ & $\{[1],[0] \}$ &
   $1\times1\times2$ & $1.$ & $\tfrac{1}{\sqrt{2}} [ 0\ -1;\ 0\ 1 ]$ &
   \multicolumn{1}{l}{$2\times2\times1$} \\
   2. & $\tfrac{1}{2};0$ & $\tfrac{1}{2};0$ & $0;0$ &
   $\left\{  [0],[1]\right\}$ & $1\times1\times2$ &
   $\tfrac{1}{\sqrt{2}}\left[  0\ -1;\ \ 0\ 1\right]$ & $1.$ &
   \multicolumn{1}{l}{$2\times2\times1$}
\\[1ex] \hline
   3. & $0;\tfrac{1}{2}$ & $0;\tfrac{1}{2}$ & $0;1$ & $1.$ & $1\times1\times1$ &
   $1.$ & $%
   \begin{array}[c]{c} \bigl\{
      [ 1\ 0;\ 0\ 0], \\
      \tfrac{1}{\sqrt{2}} [ 0\ 1;\ 0\ 1 ], \\
       {[} 0\ 0;\ 0\ 1] \bigr\}
   \end{array}
   $ & \multicolumn{1}{l}{$2\times2\times3$}
\\[4.5ex] \hline
   4. & $0;\tfrac{1}{2}$ & $\tfrac{1}{2};0$ & $\tfrac{1}{2};\tfrac{1}{2}$ &
   $\left\{[1],[0]\right\}$ & $1\times1\times2$ &
   $\left\{ [1\ 0],[0\ 1]\right\}$ & $\left\{  [1;0],[0;1]\right\}$ &
   \multicolumn{1}{l}{$2\times2\times4$} \\
   5. & $\tfrac{1}{2};0$ & $0;\tfrac{1}{2}$ & $\tfrac{1}{2};\tfrac{1}{2}$ &
   $\left\{[0],[1]\right\}  $ & $1\times1\times2$ &
   $\left\{ [1;0],[0;1]\right\}  $ & $\left\{  [1\ 0],[0\ 1]\right\}  $ &
   \multicolumn{1}{l}{$2\times2\times4$}
\\[1ex] \hline
   6. & $\tfrac{1}{2};0$ & $\tfrac{1}{2};0$ & $1;0$ & $1.$ &
   $1\times1\times1$ & $%
   \begin{array}[c]{c} \bigl\{
      {[} 1\ 0;\ 0\ 0], \\
      \tfrac{1}{\sqrt{2}} [ 0\ 1;\ 0\ 1 ], \\
      {[} 0\ 0;\ 0\ 1] \bigr\}
   \end{array}
   $ & $1.$ & \multicolumn{1}{l}{$2\times2\times3$}
%% \\[4.5ex] \hline
   \end{tabular}
   \right\}
\label{QSpace:A0_cs}%
\end{align}
\quote{\QSpace of identity \Atensor combining two fermionic sites --
the symmetry records are sorted with respect the combined symmetry
$q'' \equiv (C'',S'')$, where groups with the same $(C'',S'')$ are
separated by horizontal lines for clarity. Considering the
tensor-product of multiplet and CGC spaces, the combined space has
total dimension of $2\cdot1 + 1\cdot3 + 2\cdot4 + 1\cdot3 = 16$ as
expected for two spinful fermionic levels. Compared to the \Atensor
in \Pref{QSpace:A0_zs} with abelian charge conservation, the number
of combined symmetry sectors has been further reduced from 6 to 4
[\ie number of horizontally separated groups sharing the same
$q^{\prime\prime}$], with an overall reduction in the number of
multiplets present in the \QSpace reduced from 10 to 6 [having
$1+2+3+1+2+1=10$ in \Pref{QSpace:A0_zs}, and here $2+1+2+1=6$].
}%
\label{QSpaces:CS}
\end{table*} %% table*

Given the \AS symmetries, the abelian charge quantum number $C_{z}$
simply adds up, while for the \SU{2} spin symmetry, the usual \SU{2}
addition algebra applies. The overall number of multiplets in the
combined space $q^{\prime\prime}$ is given by the last number (index
3) in the dimensions specified with the multiplet space. The specific
input combinations entering a certain combined space
$q^{\prime\prime}$ are easily verified. The $q^{\prime\prime} = (
-\tfrac{1}{2};\tfrac{1}{2} )$ sector, for example, derives from the
two configurations $\{ q,q^{\prime} \} = \{ (-\tfrac{1}
{2};0),(0;\tfrac{1}{2}) \} $ and $\{ ( 0;\tfrac{1} {2} ) , (
-\tfrac{1}{2};0 ) \} $. Therefore the dimension of the reduced
multiplet space for this $q^{\prime\prime}$ is $2$. Each of these
multiplets has an internal z-space which is itself of dimension $2$
[last column]. The combined total dimension of the $q^{\prime\prime}
= ( \tfrac{-1}{2}; \tfrac{1}{2} )$ sector is therefor given by the
product $2\cdot2=4$. Consistently, the dimension of the two
4-dimensional sites combined add up correctly to 16 states total.
That is, multiplying the last dimension in the reduced multiplet
space with the last dimension in the combined CGC spaces for each
block separated by horizontal lines, bearing in mind that the
multiplet space and the CGC spaces are to be combined in a
tensor-product, yields the overall dimension of the combined space,
$1\cdot1 + 2\cdot2 + 3\cdot1 + 1\cdot3 + 2\cdot2 + 1\cdot1 = 16$.

The \Atensor in \Pref{QSpace:A0_zs} is an identity \Atensor, in that
up to permutations, plain identity matrices are split-up on the
reduced multiplet level. By considering, for example, the $q^{\prime
\prime} = \left( \pm\tfrac{1}{2};\tfrac{1}{2} \right)$ symmetry
sector in records $2-3$ or $8-9$ of the \QSP{QSpace:A0_zs}, the
multiplet space when viewed together, \ie ignoring all brackets,
resemble the structure of a $2$-dimensional identity matrix. Similar
so for the $q^{\prime\prime }=\left( 0;0\right) $ space in records
$4-6$, having essentially a $3$-dimensional identity matrix in the
multiplet space. Allowing for arbitrary unitaries in the multiplet
space in \QSpace \Pref{QSpace:A0_zs}, this then becomes the most
general unitary transformation of the product space of two fermionic
sites that also respects the symmetries considered.

\subsection{Fermionic sites in the presence of particle-hole symmetry}

The tensors introduced in the previous section for \AS symmetry will
now be written more compactly still by assuming the stronger
particle-hole \SU[charge]{2} symmetry instead of the plain abelian
\UA{charge}. The symmetry considered in the following is therefore
$\SC \equiv \SS$. The z-operator for charge, $\hat {C}_{z}$, is now
complemented by the raising operator $\hat{C}_{+}$ for charge \SU{2}
[\cf \Eqp{op:C+}]. The combined symmetries are given by the multiplet
label for both charge and spin \SU{2}, \ie the non-negative labels
$q=(C,S)$ with the z-labels of the charge symmetry now also taken
care of by the CGC spaces.

The basis for a single fermionic level given given \SC symmetry has
been introduced in \Eq{fstates:SC4}. Therefore the full space of the
four states $\{0, \uparrow, \downarrow, \uparrow\downarrow\}$ can be
reduced to the two symmetry multiplets
\begin{align}
   \begin{tabular}[c]{ll|c}
   \multicolumn{2}{l|}{multiplet space} & dimension \\
   \multicolumn{2}{l|}{$\left\vert C,S\right\rangle $} & $d_{C}\times d_{S}=d_{tot}$
\\[0.5ex] \hline
   $\left\vert \tfrac{1}{2};0\right\rangle $ &
   $\equiv \{ \left\vert 0\right\rangle ,s \left\vert \uparrow\downarrow\right\rangle \}$ &
   \multicolumn{1}{|l}{$2\times1=2$}
\\
   $\left\vert 0;\tfrac{1}{2}\right\rangle $ &
   $\equiv \{  \left\vert\uparrow\right\rangle ,\left\vert \downarrow\right\rangle \}$ &
   \multicolumn{1}{|l}{$1\times2=2$}
%% \\[0.5ex]\hline
   \end{tabular}
\label{QS:CZ}
\end{align}
The most general scalar operator such as the Hamiltonian is given by
the \QSpace,
\begin{equation}
   H \equiv \left\{
   \begin{tabular}[c]{cc|c|cc}
   $\left(  C;S\right)  $ & $\left(  C^{\prime};S^{\prime}\right)  $ & $\Vert
   H\Vert$ & \multicolumn{2}{|l}{CGC spaces}
   \\ \hline\hline
   $\tfrac{1}{2};0$ & $\tfrac{1}{2};0$ & $h_{\tfrac{1}{2},0}$ & $\Id^{(2)}$ & $1.$\\
   $0;\tfrac{1}{2}$ & $0;\tfrac{1}{2}$ & $h_{0,\tfrac{1}{2}}$ & $1.$ & $\Id^{(2)}$
   \end{tabular}
   \right\}
\text{.}\label{QSpace:H0_cs}%
\end{equation}
Thus only the two reduced matrix elements $h_{1/2,0}$ and $h_{0,1/2}$
are left free to choose without compromising \SC symmetry.

A non-scalar \IROP is given in terms of the 4-element spinor
$\hat{\psi}_{\CS}$ in \Eq{eq:Psi:CS4}, which combines two creation
and two annihilation operators. Its symmetries have been identified
in \Eq{eq:Psi:CS4:zlabels}, leading to the \IROP $\hat{\psi}_{\CS}$
with the \QSpace presented in \Pref{QSpace:F3_cs}. Note that the size
of the third dimension is 4 [see combined CGC dimension in the last
column], consistent with the four operators that constitute the
\IROP.

The \QSpace representation for the \Atensor combining two fermionic
levels is given in \Pref{QSpace:A0_cs}. The standard \SU{2} addition
rules are quickly confirmed. For example, the combined symmetry
$q''=(0,0)$ [records 1-2] can result from two combinations, namely
$(\tfrac{1}{2},0) \otimes(\tfrac{1}{2},0)$, or $(0,\tfrac{1}{2})
\otimes(0,\tfrac{1}{2})$, leading to a two dimensional multiplet
space. All of this is transparently encoded in the \QSpace. Here,
\QSP{QSpace:A0_cs} again shows an identity \Atensor, as seen, for
example, in the combined reduced multiplet space of records $1-2$, or
$4-5$).

\begin{table}[t!]%
   \begin{tabular}[c]{c|cccclc|cr}\hline
   & \multicolumn{6}{|l|}{multiplet space} & CG dimension & \\
   index & $\vert S$, & $C_{1}$, & $C_{2}$, & $C_{3} $ & $\rangle $ & dim. &
   $d_{S} d_{C_{1}}d_{C_{2}}d_{C_{3}}=$ & $d_{tot}$
   \\[0.5ex]\hline
    1. & $\vert\, 0$; & $0$; & $0$; & $\tfrac{1}{2}$ & $\rangle $ & $1$
       & $1\times1\times1\times2=$ & $2$ \\
    2. & $\vert\, 0$; & $0$; & $\tfrac{1}{2}$; & $0$ & $\rangle $ & $1$
       & $1\times1\times2\times1=$ & $2$ \\
    3. & $\vert\, 0$; & $\tfrac{1}{2}$; & $0$; & $0$ & $\rangle $ & $1$
       & $1\times2\times1\times1=$ & $2$\\
    4. & $\vert\, 0$; & $\tfrac{1}{2}$; & $\tfrac{1}{2}$; & $\tfrac{1}{2}$ & $\rangle $ & $1$ &
         $1\times2\times2\times2=$ & $8$ \\
    5. & $\vert\, \tfrac{1}{2}$; & $0$; & $0$; & $0$ & $\rangle $ & $2$
       & $2\times1\times1\times1=$ & $2$\\
    6. & $\vert\, \tfrac{1}{2}$; & $0$; & $\tfrac{1}{2}$; & $\tfrac{1}{2}$ & $\rangle $ & $1$ &
         $2\times1\times2\times2=$ & $8$ \\
    7. & $\vert\, \tfrac{1}{2}$; & $\tfrac{1}{2}$; & $0$; & $\tfrac{1}{2}$ & $\rangle $ & $1$ &
         $2\times2\times1\times2=$ & $8$ \\
    8. & $\vert\, \tfrac{1}{2}$; & $\tfrac{1}{2}$; & $\tfrac{1}{2}$; & $0$ & $\rangle $ & $1$ &
         $2\times2\times2\times1=$ & $8$ \\
    9. & $\vert\, 1$; & $0$; & $0$; & $\tfrac{1}{2}$ & $\rangle $ & $1$
       & $3\times1\times1\times2=$ & $6$\\
   10. & $\vert\, 1$; & $0$; & $\tfrac{1}{2}$; & $0$ & $\rangle $ & $1$
       & $3\times1\times2\times1=$ & $6$\\
   11. & $\vert\, 1$; & $\tfrac{1}{2}$; & $0$; & $0$ & $\rangle $ & $1$
       & $3\times2\times1\times1=$ & $6$\\
   12. & $\vert\, \tfrac{3}{2}$; & $0$; & $0$; & $0$ & $\rangle $ & $1$
       & $4\times1\times1\times1=$ & $4$
   \\[0.5ex]\hline
   \end{tabular}
\caption{ State space of 3-channel site with $\Sfour \equiv \SSSS$
symmetry. }
\label{64states:SU2x4}%
%% \end{table}

%% \begin{table}%
   \begin{tabular}[c]{c|crrrc|cr}
   \multicolumn{8}{l}{} \\ \hline %% spacer
   & \multicolumn{5}{|l|}{multiplet space} & CG dimension & \\
   index & $\vert\, S$ & $C_{z}$ & $T $ & $\rangle$ & dim. & $d_{S}{\cdot}d_{C_{z}}{\cdot}d_{T} =$ & $d_{tot}$
   \\[0.5ex]\hline
    1. & $\vert\, 0$; & $-\tfrac{3}{2}$; & $\phantom{-}0\ 0$ & $\rangle$ & $1$ & $1\times1\times1 =$ & $1$ \\
    2. & $\vert\, 0$; & $-\tfrac{1}{2}$; & $2\ 0$ & $\rangle$ & $1$ & $1\times1\times6 =$ & $6$ \\
    3. & $\vert\, 0$; & $\tfrac{1}{2}$; &  $0\ 2$ & $\rangle$ & $1$ & $1\times1\times6 =$ & $6$ \\
    4. & $\vert\, 0$; & $\tfrac{3}{2}$; &  $0\ 0$ & $\rangle$ & $1$ & $1\times1\times1 =$ & $1$ \\
    5. & $\vert\, \tfrac{1}{2}$; & $-1$; & $1\ 0$ & $\rangle$ & $1$ & $2\times1\times3 =$ & $6$ \\
    6. & $\vert\, \tfrac{1}{2}$; & $0$; &  $1\ 1$ & $\rangle$ & $1$ & $2\times1\times8 =$ & $16$ \\
    7. & $\vert\, \tfrac{1}{2}$; & $1$; &  $0\ 1$ & $\rangle$ & $1$ & $2\times1\times3 =$ & $6$ \\
    8. & $\vert\, 1$; & $-\tfrac{1}{2}$; & $0\ 1$ & $\rangle$ & $1$ & $3\times1\times3 =$ & $9$ \\
    9. & $\vert\, 1$; & $\tfrac{1}{2}$;  & $1\ 0$ & $\rangle$ & $1$ & $3\times1\times3 =$ & $9$ \\
   10. & $\vert\, \tfrac{3}{2}$; & $0$;  & $0\ 0$ & $\rangle$ & $1$ & $4\times1\times1 =$ & $4$
   \\[0.5ex]\hline
   \end{tabular}
\caption{ State space of 3-channel site with \ASC symmetry. }
\label{64states:SU123}%
%% \end{table}

%% \begin{table}%
   \begin{tabular}[c]{c|crrc|cc}
   \multicolumn{7}{l}{} \\ \hline %% spacer
   & \multicolumn{4}{|l|}{multiplet space} & CG dimension & \\
   index & $\vert\, S$ & $\Sp{6} $ & $\rangle$ & dim. & $d_{S}\cdot d_{C_{z}}\cdot d_{T} =$ & $d_{tot}$
   \\[0.5ex]\hline
    1. & $\vert\, 0$;  & $\phantom{-}0\ 0\ 1$ & $\rangle$ & $1$ & $1\times1\times14 =$ & $14$ \\
    2. & $\vert\, \tfrac{1}{2}$;  & $0\ 1\ 0$ & $\rangle$ & $1$ & $2\times1\times14 =$ & $28$ \\
    3. & $\vert\, 1$;             & $1\ 0\ 0$ & $\rangle$ & $1$ & $3\times1\times\phantom{1}6 =$ & $18$ \\
    4. & $\vert\, \tfrac{3}{2}$;  & $0\ 0\ 0$ & $\rangle$ & $1$ & $4\times1\times\phantom{1}1 =$ & $ \phantom{1}4$
   \\[0.5ex]\hline
   \end{tabular}
\caption{ State space of 3-channel site with \SSP symmetry. }
\label{64states:Sp6}%
\end{table}

\begin{table*}[tb]%
   \begin{tabular}[c]{|c|c|ccccc|ccccc|ccccc|}\hline
   sites & abelian dim. & \multicolumn{5}{|c|}{\SSSS}
   & \multicolumn{5}{|c|}{\ASC}
   & \multicolumn{5}{|c|}{\SSP }
\\ %\cline{3-10}%
   $n$ & $D^\ast=64^n$ &
   $N_S$ & $D$ & \!\!\!$D^{\ast}/D$\! & memory & MEM$^\ast$ &
   $N_S$ & $D$ & \!\!\!$D^{\ast}/D$\! & memory & MEM$^\ast$ &
   $N_S$ & $D$ & \!\!\!$D^{\ast}/D$\! & memory & MEM$^\ast$
\\ \hline
   1 & \multicolumn{1}{l|}{$64$}
     & 12 & 13 & 4.9 & $<18$\,K &
     & 10 & 10 & 6.4 & $<13$\,K &
     &  4 &  4 &  16 & $<6$\,K  &
   \\
   2 & \multicolumn{1}{l|}{$4,096$}
     & 61 & 388 & 10.6 & 528\,K & $>$8.7\,M
     & 69 & 260 & 15.8 & 359\,K & $>$12\,M
     & 23 &  61 & 67.1 & 162\,K & $>$34\,M
   \\
   3 & \multicolumn{1}{l|}{$262,144$}
     & 192 & 14,229 & 18.4 & 27\,M & $>$23\,G
     & 226 & 9,086 & 28.9 & 11\,M & $>$31\,G
     &  60 & 1,232 & 213 & 7\,M & $>$112\,G
   \\
   4 & \multicolumn{1}{l|}{$16,777,216$}
     & 469 & 590,856 & 28.4 & 24\,G & $>$65\,T
     & 565 & 366,744 & 45.7 & 6.8\,G & $>$85\,T
     & 132 &  31,640 & 530 & 334\,M & $>$355\,T
\\ \hline
\end{tabular}
\caption{ Comparison of different symmetry scenarios for the same
underlying physical system of a symmetric 3-channel setup, analyzing
the product spaces of up to $n=4$ sites. Each site represents a
Hilbert space of dimension $64$, thus $n$ sites amounts to an overall
Hilbert space of dimension $D^{\ast}=64^{n}$ [second column]. This
state space can be decomposed into $D$ multiplets in $N_S$ symmetry
sectors using an \Atensor for the addition of every new site. These
\Atensors are encoded in terms of \QSpaces. The total memory
requirement for each \Atensor is listed, given sparse CGC
representation. In addition, as a comparison to a fully abelian
setting, MEM$^\ast$ indicates the memory that had been required if
the tensor products between reduced multiplets and CGC spaces was
carried out explicitly [K,M,G,T for kilo-, mega-,
giga-, and tera-bytes, respectively].\\[2ex]%%
} \label{tbl:3channel_Dtotal}

\begin{tabular}[c]{|c|cccccc|ccccc|cccc|}\hline
   sites & \multicolumn{6}{|c|}{\SSSS}
   & \multicolumn{5}{|c|}{\ASC}
   & \multicolumn{4}{|c|}{\SSP }
\\ %\cline{3-10}%
   $n$ &
   $d_S$ & $d_C$ & $d_C$ & $d_C$ & sparsity & CGS/A &
   $d_S$ & $d_{U(1)}$ & $d_C$ & sparsity& CGS/A &
   $d_S$ & $d_{\Sp{6}}$ & sparsity & CGS/A
\\ \hline
   1 & 4 & 2 & 2 & 2 & 0.36 & 0.8
     & 4 & 1 & 8 & 0.12 & 0.8
     & 4 & 14 & 0.027 &  0.90
   \\
   2 & 7 & 3 & 3 & 3 & 0.36 & 0.8
     & 7 & 1 & 27 & 0.12 & 0.8
     & 7 & 126 & 0.027 & 0.90
   \\
   3 & 10 & 4 & 4 & 4 & 0.28 & 0.11
     & 10 & 1 & 64 & 0.064 & 0.33
     & 10 & 616 & 0.011 & 0.94
   \\
   4 & 13 & 5 & 5 & 5 & 0.22 & $<10^{-3}$
     & 13 & 1 & 125 & 0.039 &  0.003
     & 13 & 2457 & 0.006 & 0.55
\\ \hline
\end{tabular}
\caption{ Comparison of different symmetry scenarios as in
\Tbl{tbl:3channel_Dtotal} in terms of (i) largest multiplet dimension
$d$ for each individual symmetry, and (ii) overall average sparsity
of the CGC spaces, \ie the number of non-zero elements divided by the
total number of matrix elements. The last columns for each symmetry
(CGS/A) shows the memory requirement of all sparse CGC spaces in a
given \QSpace $A_n$ relative to the entire \QSpace.
}%
\label{tbl:3channel_sparse}%
\end{table*}

\subsection{
Three channels with \SU{3} channel symmetry
\label{sec:SU123}}

Consider a system with three spinful particle-hole symmetric
channels, as introduced in \Secp{sec:3channel}. A single site then
has a full Hilbert space of dimension $4^{3}=64$. Three symmetry
settings are analyzed: a set of plain \SU{2} symmetries, a
combination with the \SU{3} channel symmetry, and finally the largest
symmetry present which includes the enveloping symplectic symmetry
\Sp{6}. Since the \QSpaces in given context are extensive, a more
compact comparison of these symmetry settings in the numerical
context is given, instead.

The first setting, $\Sfour\equiv\SSSS$, is based on four independent
\SU{2} symmetries. The 64-dimensional state space of single site
decomposes into 13 multiplets in 12 symmetry sectors, as listed in
\Tbl{64states:SU2x4}. All of these contain a single representative
multiplet, except for the space $\vert \tfrac{1}{2}; 0;0;0 \rangle $
in row 5, which contains two multiplets.
In contrast, using the \ASC symmetry, instead, the $64$ states of the
three fermionic levels decomposes into the 10 multiplet spaces listed
in \Tbl{64states:SU123}. Thus compared to the $\SU[CS]{2}^{\otimes4}$
symmetry setting in \Tbl{64states:SU2x4}, the number of multiplet
spaces is further reduced with all multiplet spaces containing a
single multiplet only. This suggests that the latter symmetry
including the channel \SU{3} is somewhat more efficient as it allows
to compactify multiplet spaces more strongly. Given the multiplet
space in \Tbl{64states:SU123}, for example, the most general
Hamiltonian in the $64\times64$ dimensional Hilbert space compatible
with given symmetry consists of the 10 reduced matrix elements
appearing in the multiplet space only.

A further strong boost in numerical efficiency can be obtained, if
the Hamiltonian supports it, by combining the particle-hole symmetry
of \SSSS with the channel symmetry in \ASC to their enveloping \Sp{6}
symmetry. The resulting state space for the state space of a single
3-channel site is given in \Tbl{64states:Sp6}. The $64\times64$
dimensional Hilbert space has thus been reduced to a total of four
multiplets only.

All three symmetry settings have been successfully implemented within
the NRG framework. By starting with a single site [\ie the basic
fermionic three-level unit as introduced in \Secp{sec:3channel}], and
iteratively adding a site within the NRG, new multiplet spaces are
quickly explored and built up within the first few NRG iterations. In
practice, the CGC spaces of newly generated multiplets are also
stored for latter retrieval. Once truncation of the state space
within NRG sets in, however, the generation of \IREPs eventually
saturates to within a finite range of multiplet spaces.

The resulting space for adding up to three further sites without
truncation is indicated in \Tbl{tbl:3channel_Dtotal}. With reasonable
numerical resources, it is feasible within the \ASC (second) or \SSP
(third) setting, to keep all states up to three sites total within
the NRG, first truncating only when a fourth site is added. This
leads to $\NK=9,086$ [$\NK=1,232$] kept multiplets for the second and
third symmetry setting, respectively. The corresponding memory
requirements for a general basis transformation for adding another
site (\Atensor{}) then amounts to about $7\,\mathrm{G}$
[$0.3\,\mathrm{G}$]. The corresponding full NRG iteration for adding
another site then takes several hours on a state-of-the-art 8-core
workstation.

The same calculation, however, gets quickly impossible as fewer
symmetries are available or used in the actual computation as can be
seen from \Tbl{tbl:3channel_Dtotal}. For example, if only the abelian
part of the symmetry had been accounted for in the computation, the
corresponding memory requirement can be estimated by considering the
explicit tensor product of the multiplet space with the CGC spaces,
leading in terms of the \Sfour setting to about $23\,\mathrm{G}$ and
$65\,\mathrm{T}$(!) for $n=3$ and $n=4$, respectively, the latter
being completely hopeless in practice. The explicit treatment of
non-abelian symmetries, however, clearly makes the latter case
feasible with a reasonable amount of numerical resources.

\begin{figure*}[ptb]
\begin{center}
\includegraphics[width=\linewidth]{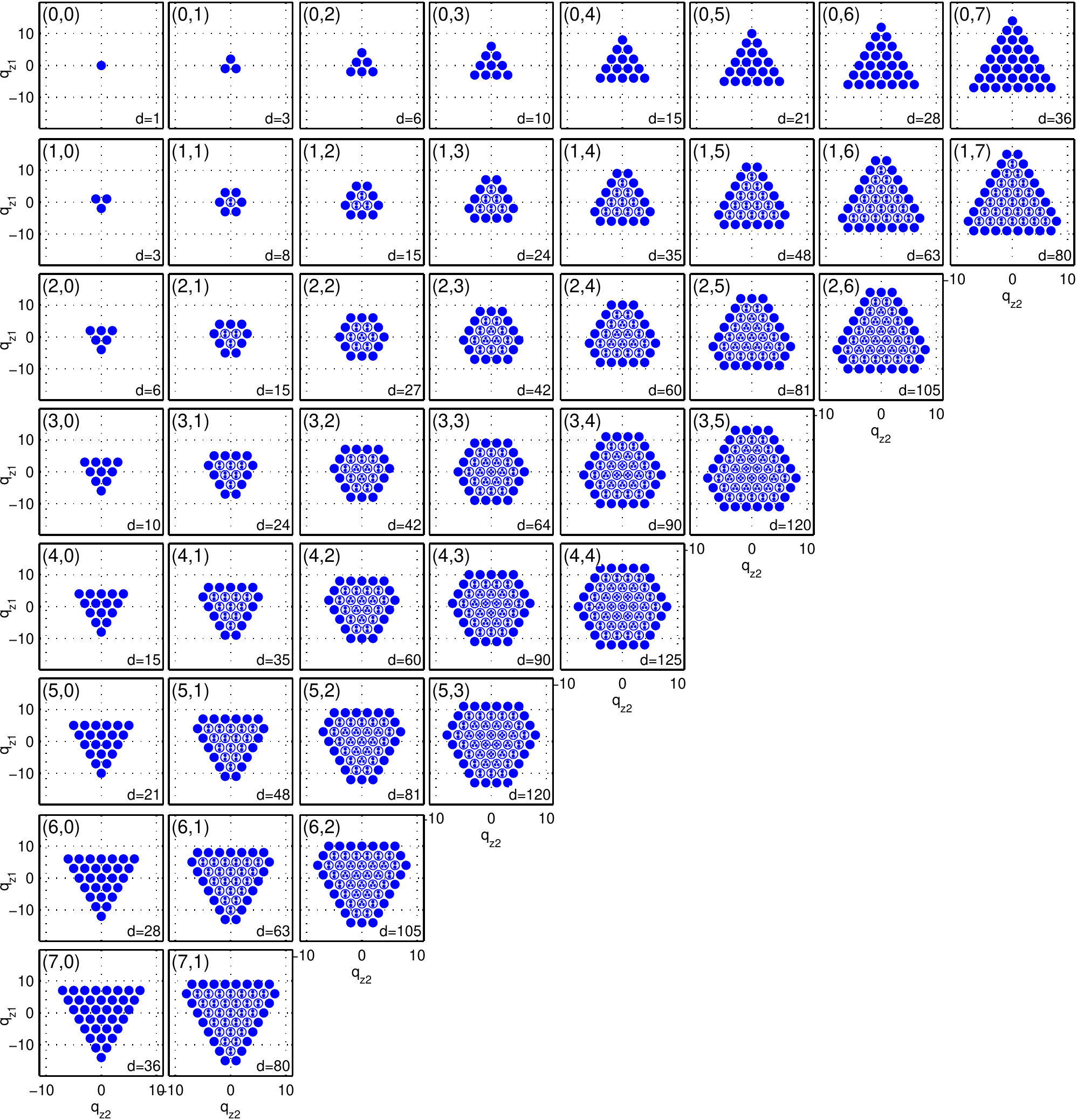}
\end{center}
\caption{ Weight diagrams of \SU{3} multiplets generated in a typical
NRG run for the symmetric 3-channel system including \ASC symmetries
($\Lambda=4$, $\Etrunc=7$). The multiplet label $(q_{1},q_{2})$, as
defined in \Eq{SU3-labels}, is specified with each multiplet in the
upper left corner of its panel. For a weight diagram of a specific
\IREP $(q_{1},q_{2})$, the corresponding z-labels $(q_{z1},q_{z2})$
of \SU{3} for each individual state within the multiplet are depicted
as points in a two-dimensional plot. In case of inner multiplicity,
\ie that several states within the same \IREP share exactly the same
z-labels, these states are shown as an encircled group of smaller
points. The dimension for every multiplet (number of points drawn
within a panel) is indicated to the lower right of each panel. The
first panel [multiplet $(0,0)$] represents the scalar representation
with multiplet dimension $d=1$. Multiplet $(1,0)$ represents the
defining three-dimensional representation [\cf
\Sec{sec:SU3:symlabels}{}], and $(1,1)$ the regular representation of
dimension $8$ equal to the dimension of \SU{3}, \ie the number of its
generators. The largest multiplet encountered in given NRG run is the
multiplet $(4,4)$ with an irreducible dimension of $d=125$.
}%
\label{fig:SU3diags}%
\end{figure*}

\subsubsection{\SU{3} symmetry \label{Sec:SU3:applic}}

The irreducible \SU{3} multiplets generated in the actual NRG run
using \ASC symmetry as presented in the main text [\cf
\Secp{sec:NRGresults}], are shown in terms of their weight diagrams
in \Fig{fig:SU3diags}. For comparison, the \SU{3} \IREPs present in
the description of a single site are $q^{\SU{3}} \equiv (q_1,q_2) \in
\{ (0,0),\ (0,1),\ (0,2),\ (1,0),\ (2,0) \}$, \cf
\Tbl{64states:SU123}. The apparent symmetry \wrt to flipping the
quantum numbers in $(q_1,q_2)$ is also reflected in the overall set
of multiplets generated within the NRG. As seen in
\Fig{fig:SU3diags}, all \IREPs $(q_1,q_2)$ with $q_1+q_2 \le 8$ are
present, except for $(0,8)$ and $(8,0)$.

Inner multiplicity, as expected for \SU{3}, is clearly present and
depicted in the weight diagrams of \Fig{fig:SU3diags} by the
encircled set of points. There the number of points inside a circle
stands for the multiplicity of the corresponding z-labels in the
multiplet. Inner multiplicity decreases in shells as one moves
outward the multiplet, which is seen particularly well for the
multiplets $q_{1}=q_{2}$. The states on the outer circumference have
no multiplicity, \ie have unique z-labels, as expected. This
demonstrates the uniqueness of the maximum weight state [cf.
\Secp{sec:maxweight}{}], which was required for the numerical state
space decomposition in \Sec{Sec:NumDeComp}.

Due to the two-dimensional label structure of the \SU{3} multiplets
together with inner multiplicity, their internal dimensions can get
significantly larger as compared to \SU{2} multiplets. The largest
\SU{3} multiplet $(4,4)$ encountered in the actual NRG run presented
in \Fig{fig:SU3diags}, for example, has an internal irreducible
dimension of $d=125$ [see also \Tbl{tbl:3channel_sparse}{}]. This
implies, for example, that with respect to the diagonalization of a
Hamiltonian, a 125-fold degeneracy has been reduced to a
\emph{single} multiplet. In contrast, the multiplet structure for
\Sfour is clearly weaker as it only includes \SU{2} symmetries. There
the largest quantum numbers encountered in an NRG run with comparable
number of kept states includes $S \le6$ in the spin sector, leading
to an individual multiplet dimension of at most $13$ [see also
\Tbl{tbl:3channel_sparse}{}]. In the overall combination of the
symmetries, this implies that for comparable number of states, \ie
for a comparable accuracy within the NRG, on average about 50\% more
multiplets need to be kept within the \Sfour setting as compared to
the case when \SU{3} is included [see \Fig{fig:nrg_anal} in the main
text].

Finally, the individual weight diagrams in \Fig{fig:SU3diags} show
well-known symmetries, such as a reflection symmetry of each diagram
around the vertical y-axes, or the reflection symmetry between the
multiplets $(q_{1},q_{2})$ and $(q_{2},q_{1})$ around the horizontal
axis. These \emph{Weyl symmetries} may be used to evaluate or encode
CGC spaces more efficiently.\cite{Alex11} For the purpose of this
paper, however, these symmetries were not exploited, given also that
the pure numerical evaluation of the CGC spaces as outlined earlier
was already sufficiently fast.

\begin{table}
   \begin{tabular}[c]{|c|c||c|c|}
   \hline
   \Sp{6} multiplet & dimension & \Sp{6} multiplet & dimension \\
    $q$ & $d$ & [cont'd] & \\ \hline
    $( 0\  0\  0 )$ &     1 &  $( 1\  2\  0 )$ &   350 \\
    $( 1\  0\  0 )$ &     6 &  $( 1\  0\  2 )$ &   378 \\
    $( 0\  0\  1 )$ &    14 &  $( 0\  3\  0 )$ &   385 \\
    $( 0\  1\  0 )$ &    14 &  $( 3\  1\  0 )$ &   448 \\
    $( 2\  0\  0 )$ &    21 &  $( 1\  1\  1 )$ &   512 \\
    $( 3\  0\  0 )$ &    56 &  $( 3\  0\  1 )$ &   525 \\
    $( 1\  1\  0 )$ &    64 &  $( 0\  1\  2 )$ &   594 \\
    $( 1\  0\  1 )$ &    70 &  $( 0\  2\  1 )$ &   616 \\
    $( 0\  0\  2 )$ &    84 &  $( 2\  2\  0 )$ &   924 \\
    $( 0\  2\  0 )$ &    90 &  $( 2\  0\  2 )$ &  1078 \\
    $( 0\  1\  1 )$ &   126 &  $( 1\  3\  0 )$ &  1344 \\
    $( 4\  0\  0 )$ &   126 &  $( 2\  1\  1 )$ &  1386 \\
    $( 2\  1\  0 )$ &   189 &  $( 1\  2\  1 )$ &  2205 \\
    $( 2\  0\  1 )$ &   216 &  $( 1\  1\  2 )$ &  2240 \\
    $( 0\  0\  3 )$ &   330 &  & \\[0.5ex]\hline
    \end{tabular}
\caption{ \Sp{6} multiplets generated in a fully converged NRG run
for the symmetric 3-channel system using \SSP ($\Lambda=4$,
$\Etrunc=7$). Multiplet $(0,0,0)$ represents the scalar
representation of dimension $1$, multiplet $(1,0,0)$ the defining
representation of dimension $6$, and multiplet $(2,0,0)$ the regular
representation of dimension $21$ which is also equal to the number of
generators for \Sp{6}. The largest tensor-product decomposition was
between the product spaces of \IREPs of dimension $14$ and $512$,
yielding a combined product space dimension of $7,168$. Run time of
the bare NRG run was about $2\,\mathrm{hrs}$ on a state-of-the-art
8-core workstation with moderate memory requirements of $\lesssim
4.5\,\mathrm{G}$. }
\label{NRG:Sp6:multiplets}%
\end{table}

\subsubsection{\Sp{6} symmetry \label{Sec:Sp6:applic}}

The complete set of \Sp{6} symmetries generated in the fully
converged NRG run (using $\Lambda=4$ and $\Etrunc=7$ as used in the
results in the main text), is listed in \Tbl{NRG:Sp6:multiplets}.
All multiplets had been generated within the first four Wilson
shells. The fact that the symmetry \Sp{6} fully incorporates
non-abelian particle-hole and channel symmetry, manifests itself by
observing that all eigenenergies in the multiplet spaces are now
strictly \emph{non-degenerate} throughout an entire NRG calculation.
\emph{Huge} degeneracies of several thousands can be split off in
terms of tensor products with \Sp{6} multiplets.

For given model, the symmetry \Sp{6} in fact also allowed to reduce
the rather coarse discretization of $\Lambda=4$ in the NRG
calculation underlying \Tbl{NRG:Sp6:multiplets}. For comparison, if
$\Lambda=2$ is used, instead, while keeping the same $\Etrunc=7$, it
turns out, the largest multiplet generated is $(2,1,2)$ of dimension
5720. The largest intermediate product space to be decomposed into
\IREPs becomes as large as $14 \times 1386 = 19,404$. Having
$\Lambda=2$, this required twice the Wilson chain length for the
same range in energy scales, leading to an overall run time of the
entire NRG run of about $32\,\mathrm{hrs}$ with still reasonably
manageable memory requirements of $\lesssim 20\,\mathrm{G}$.

As a rough general estimate, typical multiplet dimensions, as they
occurred in practice, scale like $10^r$ where $r$ is the rank of the
symmetry. For \SU[spin]{2}, this implies multiplets of dimension
$\lesssim 10$, for the \SU[channel]{4} symmetry, indeed, one had
multiplets of dimension of $\lesssim 100$, while now for \Sp{6}, a
symmetry of rank 3, one easily reaches multiplet dimensions on the
order of a few 1000 (\cf \Tbl{NRG:Sp6:multiplets}). Therefore with
increasing rank of the symmetry, the numerical effort strongly
shifts from the multiplet space to the CGC spaces. For sets of
smaller symmetries with rank $r\le 2$ this leads to a strong gain in
numerical efficiency, while the numerical overhead for the CGC
spaces remains negligible. Reaching symmetries of rank 3, such as
\Sp{6}, the numerical effort within the CGC spaces can now become
comparable to or even larger than the operations on the higher
multiplet level.

\TBL{tbl:3channel_sparse} summarizes the situation by comparing the
maximal multiplet spaces with the corresponding sparsity and memory
requirements of the CGC spaces for the first few \Atensors, when
combining up to $n=4$ sites without truncation. As the internal
multiplet dimensions quickly grow for higher rank symmetries,
nevertheless only an ever smaller fraction of Clebsch-Gordan
coefficients are non-zero. As seen from \Tbl{tbl:3channel_sparse},
the sparsity roughly grows exponentially with the rank of the
symmetry. Nevertheless, with the memory requirement of the
\emph{sparse} CGC for \Sp{6} comparable or even larger than the
storage of the reduced matrix elements on the higher multiplet level
(see last column in \Tbl{tbl:3channel_sparse}), full storage
including also the zero CGC spaces would have extremely inflated
overall storage requirement. In this sense, sparse storage of CGC
spaces becomes mandatory for larger-rank symmetries.
With the standard CGC spaces already tensors of rank-3, sparse
storage of general CGC spaces requires the extension of standard
sparse storage and sparse operations to arbitrary-rank tensors. All
of these is achieved, in general, by proper efficient permutations
in sparse index space which requires a fast sorting scheme, together
with reshaping of higher rank-objects to standard two-dimensional
sparse objects with fused indices, since this allows to employ
efficient algorithms for standard sparse matrix multiplication.

In order to distinguish numerical noise, \ie negligible CGC matrix
elements, from actual matrix elements then, this requires an
accurate evaluation of the CGC matrix elements. Double precision
accuracy as compared to the \emph{exact theoretical} CGCs was
sufficient, in practice. In particular, this implies, that also the
matrix elements of the generators for given \IREPs of the symmetry
are known \emph{numerically exact} at any step. In the iterative
approach, however, when new multiplets are generated through tensor
products with smaller entities, numerical errors can pile up. For
large-rank symmetries then the accuracy of the matrix elements of
the generators must be better than double precision. For practical
purposes, quad precision on matrix elements of the generators turned
out sufficient. Alternatively, it is also emphasized that the matrix
elements in the numerical representation of the generators \wrt some
given \IREP \emph{can actually be corrected even for sizeable
numerical errors}. The underlying reason is that the generators for
a given \IREP are unique up to similarity transformation.
Nevertheless, the similarity transformation is largely fixed by the
construction of the z-operators: these are (i) diagonal, and (ii)
their diagonal matrix elements are integer-superpositions of the
diagonals of the z-operators in the defining representation, \eg can
be chosen to be integer or half-integer valued. Hence small
numerical errors $\varepsilon \ll 1$ can be easily corrected \wrt
the z-operators. However, knowing the z-operators exactly, also the
matrix elements of the remaining operators can be fixed, in
principle, while paying attention to conventions regarding inner
multiplicity.

\end{document}